\documentclass[useAMS,usenatbib]{mn2e}
\usepackage{journal_names}
\usepackage{graphicx,times}
\usepackage{amsmath}
\usepackage[T1]{fontenc}
\usepackage{aecompl}
\usepackage{longtable}
\usepackage{tabularx}
\usepackage{pdflscape}
\usepackage{rotating}
\usepackage{multirow}
\usepackage[caption=false]{subfig}

\def\la{\mathrel{\hbox{\rlap{\hbox{\lower4pt\hbox{$\sim$}}}\hbox{$<$}}}}
\def\ga{\mathrel{\hbox{\rlap{\hbox{\lower4pt\hbox{$\sim$}}}\hbox{$>$}}}}

\newcommand{\HI}{\mbox{\normalsize H\thinspace\footnotesize I}}

\newcommand{\K}{{$K_s$}}

\def\apjl   {{ApJL }}
\def\apjs   {{ApJS }}

\title[Deep NIR catalogue of the HIZOA galaxies]{NIR Tully-Fisher in the Zone of Avoidance.  --  III. Deep NIR catalogue of the HIZOA galaxies}

\author[K.~Said et al.]
{\parbox{\textwidth}{Khaled~Said$^{1,2,3}$\thanks{E-mail: khaled@ast.uct.ac.za},
\ Ren\'ee~C.~Kraan-Korteweg$^{1}$,
T.~H.~Jarrett$^{1}$,
Lister~Staveley-Smith$^{2,3}$, and 
Wendy L. Williams$^{4}$} \vspace{0.4cm}\\
\parbox{\textwidth}{$^{1}$Astrophysics, Cosmology and Gravity Centre (ACGC), Astronomy Department, 
University of Cape Town, Private Bag X3, Rondebosch, 7701, South Africa\\
$^{2}$International Centre for Radio Astronomy Research (ICRAR), M468, The University of Western Australia, 35 Stirling Highway, Crawley, WA 6009, Australia\\
$^{3}$ARC Centre of Excellence for All-sky Astrophysics (CAASTRO)\\
$^{4}$School of Physics, Astronomy and Mathematics, University of Hertfordshire, College Lane, Hatfield AL10 9AB, UK} }

\begin{document}

\date{Accepted 00. Received 00; in original form 00}

\pagerange{\pageref{firstpage}--\pageref{lastpage}} \pubyear{2016}

\maketitle

\label{firstpage}

\begin{abstract}
We present a deep near-infrared (NIR; $J$, $H$, and $K_s$ bands) photometric catalogue of sources from the Parkes {\HI} Zone of Avoidance (HIZOA) survey, which forms the basis for an investigation of the  matter distribution in the Zone of Avoidance. Observations were conducted between 2006 and 2013 using the Infrared Survey Facility (IRSF), a 1.4-m  telescope situated at the South African Astronomical Observatory site in Sutherland. The images cover all 1108 HIZOA detections and yield 915 galaxies. An additional 105 bright 2MASS galaxies in the southern ZOA were imaged with the IRSF, resulting in 129 galaxies. The average $K_s$-band seeing and sky background for the survey are 1.38 arcsec and 20.1 mag, respectively. The detection rate as a function of stellar density and dust extinction is found to  depend mainly on the {\HI} mass of the {\HI} detected galaxies, which in principal correlates with the NIR brightness of the spiral galaxies. The measured isophotal magnitudes are of sufficient accuracy (errors $\sim$ 0.02 mag) to be used in a Tully-Fisher analysis. In the final NIR catalogue, 285  galaxies have both IRSF and 2MASS photometry (180 HIZOA plus 105 bright 2MASX galaxies). The $K_s$-band isophotal magnitudes presented in this paper agree, within the uncertainties, with those reported in the 2MASX catalogue. Another 30 galaxies, from the HIZOA northern extension, are also covered by UKIDSS Galactic Plane Survey (GPS) images, which are one magnitude deeper than our IRSF images. A modified version of our photometry pipeline was used to derive the photometric parameters of these UKIDSS galaxies. Good agreement was found between the respective $K_s$-band isophotal magnitudes. These comparisons confirm the robustness of the isophotal parameters and demonstrate that the IRSF images do not suffer from foreground contamination, after star removal, nor under-estimate the isophotal fluxes of ZoA galaxies. 
\end{abstract}

\begin{keywords}
galaxies: spiral -- galaxies: distances and redshifts -- galaxies: photometry -- infrared: galaxies -- cosmology: observations -- cosmology: large-scale structure of Universe
\end{keywords}

\section{Introduction}
The Milky Way obscures the background extragalactic sky through dust extinction, notably within $\pm5$ degrees of the Galactic Plane, and stellar confusion. This results in the so-called `Zone of Avoidance' (ZOA), literally meaning that the region is devoid of galaxies, and accordingly, extragalactic astronomers avoid working in this region of the sky. This has led to a gap in our understanding of large scale structure (LSS) in the ZOA, especially the conspicuous features that cross the ZOA, such as the Perseus-Pisces Supercluster (PPS; \citealt{1980MNRAS.193..353E,1982AJ.....87.1355G,1984A&A...136..178F,1987A&A...184...43H}), the Great Attractor (GA; \citealt{1987ApJ...313L..37D,1988ApJ...326...19L,1999A&A...352...39W}) and the Local Void (LV; \citealt{1987ang..book.....T,2008glv..book...13K}).\\

Galaxy peculiar velocities, i.e. deviations from the isotropic Hubble expansion, are sensitive to LSS such as the GA  \citep{1987ApJ...313L..37D} and can be used alongside redshift surveys for cosmography and to study the dynamics of the Local Group (LG), cosmic flow fields, and the origin of the observed dipole in the Cosmic Microwave Background (CMB) (e.g. \citealt{2006ApJ...653..861M,2007ApJS..172..599S,2008AJ....135.1738M,2012MNRAS.427..245M,2013AJ....146...69C,2014MNRAS.439.3666M,2014MNRAS.445..402H,2014MNRAS.445.2677S,2014Natur.513...71T,2015MNRAS.450..317C,2016MNRAS.456.1886S,2016arXiv160501765T}). The use of infrared wavelengths in these studies (\citealt{2008AJ....135.1738M,2013ApJ...765...94S,2014MNRAS.444..527S,2014ApJ...792..129N,2013ApJ...771...88L}) has minimized the impact of the ZOA, but the most obscured part of the ZOA ($|b| \leq 5^\circ$) remains mostly unexplored.

This project aims to derive the peculiar velocities of galaxies in the ZOA to supplement the above studies and provide a truly all-sky peculiar velocity survey. In this series of papers, we aim to map most of the dynamically important structures in the ZOA such as the GA and LV in greater detail. The requirements to pursue such a project are: (i) a calibrated and unbiased  Tully-Fisher (TF) relation to be used as the global template relation; (ii) 21 cm observations of spiral galaxies in the ZOA from which to extract the redshift and the rotational velocity of galaxies; (iii) follow-up NIR imaging of the {\HI} sources to measure the apparent magnitude of each galaxy. With these three ingredients, the absolute magnitude can be derived using the template relation given the rotational velocity. The distance to each galaxy can then be calculated independent of the redshift using the distance modulus. 

The TF template is constructed by \cite{2015MNRAS.447.1618S} using the isophotal magnitudes of the 888 spiral galaxies from \cite{2008AJ....135.1738M}. The advantage of using isophotal and not total magnitudes is twofold. First, they are more robust and can be measured easily both in and out of the ZOA. Second, they are consistent between different surveys which means they can be used to combine data from different data-sets without corrections (see fig. 1 in \citealt{2015MNRAS.447.1618S}). A correction model for the change in the shape of galaxies due to dust extinction is also presented in \cite{2015MNRAS.447.1618S}.  \cite{2016MNRAS.457.2366S} presents the second element for this ZOA peculiar velocity survey which is the 21 cm {\HI}-line spectra of inclined, $(b/a)^\circ < 0.5$, spiral galaxies. The average signal-to-noise ratio for this {\HI} survey was 14.7 which is adequate for TF studies. Five different types of line-widths are presented to select the most robust one. Conversions between these widths are derived to allow combination of data from different surveys. This third paper is dedicated to systematic NIR follow-up observations of all galaxies in the HIZOA catalogue  (HIZOA-S; \citealt{2016AJ....151...52S}, HIZOA-N; \citealt{2005AJ....129..220D}, GB; \citealt{2008glv..book...13K}). 

At low Galactic latitude NIR wavelengths are preferred over optical wavelengths because of the ability of NIR radiation to penetrate through dust.  The NIR also provides a more stable indicator of total stellar mass \citep{2000A&ARv..10..211K,2005RvMA...18...48K}. In the last few decades many surveys have used the NIR to unveil the LSS hidden behind the Milky Way. The 2MASS extended source catalogue (2MASX) contains galaxies that have never been seen before in the ZOA   \citep{2000AJ....119.2498J}. \cite{2000AJ....120..298J} present new extended sources in the ZOA at Galactic longitude between $40^{\circ}$ and $70^{\circ}$. Deeper NIR observations have been used specifically for dedicated surveys in the ZOA. \cite{2004MNRAS.354..980N} used the same instrument that we have used in this work to conduct a NIR survey around the radio galaxy PKS1343-601. They detected 19 galaxies and another 38 galaxy candidates of which only three were known previously. \cite{2005ASPC..329..147W} also used the IRSF to obtain deep photometry for 76 galaxies which was used in the determination of the distance to the Norma cluster.  \cite{2006MNRAS.368..534N} used a deep NIR survey of a luminous cluster in the GA region and identified 111 galaxy candidates. Longer wavelengths are also used in the ZOA; \cite{2007AJ....133..979J} used mid-infrared wavelengths to unveil two galaxies in the GA region. A large deep NIR survey of the Norma Wall (NWS: \citealt{2010MNRAS.401..924R,2010PhDT........33W,2011arXiv1107.1069K}; Riad et al,. in prep) was also conducted with the IRSF. This survey resulted in a catalogue of 4360 galaxies with completeness limits of 15.6, 15.3 and 14.8 mag in the $J$, $H$, and $K_s$ bands, respectively. Given the success of these surveys, we started a follow-up NIR survey of the HIZOA galaxies in the southern ZOA using the same telescope and instrument (IRSF).\\

This paper is organized as follows. Section \ref{sect:obs} discusses the observations, calibration and observatory site conditions. The final extended source catalogue and parameter characterization are presented in Section \ref{sect:cat}. Completeness as a function of  dust extinction and stellar density is discussed in detail in Section \ref{sect:completeness}. Comparison of the resulting photometry with the 2MASS and UKIDSS surveys is presented in Section \ref{sect:counterparts}. We summarise our results in Section \ref{sect:summary}. All magnitudes are quoted in the Vega System.

\section{Observations}
\label{sect:obs}
Deep NIR follow-up observations of all HIZOA galaxies were conducted with the Infrared Survey Facility (IRSF), a 1.4-m  telescope situated at the South African Astronomical Observatory  (SAAO) site in Sutherland, South Africa. The Simultaneous 3-colour ($J$, $H$, and $K_s$) Infrared Imager for Unbiased Survey (SIRIUS; \citealt{2003SPIE.4841..459N}) camera on the IRSF has a field of view of $7.7\times7.7$ arcmin$^2$ (ideally suited for HIZOA follow-up given the 4 arcmin positional accuracy of the \HI\ detections) with a pixel scale of 0.45 arcsec pixel$^{-1}$ compared to 2.0 arcsec pixel$^{-1}$ for 2MASS  \citep{2006AJ....131.1163S}. A pilot project and the first results of the catalogue were published by \cite{2014MNRAS.443...41W} who presented photometry for 557 galaxies in the HIZOA catalogue with $cz \leq 6000$ km s$^{-1}$. For completion we have included their 578 fields in the current study.

An additional 105 2MASX fields in the southern ZOA but not in the HIZOA survey were also observed. The {\HI} spectral line data for these 105 2MASX fields are from Parkes observations and available either from the 2MASS TF Survey \citep{2013MNRAS.432.1178H} or from Said et al. (in prep.). These additional galaxies are all the bright ($K_s^{\circ} = 11.25$ mag),  edge-on ($b/a = 0.5$) 2MASX galaxies in the southern ZOA ($5^\circ \leq|b|\leq 10^\circ$).

\subsection{Data acquisition}
Data acquisition started in 2006 and was completed by 2013, resulting in deep NIR imaging of all the HIZOA targets. The images have exposure times of 10 min, and are 2 mag deeper than 2MASS in the \K-band \citep{2010PhDT........33W}. We used the dithering technique to overcome the problem of faulty pixels in the NIR detector. We repeated a 24 s exposure 25 times with a dithering step of 15 arcsec. This dithering resulted in increasing the final image size to $8.6\times8.6$ arcmin$^2$. A total of 12 weeks were allocated to this project starting in 2009. 101 fields were observed between 2006 and 2008 as part of other projects to test the feasibility of this project. Table \ref{IRSF_observers} shows the observations, number of allocated weeks, number of fields observed and the Observer In Charge (OIC). Substantial time was lost during the 2009 and 2010 runs due to bad weather and cooling system problems. 

\begin{table}
\begin{center}
\caption[]{IRSF observations and the Observer In Charge (OIC). OIC: Wendy Williams (WW), Tom Mutabazi (TM) and Khaled Said (KS).}
\begin{tabular}{c c c c c}
\hline
\hline
Year & Month(s) & Allocated weeks & No. of fields & OIC\\ 
\hline
2006-2008 & & & 101 & \\
2009 & March/April & 2 & 249 & WW\\
2009 & March/April & 2 & 67 & WW\\
2010 & June/July & 3 & 138 & WW\\
2012 & May & 2 & 231 & TM\\
2013 & April & 3 & 430 & KS\\
\hline
\end{tabular}
\label{IRSF_observers}
\end{center}
\end{table} 
 
\subsection{Data reduction and calibration}
The primary data reduction, including dark frame subtraction, flat field correction, sky-subtraction, dither combination and astrometric and photometric calibration, was carried out using the pipeline software for SIRIUS\footnote{http://irsf-software.appspot.com/yas/nakajima/sirius.html}.

\subsection{Observatory site conditions and quality control}
\begin{figure}
\begin{center}
\includegraphics[scale=0.31]{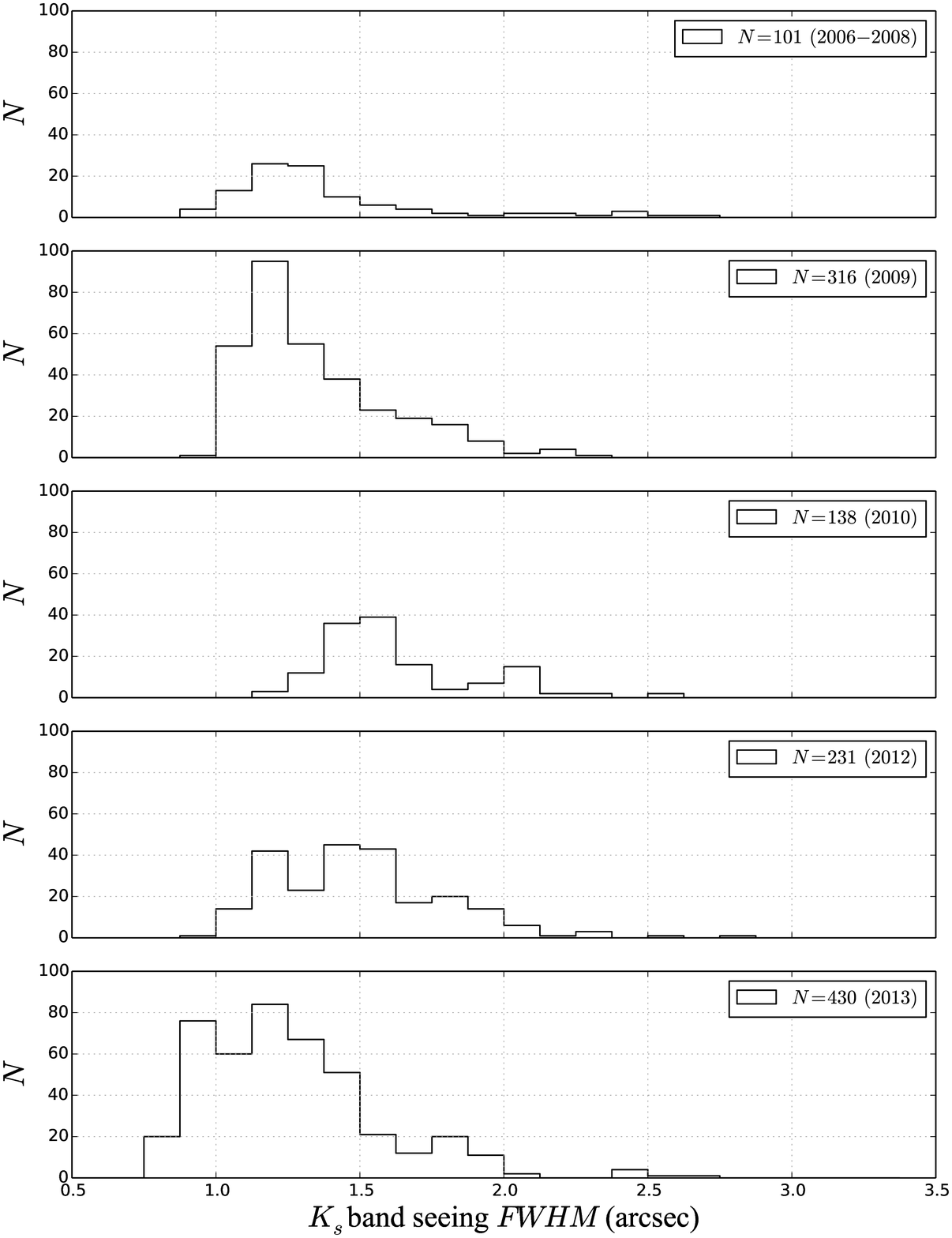}
\caption{Distributions of the measured $K_s$-band seeing \textit{FWHM} from 2006 to 2013.}
\label{seeing}
\end{center}
\end{figure}

\begin{figure}
\begin{center}
\includegraphics[scale=0.31]{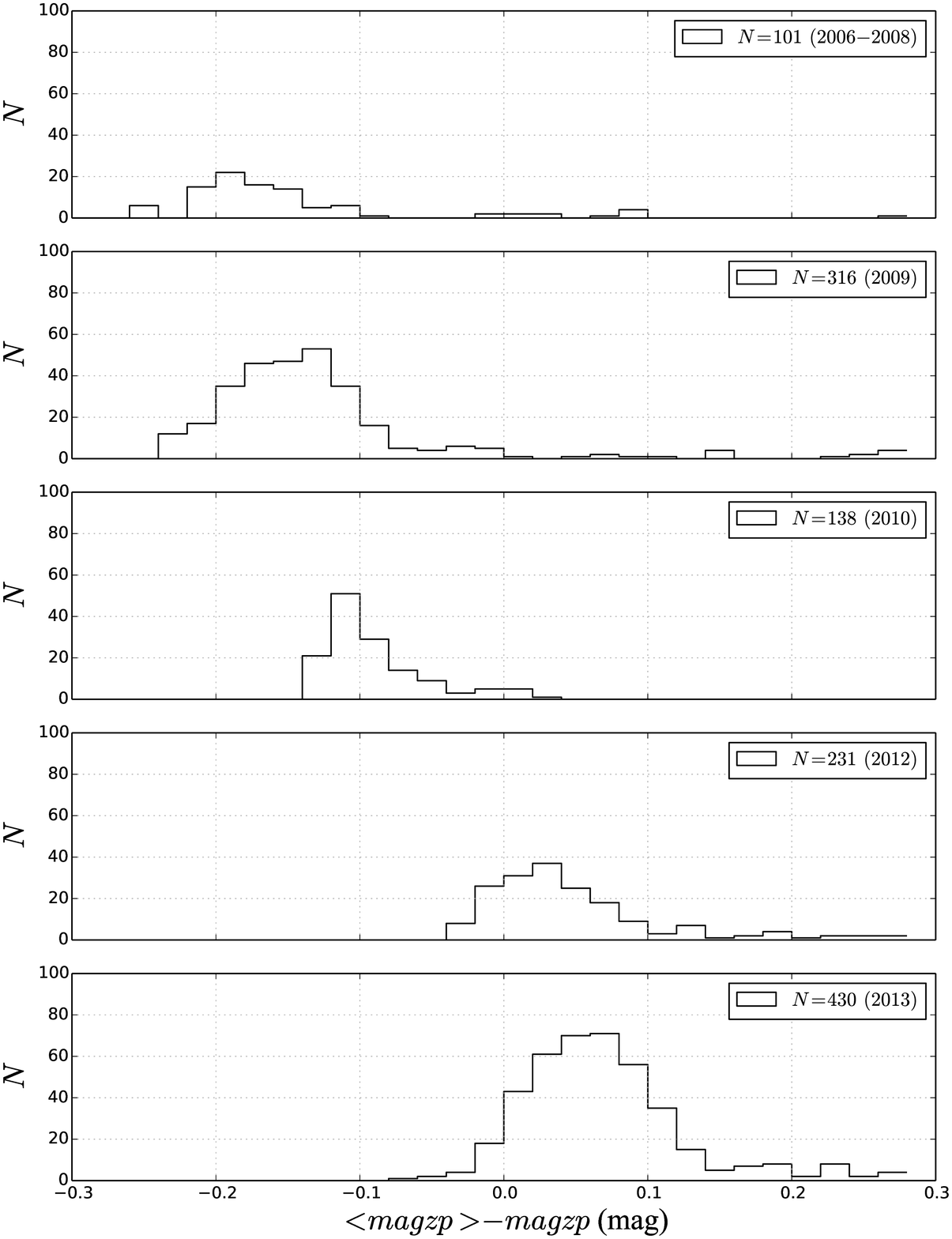}
\caption{Distributions of the $K_s$-band photometric zero point magnitude from 2006 to 2013.}
\label{magzp}
\end{center}
\end{figure}

Figs \ref{seeing} and \ref{magzp} show the distributions of the $K_s$-band seeing and the deviation of magnitude zero point from the mean, respectively, for the 1229 observed fields as a function of observation date. An average $K_s$-band seeing of 1.38 arcsec and an average $K_s$-band magnitude zero point of 20.1 mag were found for the entire survey. These values agree with those found for the Norma wall survey \citep{2010PhDT........33W,2011arXiv1107.1069K}. Fig. \ref{seeing} shows that 95 per cent of the survey has seeing values below 2.0 arcsec. Fig. \ref{magzp} shows that 85 per cent of the survey lies within $\pm$ 0.2 mag of the mean magnitude zero point. While the deviations of magnitude zero point from the mean are not significant, Fig. \ref{magzp} shows a clear trend that the sky is getting brighter with time. 
All fields with either seeing or zero point magnitude outside of these two ranges were inspected visually to check their image quality. Poor quality fields were re-observed under photometric conditions\footnote{These poor fields were removed and replaced with the photometric-quality ones for the final catalogue but we have made both FITS files available for comparison purposes}.

\begin{figure*}
\begin{center}
\includegraphics[scale=0.42]{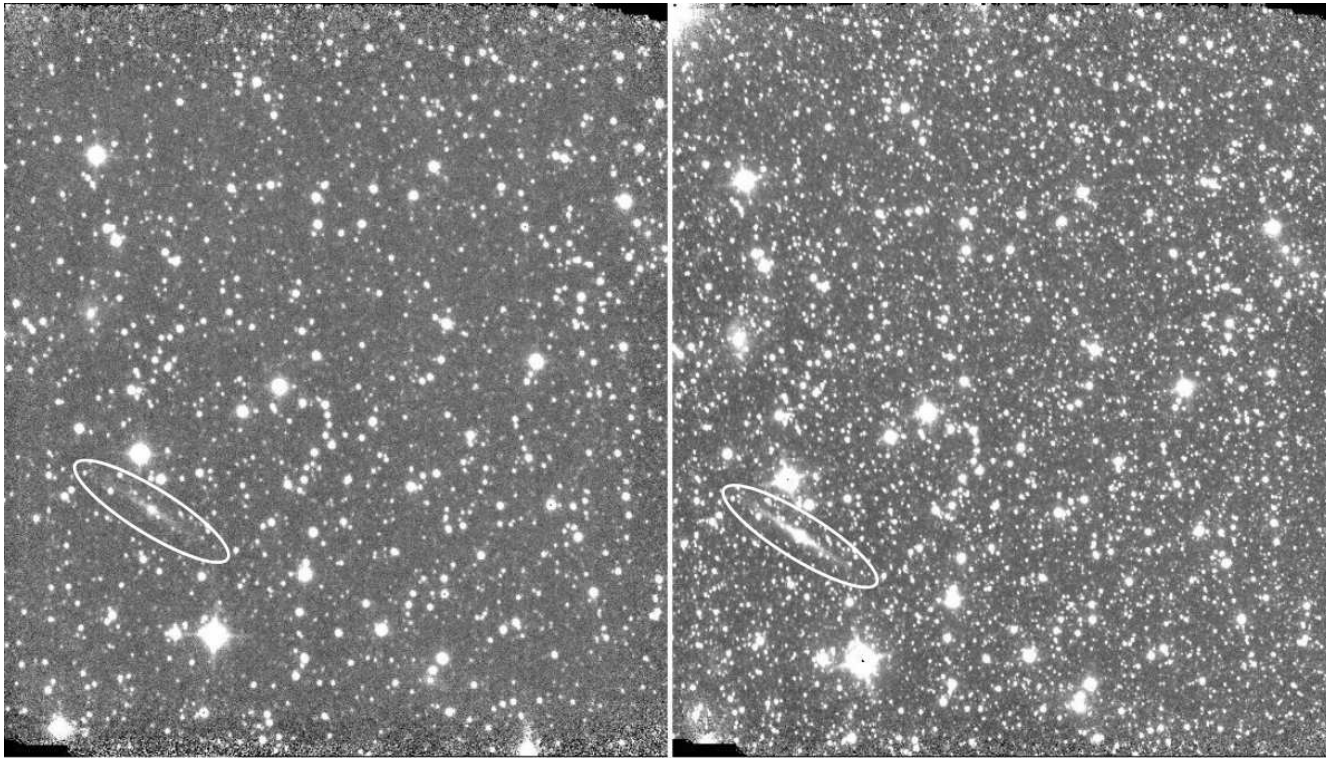}
\caption{$K_s$-band photometric vs non-photometric images of the same field. Both images are $8.6 \times 8.6$ arcmin; the image on the left was observed under non-photometric conditions while the image on the right was re-observed under photometric conditions.  The $K_s$-band seeing and zero point magnitude for the image on the left are 2.6 arcsec and 18.1 mag, respectively, while the $K_s$-band seeing and zero point magnitude for the image on the right are 0.9 arcsec and 20.1 mag, respectively. Note also the increase in resolved stars in the right image.}
\label{kJ1624-45A_comp}
\end{center}
\end{figure*}

Fig. \ref{kJ1624-45A_comp} shows one field that was re-observed, where the left panel shows the field observed under non-photometric conditions and the right panel shows that same field re-observed under photometric conditions.
The image on the left hand side was taken under a partially cloudy sky which affects both the seeing and magnitude zero point. The $K_s$-band seeing for the left hand side image is 2.6 arcsec. The $K_s$-band zero point magnitude for the image on the left hand side was found to be 18.1 mag. In contrast, the right hand side image shows the field observed under photometric conditions. The $K_s$-band seeing and zero point magnitude for the right hand side image are 0.9 arcsec and 20.1 mag, respectively. The counterpart of the HIZOA galaxy, J1624-45A, in this field is an edge-on galaxy and is marked with the white ellipse in the bottom-left corner of each image. This example demonstrates how important photometric conditions are for this kind of survey. The extent of the galaxy on the left hand side is underestimated while the real size of the galaxy is apparent in the right hand image.




\section{Final extended source catalogue}
\label{sect:cat}
The procedures for source identification, star subtraction and photometry are described in detail by \cite{2014MNRAS.443...41W}. In this section we will therefore only present the final extended source catalogue. This catalogue contains photometry in the $J$, $H$, and $K_s$ bands for 1044 galaxies. We divide the final catalogue into two separate catalogues, one for the HIZOA galaxies and the second for the bright 2MASX galaxies, full catalogues are available electronically. FITS files for the three bands are also available upon request. An example of the catalogue is given here for the brightest 100 galaxies in the catalogue.

\subsection{NIR Parametrization and catalogue}
The main goal of this paper is to provide accurate NIR photometric parameters for galaxies hidden behind the ZOA. These parameters will be used in the forthcoming NIR TF analysis. In this section we provide the required parameters. For consistency with \cite{2014MNRAS.443...41W}, we employ the same methods and naming for the whole survey  as that  described by \cite{2003tmc..book.....C}. Table \ref{all_NIR_par} and Fig. \ref{stamps} present the photometry for and the postage stamp images of the brightest 100 galaxies in the catalogue, respectively, in order of isophotal $K_{s20}$ fiducial elliptical aperture magnitude. The catalogue is presented in its entire form online. The NIR parameters listed in the catalogue are as follows:

Column (1) -- unique ZOA ID formed from sexigesimal coordinates [ZOAhhmmss.sss$\pm$ddmmss.ss].

Column (2) -- HIZOA ID as reported in the HIZOA survey publications (\citealt{2005AJ....129..220D}; Staveley-Smith et al., 2015; Kraan-Korteweg et al., in prep.)\footnote{If the galaxy is not in the HIZOA catalogue, the 2MASX ID is provided instead.}.

Column (3) -- Survey name.

Columns (4 and 5) -- Right Ascension (RA) and Declination (Dec.) in the J2000.0 epoch.

Columns (6 and 7) -- Galactic coordinates [degree].

Column (8) -- $J$-band ellipticity ($\epsilon=1-b/a$) measured as the mean value of the ellipticities of the ellipses fitted between the $1 \sigma$ and $2 \sigma$, where $\sigma$ is the sky rms.

Column (9) -- Isophotal $K_{s20}$ fiducial elliptical aperture semi-major axis [arcsec].

Columns (10--12) -- $J$-, $H$-, and $K_s$-band $K_{s20}$ fiducial elliptical aperture magnitudes and associated errors [mag].

Columns (13--15) -- $J$-, $H$-, and $K_s$-band central surface brightness [mag arcsec$^{-2}$].

Column (16) -- E($B-V$), Galactic extinction as reported by \cite{2011ApJ...737..103S}.

Column (17) -- $SD$, IRSF stellar density $\log(N_{K_s<14}/\mathrm{deg}^{2})$ where $N_{K_s<14}$ is the number density of stars brighter than 14 mag in the $K_s$ band.

\onecolumn
\begin{landscape}
\renewcommand{\thefootnote}{\fnsymbol{footnote}}
\renewcommand{\arraystretch}{1.1}
\scriptsize
\begin{longtable}{c c c  c c c c  c c c c c  c c c c c}
\caption[]{NIR coordinates, sizes, shapes and photometry of the brightest 100 galaxies in the catalogue in order of isophotal $K_{s20}$ fiducial elliptical aperture magnitude. The full table is available online.}\\
 \hline
 \hline
ZOA ID & HI Name & Survey & RA & DEC & $l$ & $b$ & $\epsilon_J$ & $r_{K_{s20fe}}$ & $J_{K_{s20fe}}$ & $H_{K_{s20fe}}$ & $K_{sK_{s20fe}}$ & $\mu_{cJ}$ & $\mu_{cH}$ & $\mu_{cKs}$ & $E(B-V)$ & $SD$\\
 & & & \multicolumn{2}{c}{(J2000)} & \multicolumn{2}{c}{[deg]} &   & [arcsec] & [mag] & [mag] & [mag]  & \multicolumn{3}{c}{[mag arcsec$^{-2}$]} & [mag] &\\
(1) & (2) & (3) & (4) & (5) & (6)  & (7) & (8) & (9) & (10) & (11) & (12) & (13) & (14) & (15) & (16) & (17)\\
\hline
ZOA141309.873-652020.76 &        J1413-65 &  HIZOA &  213.291 &     $-$65.339 &    311.326 &     $-$3.808 &     0.44 &    87.05 &     7.50 $\pm$     0.02 &     6.49 $\pm$     0.02 &     6.09 $\pm$     0.02 &    12.67 &    11.69 &    11.18 &    1.264 &    4.374 \\ 
ZOA151434.147-525921.52 &        J1514-53 &  HIZOA &  228.642 &    $-$52.989 &    323.594 &      4.043 &     0.77 &   136.09 &     8.86 $\pm$     0.02 &     7.91 $\pm$     0.02 &     7.42 $\pm$     0.02 &    14.44 &    13.49 &    13.01 &    0.851 &    4.299 \\ 
ZOA085728.473-391605.66 &        J0857-39 &  HIZOA &  134.369 &    $-$39.268 &    261.500 &       4.100 &     0.10 &    45.77 &     9.18 $\pm$     0.02 &     8.32 $\pm$     0.02 &     7.95 $\pm$     0.02 &    14.09 &    13.22 &    12.85 &    0.619 &    3.611 \\ 
ZOA150928.962-523320.67 &        J1509-52 &  HIZOA &  227.371 &    $-$52.556 &    323.155 &      4.810 &     0.79 &    98.73 &     9.24 $\pm$     0.02 &     8.38 $\pm$     0.02 &     7.97 $\pm$     0.02 &    14.97 &    14.05 &    13.60 &    0.662 &    4.090 \\ 
ZOA081706.065-272720.64 &        J0817-27 &  HIZOA &  124.275 &    $-$27.456 &    246.973 &      4.481 &     0.41 &    59.84 &     9.17 $\pm$     0.02 &     8.41 $\pm$     0.02 &     8.11 $\pm$     0.02 &    14.83 &    14.09 &    13.81 &    0.220 &    3.746 \\ 
ZOA181427.987-022505.11 &        J1814-02 &  HIZOA &  273.617 &     $-$2.418 &     26.526 &      7.099 &     0.47 &    47.82 &    10.04 $\pm$     0.02 &     8.90 $\pm$     0.02 &     8.50 $\pm$     0.02 &    14.72 &    13.70 &    13.40 &    1.980 &    3.986 \\ 
ZOA122238.290-583657.66 &        J1222-58 &  HIZOA &  185.659 &    $-$58.616 &    299.180 &      4.046 &     0.69 &    60.30 &     9.71 $\pm$     0.02 &     8.92 $\pm$     0.02 &     8.60 $\pm$     0.02 &    14.62 &    13.83 &    13.55 &    0.499 &    4.121 \\ 
ZOA151413.880-464827.22 &   2MASX1514-464 &  2MASS &  228.558 &    $-$46.808 &    326.808 &      9.338 &     0.53 &   114.03 &    10.75 $\pm$     0.02 &    10.16 $\pm$     0.02 &     8.60 $\pm$     0.02 &    18.23 &    17.64 &    16.89 &    0.217 &    3.805 \\ 
ZOA094916.505-475511.27 &       J0949-47A &  HIZOA &  147.319 &     $-$47.919 &    274.257 &      4.549 &     0.10 &    38.65 &     9.67 $\pm$     0.02 &     8.97 $\pm$     0.02 &     8.69 $\pm$     0.02 &    14.37 &    13.62 &    13.36 &    0.301 &    3.797 \\ 
ZOA145709.815-542331.46 &        J1457-54 &  HIZOA &   224.291 &    $-$54.392 &      320.654 &      4.0957 &     0.51 &    53.80 &    10.04 $\pm$     0.02 &     9.14 $\pm$     0.02 &     8.69 $\pm$     0.02 &    14.96 &    14.07 &    13.64 &    0.731 &    4.210 \\ 
ZOA114606.371-562326.95 &        J1145-56 &  HIZOA &  176.526 &    $-$56.391 &    293.937 &      5.336 &     0.27 &    54.68 &     9.99 $\pm$     0.02 &     9.26 $\pm$     0.02 &     8.82 $\pm$     0.02 &    16.05 &    15.26 &    14.82 &    0.335 &    3.844 \\ 
ZOA135138.534-583515.22 &        J1351-58 &  HIZOA &  207.910 &    $-$58.588 &    310.724 &       3.370 &     0.52 &    47.27 &    10.23 $\pm$     0.02 &     9.31 $\pm$     0.02 &     8.87 $\pm$     0.02 &    14.80 &    13.88 &    13.41 &    0.834 &    4.287 \\ 
ZOA074752.048-184453.18 &        J0747-18 &  HIZOA &  116.967 &    $-$18.748 &    236.009 &      3.374 &     0.82 &    84.90 &    10.06 $\pm$     0.02 &     9.31 $\pm$     0.02 &     8.91 $\pm$     0.02 &    15.79 &    14.95 &    14.77 &    0.327 &    3.547 \\ 
ZOA072210.950-055547.38 &        J0722-05 &  HIZOA &  110.546 &     $-$5.930 &    221.732 &      4.088 &     0.78 &    48.64 &     9.99 $\pm$     0.02 &     9.28 $\pm$     0.02 &     8.99 $\pm$     0.02 &    14.08 &    13.41 &    13.05 &    0.250 &    3.528 \\ 
ZOA070103.346+015439.69 &        J0701+01 &  HIZOA &  105.264 &      1.911 &    212.326 &      3.011 &     0.48 &    41.91 &    10.16 $\pm$     0.02 &     9.37 $\pm$     0.02 &     9.01 $\pm$     0.02 &    15.09 &    14.34 &    13.93 &    0.520 &    3.563 \\ 
ZOA143158.829-552758.82 &        J1431-55 &  HIZOA &  217.995 &    $-$55.466 &    316.912 &      4.653 &     0.10 &    34.39 &    10.32 $\pm$     0.02 &     9.46 $\pm$     0.02 &     9.05 $\pm$     0.02 &    15.65 &    14.75 &    14.31 &    0.731 &    4.091 \\ 
ZOA080611.134-273140.86 &        J0806-27 &  HIZOA &  121.546 &    $-$27.528 &    245.709 &      2.414 &     0.58 &    43.68 &    10.26 $\pm$     0.02 &     9.51 $\pm$     0.02 &     9.09 $\pm$     0.02 &    14.84 &    14.10 &    13.67 &    0.420 &    3.731 \\ 
ZOA074843.871-261445.62 &       J0748-26B &  HIZOA &   117.183 &    $-$26.246 &    242.586 &     $-$0.239 &     0.32 &    39.43 &    10.49 $\pm$     0.02 &     9.64 $\pm$     0.02 &     9.10 $\pm$     0.02 &    15.53 &    14.56 &    14.04 &    0.482 &    3.723 \\ 
ZOA163211.878-280530.82 &        J1632-28 &  HIZOA &  248.050 &     $-$28.092 &    351.084 &     13.502 &     0.72 &    81.48 &    10.28 $\pm$     0.02 &     9.45 $\pm$     0.02 &     9.10 $\pm$     0.02 &    15.64 &    14.80 &    14.46 &    0.525 &    3.536 \\ 
ZOA141036.181-653457.76 &        J1410-65 &  HIZOA &  212.651 &    $-$65.583 &    310.997 &      $-$3.958 &     0.52 &    46.18 &    10.35 $\pm$     0.02 &     9.48 $\pm$     0.02 &     9.14 $\pm$     0.02 &    15.13 &    14.34 &    13.98 &    0.516 &    4.325 \\ 
ZOA155524.078-581431.30 &       J1555-581 &  2MASS &  238.850 &    $-$58.242 &    325.222 &     $-$3.570 &     0.37 &    65.37 &    11.32 $\pm$     0.02 &     9.98 $\pm$     0.02 &     9.16 $\pm$     0.02 &    16.32 &    15.40 &    14.86 &    0.641 &    4.404 \\ 
ZOA063554.386+110808.32 &        J0635+11 &  HIZOA &   98.977 &     11.136 &    201.262 &      1.658 &     0.27 &    31.91 &    10.78 $\pm$     0.02 &     9.63 $\pm$     0.02 &     9.19 $\pm$     0.02 &    15.67 &    14.70 &    14.29 &    1.310 &    3.518 \\ 
ZOA083439.531-400855.61 &        J0834-40 &  HIZOA &  128.665 &    $-$40.149 &    259.448 &        0.122 &     0.22 &    34.38 &    10.02 $\pm$     0.02 &     9.52 $\pm$     0.02 &     9.30 $\pm$     0.02 &    15.30 &    14.61 &    14.51 &    1.823 &    3.838 \\ 
ZOA132723.827-572922.23 &        J1327-57 &  HIZOA &  201.849 &    $-$57.490 &    307.768 &      5.044 &     0.74 &    46.73 &    10.86 $\pm$     0.02 &     9.82 $\pm$     0.02 &     9.30 $\pm$     0.02 &    15.93 &    14.75 &    14.09 &    0.697 &    3.980 \\ 
ZOA085838.795-423157.31 &        J0858-42 &  HIZOA &  134.662 &    $-$42.533 &    264.125 &      2.141 &     0.45 &    43.00 &    11.95 $\pm$     0.02 &    10.10 $\pm$     0.02 &     9.35 $\pm$     0.02 &    16.58 &    14.86 &    14.05 &    3.431 &    3.735 \\ 
ZOA072456.870-093933.95 &        J0724-09 &  HIZOA &  111.237 &     $-$9.659 &    225.354 &      2.942 &     0.29 &    32.86 &    10.48 $\pm$     0.02 &     9.71 $\pm$     0.02 &     9.41 $\pm$     0.02 &    15.54 &    14.90 &    14.39 &    0.340 &    3.529 \\ 
ZOA160349.297-605840.50 &   2MASX1603-605 &  2MASS &  240.955 &    $-$60.978 &    324.235 &     $-$6.336 &     0.81 &    58.98 &    10.71 $\pm$     0.02 &     9.85 $\pm$     0.02 &     9.45 $\pm$     0.02 &    15.62 &    15.02 &    14.67 &    0.275 &    4.036 \\ 
ZOA094952.868-563235.55 &        J0949-56 &  HIZOA &  147.470 &    $-$56.543 &    279.808 &     $-$2.054 &     0.24 &    45.57 &    11.43 $\pm$     0.02 &    10.29 $\pm$     0.02 &     9.58 $\pm$     0.02 &    16.56 &    15.33 &    14.62 &    1.806 &    4.159 \\ 
ZOA141710.092-553240.92 &   2MASX1417-553 &  2MASS &  214.292 &    $-$55.545 &    314.915 &      5.320 &     0.85 &    58.28 &    11.27 $\pm$     0.02 &    10.13 $\pm$     0.02 &     9.58 $\pm$     0.02 &    16.54 &    15.49 &    15.10 &    0.535 &    3.994 \\ 
ZOA155422.988-612025.58 &   2MASX1554-612 &  2MASS &  238.596 &    $-$61.340 &    323.130 &     $-$5.866 &     0.56 &    33.80 &    10.84 $\pm$     0.02 &    10.11 $\pm$     0.02 &     9.58 $\pm$     0.02 &    15.45 &    15.09 &    14.44 &    0.348 &    4.042 \\ 
ZOA141933.720-580850.19 &       J1419-58B &  HIZOA &   214.890 &    $-$58.147 &    314.363 &      2.755 &     0.59 &    52.89 &    11.00 $\pm$     0.02 &    10.14 $\pm$     0.02 &     9.59 $\pm$     0.02 &    16.82 &    15.82 &    15.27 &    1.307 &    4.363 \\ 
ZOA163140.118-280606.66 &        J1631-28 &  HIZOA &  247.917 &    $-$28.101 &    350.997 &     13.584 &     0.27 &    35.29 &    10.83 $\pm$     0.02 &    10.07 $\pm$     0.02 &     9.59 $\pm$     0.02 &    15.78 &    14.94 &    14.48 &    0.542 &    3.598 \\ 
ZOA161710.946-581845.49 &   2MASX1617-581 &  2MASS &  244.296 &    $-$58.313 &    327.304 &     $-$5.542 &     0.57 &    31.39 &    10.83 $\pm$     0.02 &    10.05 $\pm$     0.02 &     9.61 $\pm$     0.02 &    15.11 &    14.33 &    13.80 &    0.260 &    4.203 \\ 
ZOA101655.552-485252.32 &   2MASX1016-485 &  2MASS &  154.231 &    $-$48.881 &    278.525 &      6.529 &     0.54 &    37.76 &    10.72 $\pm$     0.02 &     9.74 $\pm$     0.02 &     9.63 $\pm$     0.02 &    16.01 &    15.29 &    14.91 &    0.183 &    3.659 \\ 
ZOA160425.042-604415.93 &   2MASX1604-604 &  2MASS &  241.104 &    $-$60.738 &    324.450 &     $-$6.205 &     0.77 &    40.77 &    10.73 $\pm$     0.02 &    10.01 $\pm$     0.02 &     9.65 $\pm$     0.02 &    15.08 &    14.63 &    14.42 &    0.257 &    4.039 \\ 
ZOA090033.110-392626.93 &        J0900-39 &  HIZOA &  135.138 &    $-$39.441 &    262.020 &      4.438 &     0.39 &    37.83 &    10.79 $\pm$     0.02 &    10.15 $\pm$     0.02 &     9.67 $\pm$     0.02 &    16.21 &    15.45 &    15.04 &    0.576 &    3.611 \\ 
ZOA153603.059-594449.42 &    2MASXJ1536-5944 & 2MASS &   234.013 &    $-$59.747 &    322.327 &     $-$3.226 &     0.46 &    41.50 &    10.92 $\pm$     0.02 &    10.04 $\pm$     0.02 &     9.67 $\pm$     0.02 &    15.87 &    14.93 &    14.40 &    0.833 &    4.435 \\ 
ZOA131033.767-580021.18 &        J1310-57 &  HIZOA &  197.641 &    $-$58.006 &    305.472 &      4.772 &     0.27 &    27.20 &    10.81 $\pm$     0.02 &    10.02 $\pm$     0.02 &     9.68 $\pm$     0.02 &    15.19 &    14.52 &    14.22 &    0.480 &    4.080 \\ 
ZOA161319.695-562349.17 &        J1613-56 &  HIZOA &  243.332 &    $-$56.397 &    328.261 &     $-$3.802 &     0.35 &    40.49 &    10.93 $\pm$     0.02 &    10.69 $\pm$     0.02 &     9.68 $\pm$     0.02 &    16.15 &    15.39 &    14.87 &    0.464 &    4.425 \\ 
ZOA074252.007-315959.79 &        J0742-31 &  HIZOA &   115.717 &    $-$32.000 &     246.932 &     $-$4.2257 &     0.52 &    26.61 &    11.15 $\pm$     0.02 &    10.38 $\pm$     0.02 &     9.71 $\pm$     0.02 &    15.39 &    14.72 &    14.15 &    0.714 &    3.757 \\ 
ZOA101212.032-623159.40 &        J1012-62 &  HIZOA &  153.050 &    $-$62.533 &    285.686 &     $-$5.121 &     0.62 &    39.84 &    10.84 $\pm$     0.02 &    10.08 $\pm$     0.02 &     9.71 $\pm$     0.02 &    15.80 &    15.03 &    14.51 &    0.249 &    4.036 \\ 
ZOA162101.624-360831.49 &        J1621-36 &  HIZOA &  245.257 &    $-$36.142 &    343.413 &      9.765 &     0.55 &    38.44 &    10.84 $\pm$     0.02 &    10.05 $\pm$     0.02 &     9.71 $\pm$     0.02 &    15.25 &    14.44 &    14.09 &    0.611 &    3.789 \\ 
ZOA070056.215-114734.32 &        J0700-11 &  HIZOA &  105.234 &    $-$11.793 &    224.511 &     $-$3.273 &     0.70 &    42.70 &    11.39 $\pm$     0.02 &    10.29 $\pm$     0.02 &     9.72 $\pm$     0.02 &    16.21 &    15.12 &    14.61 &    0.680 &    3.595 \\ 
ZOA072457.474-271516.87 &        J0724-27 &  HIZOA &  111.240 &    $-$27.255 &    240.882 &     $-$5.366 &     0.36 &    41.90 &    11.78 $\pm$     0.02 &    11.03 $\pm$     0.02 &     9.72 $\pm$     0.02 &    16.64 &    15.89 &    15.49 &    0.370 &    3.668 \\ 
ZOA181530.143-025348.41 &        J1815-02 &  HIZOA &  273.876 &     $-$2.897 &     26.220 &      6.647 &     0.10 &    32.73 &    11.41 $\pm$     0.02 &    10.26 $\pm$     0.02 &     9.72 $\pm$     0.02 &    16.32 &    15.29 &    14.62 &    2.230 &    3.973 \\ 
ZOA114948.692-640006.93 &        J1149-64 &  HIZOA &  177.453 &    $-$64.002 &    296.241 &      $-$1.934 &     0.57 &    42.66 &    11.73 $\pm$     0.02 &    10.42 $\pm$     0.02 &     9.73 $\pm$     0.02 &    17.05 &    15.76 &    15.05 &    2.098 &    4.382 \\ 
ZOA141710.099-553238.77 &       J1416-55B &  HIZOA &  214.292 &     $-$55.544 &    314.916 &      5.320 &     0.86 &    68.61 &    11.23 $\pm$     0.02 &    10.10 $\pm$     0.02 &     9.73 $\pm$     0.02 &    16.58 &    15.47 &    14.96 &    0.533 &    3.982 \\ 
ZOA074141.201-223112.25 &        J0741-22 &  HIZOA &  115.422 &    $-$22.520 &    238.558 &      0.239 &     0.80 &    48.26 &    11.31 $\pm$     0.02 &    10.24 $\pm$     0.02 &     9.74 $\pm$     0.02 &    16.54 &    15.35 &    14.64 &    0.568 &    3.737 \\ 
ZOA161710.749-581844.59 &        J1617-58 &  HIZOA &  244.295 &    $-$58.313 &    327.304 &     $-$5.542 &     0.53 &    33.09 &    10.82 $\pm$     0.02 &    10.04 $\pm$     0.02 &     9.75 $\pm$     0.02 &    15.19 &    14.39 &    14.00 &    0.258 &    4.159 \\ 
ZOA101220.012-471741.58 &   2MASX1012-471 &  2MASS &  153.083 &    $-$47.295 &    276.983 &      7.402 &     0.81 &    68.92 &    11.15 $\pm$     0.02 &    10.21 $\pm$     0.02 &     9.76 $\pm$     0.02 &    16.72 &    15.89 &    15.26 &    0.164 &    3.478 \\   
\hline
\multicolumn{15}{r}{{Continued on Next Page\ldots}} \\

\multicolumn{16}{c}{{\tablename} \thetable{} -- Continued}\\[0.5ex]
 \hline
 \hline
ZOA ID & HI Name & Survey & RA & DEC & $l$ & $b$ & $\epsilon_J$ & $r_{K_{s20fe}}$ & $J_{K_{s20fe}}$ & $H_{K_{s20fe}}$ & $K_{sK_{s20fe}}$ & $\mu_{cJ}$ & $\mu_{cH}$ & $\mu_{cKs}$ & $E(B-V)$ & $SD$\\
 & & & \multicolumn{2}{c}{(J2000)} & \multicolumn{2}{c}{[deg]} &   & [arcsec] & [mag] & [mag] & [mag]  & \multicolumn{3}{c}{[mag arcsec$^{-2}$]} & [mag] &\\
(1) & (2) & (3) & (4) & (5) & (6)  & (7) & (8) & (9) & (10) & (11) & (12) & (13) & (14) & (15) & (16) & (17)\\
\hline
ZOA073008.083-220105.84 &        J0730-22 &  HIZOA &  112.534 &    $-$22.018 &    236.817 &     $-$1.851 &     0.77 &   104.61 &    11.51 $\pm$     0.02 &    10.04 $\pm$     0.02 &     9.79 $\pm$     0.02 &    17.94 &    16.76 &    16.38 &    1.557 &    3.695 \\ 
ZOA075220.625-250840.47 &       J0752-25A &  HIZOA &  118.086 &    $-$25.145 &    242.052 &      1.022 &     0.56 &    33.38 &    10.92 $\pm$     0.02 &    10.17 $\pm$     0.02 &     9.79 $\pm$     0.02 &    15.47 &    14.70 &    14.35 &    0.327 &    3.678 \\ 
ZOA065010.633-111513.52 &        J0650-11 &  HIZOA &  102.544 &    $-$11.254 &    222.835 &     $-$5.381 &     0.18 &    26.46 &    11.27 $\pm$     0.02 &    10.36 $\pm$     0.02 &     9.83 $\pm$     0.02 &    15.75 &    14.90 &    14.42 &    0.960 &    3.493 \\ 
ZOA182226.663-354035.70 &        J1822-35 &  HIZOA &   275.611 &    $-$35.677 &    357.859 &    $-$10.062 &     0.65 &    37.90 &    10.98 $\pm$     0.02 &    10.14 $\pm$     0.02 &     9.84 $\pm$     0.02 &    16.10 &    15.22 &    14.79 &    0.120 &    4.113 \\ 
ZOA164634.204-390308.21 &        J1646-39 &  HIZOA &  251.642 &    $-$39.052 &    344.676 &      4.067 &     0.24 &    39.69 &    11.94 $\pm$     0.02 &    10.98 $\pm$     0.02 &     9.85 $\pm$     0.02 &    16.67 &    15.79 &    15.34 &    0.990 &    4.446 \\ 
ZOA141604.868-651502.53 &        J1416-65 &  HIZOA &  214.020 &     $-$65.250 &    311.644 &     $-$3.821 &     0.48 &    36.60 &    11.23 $\pm$     0.02 &    10.32 $\pm$     0.02 &     9.89 $\pm$     0.02 &    15.77 &    14.86 &    14.45 &    0.662 &    4.314 \\ 
ZOA100318.769-645803.19 &   2MASX1003-645 &  2MASS &  150.828 &    $-$64.968 &    286.322 &     $-$7.668 &     0.58 &    29.86 &    10.93 $\pm$     0.02 &     9.98 $\pm$     0.02 &     9.90 $\pm$     0.02 &    15.68 &    14.93 &    14.73 &    0.197 &    3.759 \\ 
ZOA105345.693-625013.17 &        J1053-62 &  HIZOA &  163.440 &    $-$62.837 &    289.956 &     $-$2.968 &     0.67 &    65.48 &    11.60 $\pm$     0.02 &    10.68 $\pm$     0.02 &     9.90 $\pm$     0.02 &    17.78 &    16.80 &    16.23 &    0.714 &    4.234 \\ 
ZOA080708.583-280309.50 &        J0807-28 &  HIZOA &  121.786 &    $-$28.053 &    246.266 &      2.310 &     0.61 &    42.80 &    11.01 $\pm$     0.02 &    10.29 $\pm$     0.02 &     9.92 $\pm$     0.02 &    16.69 &    16.02 &    15.56 &    0.460 &    3.676 \\ 
ZOA133732.784-585414.06 &       J1337-58B &  HIZOA &   204.387 &    $-$58.904 &    308.867 &      3.436 &     0.28 &    33.07 &    11.33 $\pm$     0.02 &    10.31 $\pm$     0.02 &     9.92 $\pm$     0.02 &    16.20 &    15.11 &    14.48 &    0.937 &    4.303 \\ 
ZOA143927.759-552503.43 &        J1439-55 &  HIZOA &  219.866 &    $-$55.418 &    317.910 &      4.281 &     0.22 &    30.55 &    11.17 $\pm$     0.02 &    10.41 $\pm$     0.02 &     9.97 $\pm$     0.02 &    16.75 &    15.92 &    15.53 &    0.550 &    4.110 \\ 
ZOA105859.839-502155.66 &   2MASX1058-501 &  2MASS &  164.749 &    $-$50.365 &    285.236 &      8.603 &     0.13 &    24.61 &    11.03 $\pm$     0.02 &    10.11 $\pm$     0.02 &     9.98 $\pm$     0.02 &    15.41 &    14.64 &    14.33 &    0.246 &    3.651 \\ 
ZOA074901.358-261442.69 &       J0748-26A &  HIZOA &  117.256 &    $-$26.245 &    242.618 &     $-$0.182 &     0.27 &    26.65 &    11.49 $\pm$     0.02 &    10.33 $\pm$     0.02 &     9.99 $\pm$     0.02 &    15.56 &    14.61 &    14.20 &    0.619 &    3.837 \\ 
ZOA113728.729-644822.59 &       J1137-644 &  2MASS &  174.370 &    $-$64.806 &    295.162 &     $-$3.056 &     0.21 &    26.51 &    11.96 $\pm$     0.02 &    10.63 $\pm$     0.02 &    10.00 $\pm$     0.02 &    16.35 &    15.00 &    14.46 &    1.363 &    4.252 \\ 
ZOA082837.437-371316.76 &        J0828-37 &  HIZOA &  127.156 &    $-$37.221 &    256.391 &      0.904 &     0.66 &    36.26 &    11.42 $\pm$     0.02 &    10.46 $\pm$     0.02 &    10.03 $\pm$     0.02 &    16.18 &    15.34 &    14.99 &    0.920 &    3.855 \\ 
ZOA134456.207-654051.40 &        J1344-65 &  HIZOA &  206.234 &    $-$65.681 &    308.408 &     $-$3.380 &     0.54 &    34.10 &    11.38 $\pm$     0.02 &    10.47 $\pm$     0.02 &    10.03 $\pm$     0.02 &    16.09 &    15.23 &    14.68 &    0.870 &    4.326 \\ 
ZOA132159.534-543645.62 &   2MASX1321-543 &  2MASS &  200.498 &    $-$54.613 &    307.390 &      7.994 &     0.73 &    41.15 &    11.28 $\pm$     0.02 &    10.45 $\pm$     0.02 &    10.04 $\pm$     0.02 &    16.63 &    15.84 &    15.52 &    0.349 &    3.786 \\ 
ZOA164421.521-552937.33 &        J1644-55 &  HIZOA &  251.090 &     $-$55.494 &    331.917 &     $-$6.333 &     0.83 &    41.24 &    11.41 $\pm$     0.02 &    10.41 $\pm$     0.02 &    10.04 $\pm$     0.02 &    16.52 &    15.54 &    15.03 &    0.292 &    4.129 \\ 
ZOA140835.888-532111.29 &   2MASX1408-532 &  2MASS &  212.150 &    $-$53.353 &    314.411 &      7.790 &     0.33 &    33.68 &    11.00 $\pm$     0.02 &    10.35 $\pm$     0.02 &    10.05 $\pm$     0.02 &    15.50 &    15.31 &    15.22 &    0.424 &    3.829 \\ 
ZOA105842.989-501930.65 &   2MASX1058-501 &  2MASS &  164.679 &    $-$50.325 &    285.177 &      8.620 &     0.82 &    45.60 &    11.40 $\pm$     0.02 &    10.53 $\pm$     0.02 &    10.08 $\pm$     0.02 &    16.09 &    15.27 &    14.73 &    0.246 &    3.651 \\ 
ZOA064400.636+122407.13 &        J0644+12 &  HIZOA &  101.003 &     12.402 &    201.040 &      4.001 &     0.65 &    38.29 &    11.28 $\pm$     0.02 &    10.55 $\pm$     0.02 &    10.09 $\pm$     0.02 &    15.70 &    15.07 &    14.46 &    0.440 &    3.409 \\ 
ZOA182423.339-341054.15 &        J1824-34 &  HIZOA &  276.097 &    $-$34.181 &     359.399 &     $-$9.758 &     0.36 &    25.92 &    11.08 $\pm$     0.02 &    10.38 $\pm$     0.02 &    10.09 $\pm$     0.02 &    15.80 &    15.05 &    14.72 &    0.112 &    4.079 \\ 
ZOA154526.828-605931.93 &        J1545-61 &  HIZOA &  236.362 &    $-$60.992 &    322.508 &     $-$4.918 &     0.49 &    27.22 &    11.49 $\pm$     0.02 &    10.58 $\pm$     0.02 &    10.10 $\pm$     0.02 &    16.00 &    14.97 &    14.41 &    0.611 &    4.140 \\ 
ZOA123157.581-595058.07 &       J1231-595 &  2MASS &  187.990 &    $-$59.850 &    300.485 &      2.930 &     0.28 &    28.59 &    11.50 $\pm$     0.02 &    10.53 $\pm$     0.02 &    10.11 $\pm$     0.02 &    16.27 &    15.32 &    15.10 &    0.803 &    4.215 \\ 
ZOA151113.653-535743.36 &       J1511-535 &  2MASS &  227.807 &    $-$53.962 &    322.664 &      3.466 &     0.12 &    21.61 &    11.39 $\pm$     0.02 &    10.52 $\pm$     0.02 &    10.11 $\pm$     0.02 &    15.70 &    15.03 &    14.87 &    0.858 &    4.316 \\ 
ZOA085809.386-454812.51 &       J0858-45A &  HIZOA &  134.539 &    $-$45.803 &    266.544 &     $-$0.062 &     0.30 &    28.25 &    12.14 $\pm$     0.02 &    10.95 $\pm$     0.02 &    10.12 $\pm$     0.02 &    17.26 &    15.87 &    15.08 &    2.356 &    3.890 \\ 
ZOA085828.676-451630.99 &       J0858-45B &  HIZOA &  134.619 &    $-$45.275 &    266.181 &      0.325 &     0.29 &    35.29 &    12.29 $\pm$     0.02 &    10.94 $\pm$     0.02 &    10.12 $\pm$     0.02 &    17.97 &    16.54 &    15.77 &    3.148 &    3.819 \\ 
ZOA133724.550-585221.57 &       J1337-58B &  HIZOA &  204.35229 &    $-$58.873 &    308.855 &      3.469 &     0.68 &    34.33 &    11.48 $\pm$     0.02 &    10.51 $\pm$     0.02 &    10.12 $\pm$     0.02 &    15.54 &    14.67 &    14.28 &    0.937 &    4.303 \\ 
ZOA154710.889-590408.56 &        J1547-59 &  HIZOA &  236.795 &    $-$59.069 &    323.868 &      $-$3.538 &     0.39 &    47.97 &    11.32 $\pm$     0.02 &    10.70 $\pm$     0.02 &    10.13 $\pm$     0.02 &    16.82 &    15.96 &    15.48 &    0.550 &    4.345 \\ 
ZOA163617.005-421325.00 &       J1636-421 &  2MASS &  249.071 &    $-$42.224 &    340.998 &      3.445 &     0.39 &    28.61 &    11.15 $\pm$     0.02 &    10.54 $\pm$     0.02 &    10.13 $\pm$     0.02 &    16.10 &    15.30 &    14.72 &    1.669 &    4.401 \\ 
ZOA165408.098-353438.65 &        J1653-35 &  HIZOA &  253.534 &    $-$35.577 &    348.325 &      5.120 &     0.13 &    18.68 &    11.47 $\pm$     0.02 &    10.68 $\pm$     0.02 &    10.17 $\pm$     0.02 &    15.53 &    15.12 &    14.29 &    0.930 &    4.395 \\ 
ZOA140627.300-575142.26 &        J1406-57 &  HIZOA &  211.614 &    $-$57.862 &    312.799 &      3.566 &     0.27 &    32.83 &    11.57 $\pm$     0.02 &    10.62 $\pm$     0.02 &    10.18 $\pm$     0.02 &    16.46 &    15.50 &    15.13 &    0.628 &    4.171 \\ 
ZOA160441.177-413947.62 &        J1604-41 &  HIZOA &  241.172 &    $-$41.663 &     337.221 &      7.994 &     0.35 &    28.11 &    11.25 $\pm$     0.02 &    10.55 $\pm$     0.02 &    10.18 $\pm$     0.02 &    16.32 &    15.53 &    15.17 &    0.559 &    3.887 \\ 
ZOA090240.287-413502.67 &       J0902-413 &  2MASS &  135.668 &    $-$41.584 &    263.902 &      3.328 &     0.52 &    28.06 &    11.53 $\pm$     0.02 &    10.61 $\pm$     0.02 &    10.19 $\pm$     0.02 &    16.09 &    15.18 &    14.80 &    1.000 &    3.659 \\ 
ZOA155335.142-614059.08 &   2MASX1553-614 &  2MASS &  238.396 &    $-$61.683 &    322.836 &     $-$6.068 &     0.74 &    38.79 &    11.32 $\pm$     0.02 &    10.51 $\pm$     0.02 &    10.20 $\pm$     0.02 &    16.10 &    15.50 &    15.47 &    0.287 &    4.020 \\ 
ZOA182700.997-203159.00 &        J1826-20 &  HIZOA &  276.754 &    $-$20.533 &     11.924 &     $-$4.088 &     0.70 &    41.25 &    11.59 $\pm$     0.02 &    10.62 $\pm$     0.02 &    10.20 $\pm$     0.02 &    16.62 &    15.62 &    15.17 &    0.714 &    4.688 \\ 
ZOA140621.248-602544.76 &       J1406-602 &  2MASS &  211.589 &    $-$60.429 &    312.051 &      1.109 &     0.42 &    22.18 &    12.47 $\pm$     0.02 &    10.97 $\pm$     0.02 &    10.21 $\pm$     0.02 &    16.55 &    15.26 &    14.53 &    2.962 &    4.638 \\ 
ZOA134423.990-522211.32 &   2MASX1344-522 &  2MASS &  206.100 &    $-$52.370 &    311.087 &      9.661 &     0.75 &    40.16 &    11.57 $\pm$     0.02 &    10.64 $\pm$     0.02 &    10.22 $\pm$     0.02 &    16.65 &    15.79 &    15.43 &    0.387 &    3.632 \\ 
ZOA170643.848-482357.97 &    2MASXJ1706-4823 &  2MASS &  256.683 &    $-$48.399 &    339.642 &     $-$4.607 &     0.71 &    28.00 &    11.68 $\pm$     0.02 &    10.79 $\pm$     0.02 &    10.24 $\pm$     0.02 &    16.10 &    15.27 &    14.60 &    0.747 &    4.363 \\ 
ZOA080953.826-414136.58 &        J0809-41 &  HIZOA &  122.474 &    $-$41.694 &    258.049 &      $-$4.600 &     0.84 &    66.41 &    12.03 $\pm$     0.02 &    10.82 $\pm$     0.02 &    10.25 $\pm$     0.02 &    17.66 &    16.41 &    15.72 &    1.049 &    3.750 \\ 
ZOA151548.734-600409.37 &       J1515-60B &  HIZOA &  228.953 &    $-$60.069 &     320.037 &     $-$2.083 &     0.17 &    32.38 &    12.09 $\pm$     0.02 &    10.84 $\pm$     0.02 &    10.25 $\pm$     0.02 &    17.56 &    16.25 &    15.61 &    3.113 &    4.611 \\ 
ZOA165805.966-211622.32 &       J1658-21A &  HIZOA &  254.525 &    $-$21.273 &      0.331 &     13.190 &     0.12 &    28.09 &    11.30 $\pm$     0.02 &    10.52 $\pm$     0.02 &    10.27 $\pm$     0.02 &    16.21 &    15.52 &    15.39 &    0.300 &    3.667 \\ 
ZOA105838.725-645044.49 &       J1058-645 &  HIZOA &  164.661 &    $-$64.846 &    291.308 &     $-$4.551 &     0.16 &    22.02 &    11.53 $\pm$     0.02 &    10.65 $\pm$     0.02 &    10.28 $\pm$     0.02 &    15.82 &    14.93 &    14.84 &    0.533 &    4.122 \\ 
ZOA160449.497-414301.20 &        J1604-41 &  HIZOA &  241.206 &      $-$41.717 &    337.204 &      7.937 &     0.44 &    36.21 &    11.42 $\pm$     0.02 &    10.71 $\pm$     0.02 &    10.29 $\pm$     0.02 &    16.40 &    15.62 &    15.24 &    0.559 &    3.887 \\ 
ZOA183155.989-314742.59 &        J1831-31 &  HIZOA &  277.983 &    $-$31.795 &      2.280 &    $-$10.134 &     0.35 &    32.63 &    11.24 $\pm$     0.02 &    10.57 $\pm$     0.02 &    10.30 $\pm$     0.02 &    16.16 &    15.49 &    15.21 &    0.146 &    4.054 \\ 
ZOA063556.737+143557.75 &       J0635+14B &  HIZOA &   98.986 &     14.599 &    198.185 &      3.254 &     0.20 &    21.72 &    11.49 $\pm$     0.02 &    10.66 $\pm$     0.02 &    10.31 $\pm$     0.02 &    15.74 &    14.92 &    14.65 &    0.570 &    3.529 \\ 
ZOA120920.790-622912.31 &        J1209-62 &  HIZOA &  182.337 &    $-$62.487 &    298.092 &     $-$0.011 &     0.36 &    22.97 &    13.08 $\pm$     0.02 &    11.21 $\pm$     0.02 &    10.31 $\pm$     0.02 &    17.35 &    15.59 &    14.70 &    3.294 &    4.553 \\ 
ZOA141232.785-563433.93 &       J1412-56A &  HIZOA &  213.137 &    $-$56.576 &    313.972 &      4.547 &     0.46 &    42.52 &    11.43 $\pm$     0.02 &    10.62 $\pm$     0.02 &    10.32 $\pm$     0.02 &    17.21 &    16.41 &    16.07 &    0.559 &    4.149 \\ 
ZOA072653.624-073252.01 &        J0726-07 &  HIZOA &  111.723 &     $-$7.548 &    223.715 &      4.362 &     0.21 &    27.41 &    11.00 $\pm$     0.02 &    10.64 $\pm$     0.02 &    10.35 $\pm$     0.02 &    16.59 &    16.08 &    15.80 &    0.240 &    3.492 \\ 
ZOA100655.897-450248.59 &   2MASX1006-450 &  2MASS &  151.733 &    $-$45.047 &    274.892 &      8.682 &     0.65 &    33.97 &    11.43 $\pm$     0.02 &    10.72 $\pm$     0.02 &    10.35 $\pm$     0.02 &    16.53 &    15.93 &    15.52 &    0.132 &    3.504 \\ 
\hline
\label{all_NIR_par}
\end{longtable}

\normalsize
\renewcommand{\thefootnote}{\arabic{footnote}}
\renewcommand{\arraystretch}{1.0}
\end{landscape}
\twocolumn

\begin{figure*}
\begin{tabular}{ccccc}
 \subfloat{\includegraphics[scale=0.17]{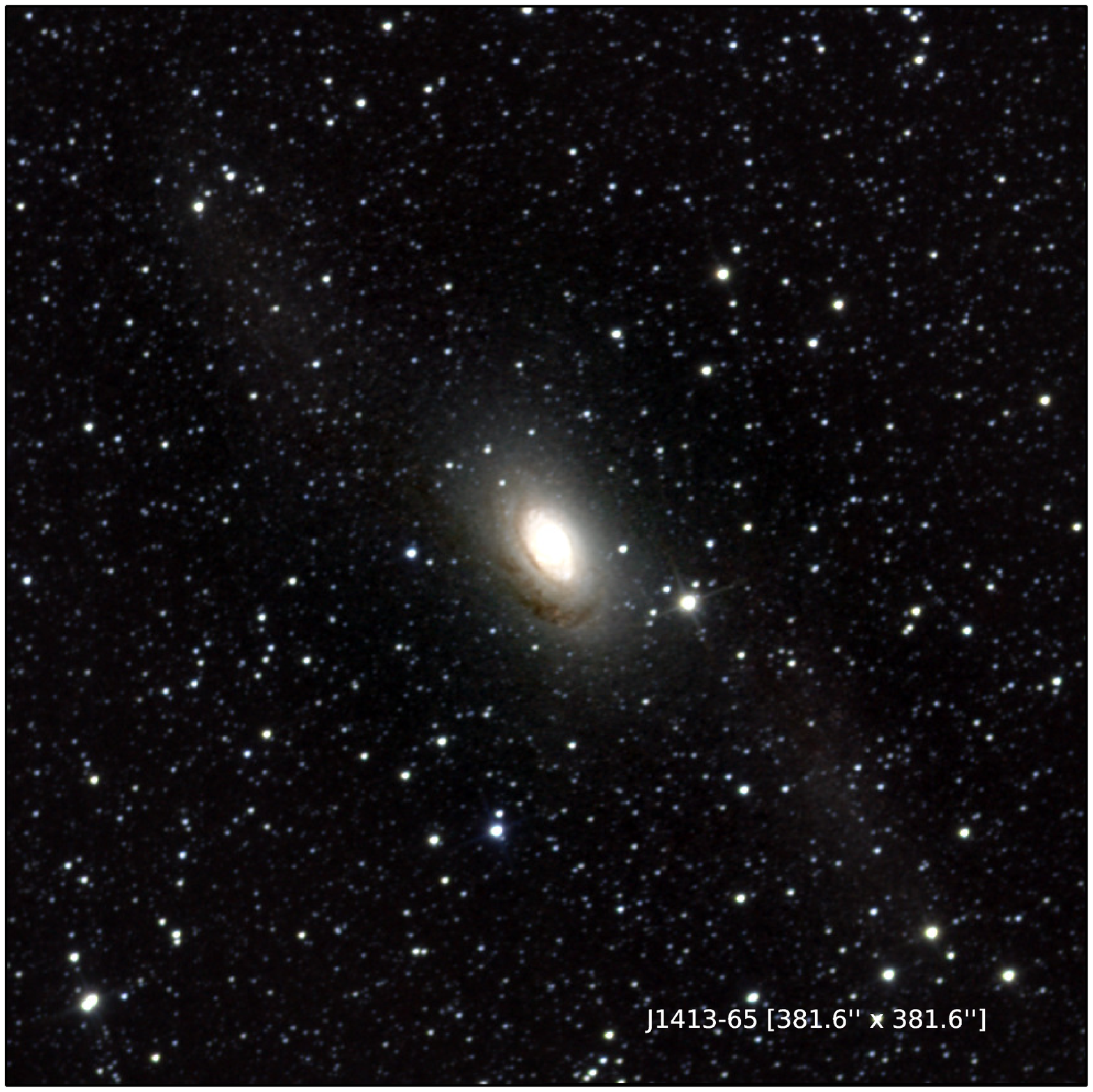}}
 & \subfloat{\includegraphics[scale=0.17]{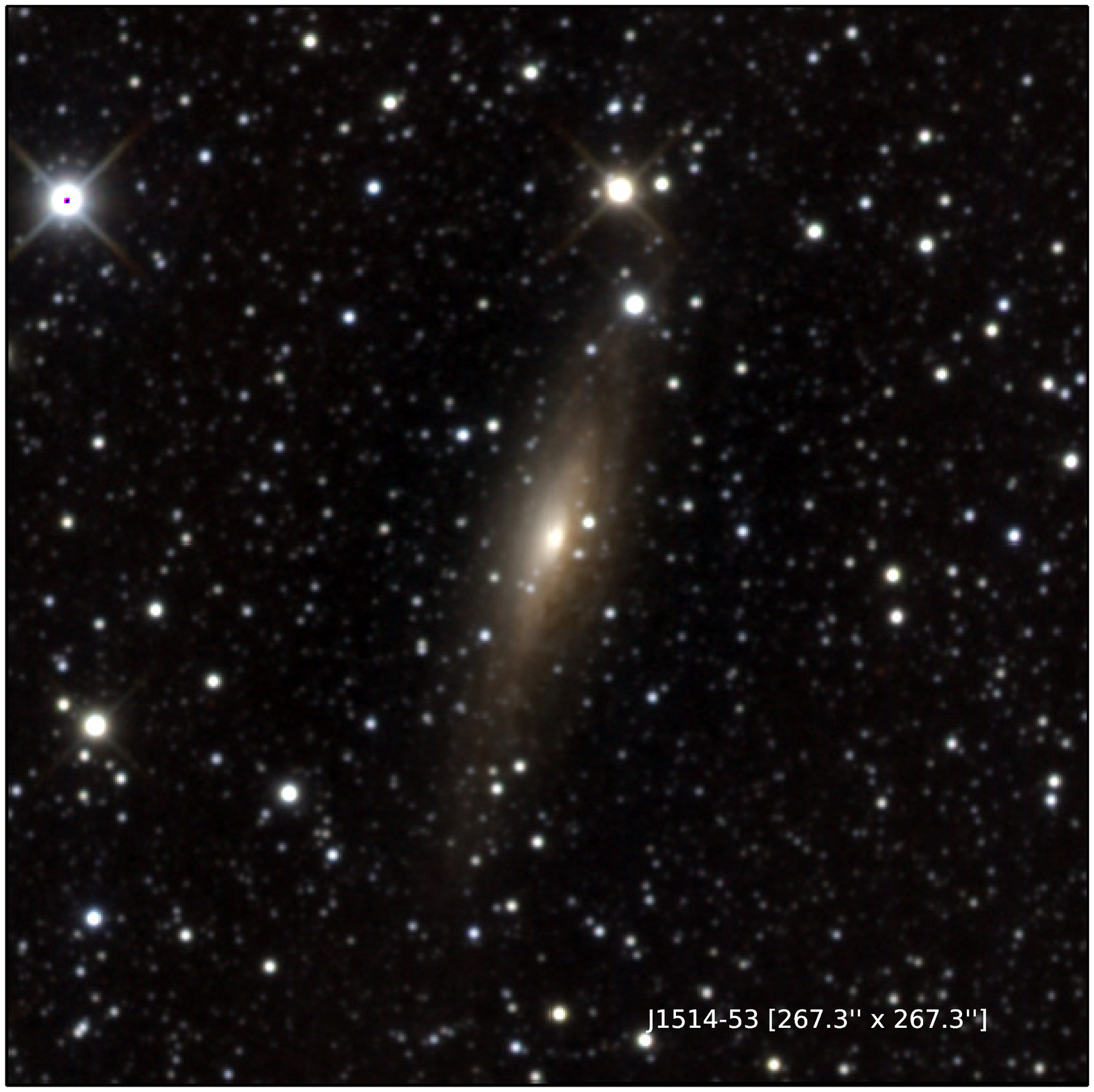}}
 & \subfloat{\includegraphics[scale=0.17]{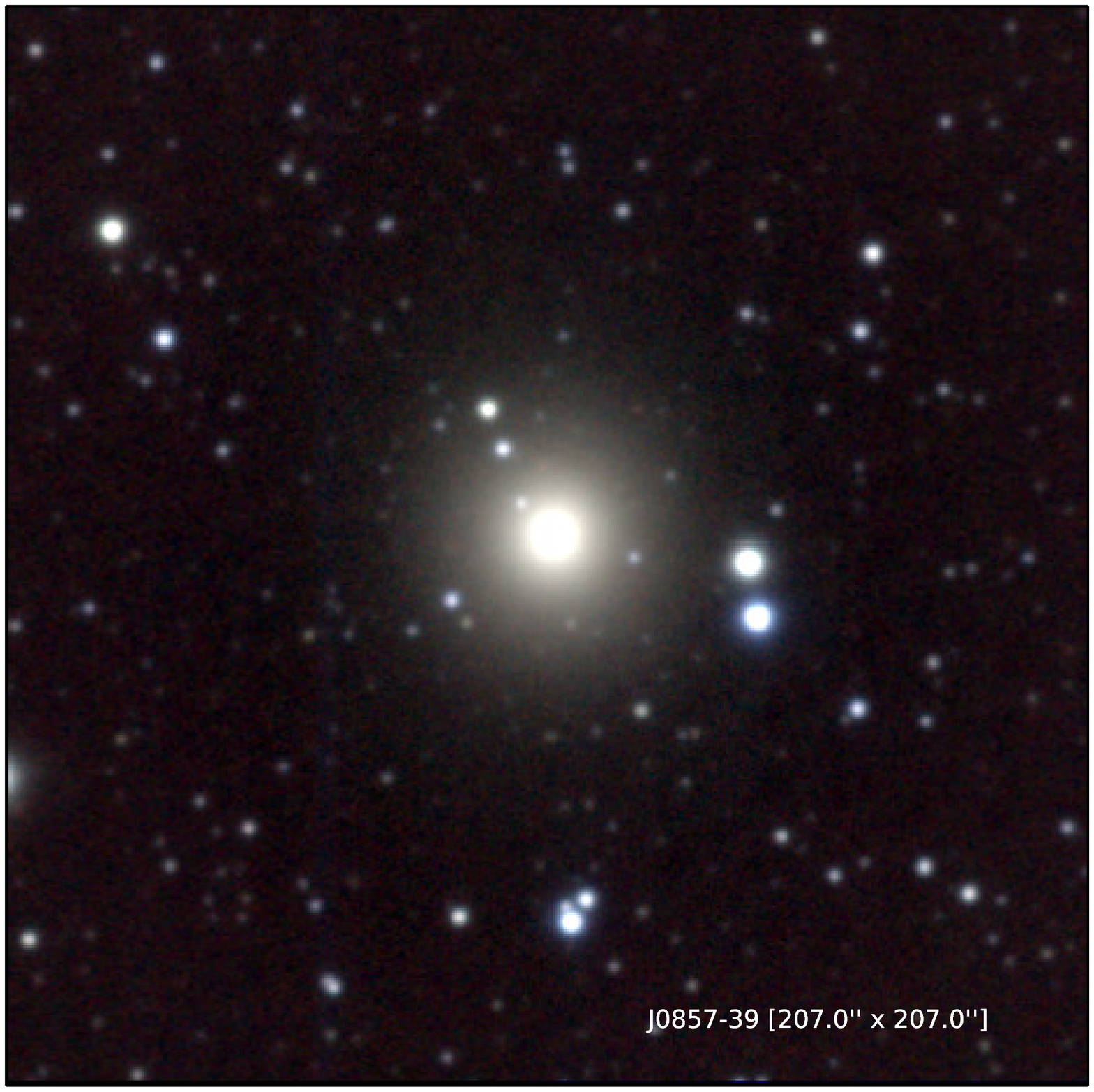}}
 & \subfloat{\includegraphics[scale=0.17]{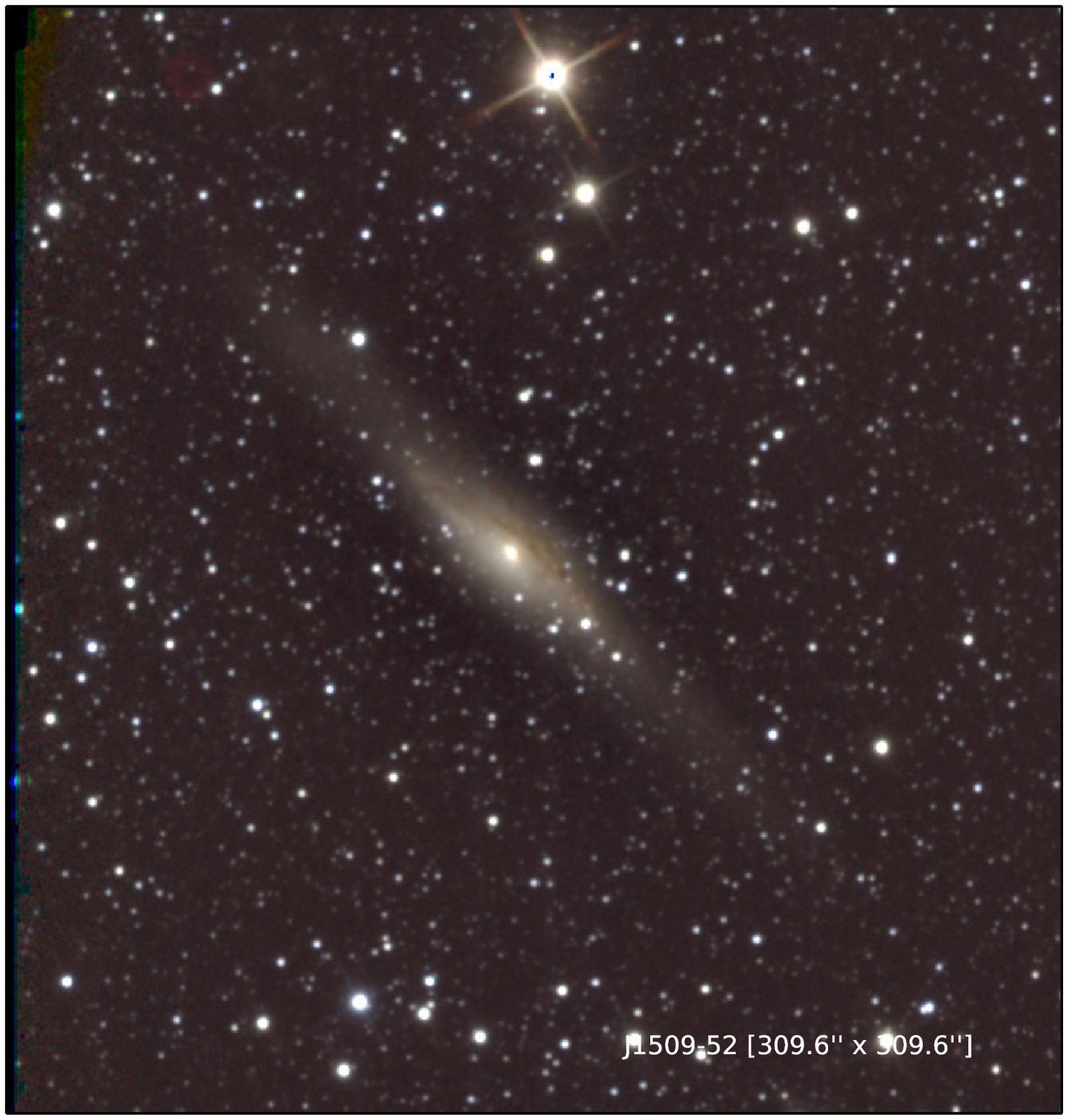}}
 & \subfloat{\includegraphics[scale=0.17]{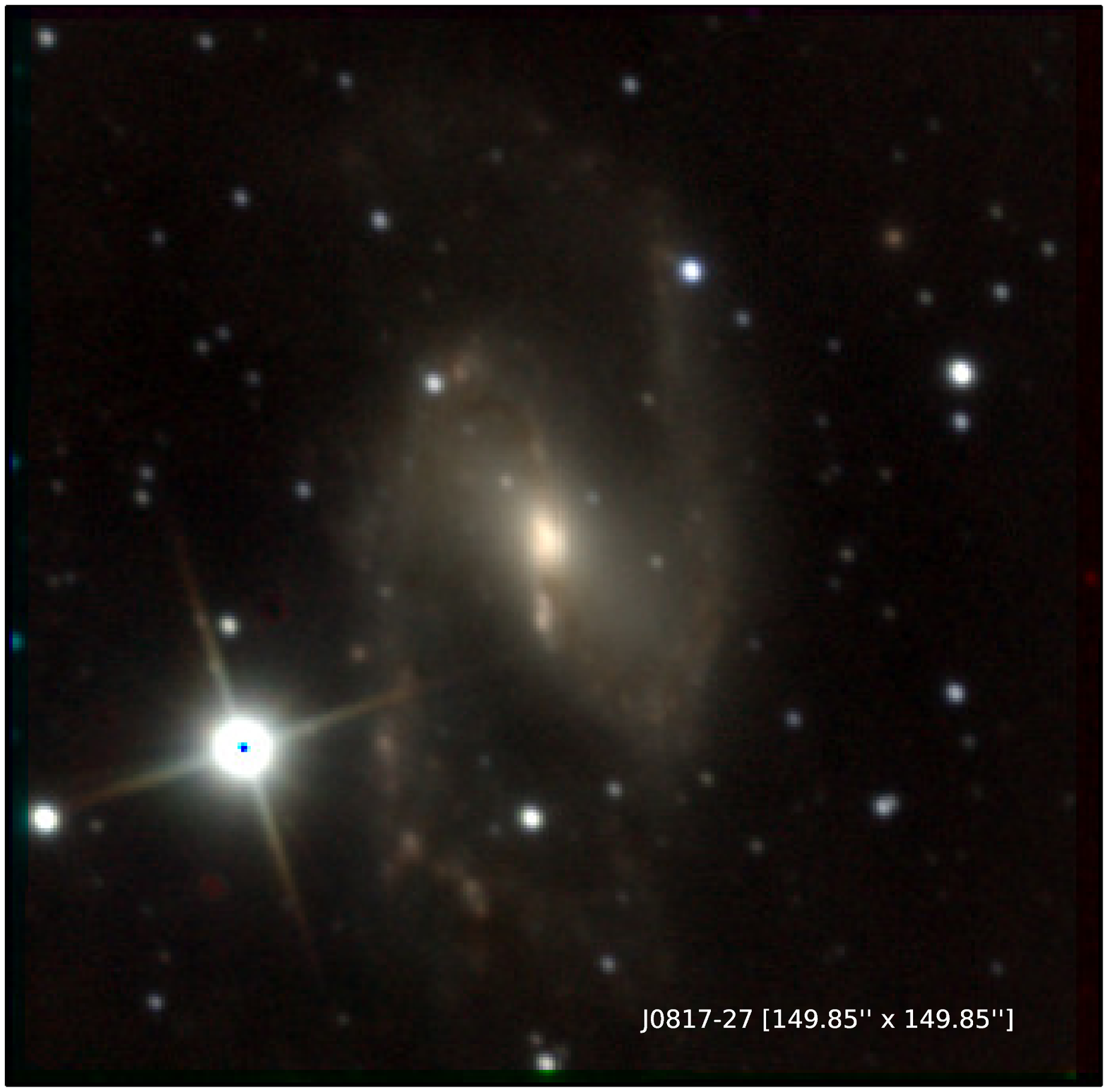}}\\
  \subfloat{\includegraphics[scale=0.17]{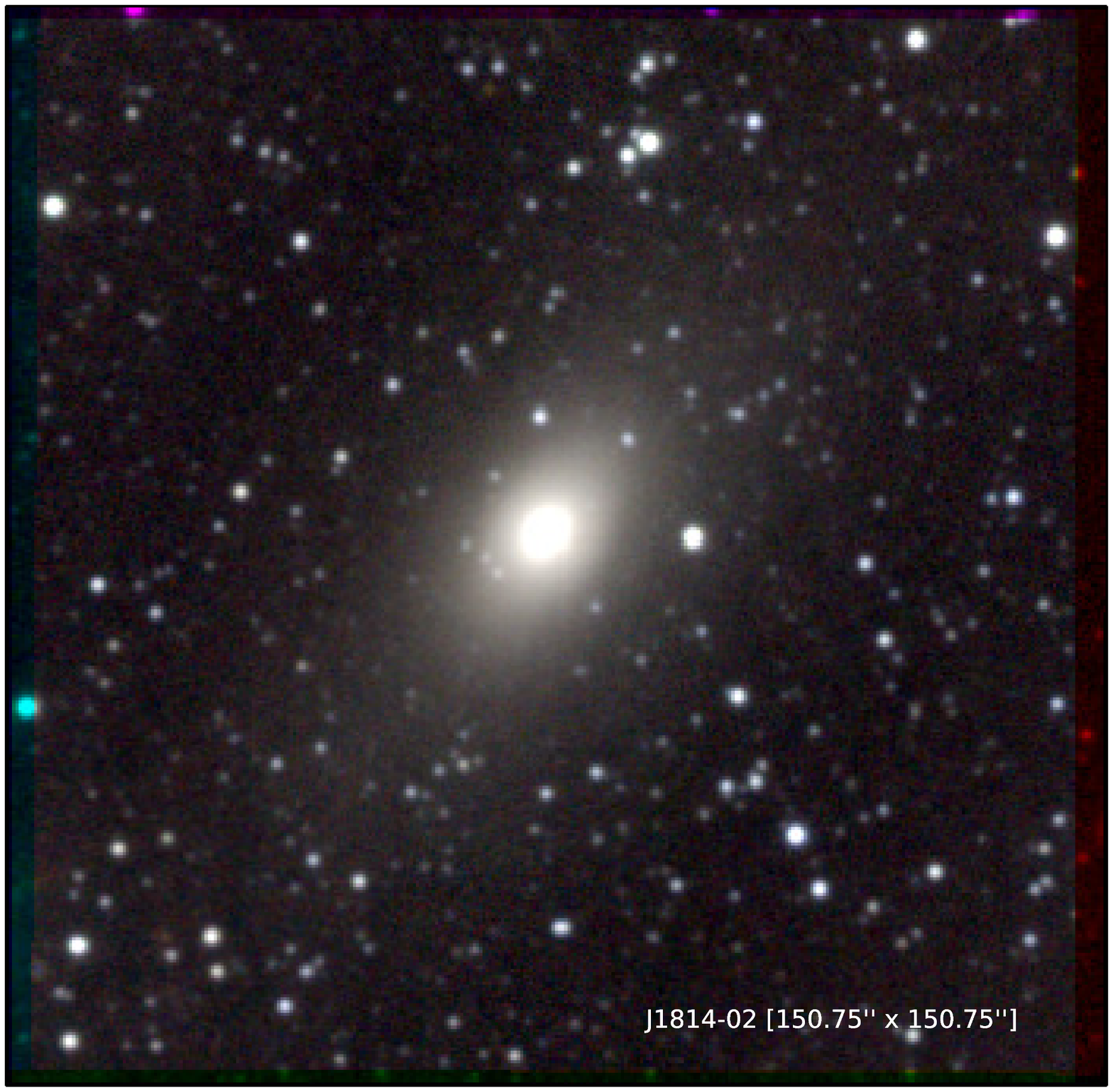}}
 & \subfloat{\includegraphics[scale=0.17]{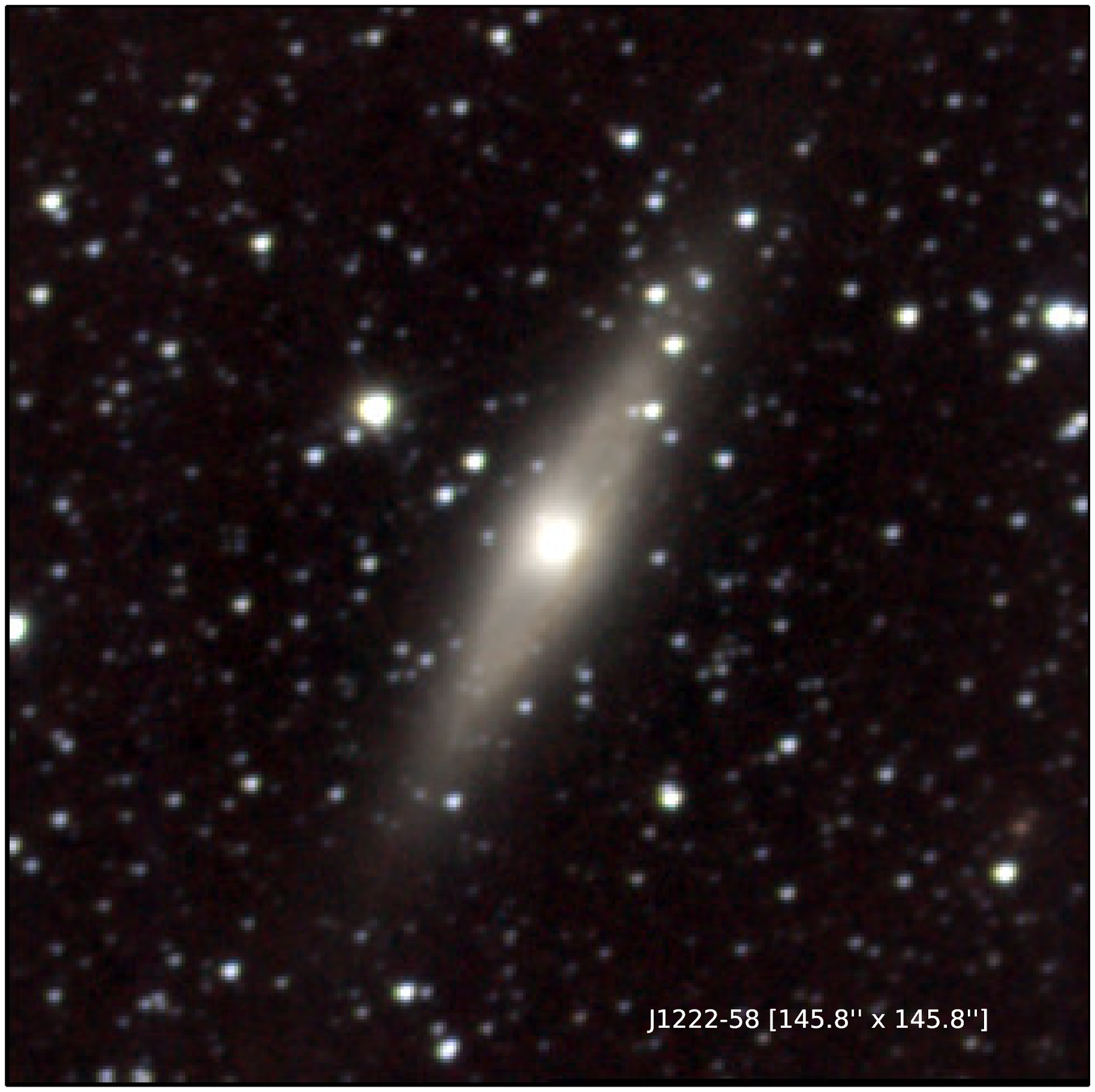}}
 & \subfloat{\includegraphics[scale=0.17]{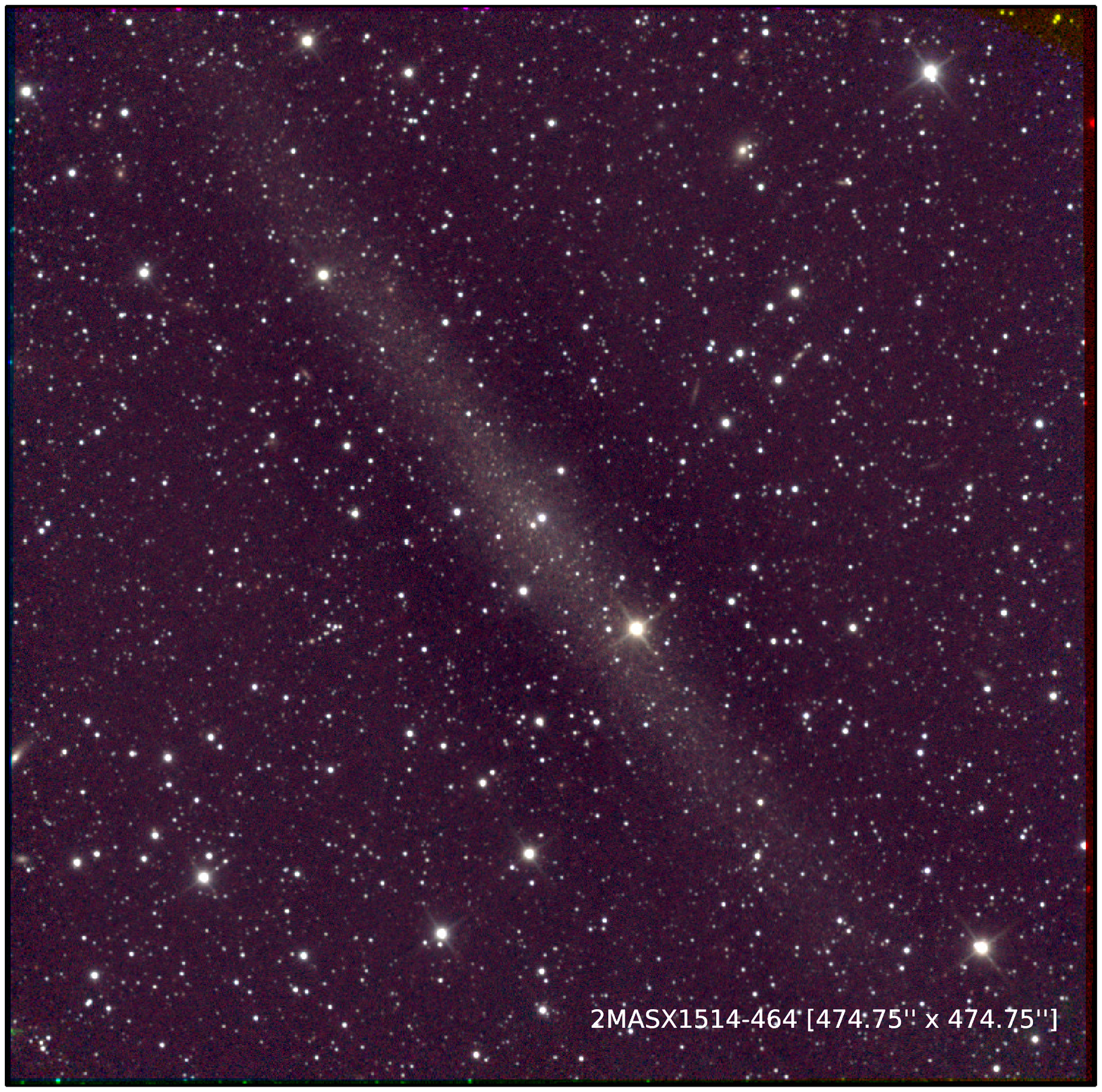}}
 & \subfloat{\includegraphics[scale=0.17]{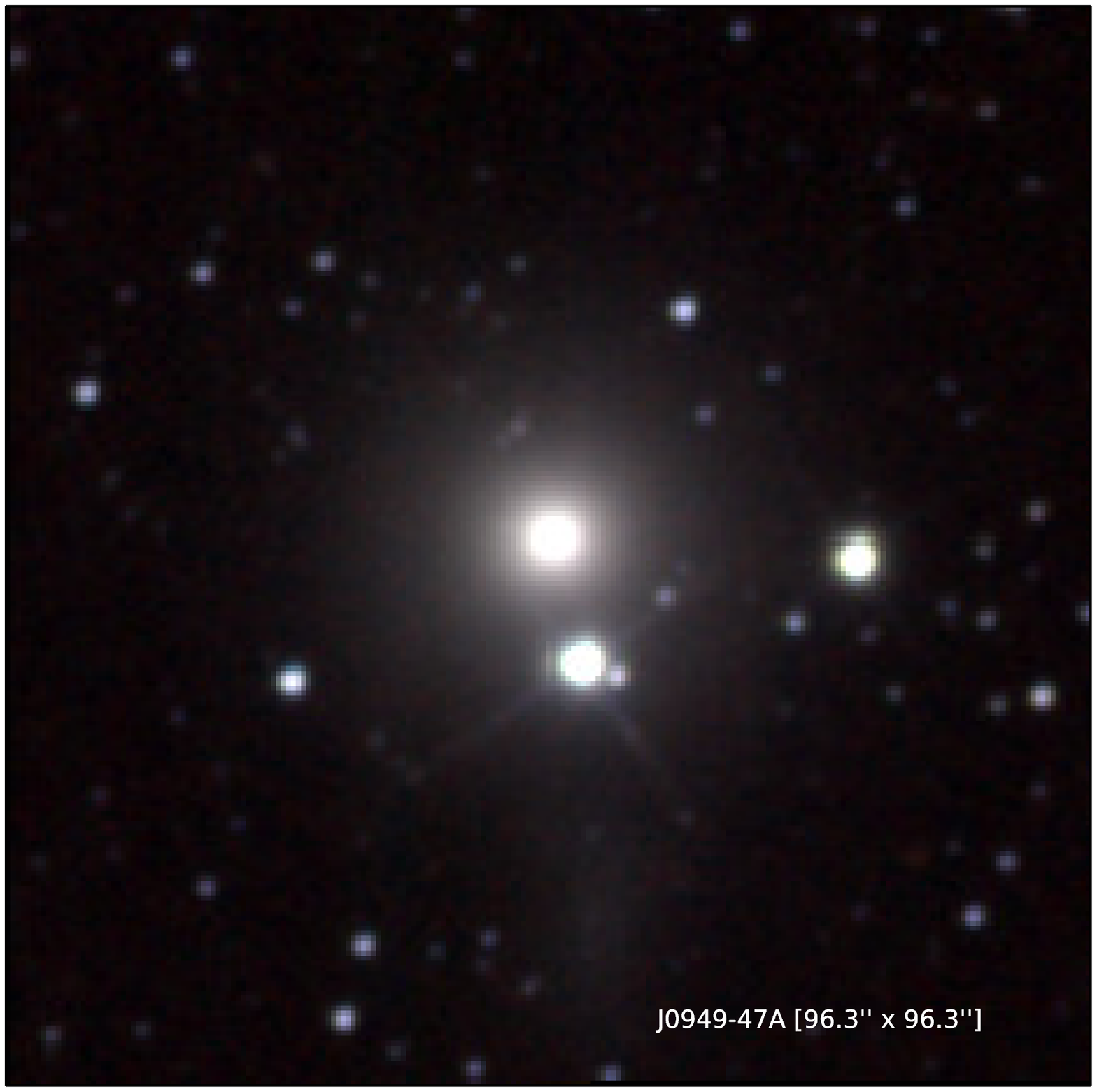}}
 & \subfloat{\includegraphics[scale=0.17]{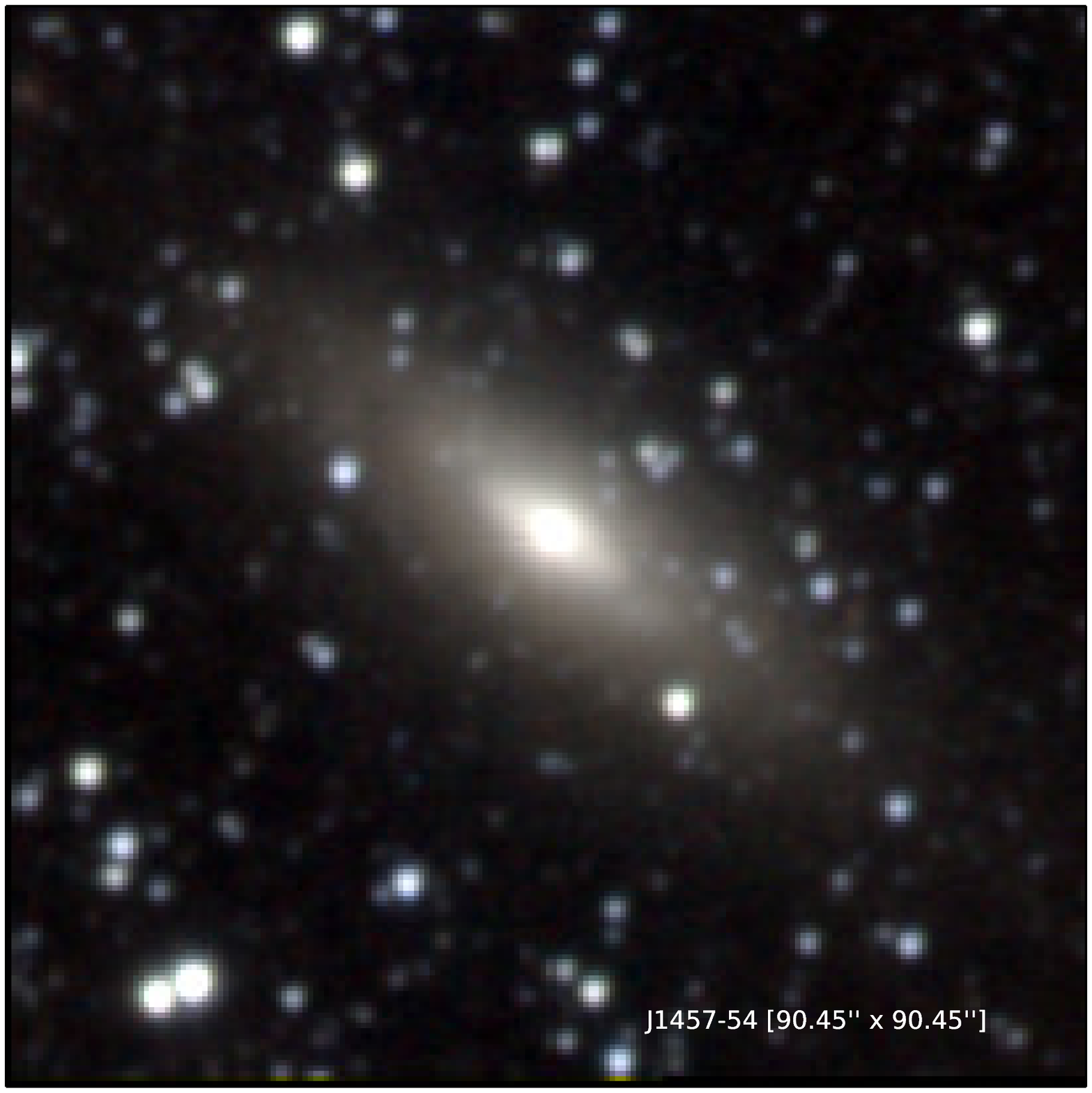}}\\
  \subfloat{\includegraphics[scale=0.17]{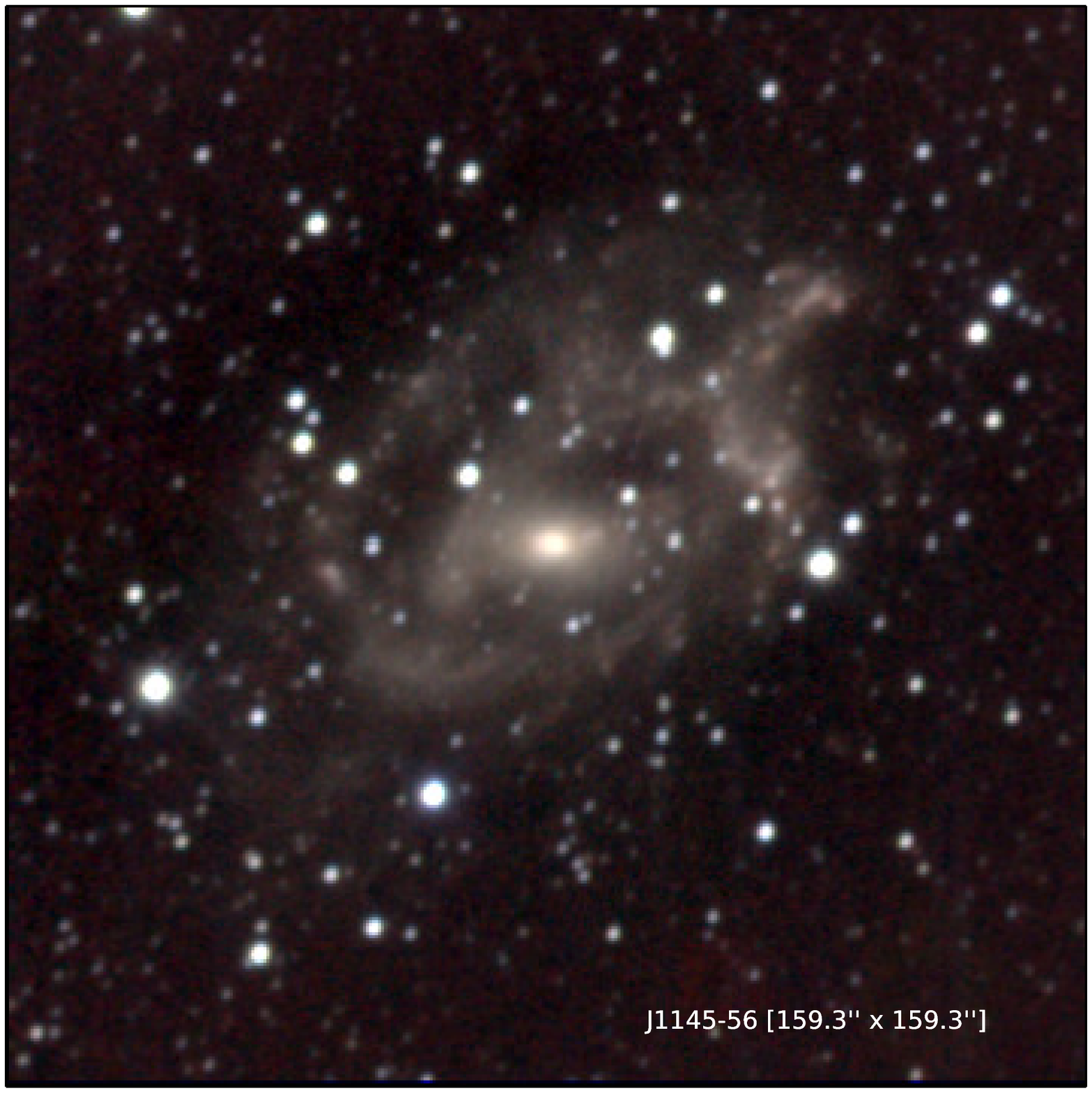}}
 & \subfloat{\includegraphics[scale=0.17]{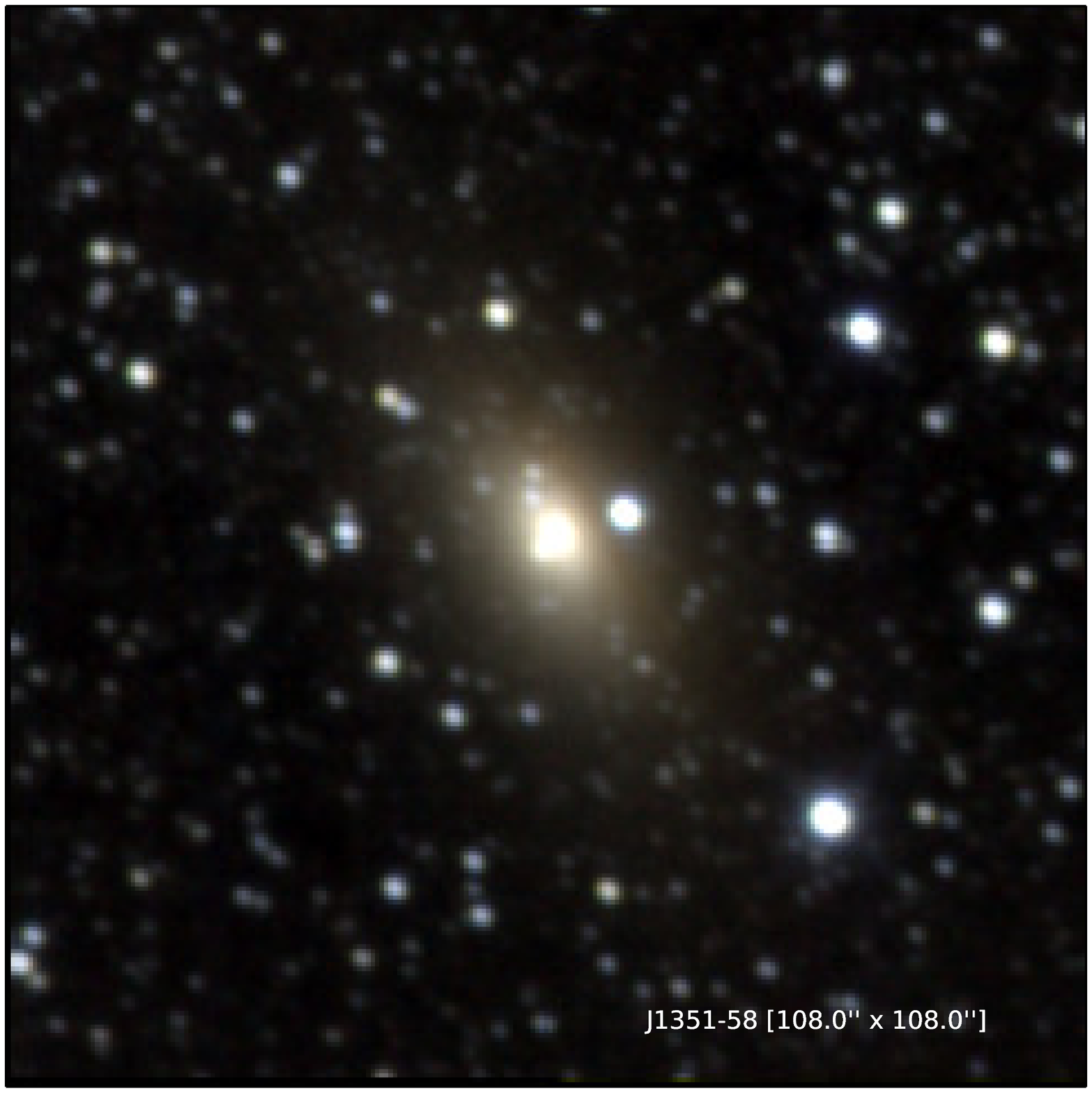}}
 & \subfloat{\includegraphics[scale=0.17]{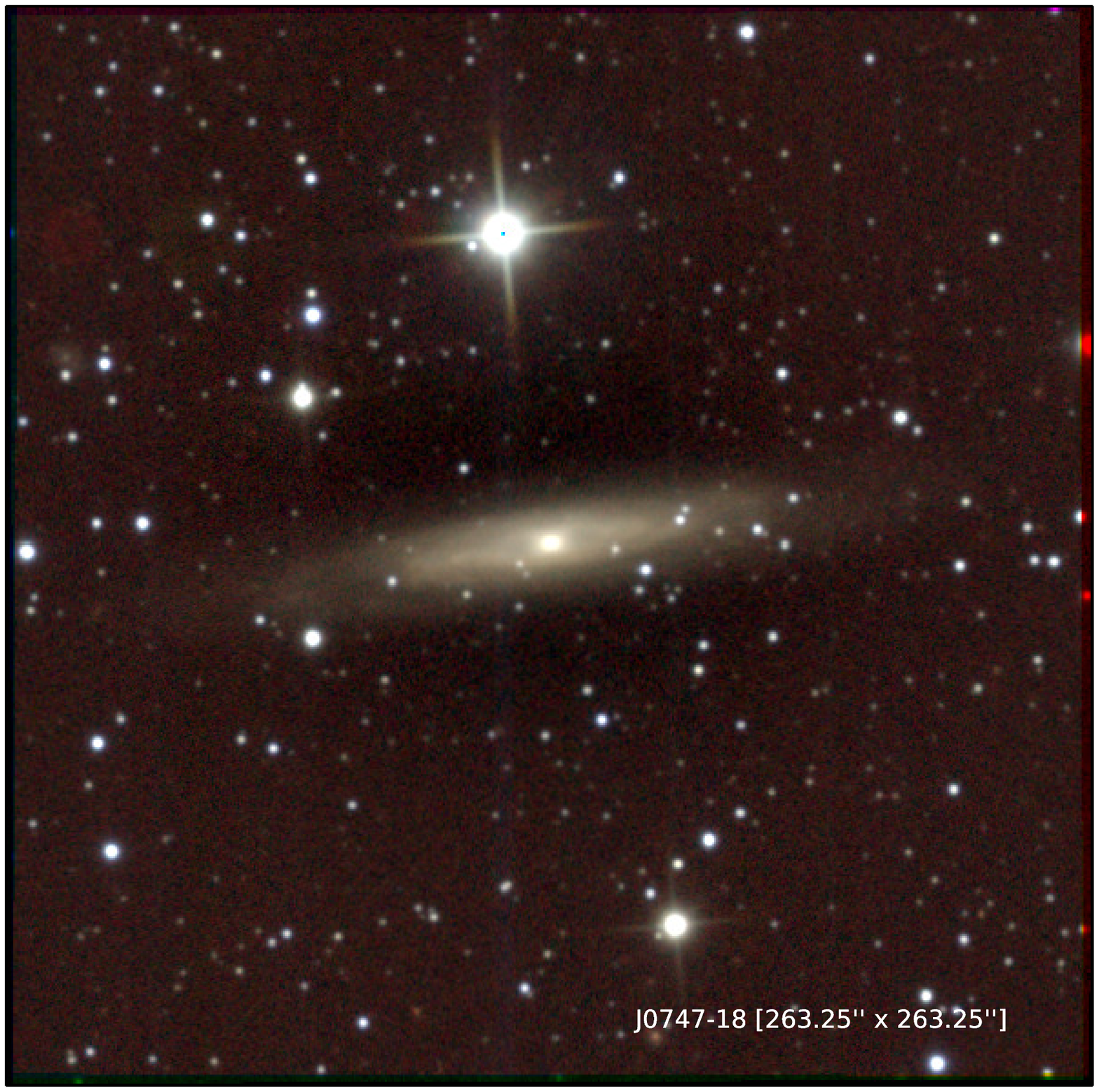}}
 & \subfloat{\includegraphics[scale=0.17]{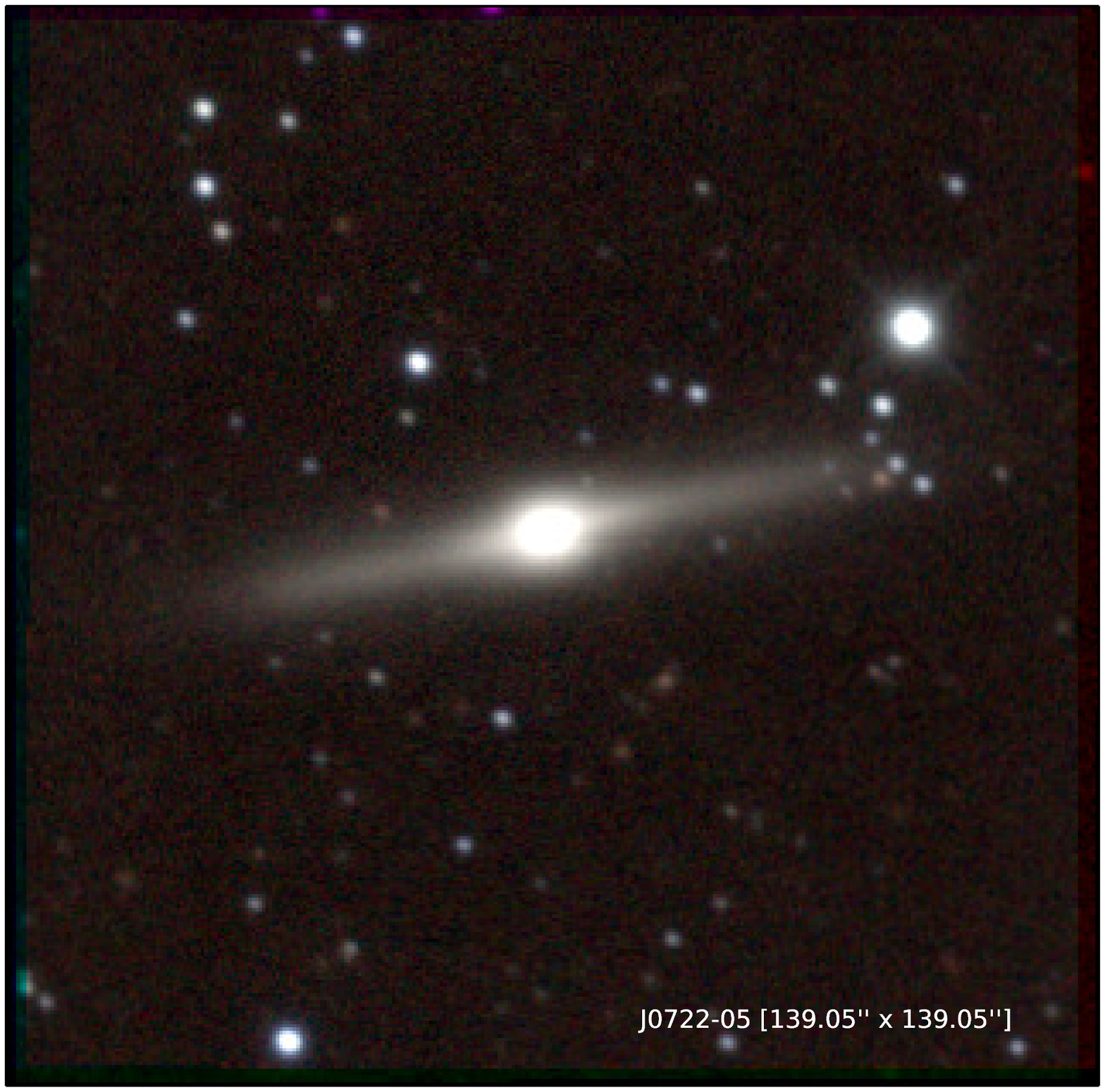}}
 & \subfloat{\includegraphics[scale=0.17]{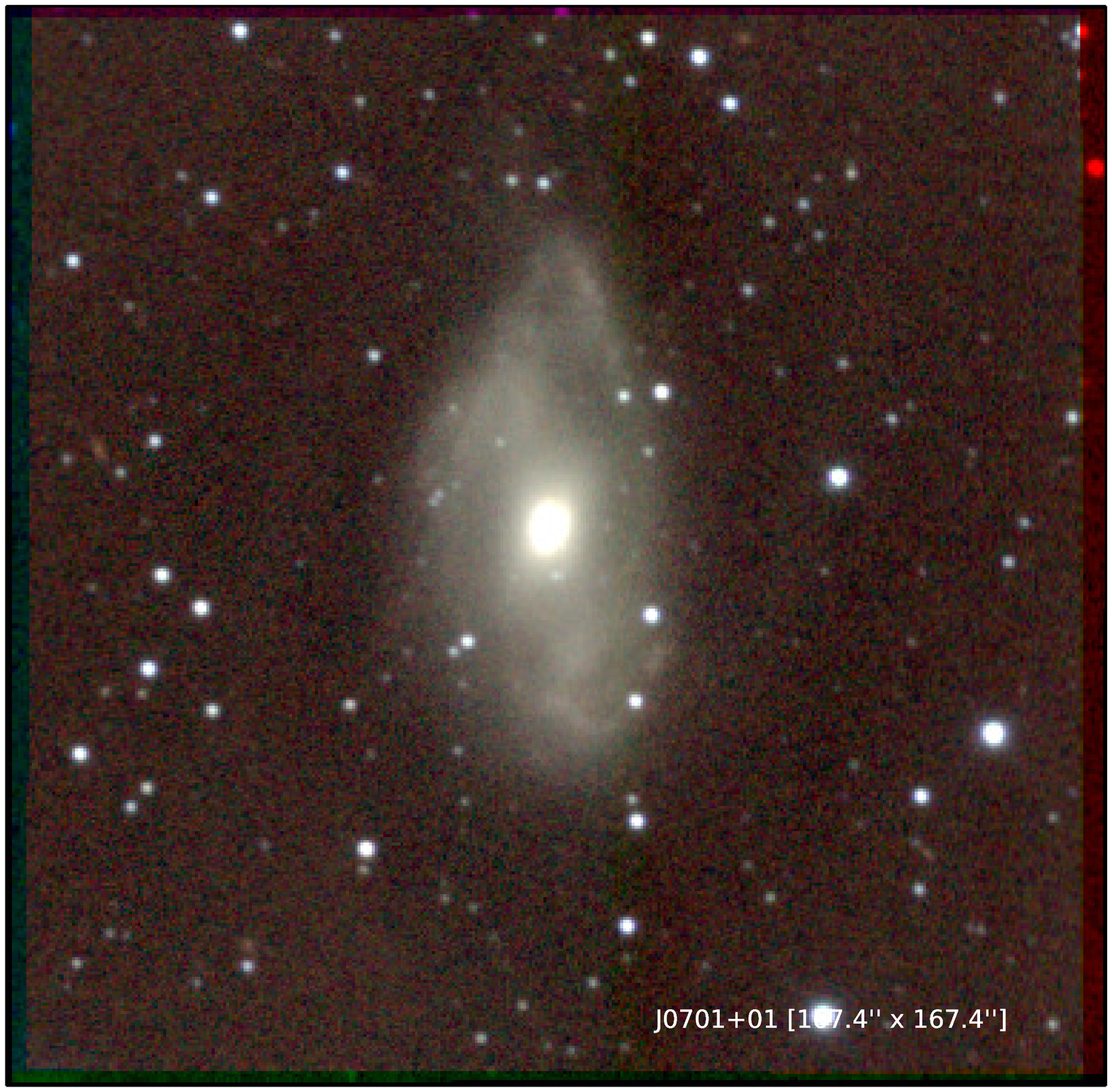}}\\
  \subfloat{\includegraphics[scale=0.17]{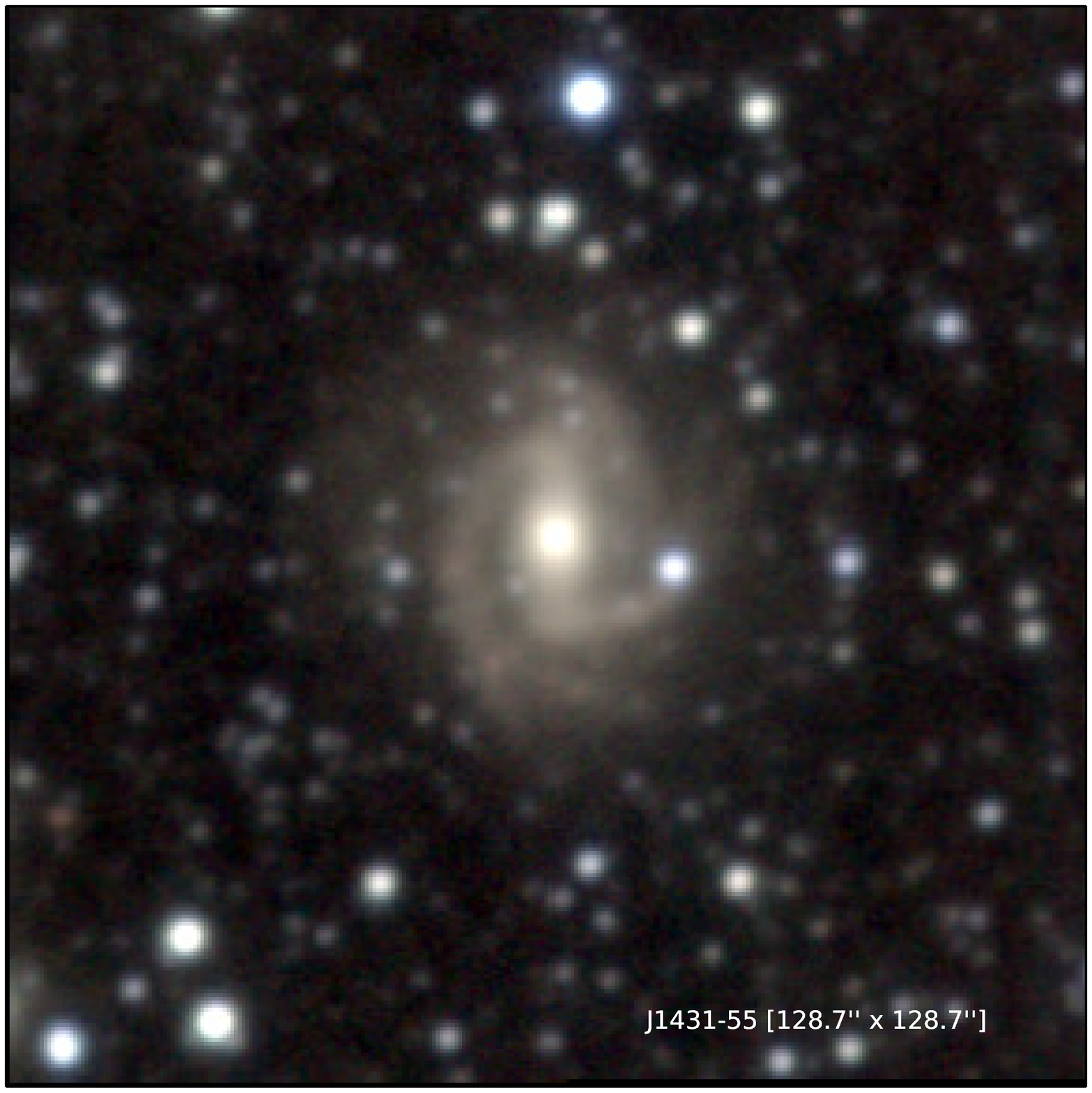}}
 & \subfloat{\includegraphics[scale=0.17]{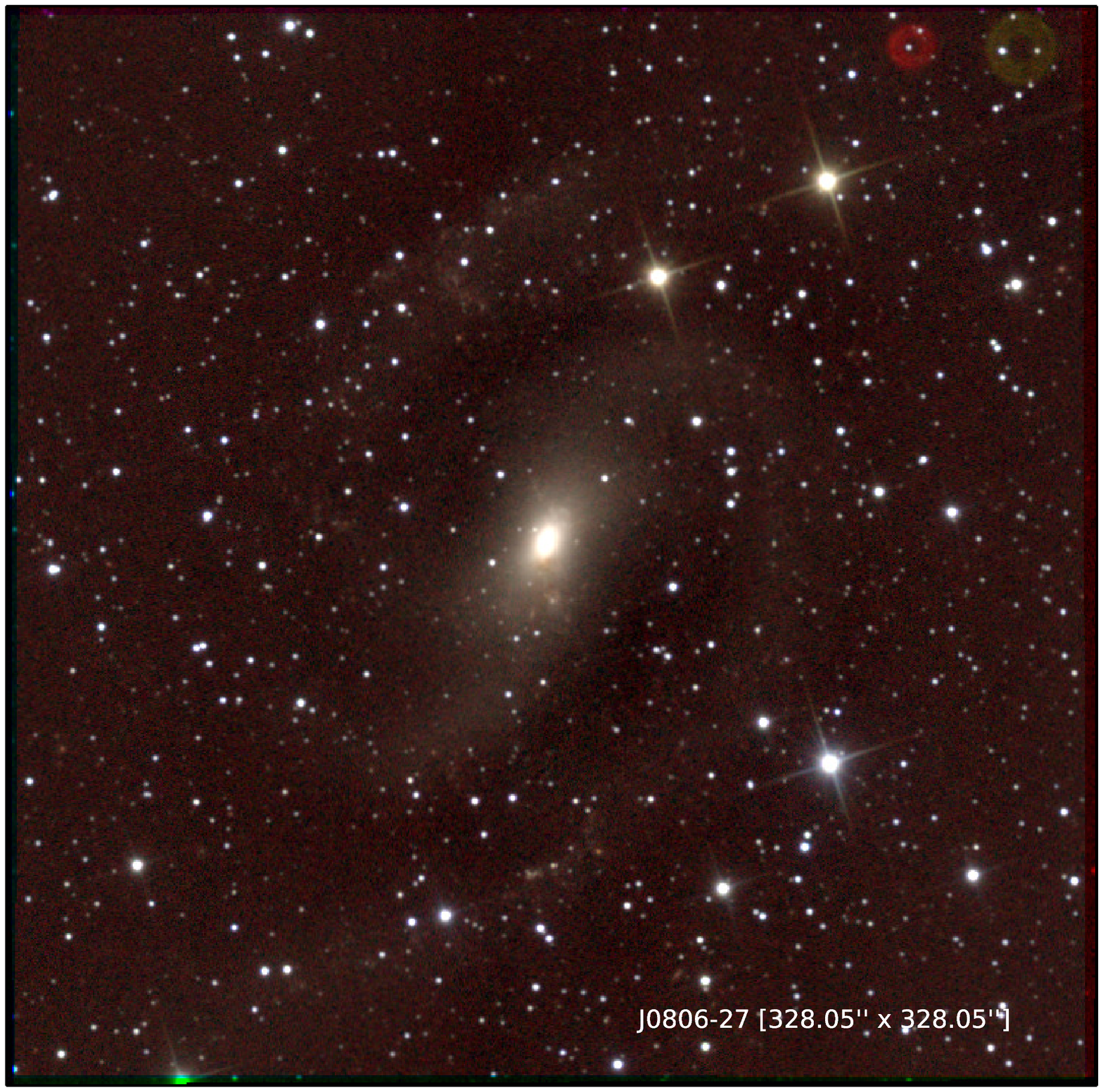}}
 & \subfloat{\includegraphics[scale=0.17]{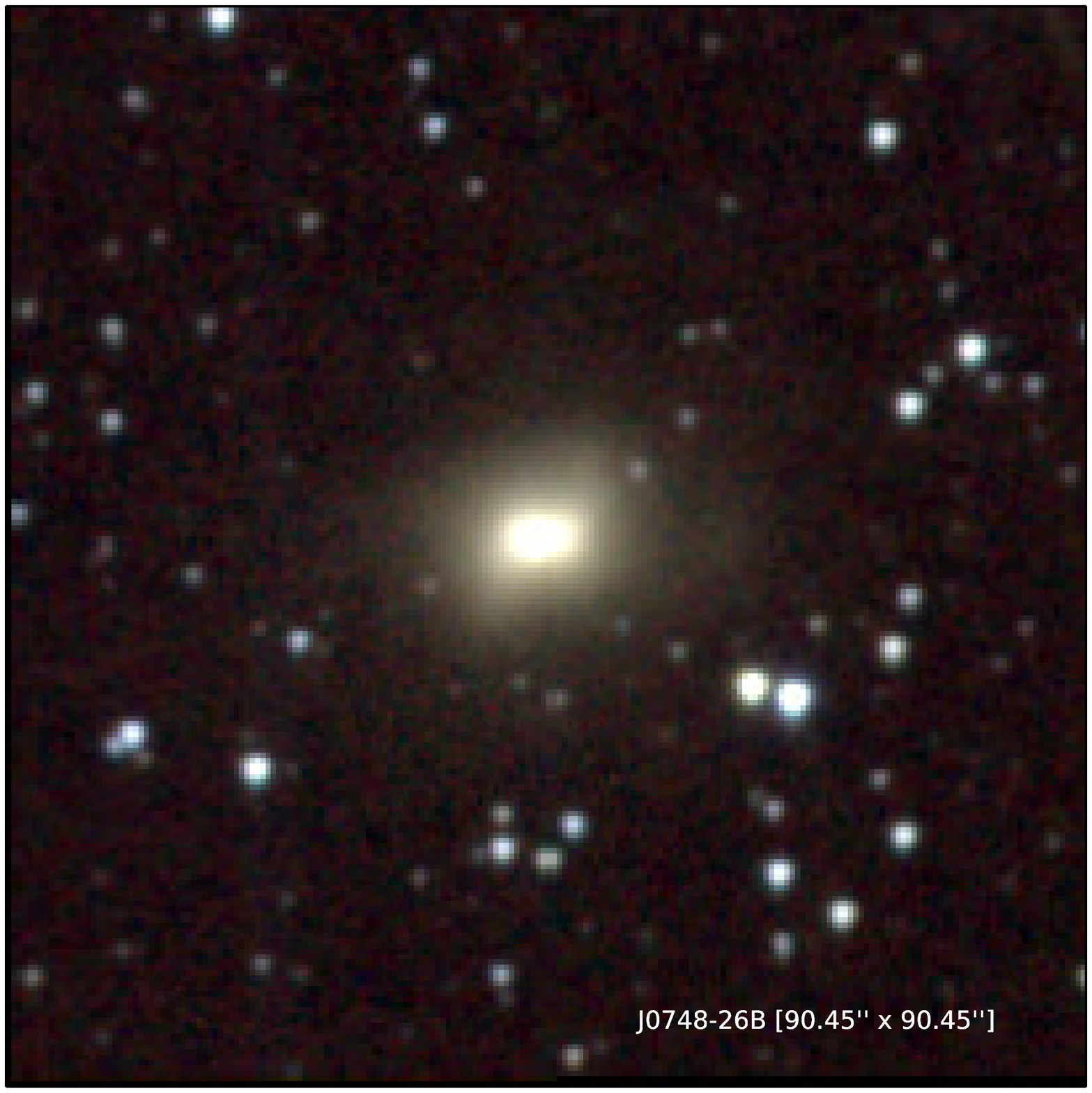}}
 & \subfloat{\includegraphics[scale=0.17]{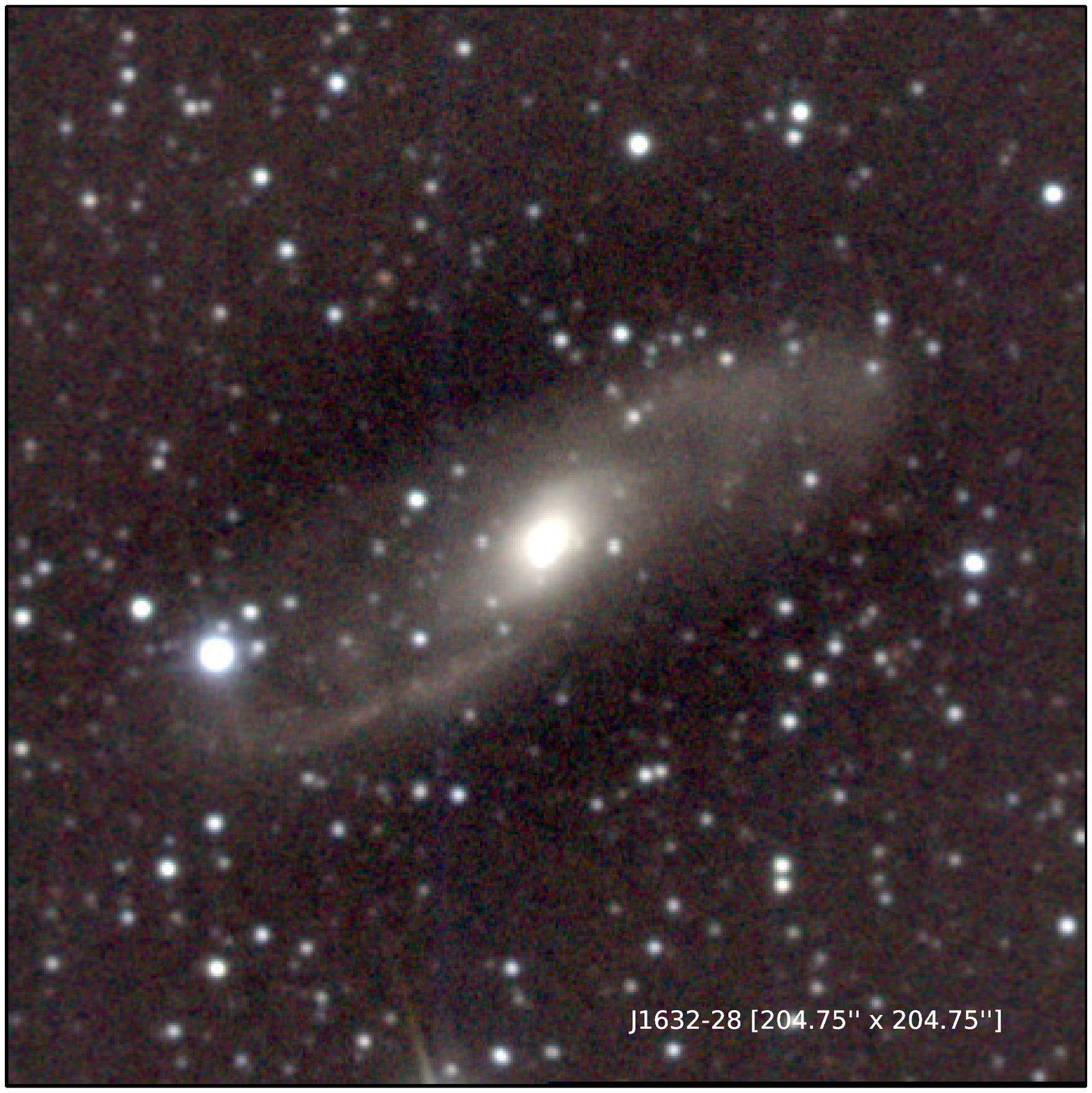}}
 & \subfloat{\includegraphics[scale=0.17]{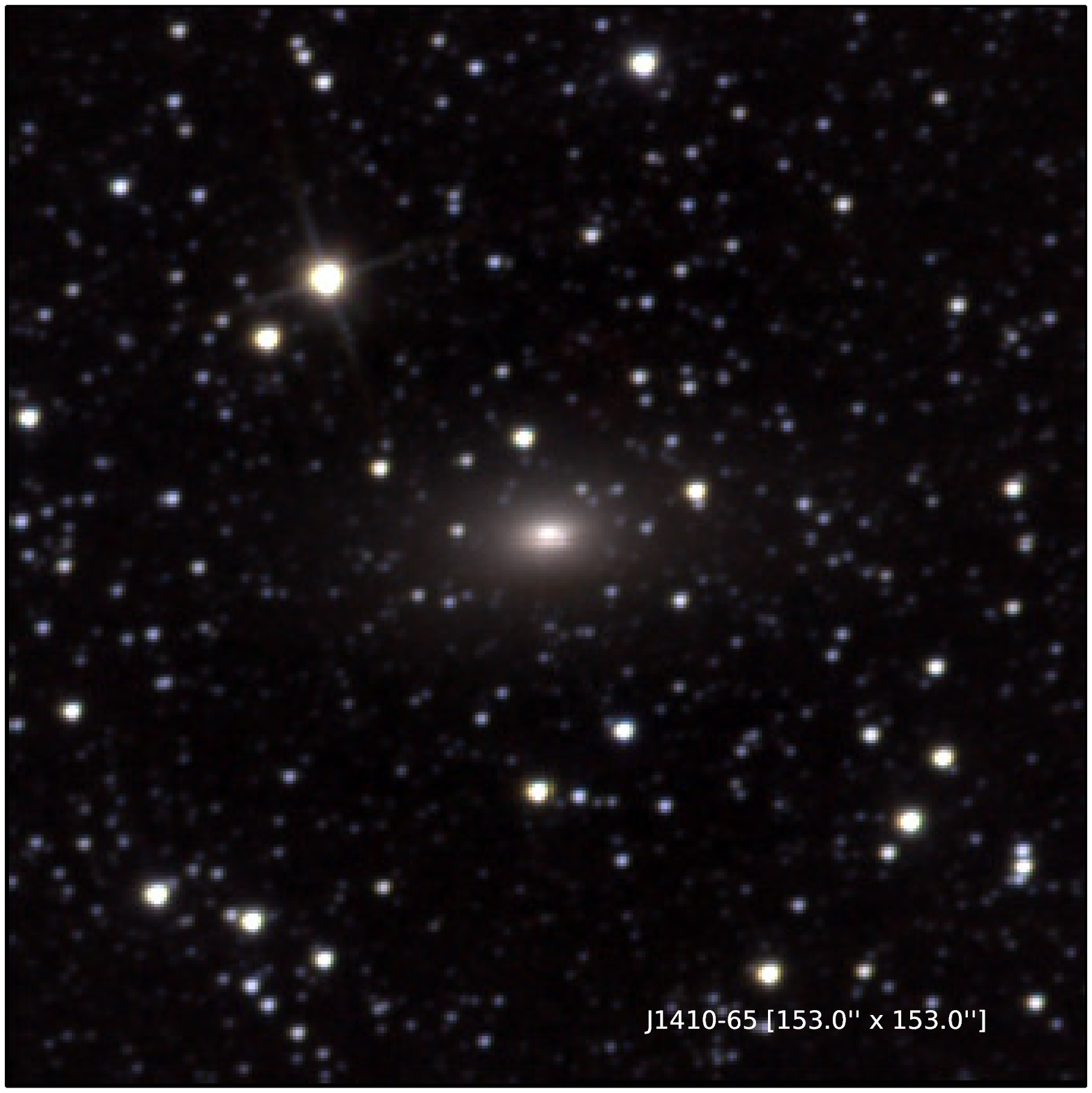}}\\
  \subfloat{\includegraphics[scale=0.17]{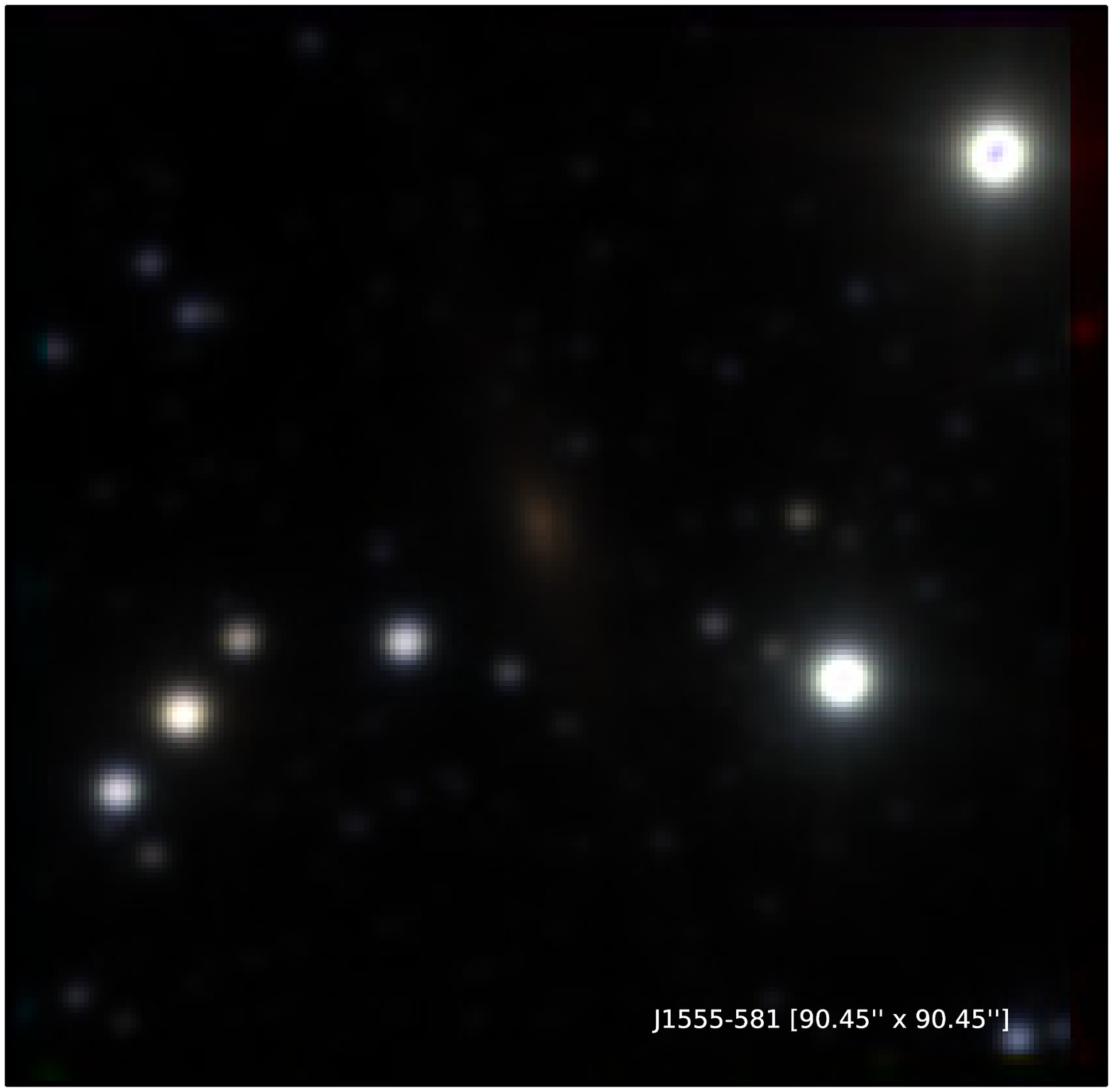}}
 & \subfloat{\includegraphics[scale=0.17]{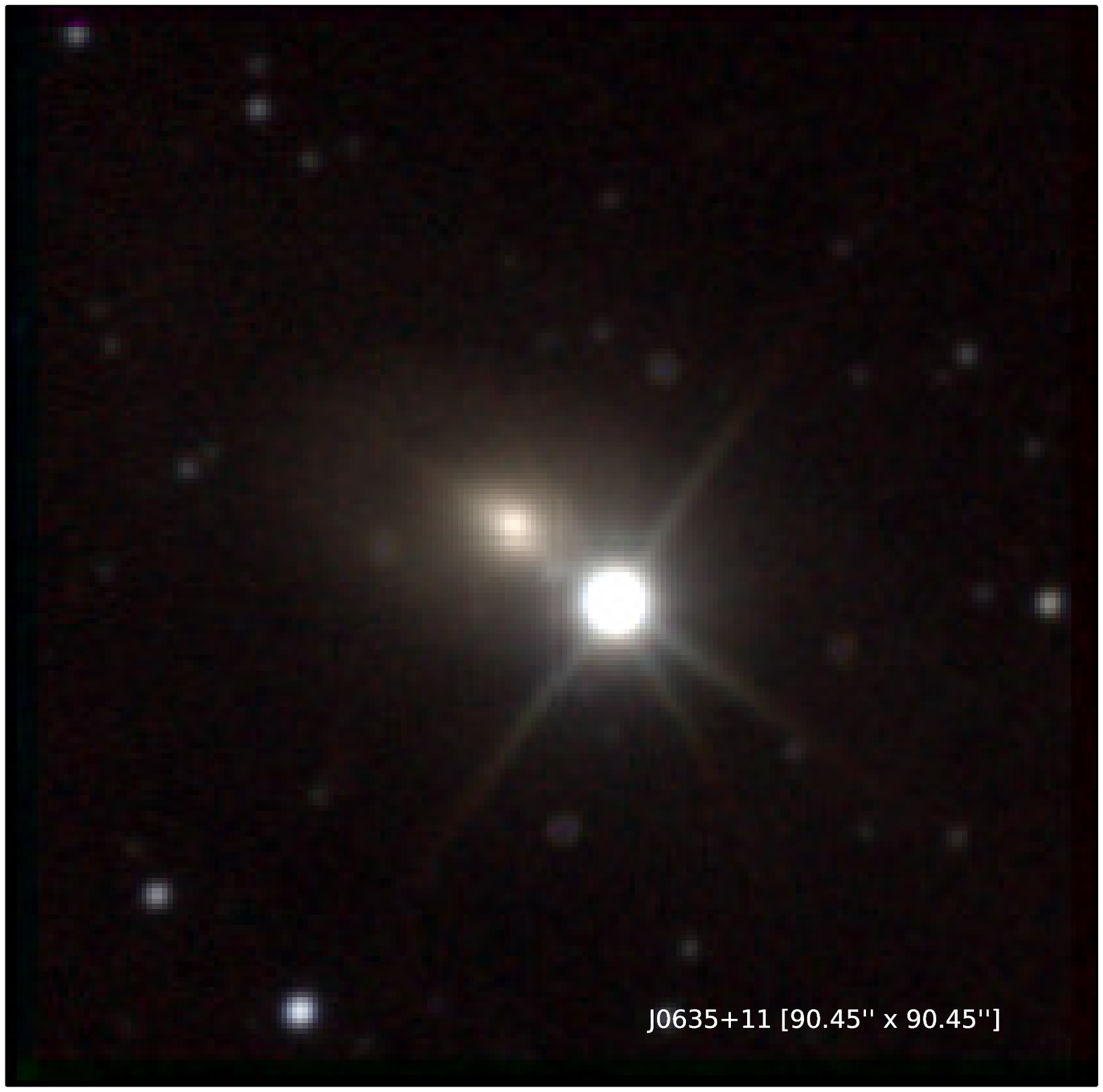}}
 & \subfloat{\includegraphics[scale=0.17]{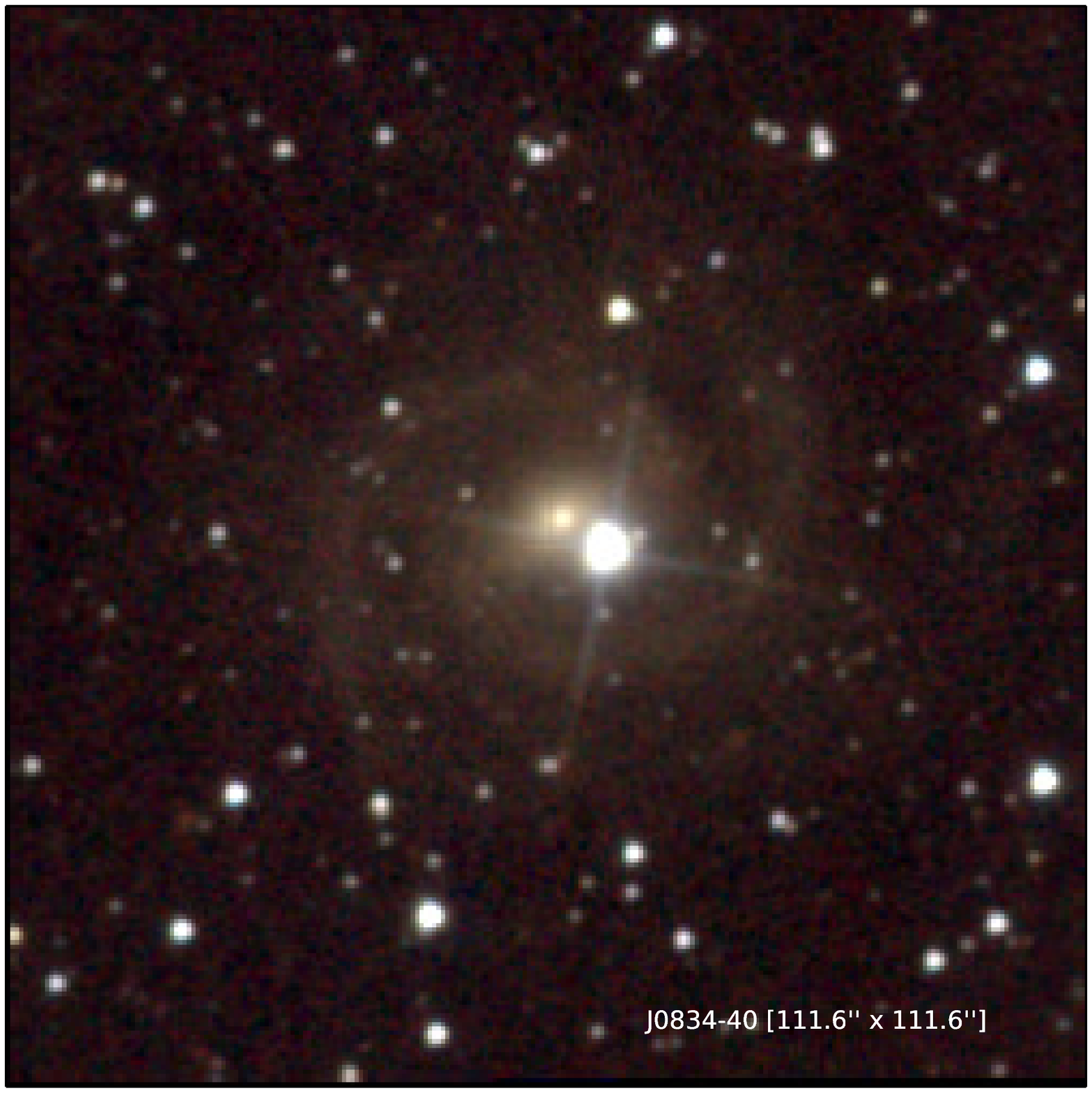}}
 & \subfloat{\includegraphics[scale=0.17]{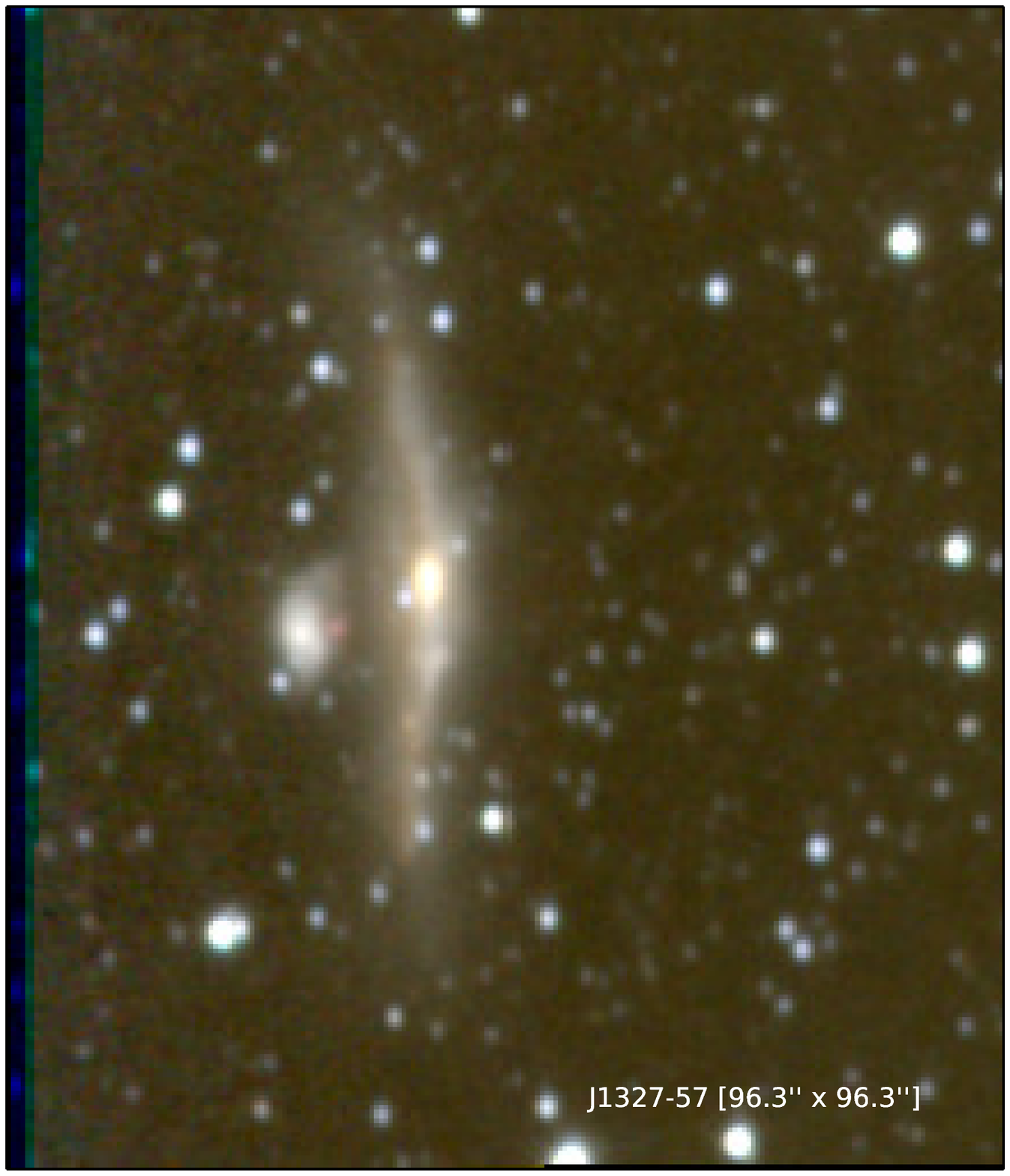}}
 & \subfloat{\includegraphics[scale=0.17]{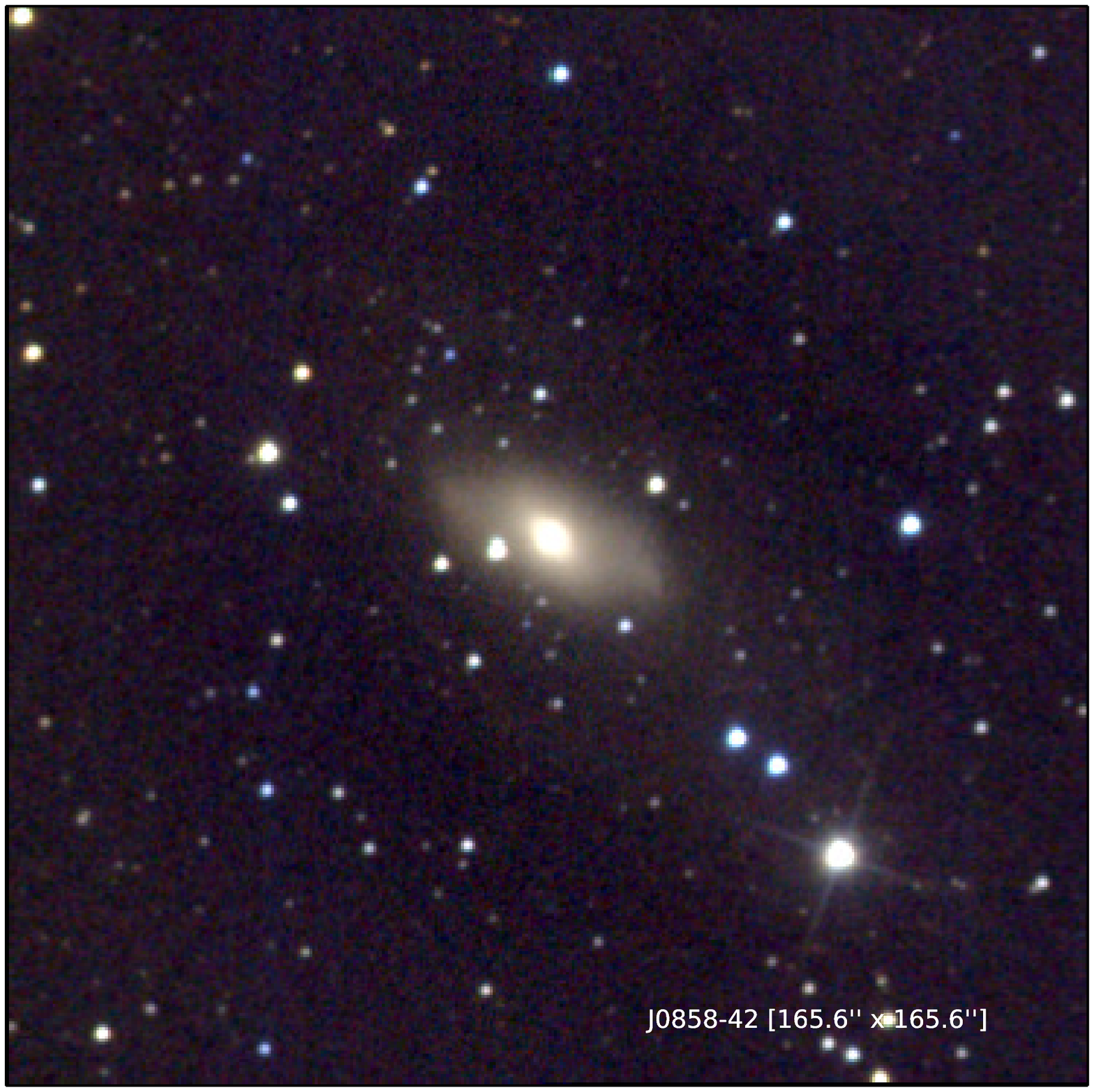}}\\
  \subfloat{\includegraphics[scale=0.17]{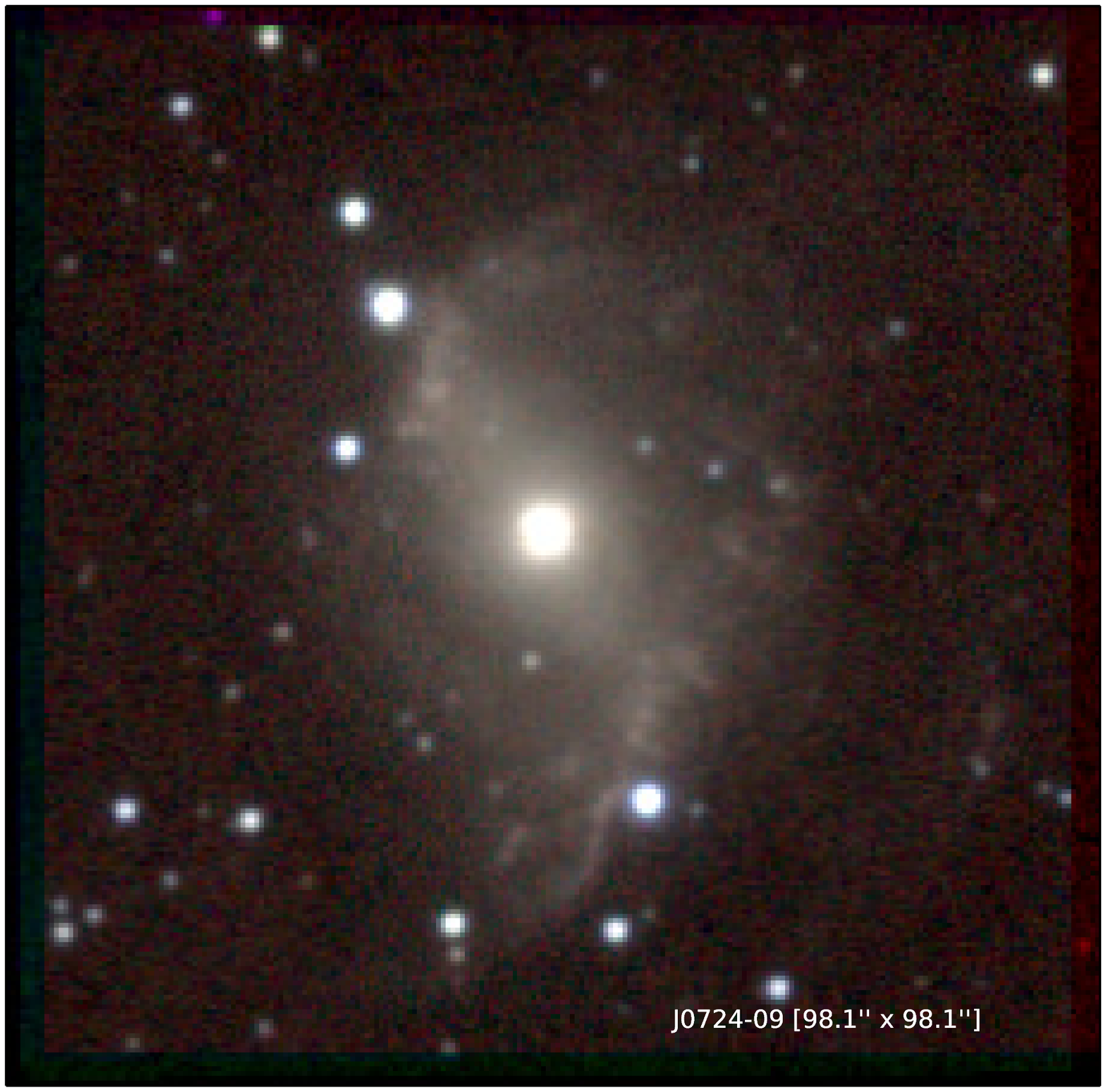}}
 & \subfloat{\includegraphics[scale=0.17]{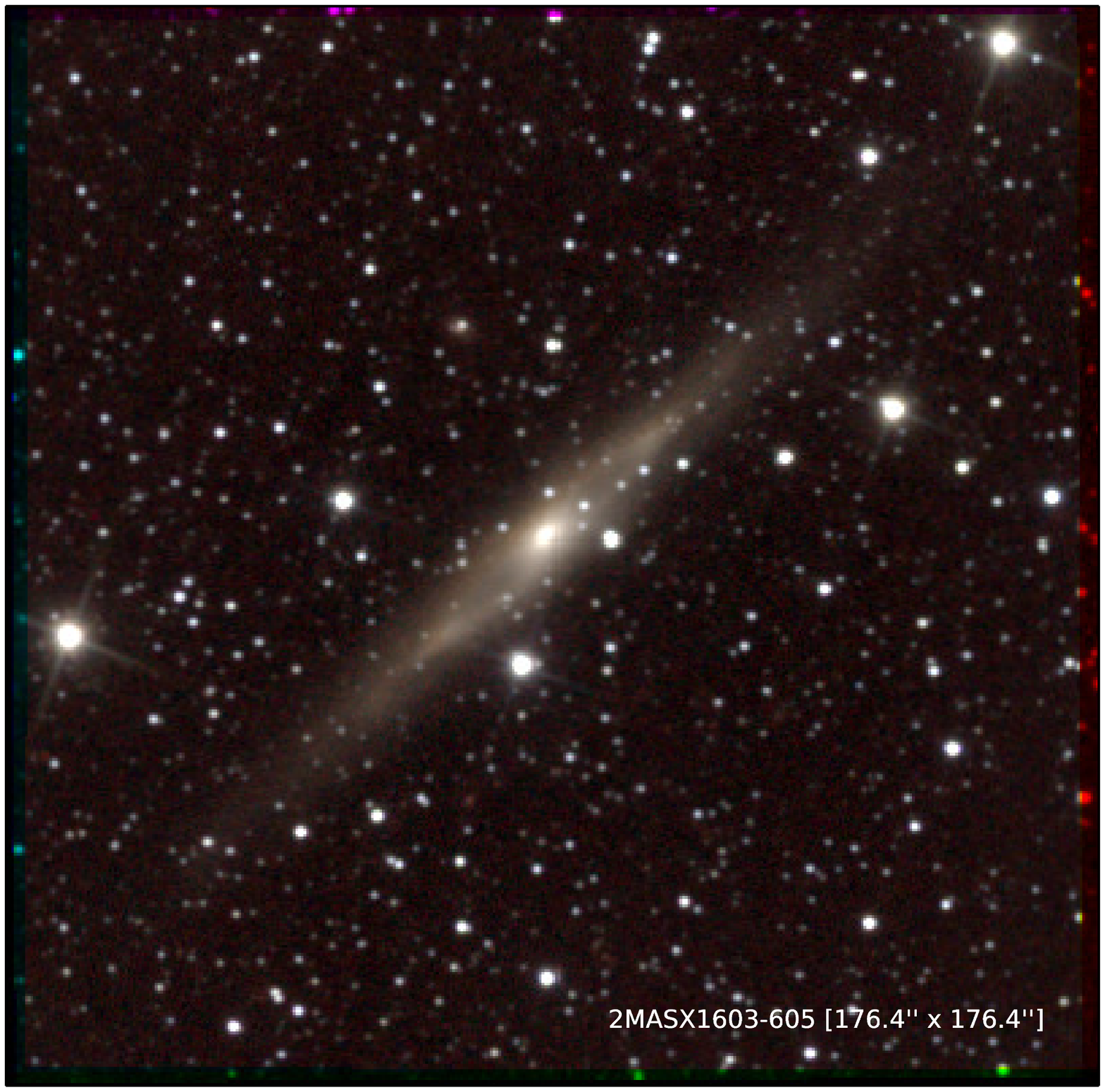}}
 & \subfloat{\includegraphics[scale=0.17]{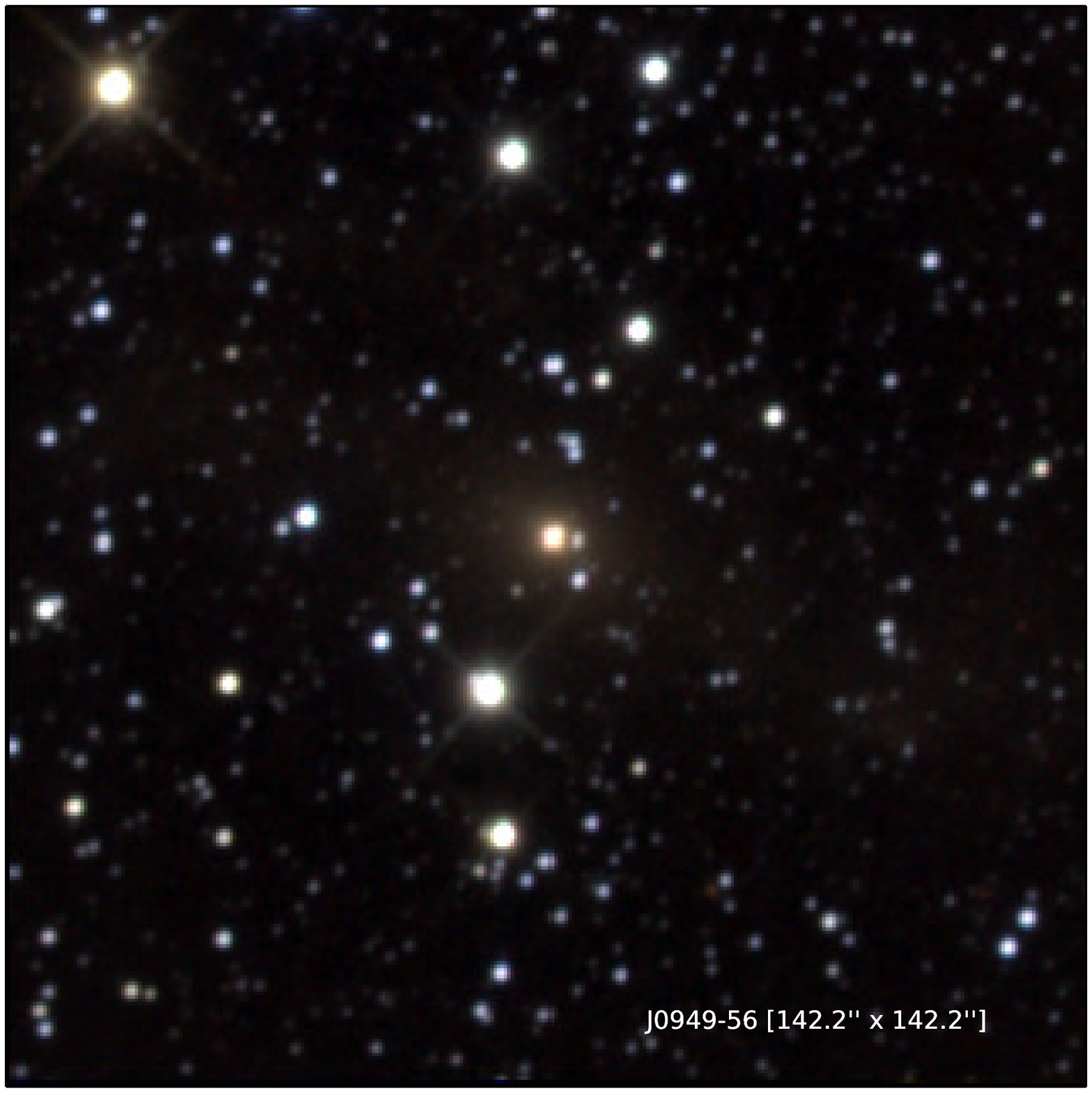}}
 & \subfloat{\includegraphics[scale=0.17]{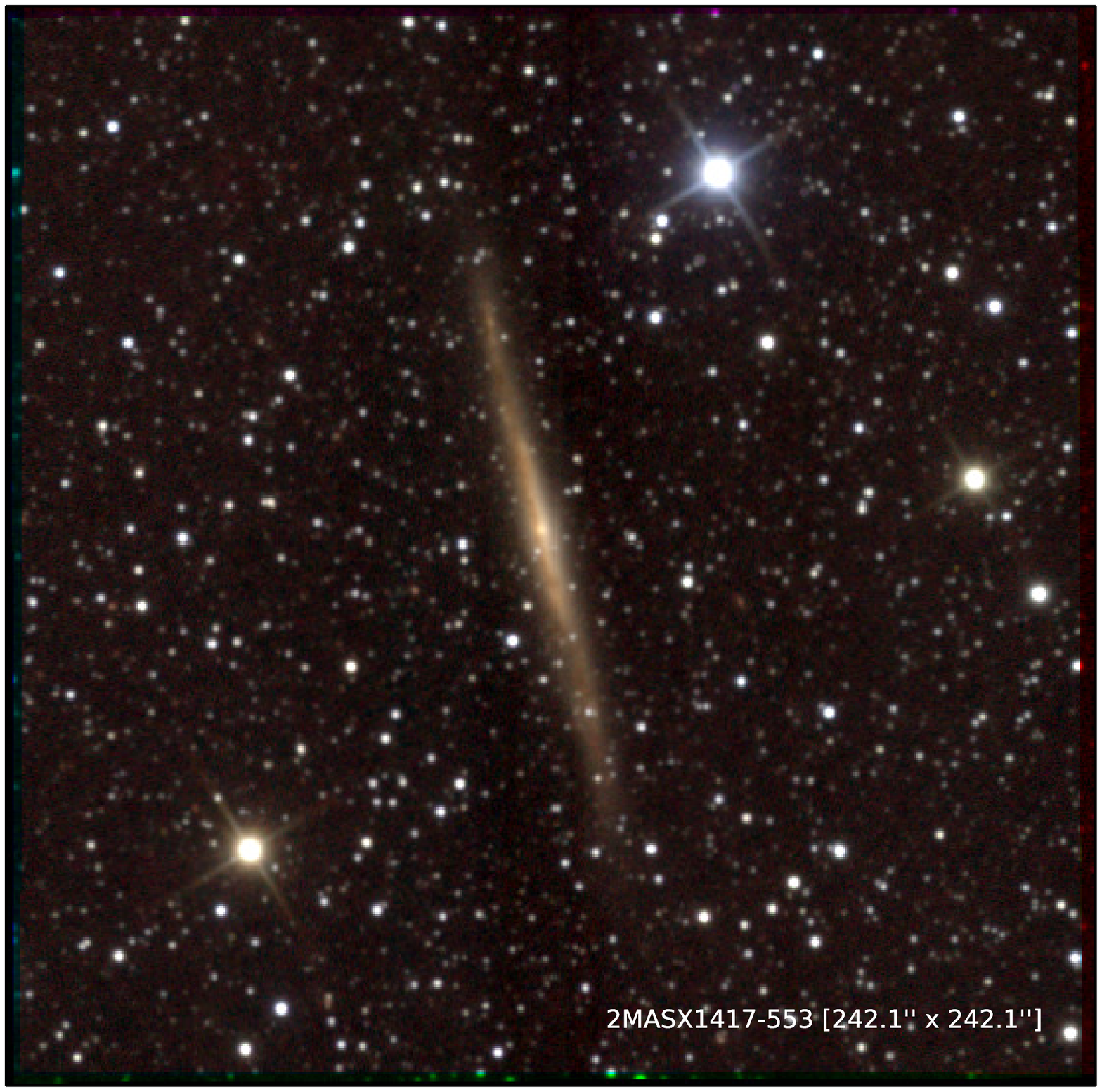}}
 & \subfloat{\includegraphics[scale=0.17]{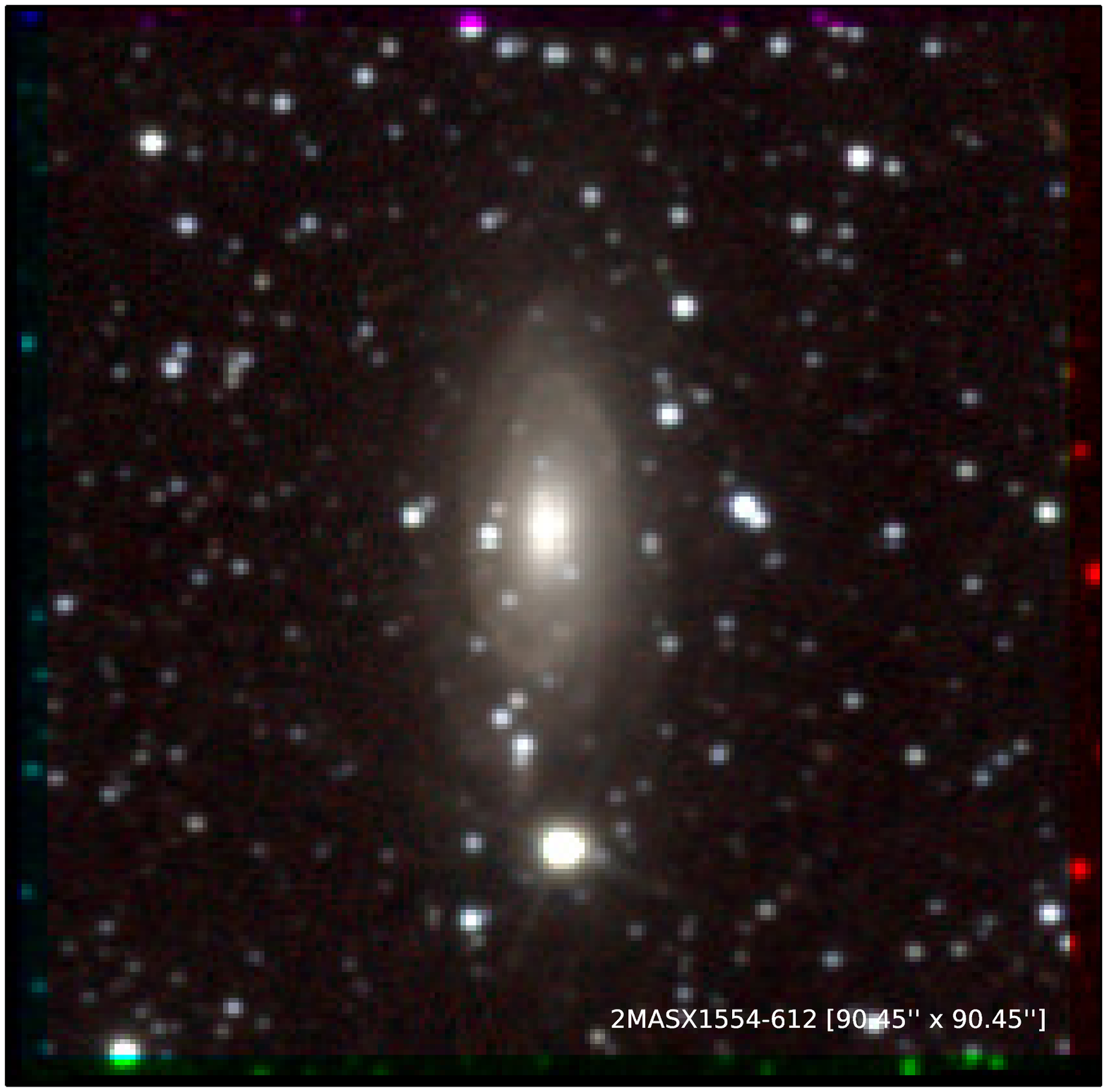}}\\
\\
\\

 \end{tabular}
 \caption{Postage stamp images of the brightest 100 galaxies in the catalogue in order of isophotal $K_{s20}$ fiducial elliptical aperture magnitude. The color composites are generated in the standard fashion: blue -- $J$ band, green -- $H$ band, and red -- $K_s$ band.}
 \label{stamps}
\end{figure*}

\begin{figure*}
\ContinuedFloat
\begin{tabular}{ccccc}
  \subfloat{\includegraphics[scale=0.17]{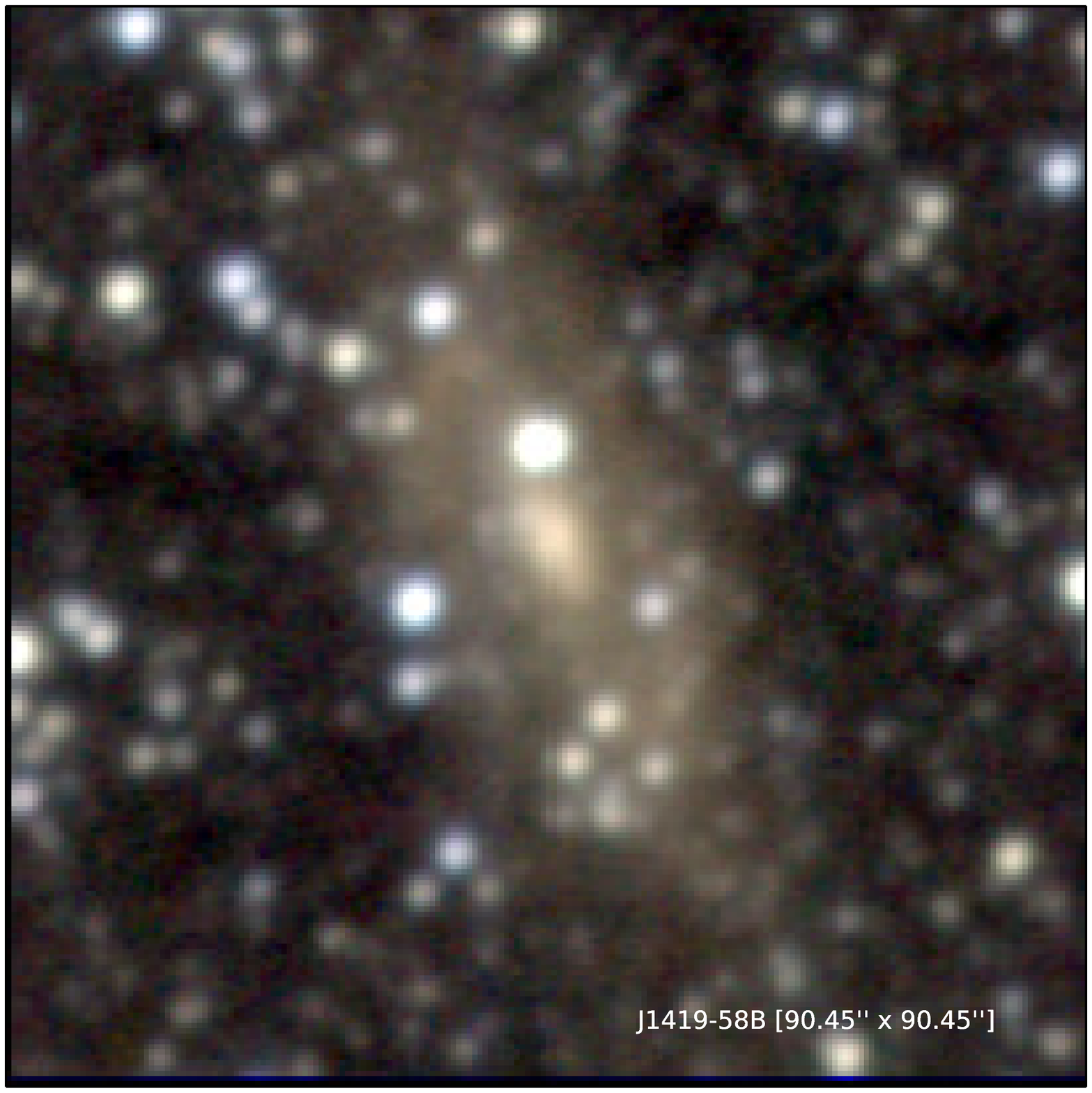}}
 & \subfloat{\includegraphics[scale=0.17]{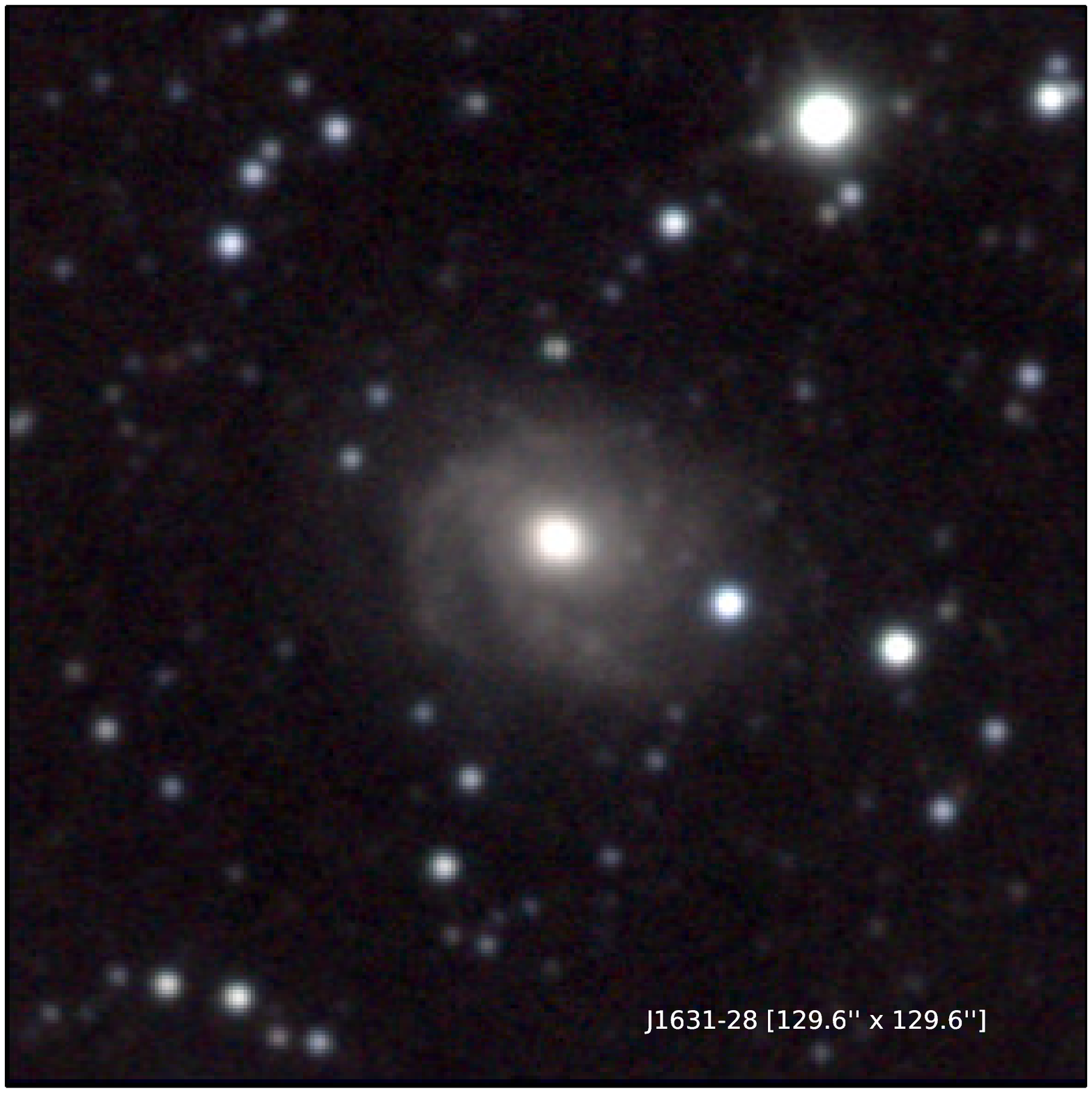}}
 & \subfloat{\includegraphics[scale=0.17]{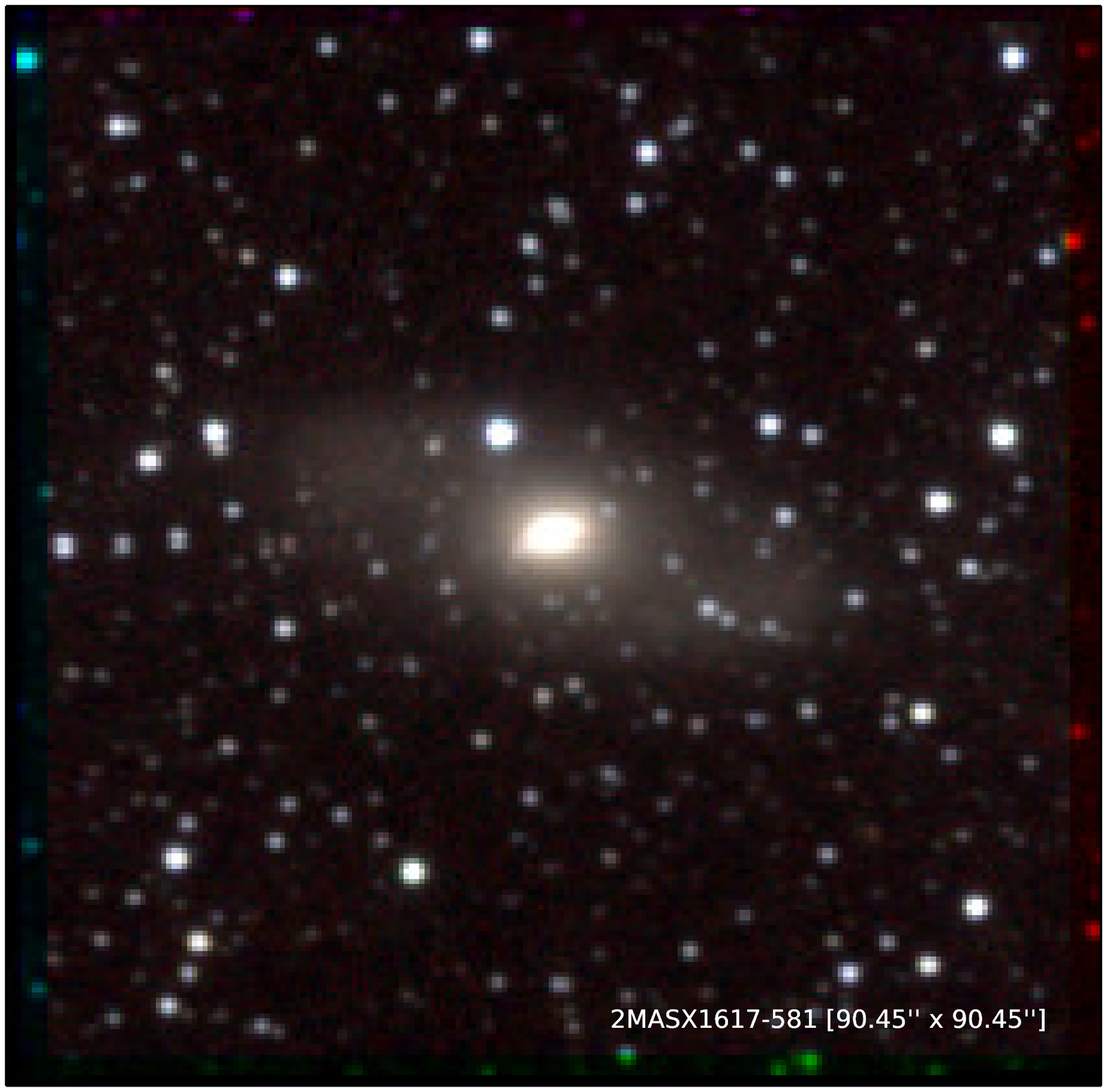}}
 & \subfloat{\includegraphics[scale=0.17]{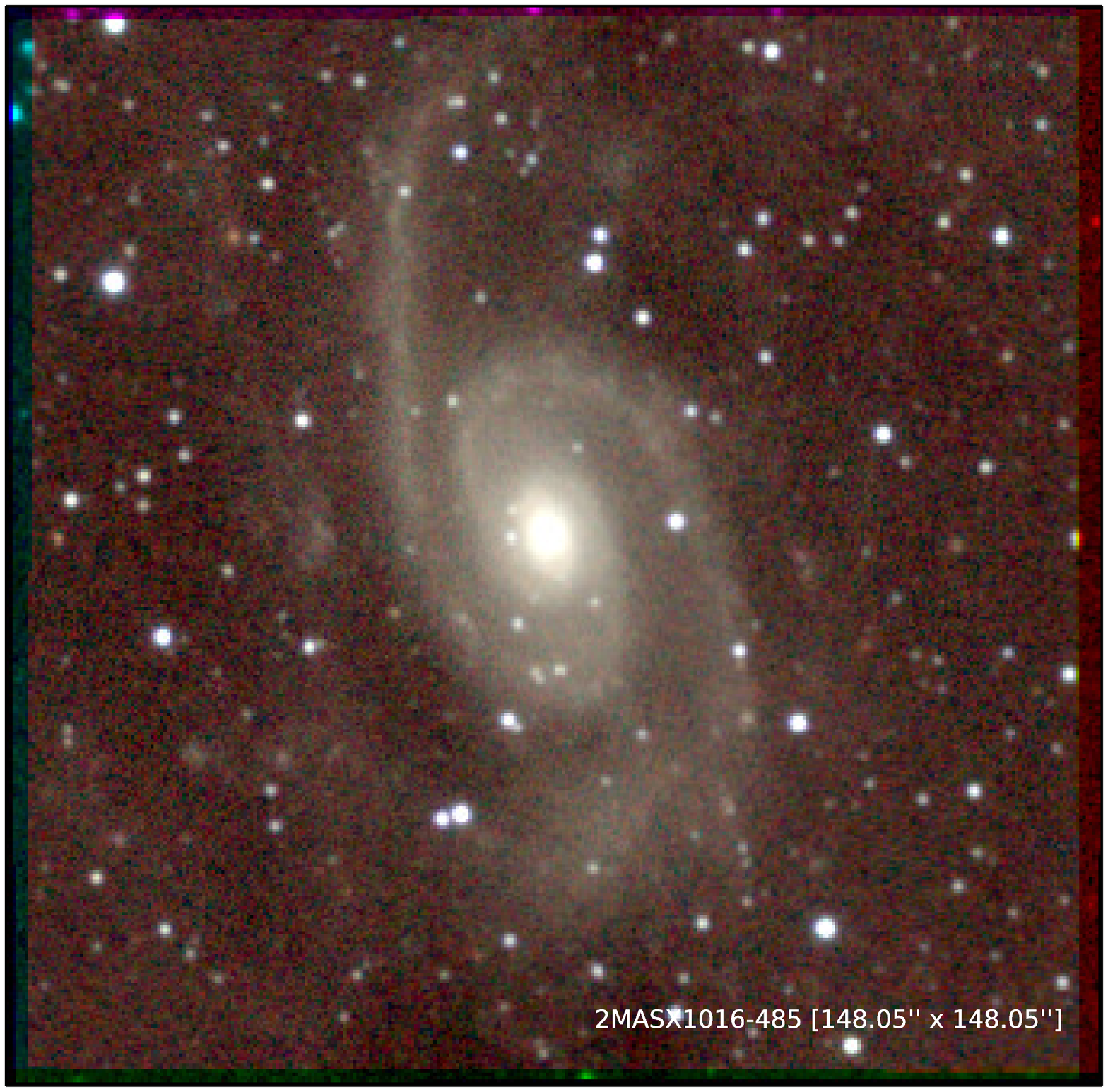}}
 & \subfloat{\includegraphics[scale=0.17]{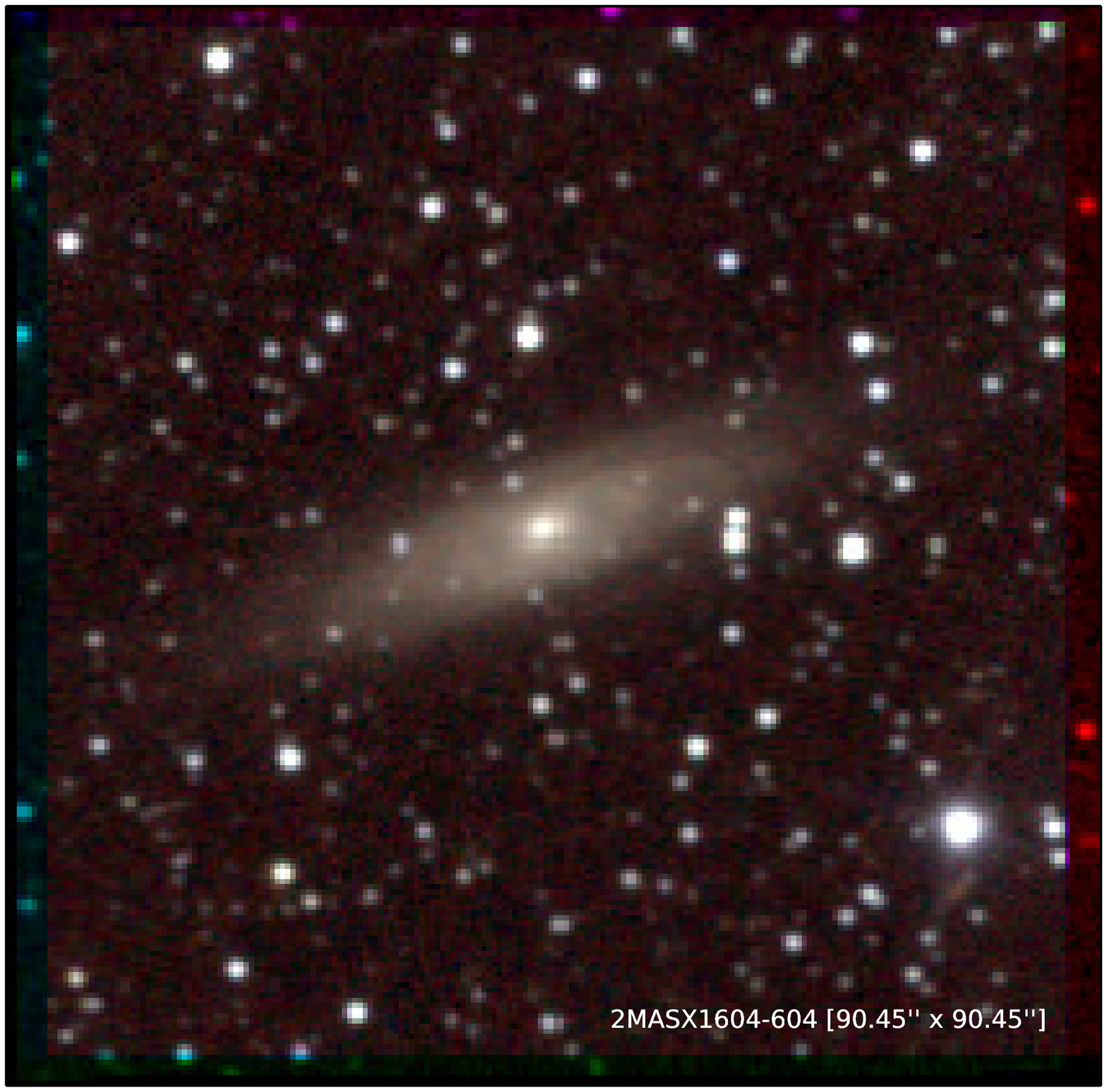}}\\
  \subfloat{\includegraphics[scale=0.17]{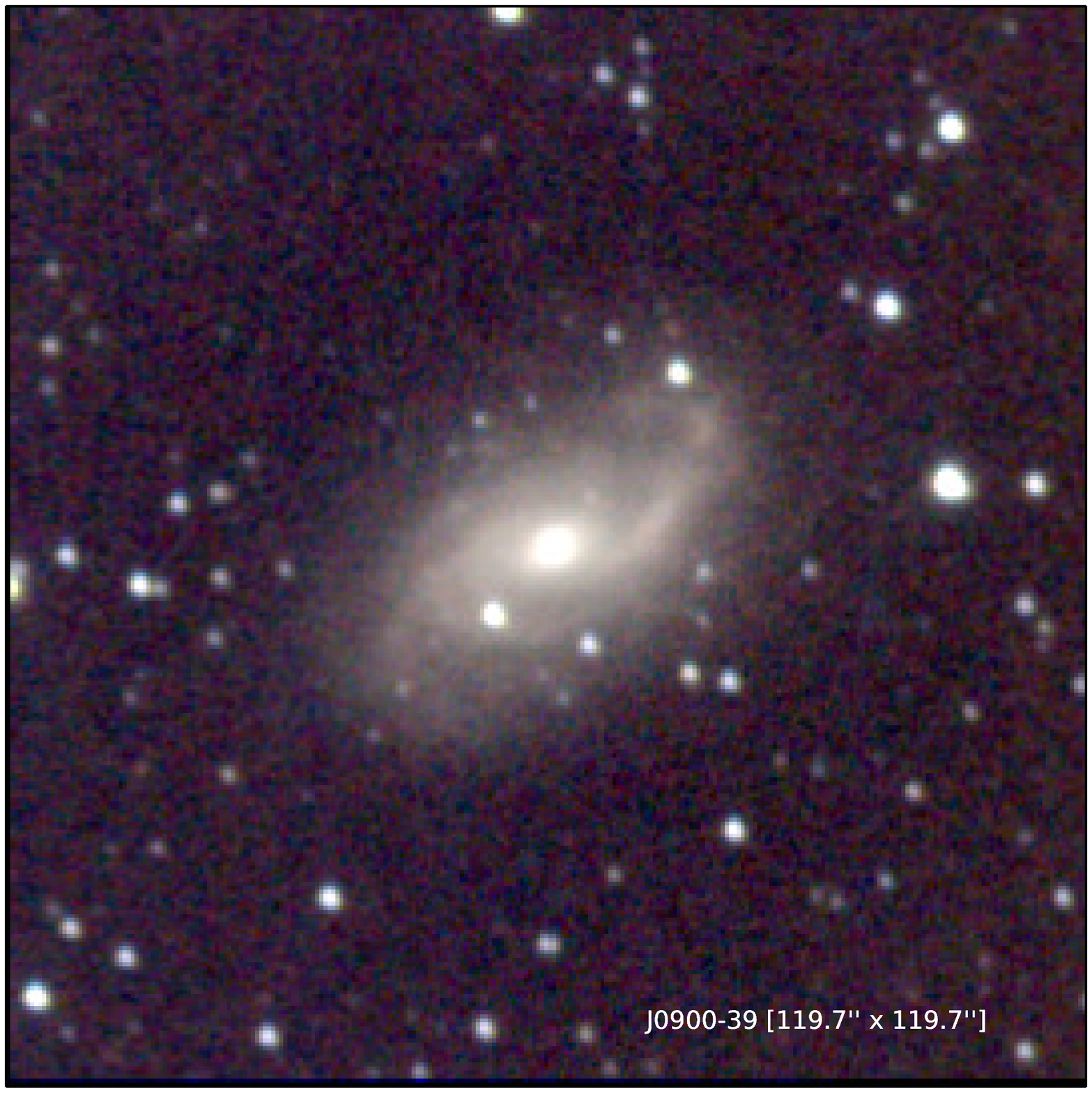}}
 & \subfloat{\includegraphics[scale=0.17]{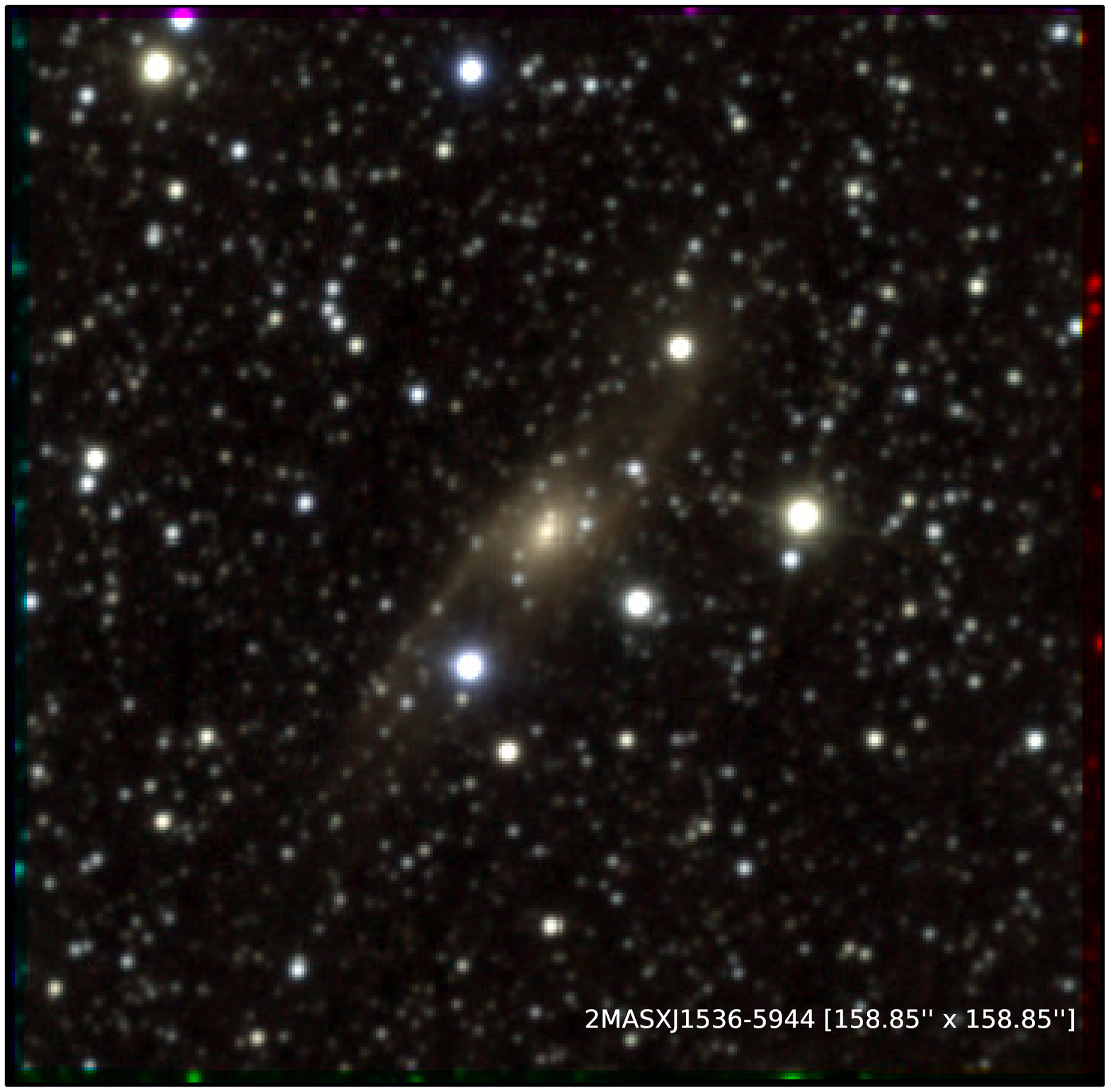}}
 & \subfloat{\includegraphics[scale=0.17]{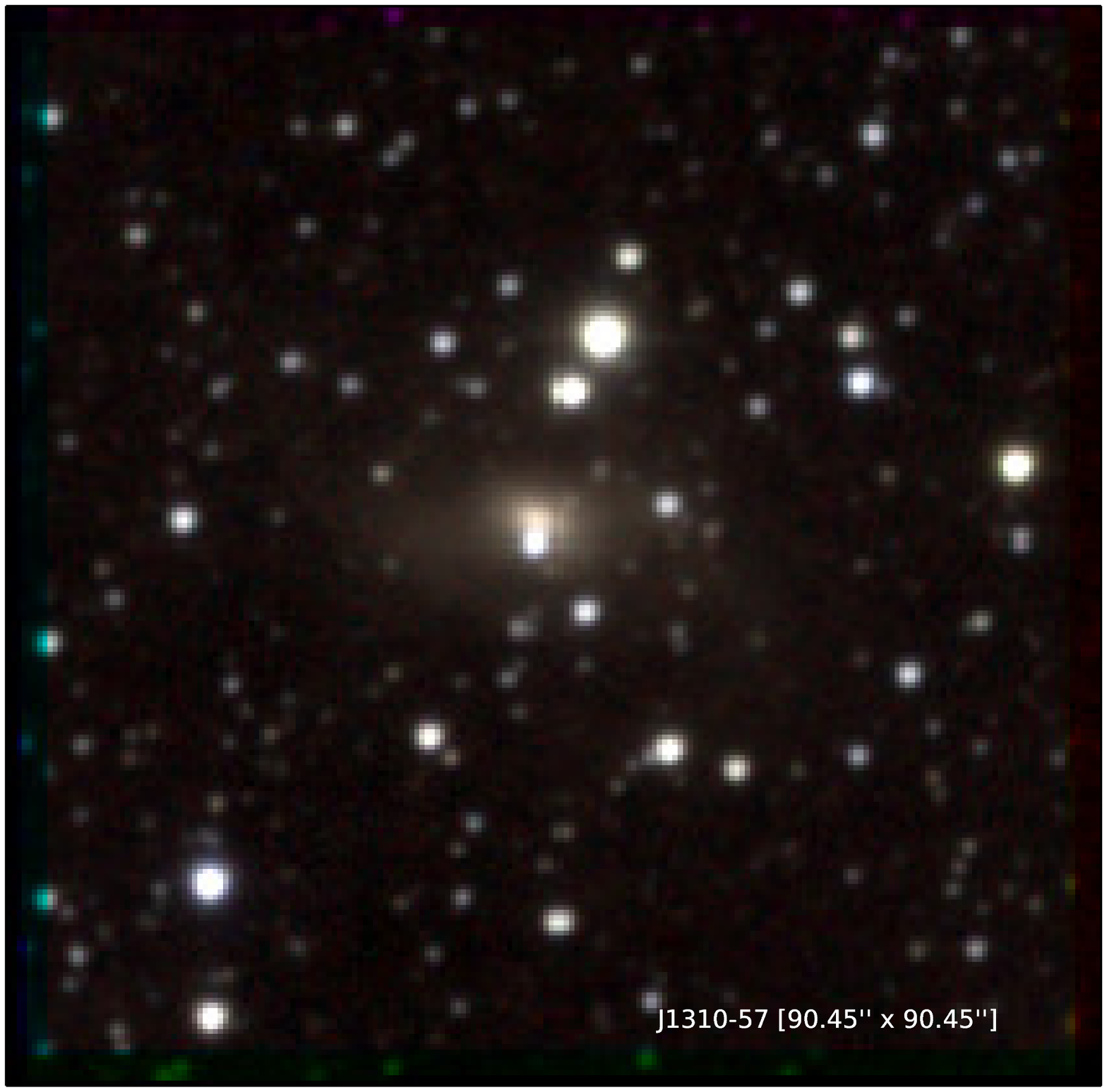}}
 & \subfloat{\includegraphics[scale=0.17]{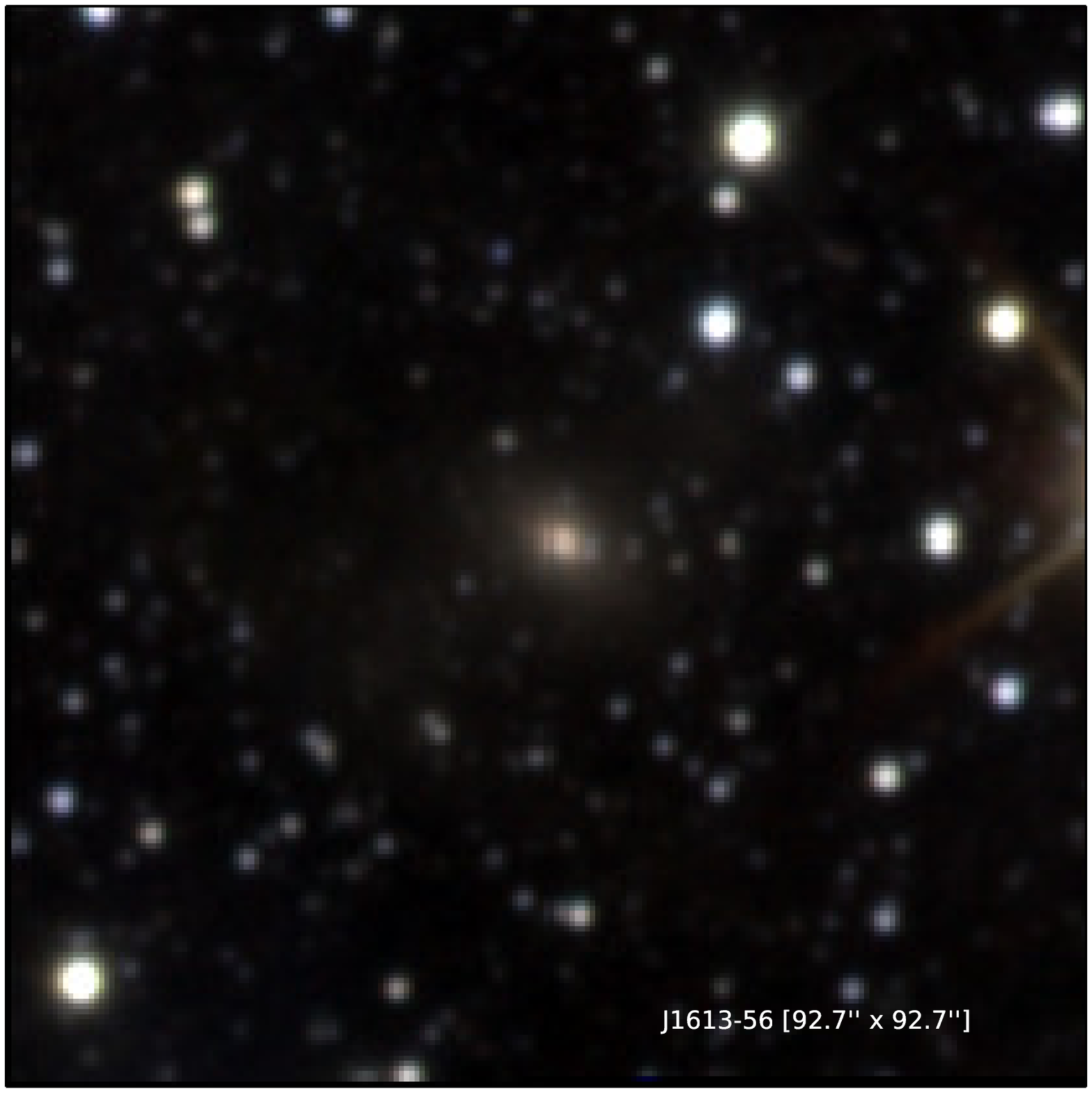}}
 & \subfloat{\includegraphics[scale=0.17]{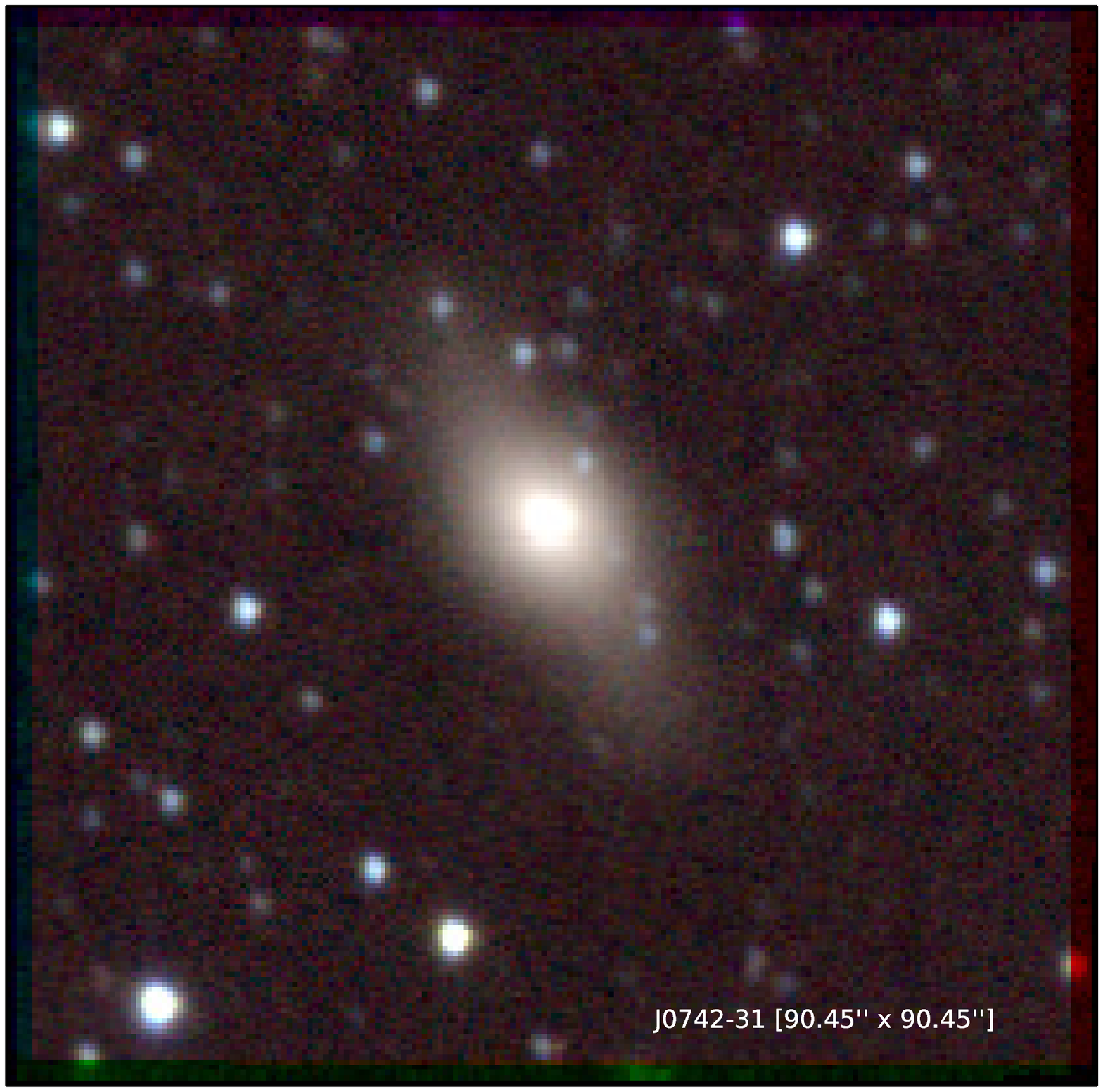}}\\
  \subfloat{\includegraphics[scale=0.17]{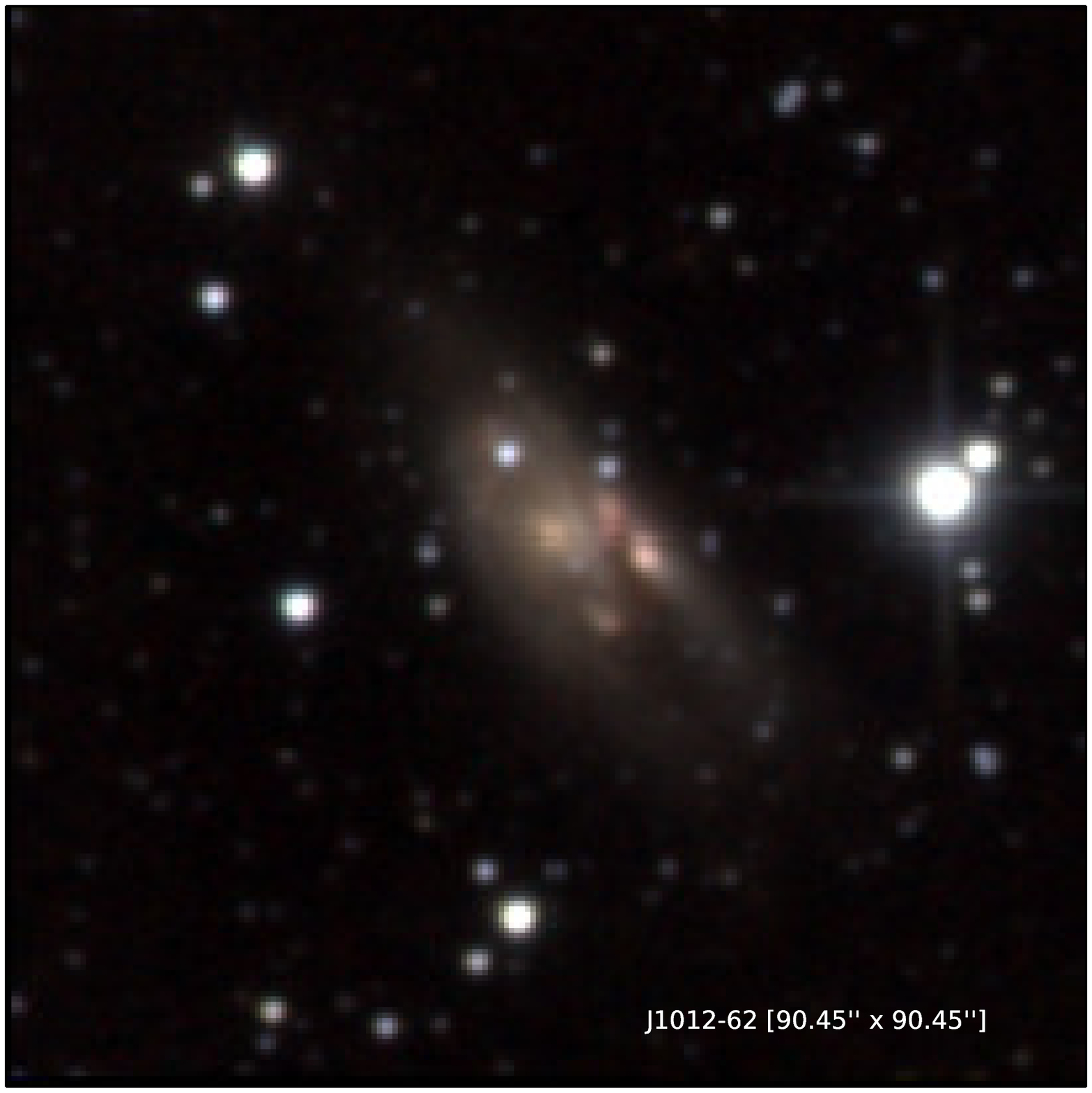}}
 & \subfloat{\includegraphics[scale=0.17]{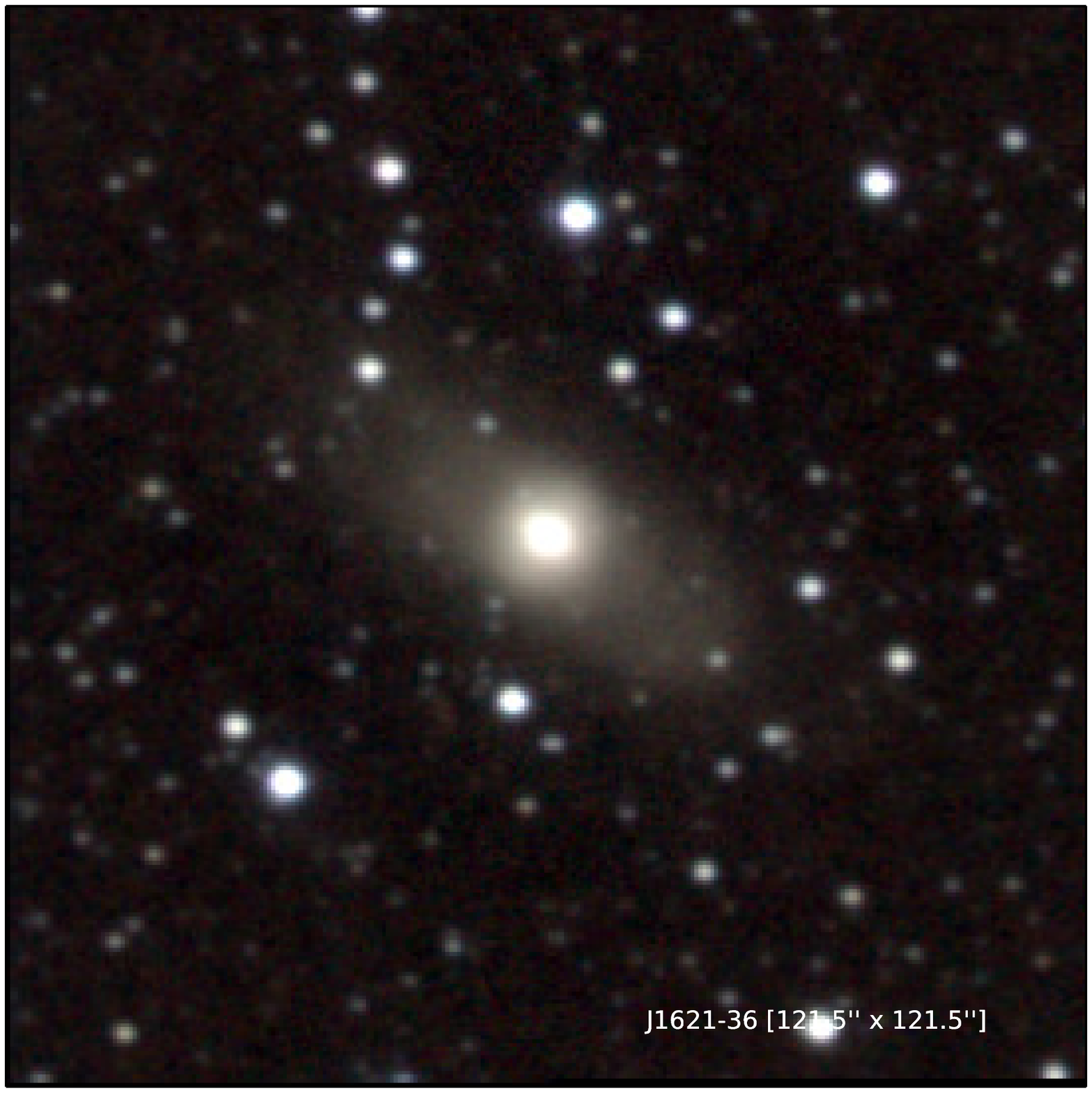}}
 & \subfloat{\includegraphics[scale=0.17]{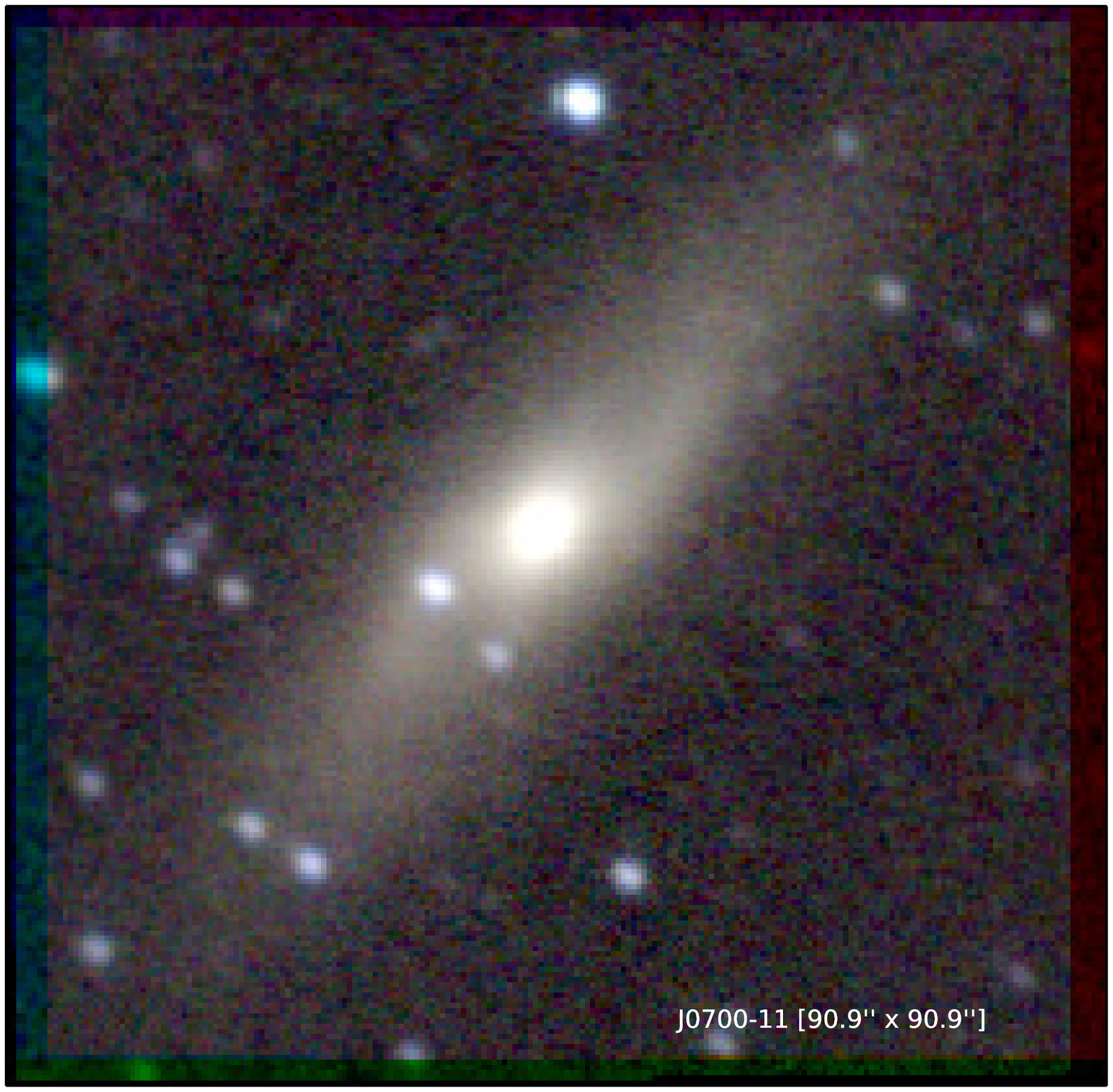}}
 & \subfloat{\includegraphics[scale=0.17]{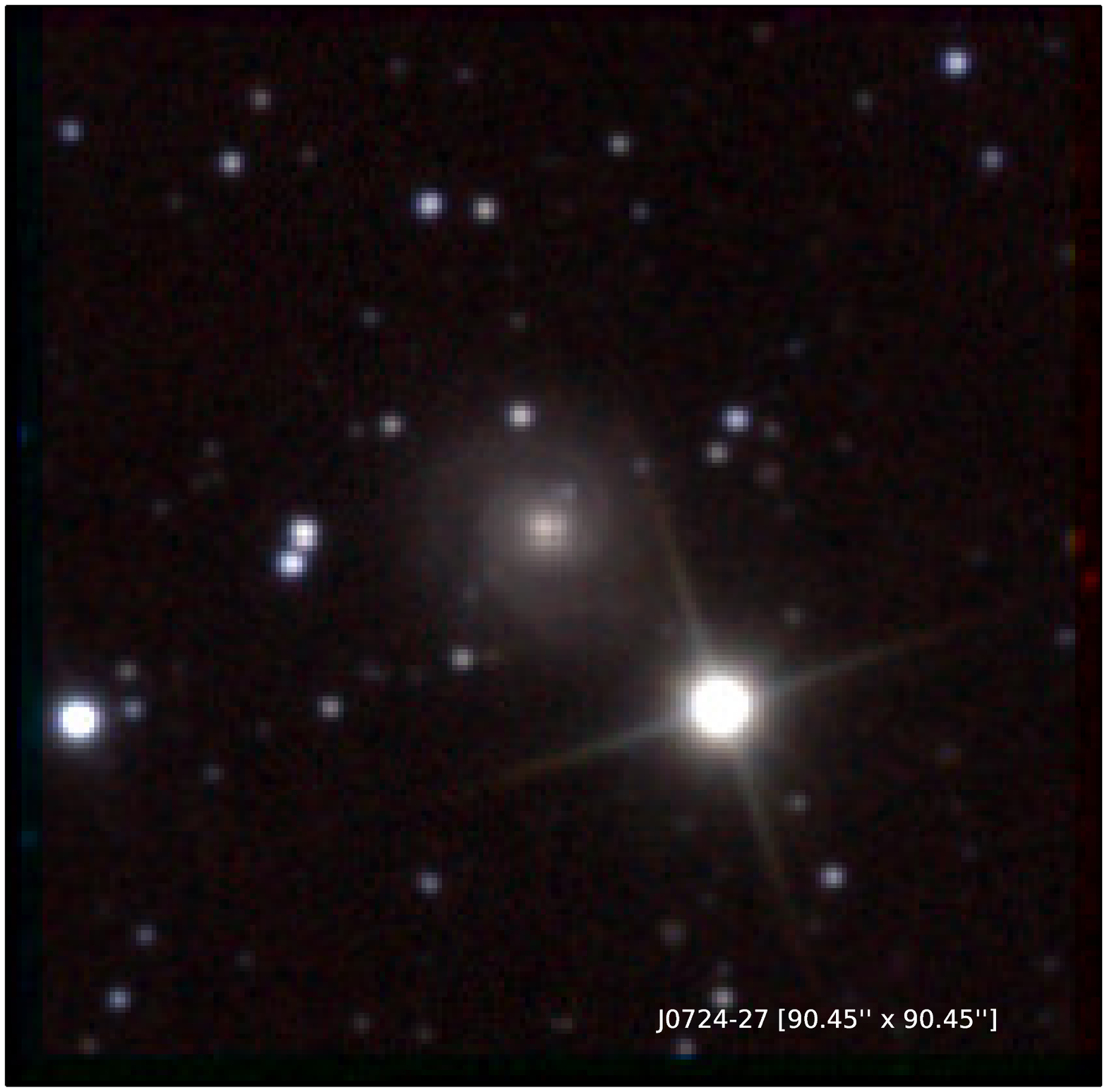}}
 & \subfloat{\includegraphics[scale=0.17]{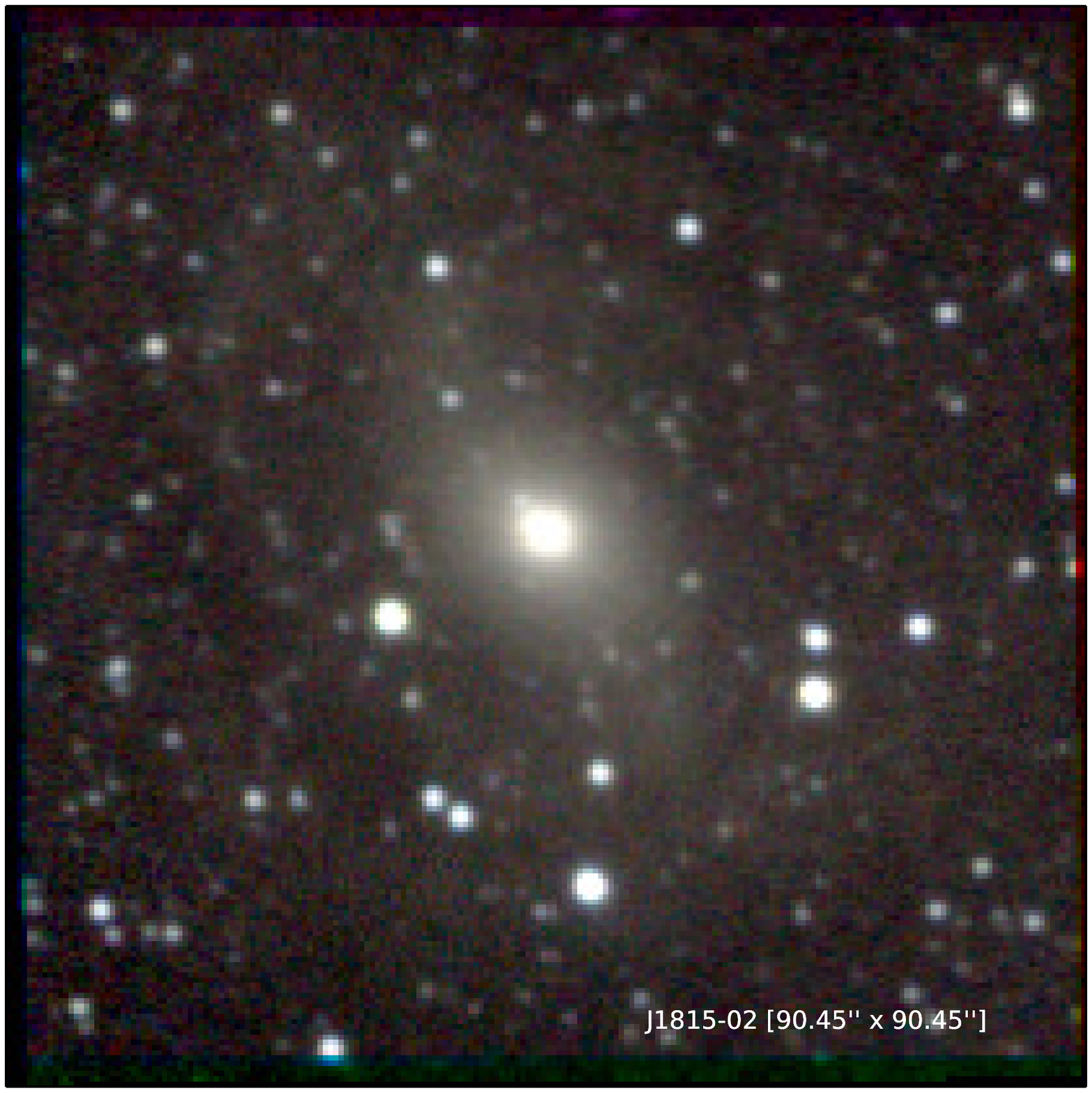}}\\
  \subfloat{\includegraphics[scale=0.17]{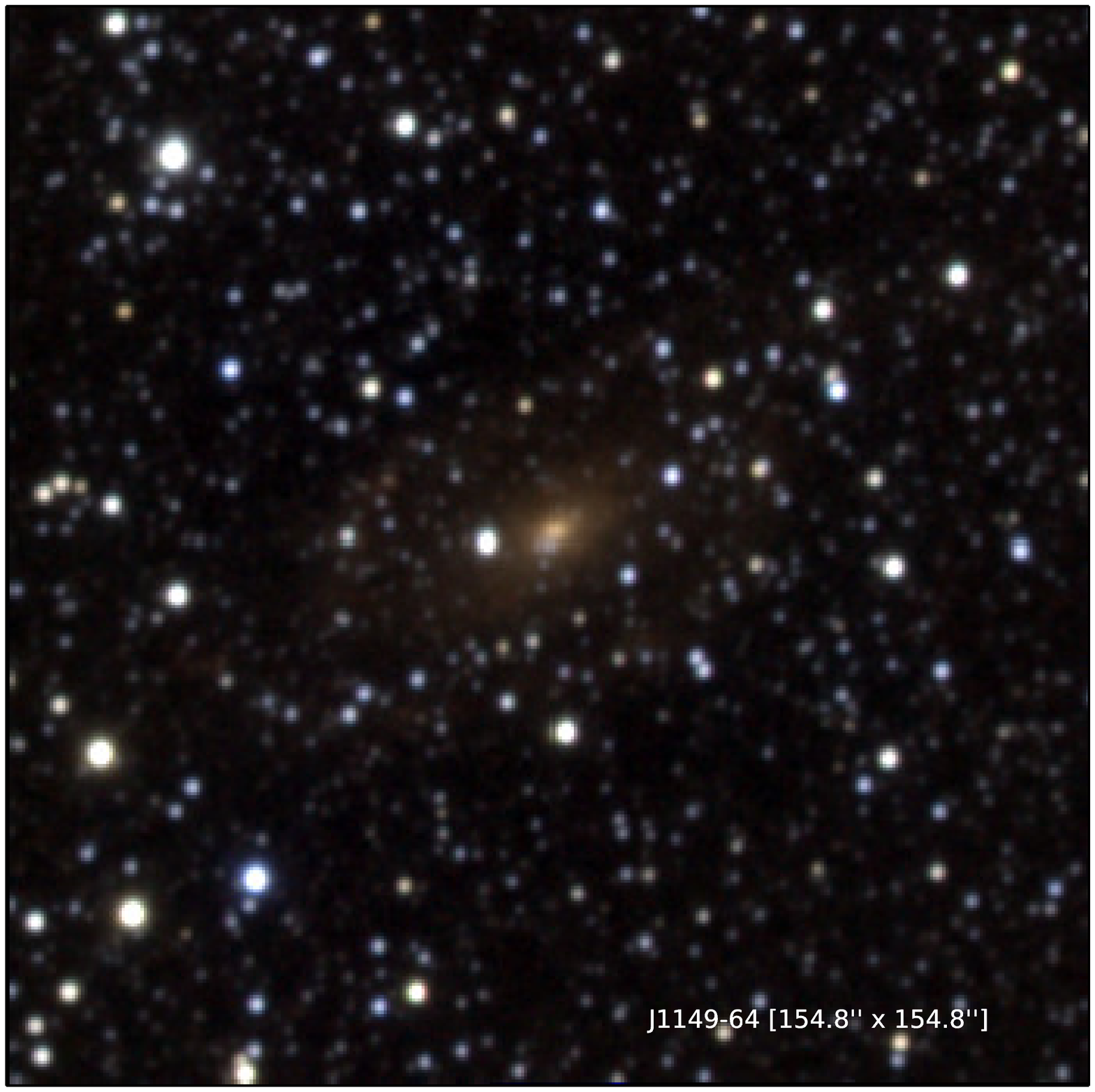}}
 & \subfloat{\includegraphics[scale=0.17]{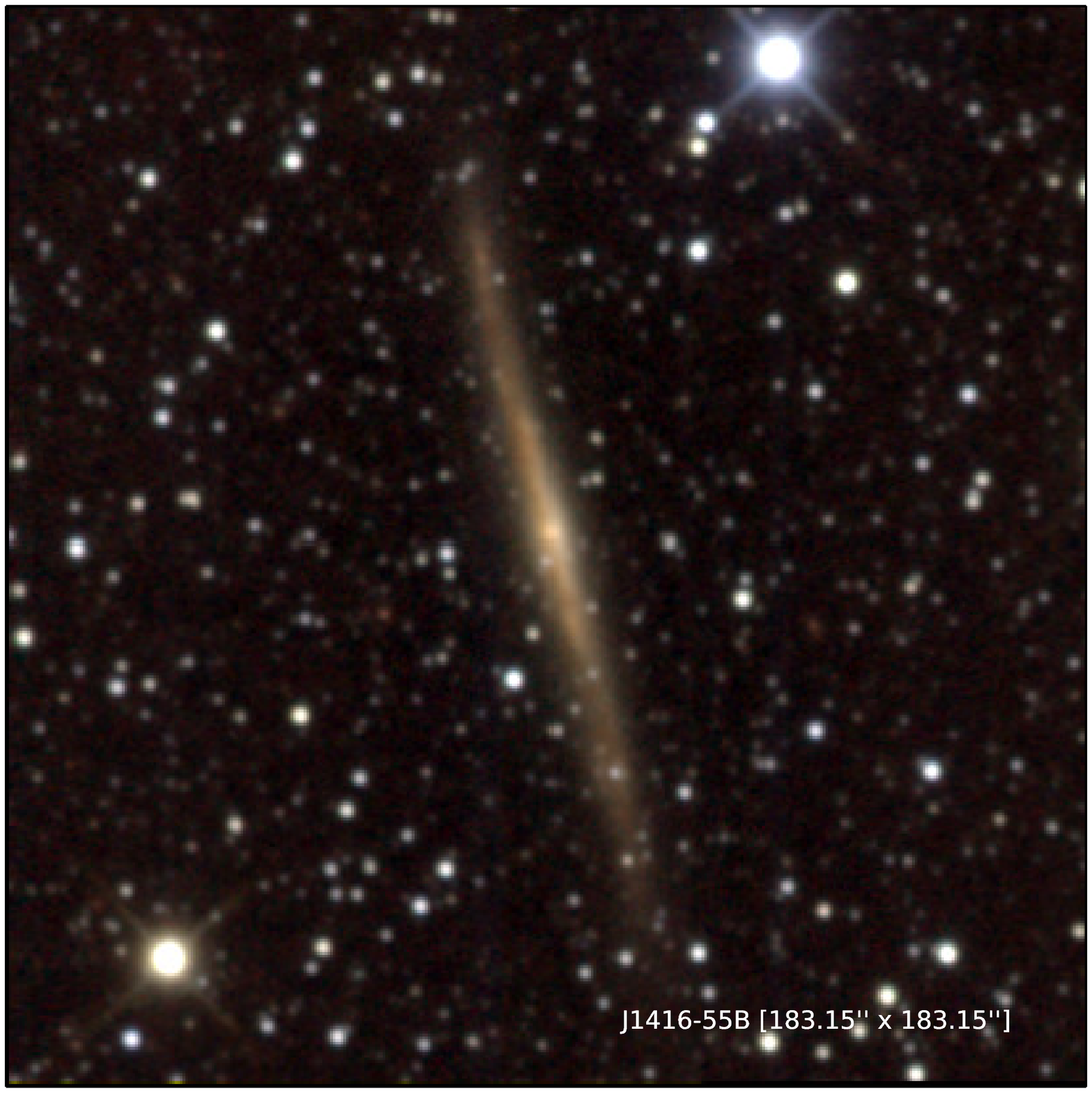}}
 & \subfloat{\includegraphics[scale=0.17]{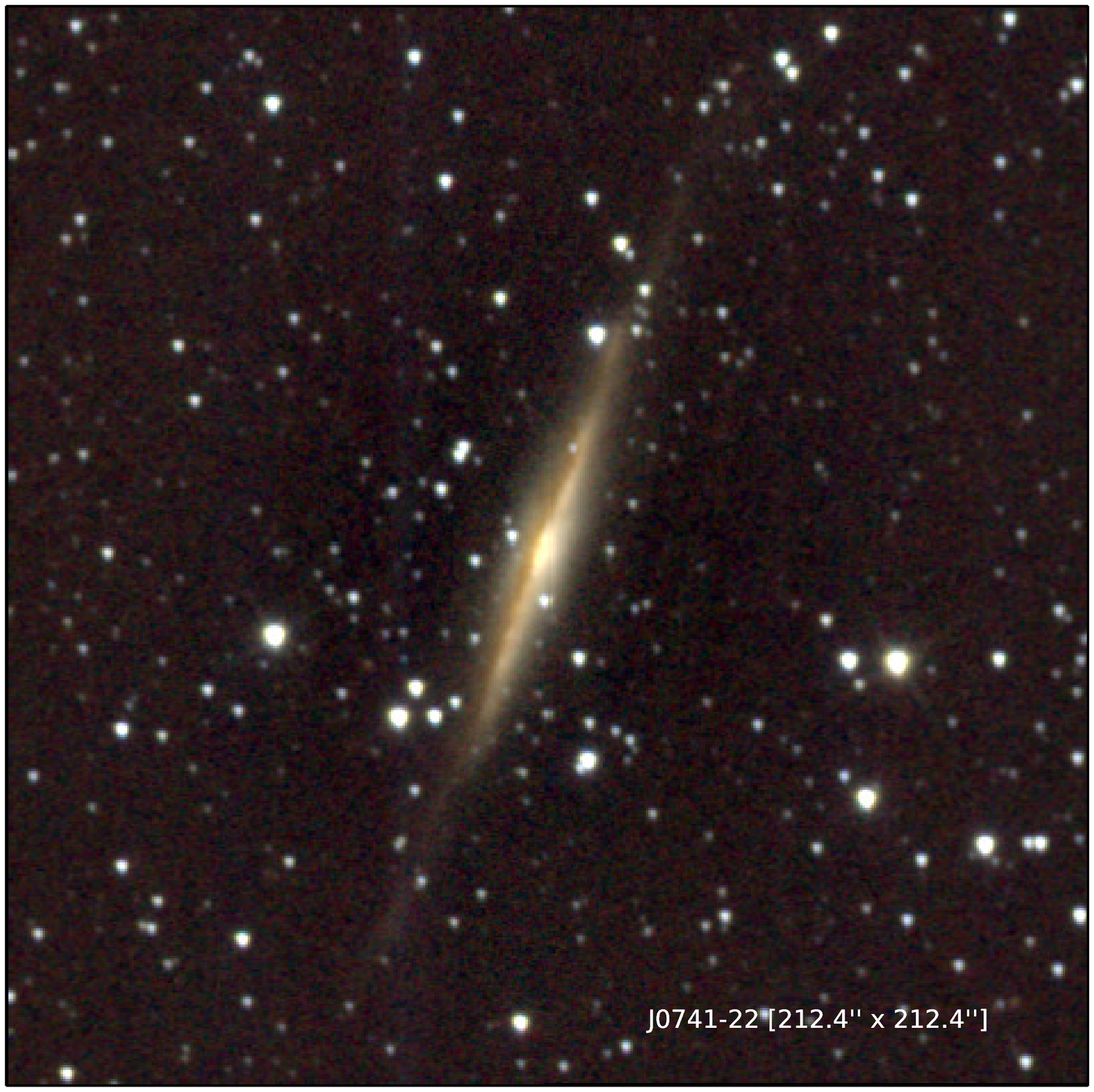}}
 & \subfloat{\includegraphics[scale=0.17]{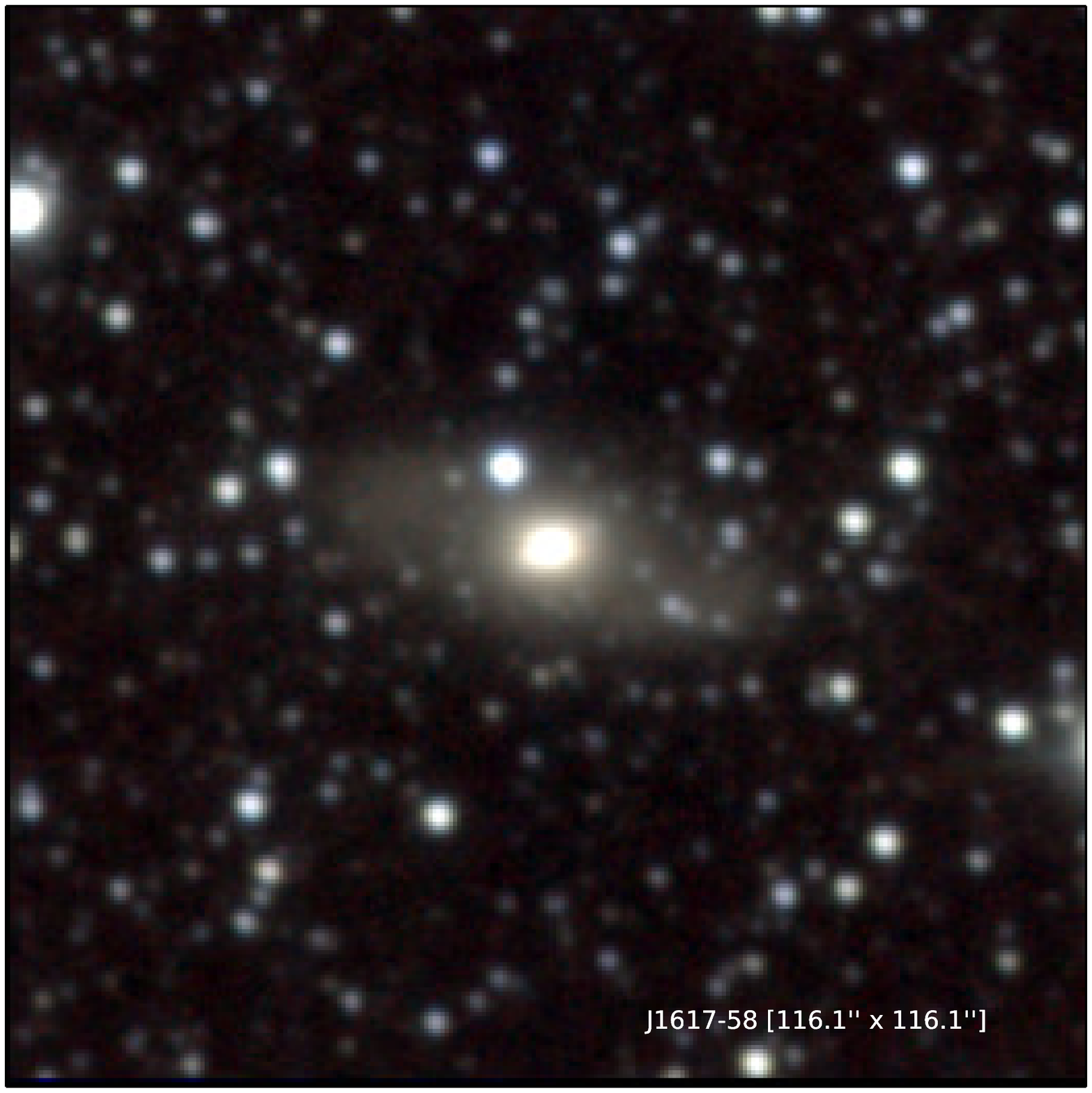}}
 & \subfloat{\includegraphics[scale=0.17]{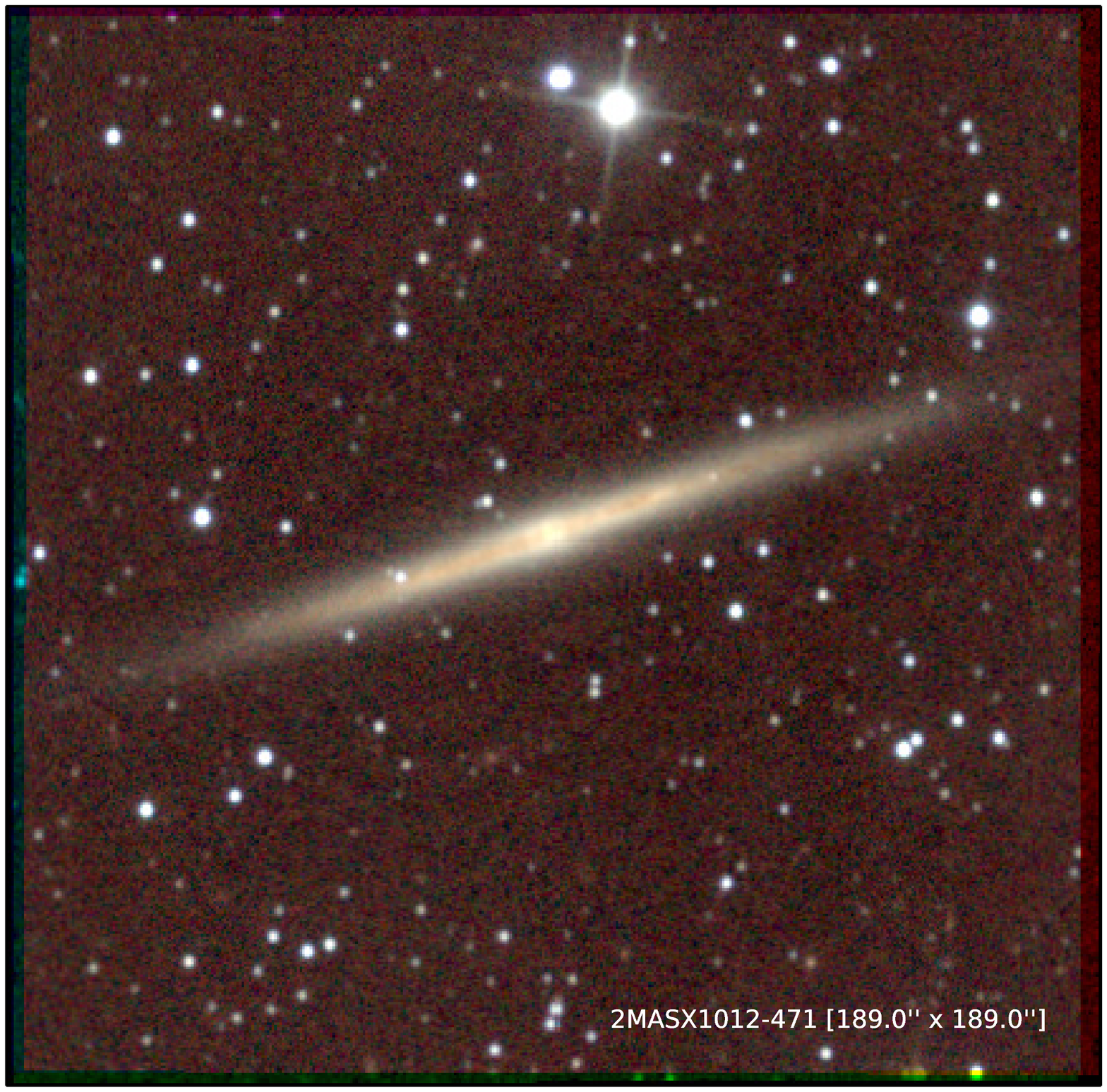}}\\
  \subfloat{\includegraphics[scale=0.17]{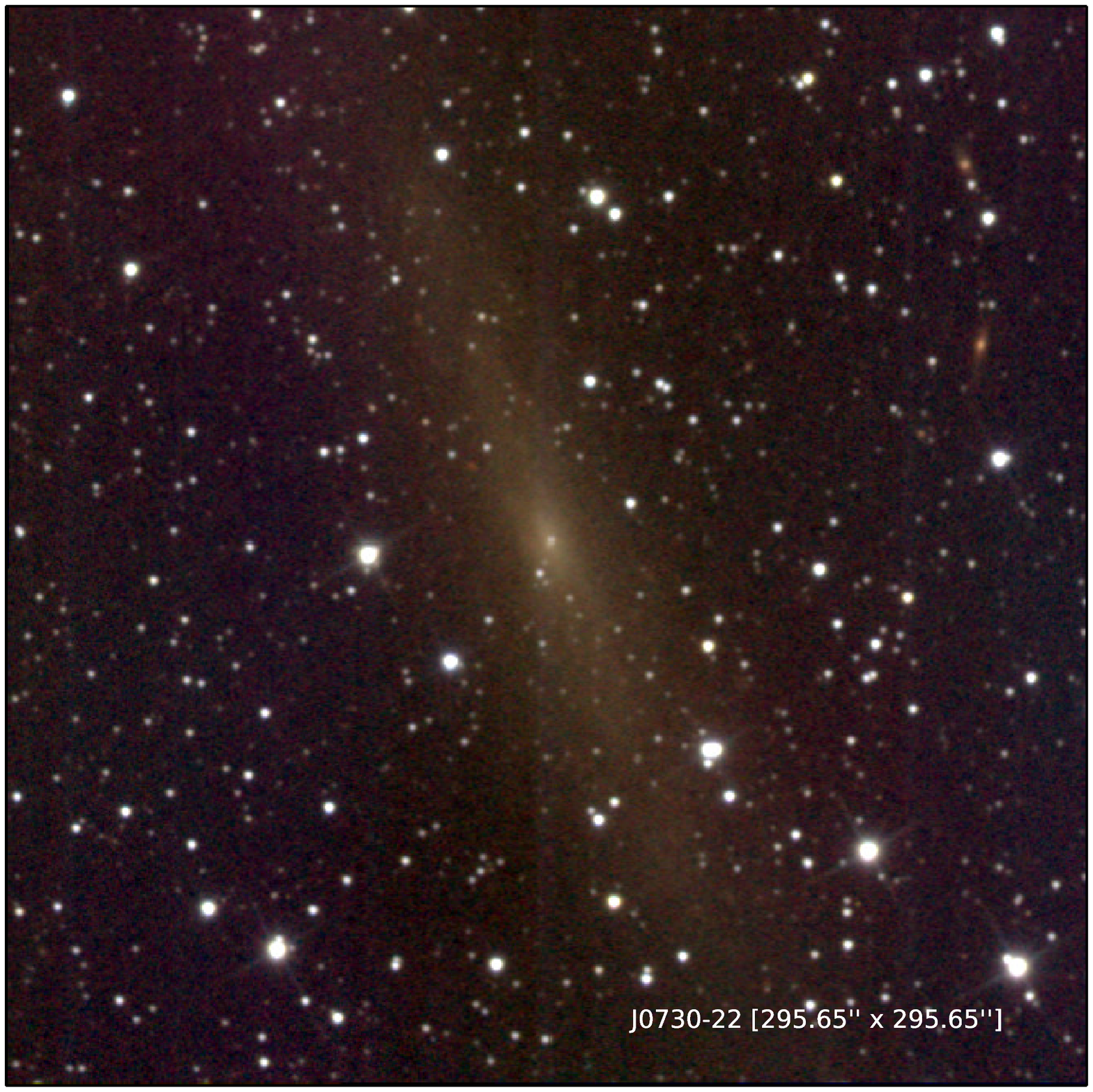}}
 & \subfloat{\includegraphics[scale=0.17]{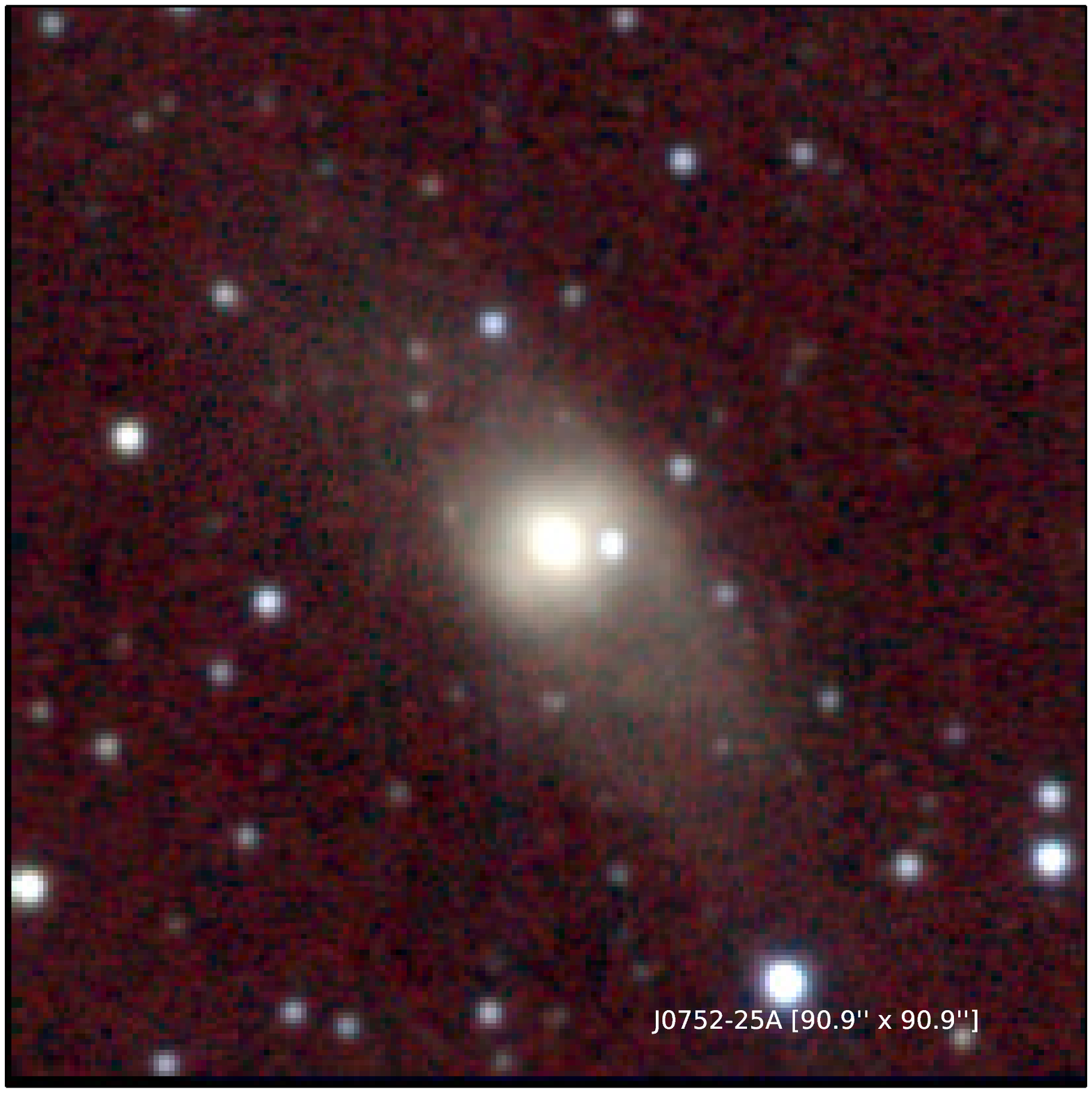}}
 & \subfloat{\includegraphics[scale=0.17]{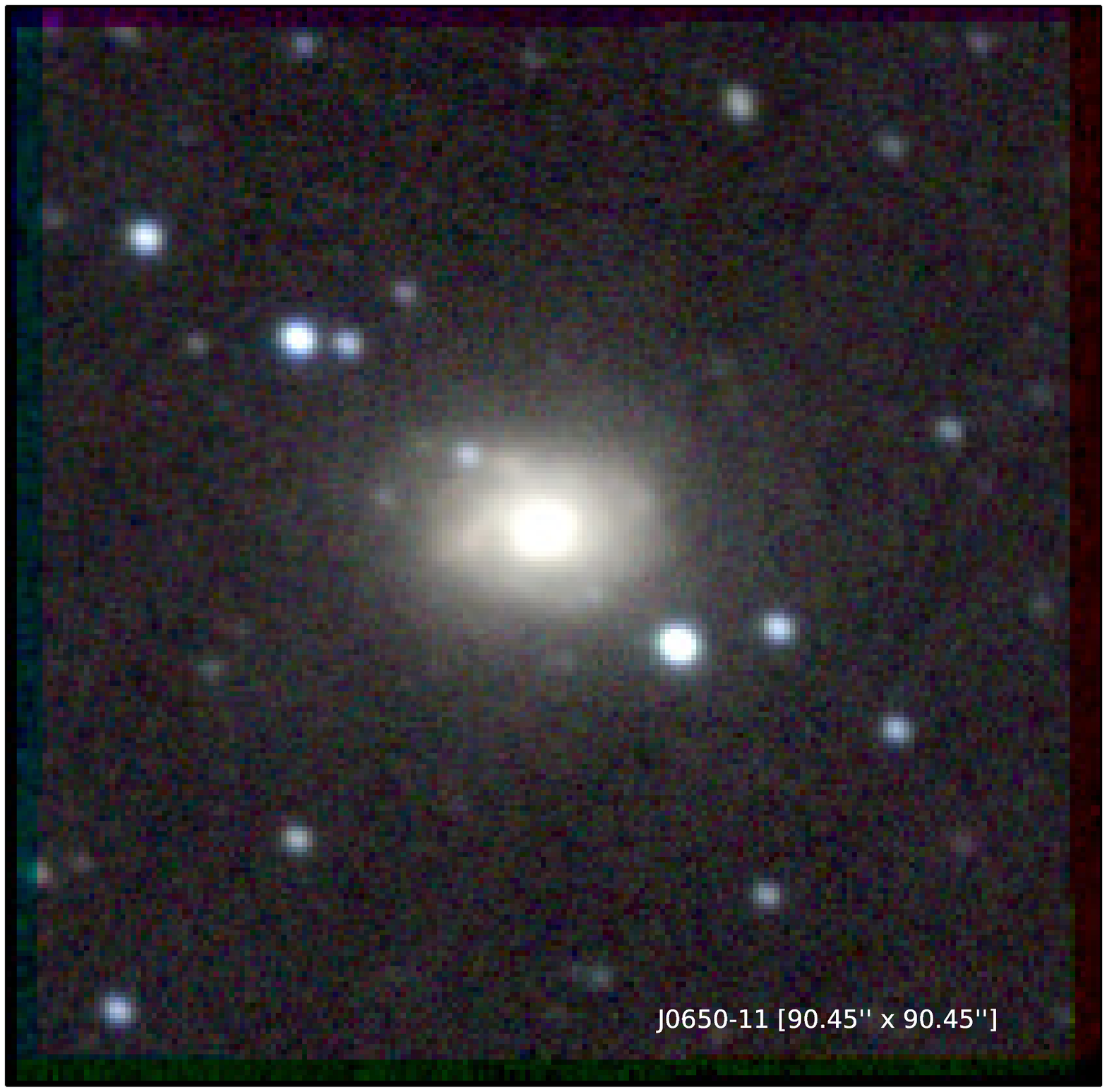}}
 & \subfloat{\includegraphics[scale=0.17]{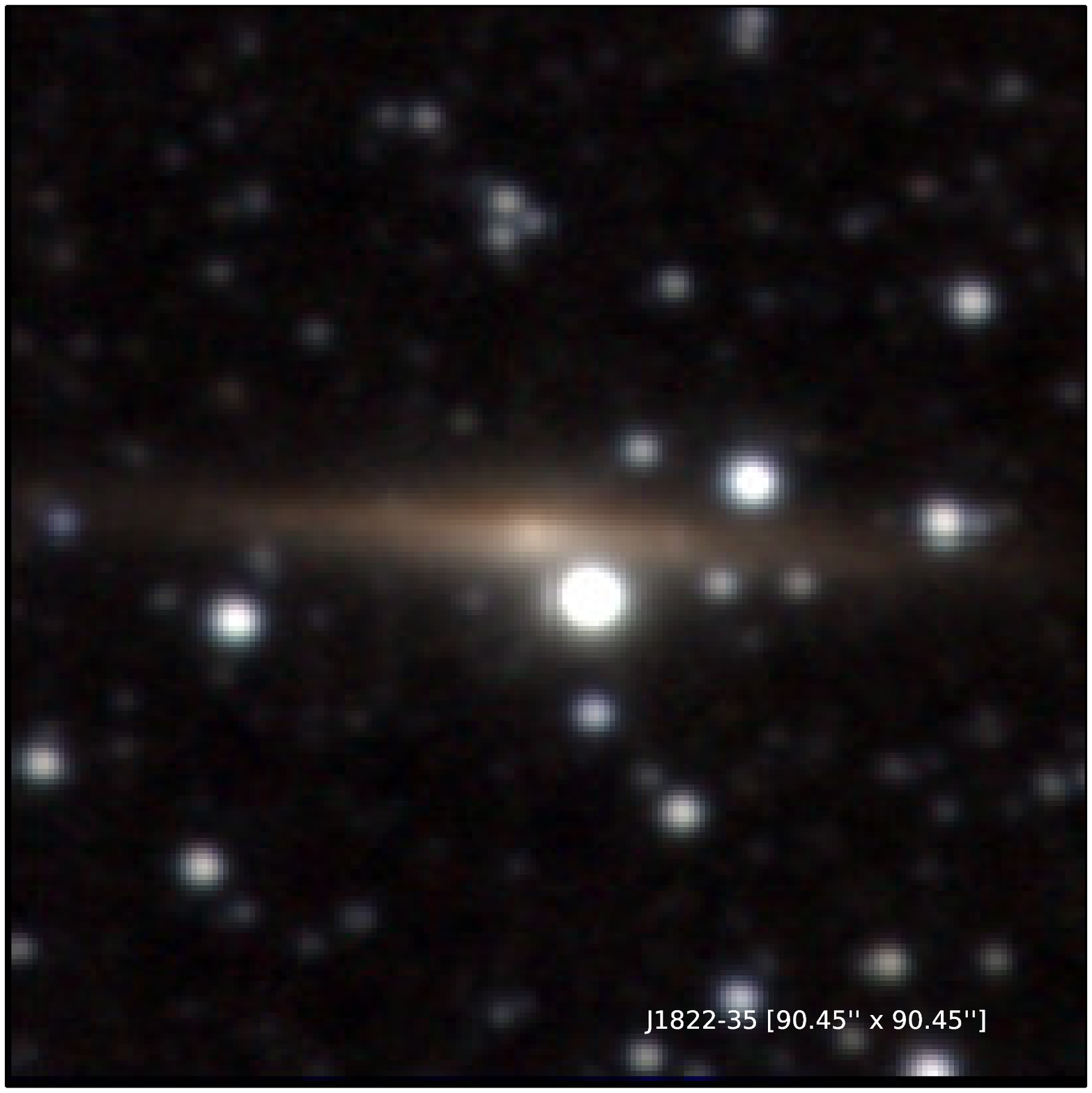}}
 & \subfloat{\includegraphics[scale=0.17]{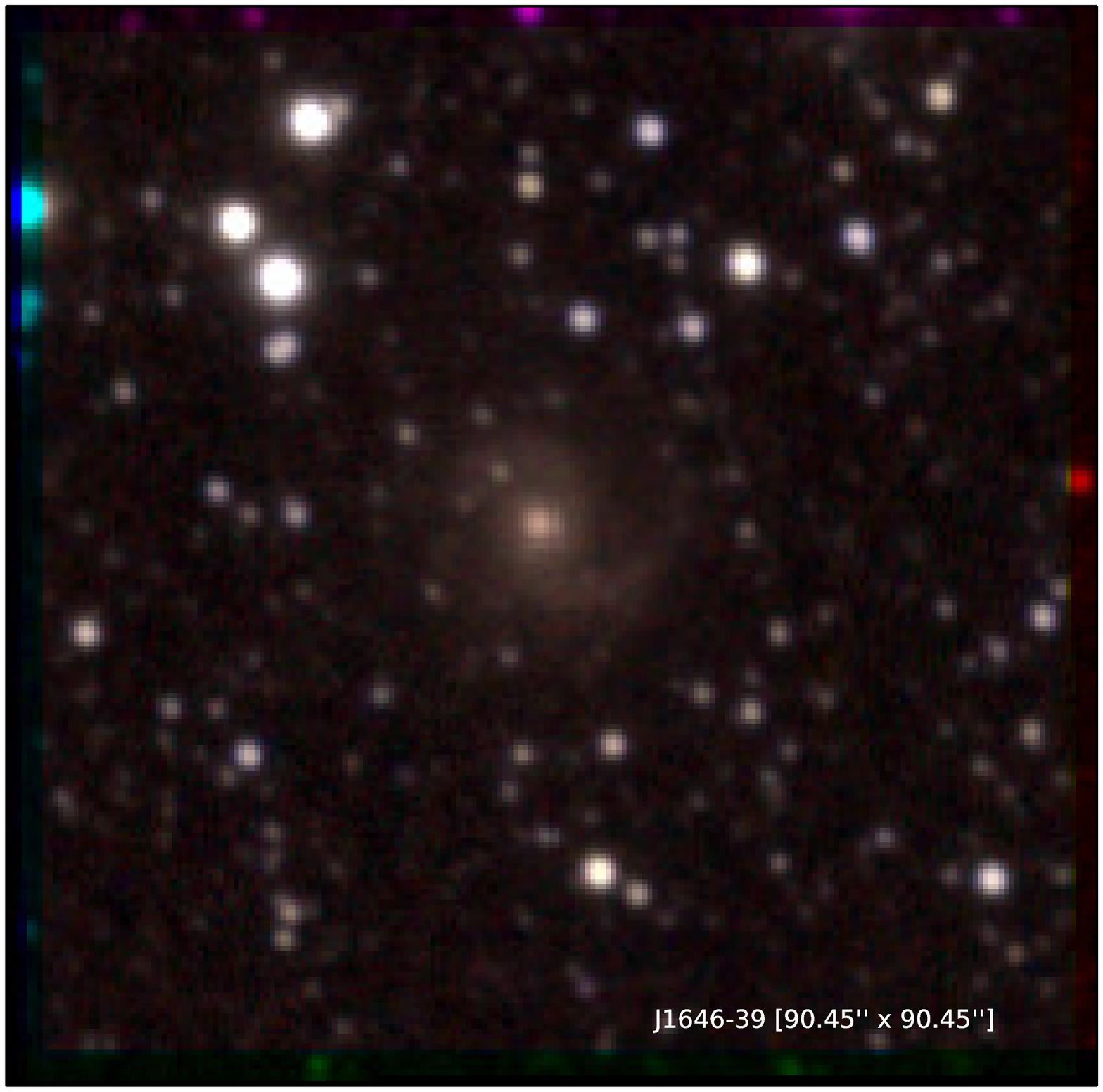}}\\
  \subfloat{\includegraphics[scale=0.17]{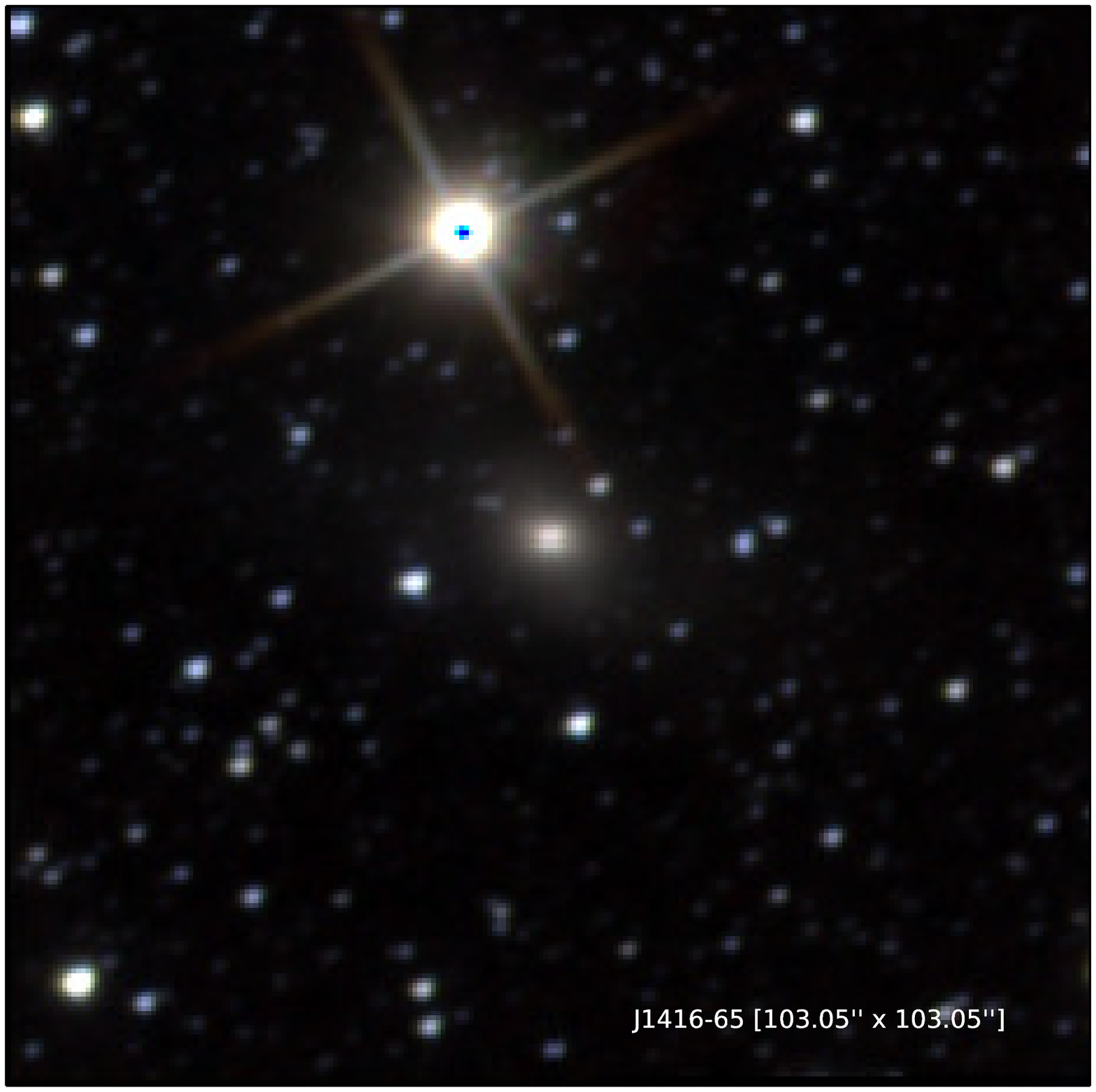}}
 & \subfloat{\includegraphics[scale=0.17]{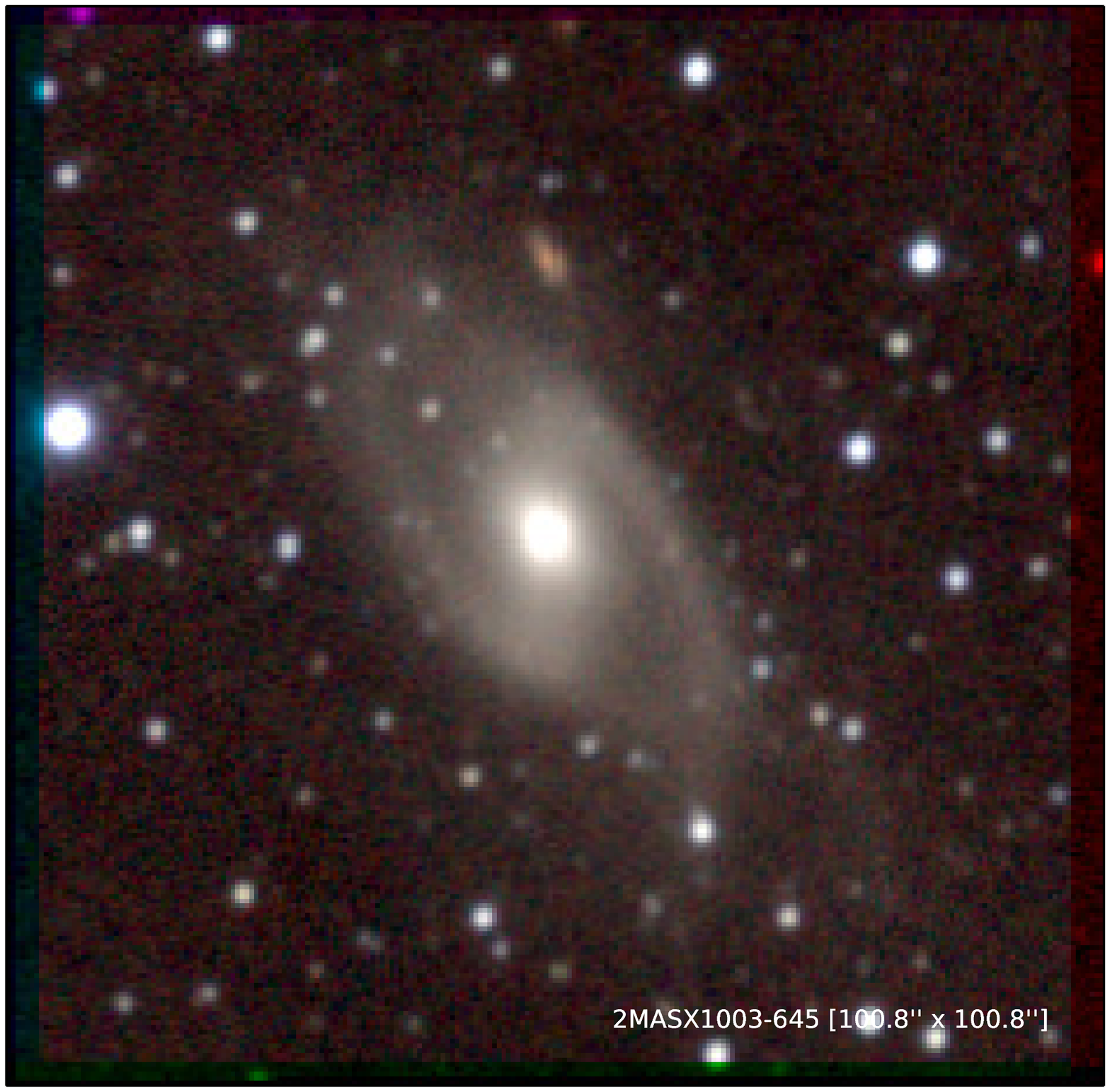}}
 & \subfloat{\includegraphics[scale=0.17]{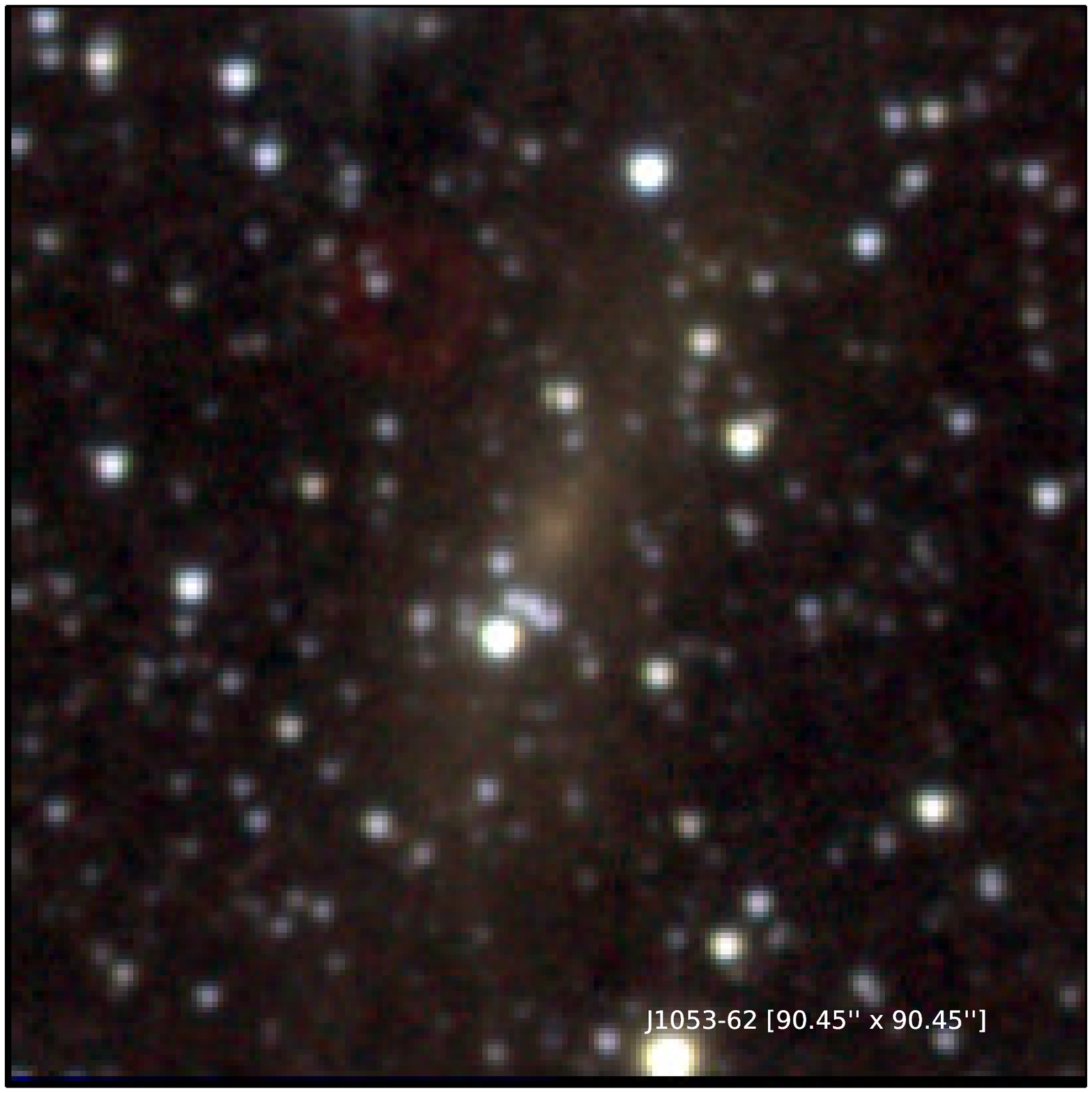}}
 & \subfloat{\includegraphics[scale=0.17]{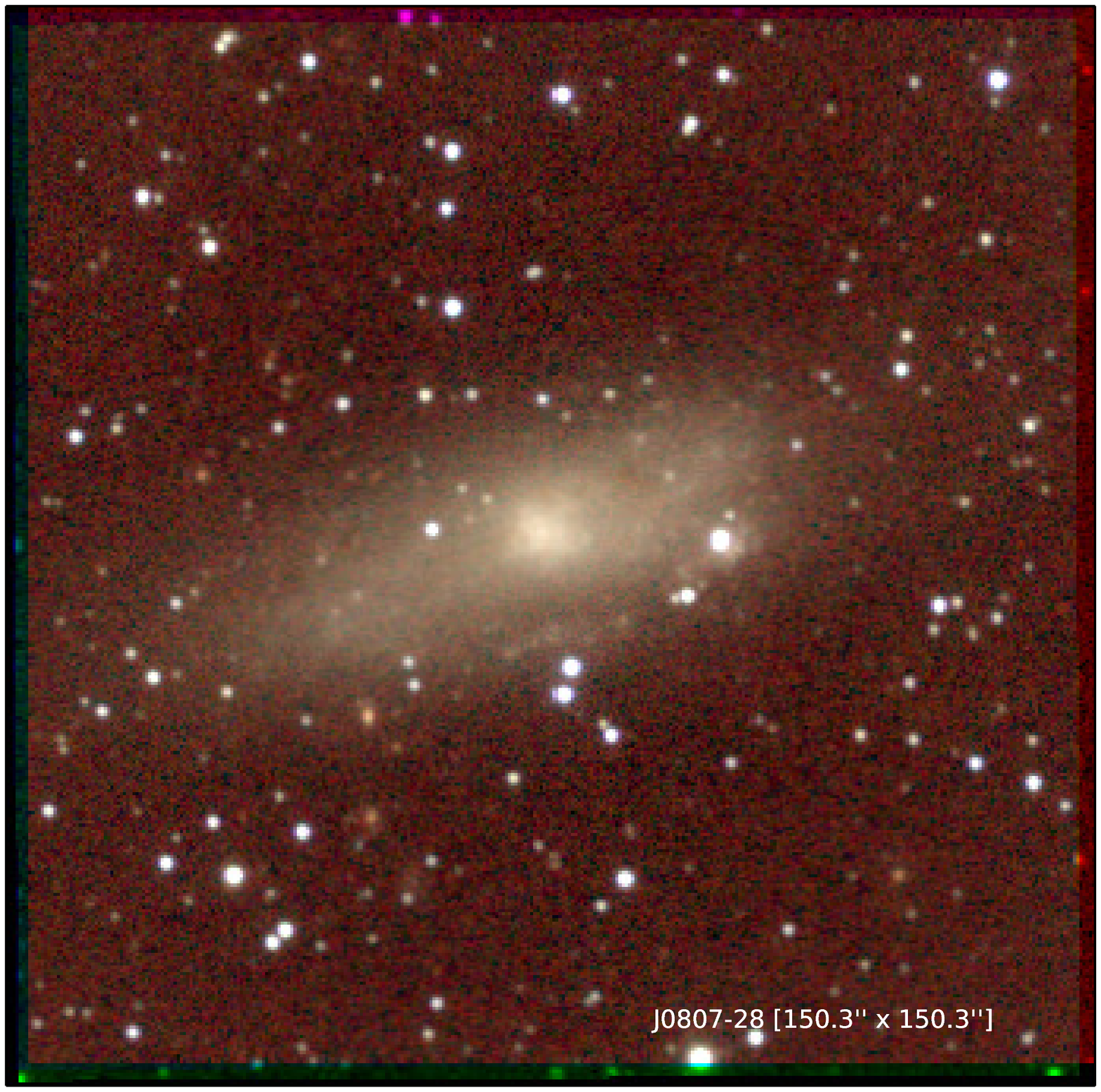}}
 & \subfloat{\includegraphics[scale=0.17]{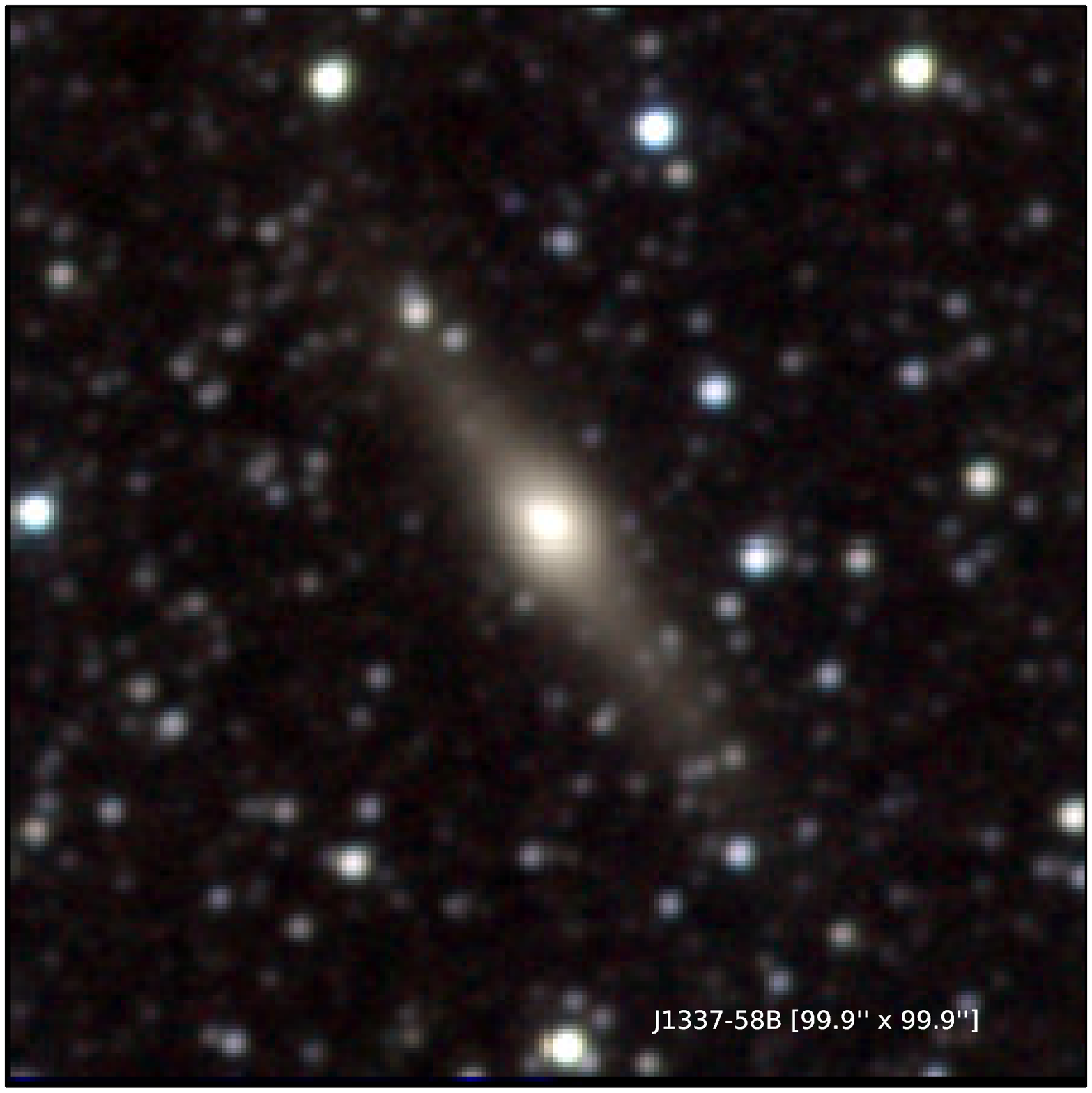}}\\
\\
\\

 \end{tabular}
 \caption{Continued}
\end{figure*}

\begin{figure*}
\ContinuedFloat
\begin{tabular}{ccccc}
  \subfloat{\includegraphics[scale=0.17]{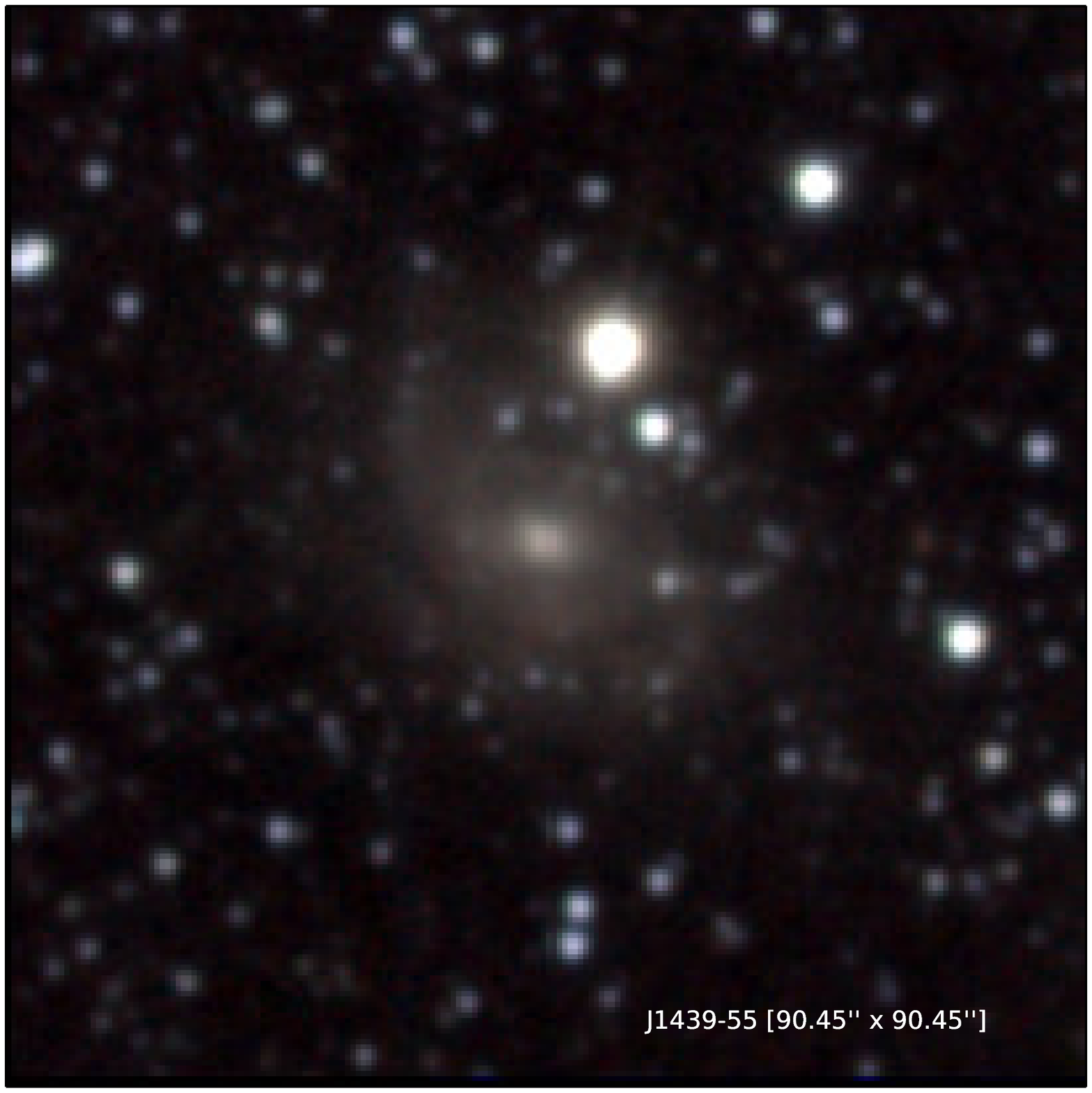}}
 & \subfloat{\includegraphics[scale=0.17]{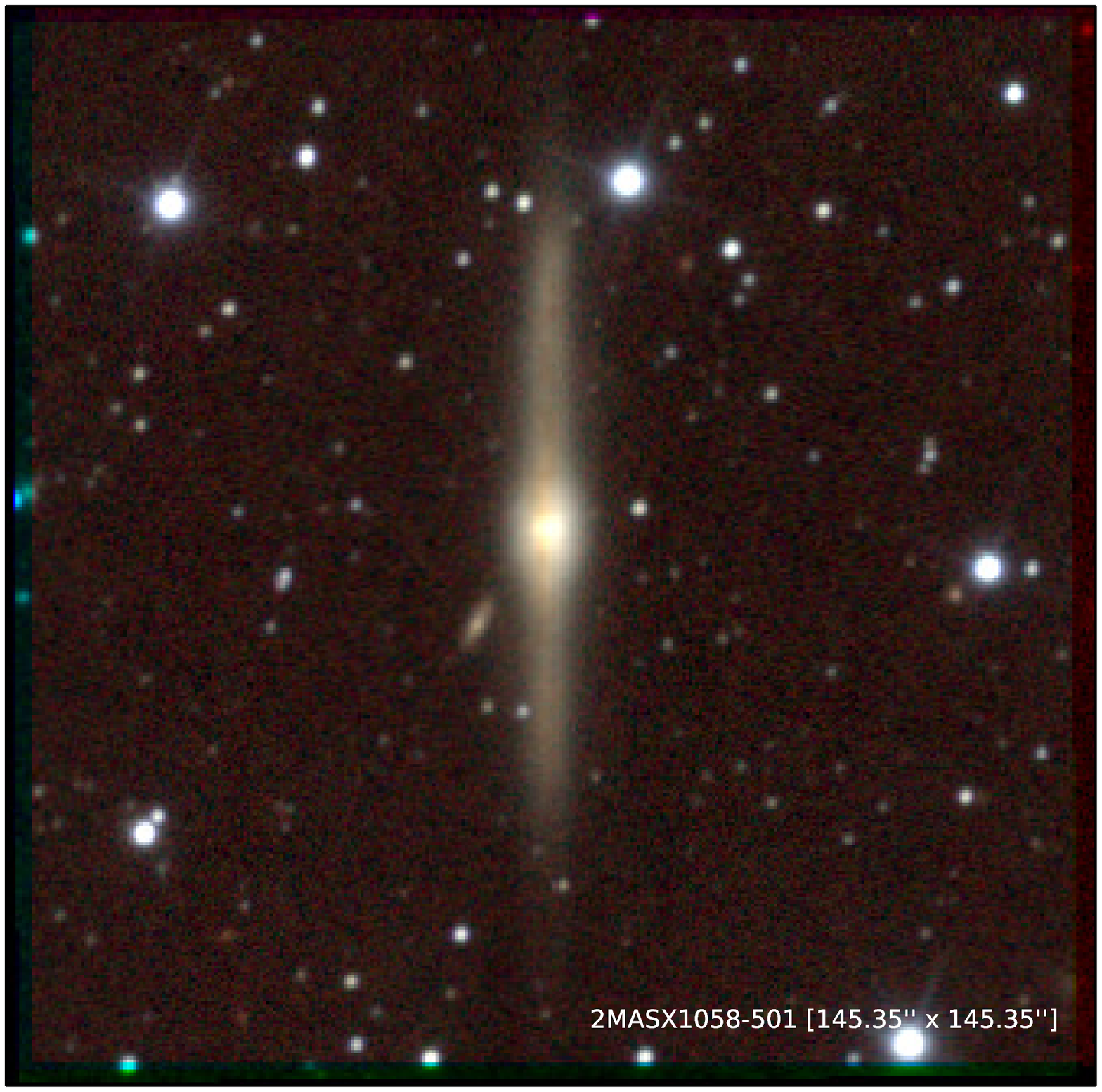}}
 & \subfloat{\includegraphics[scale=0.17]{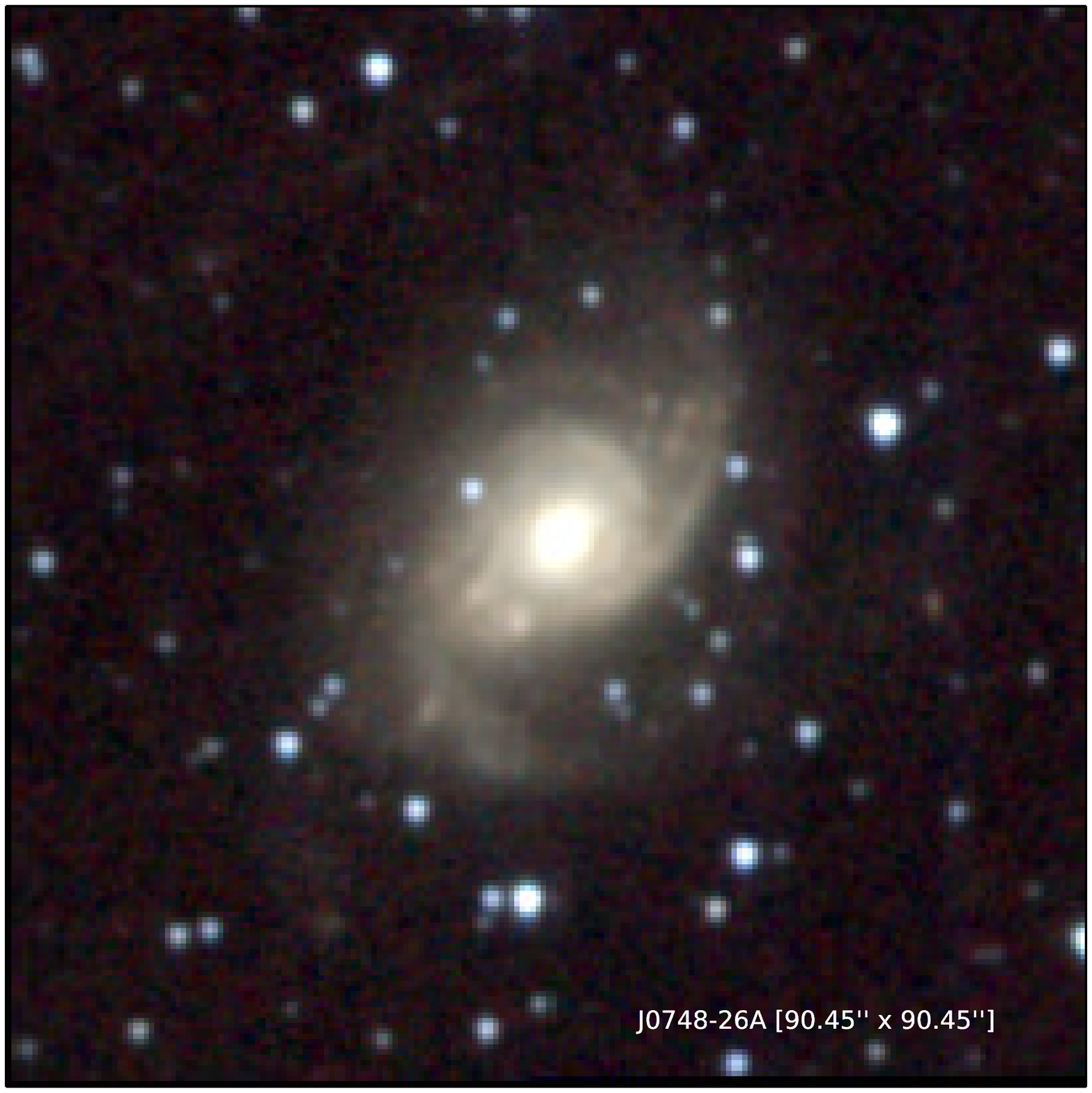}}
 & \subfloat{\includegraphics[scale=0.17]{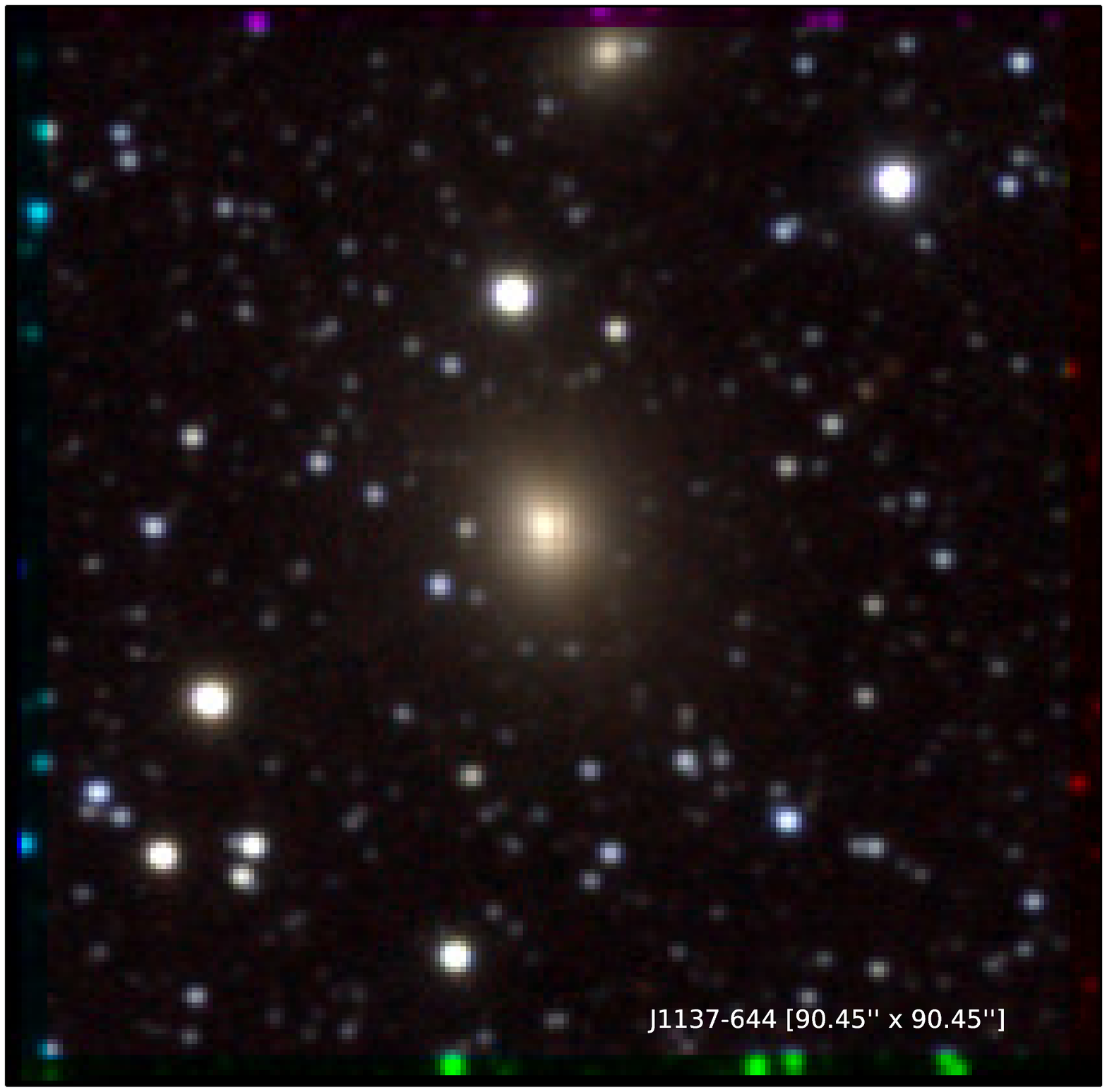}}
 & \subfloat{\includegraphics[scale=0.17]{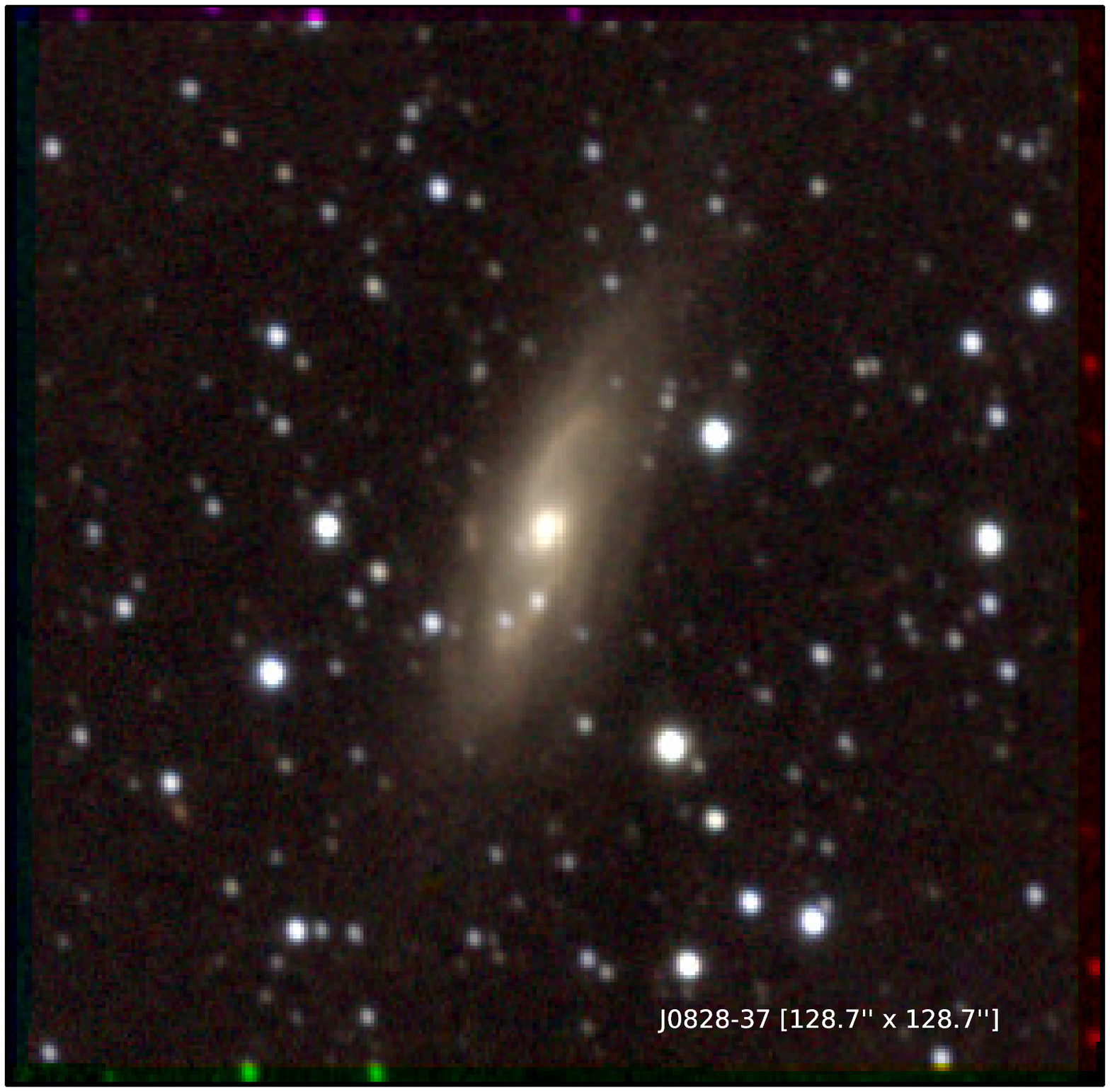}}\\
  \subfloat{\includegraphics[scale=0.17]{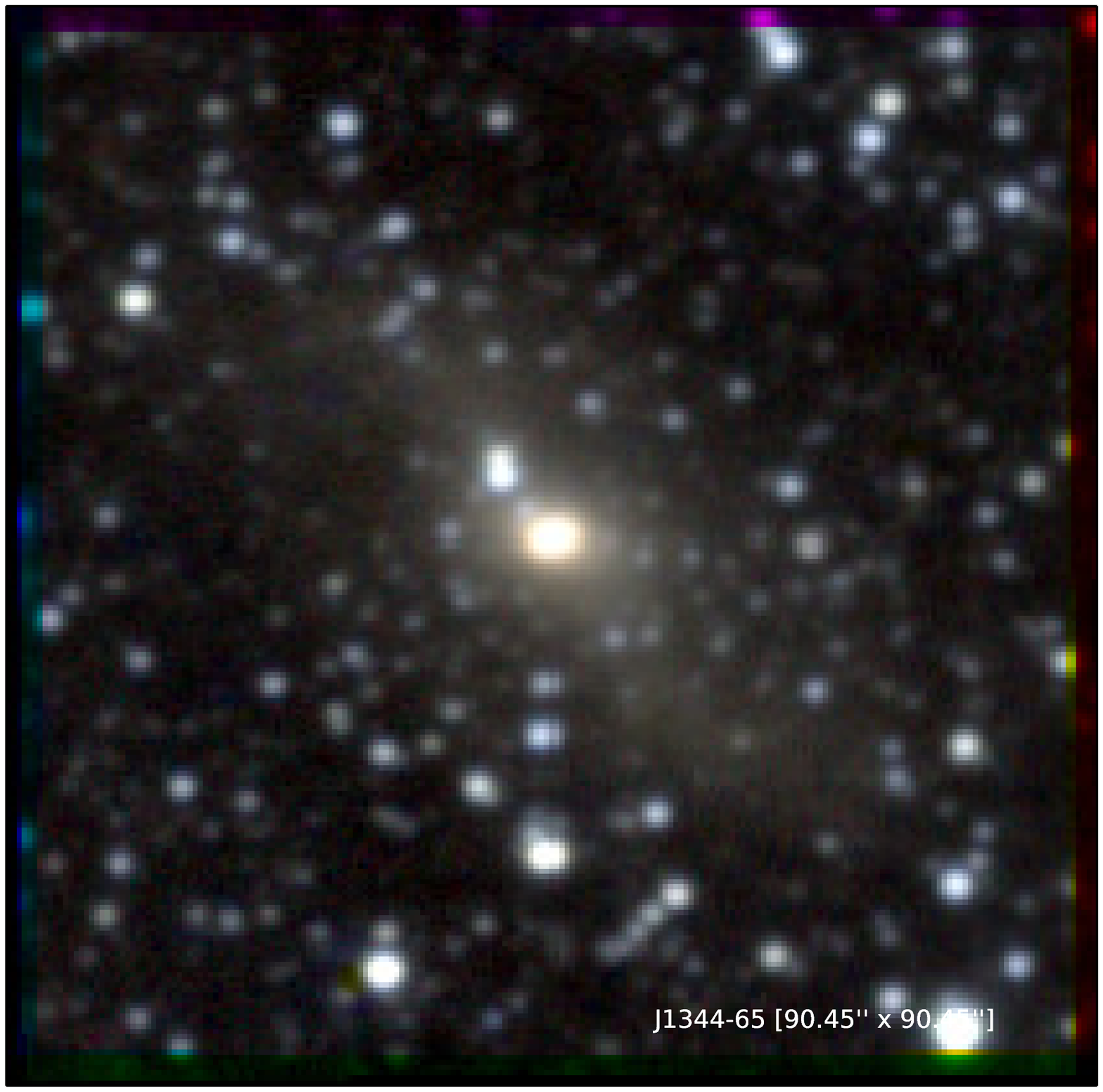}}
 & \subfloat{\includegraphics[scale=0.17]{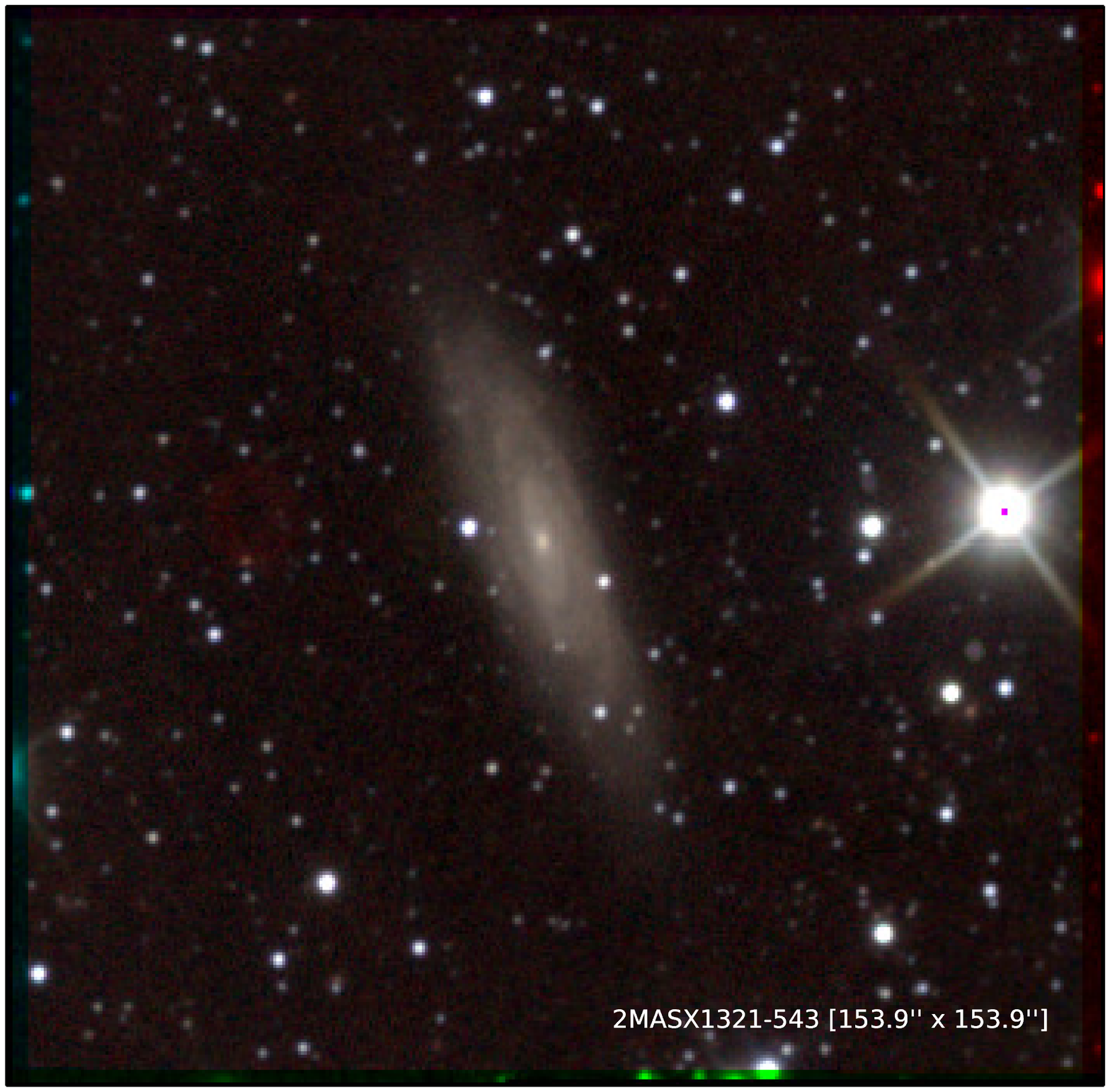}}
 & \subfloat{\includegraphics[scale=0.17]{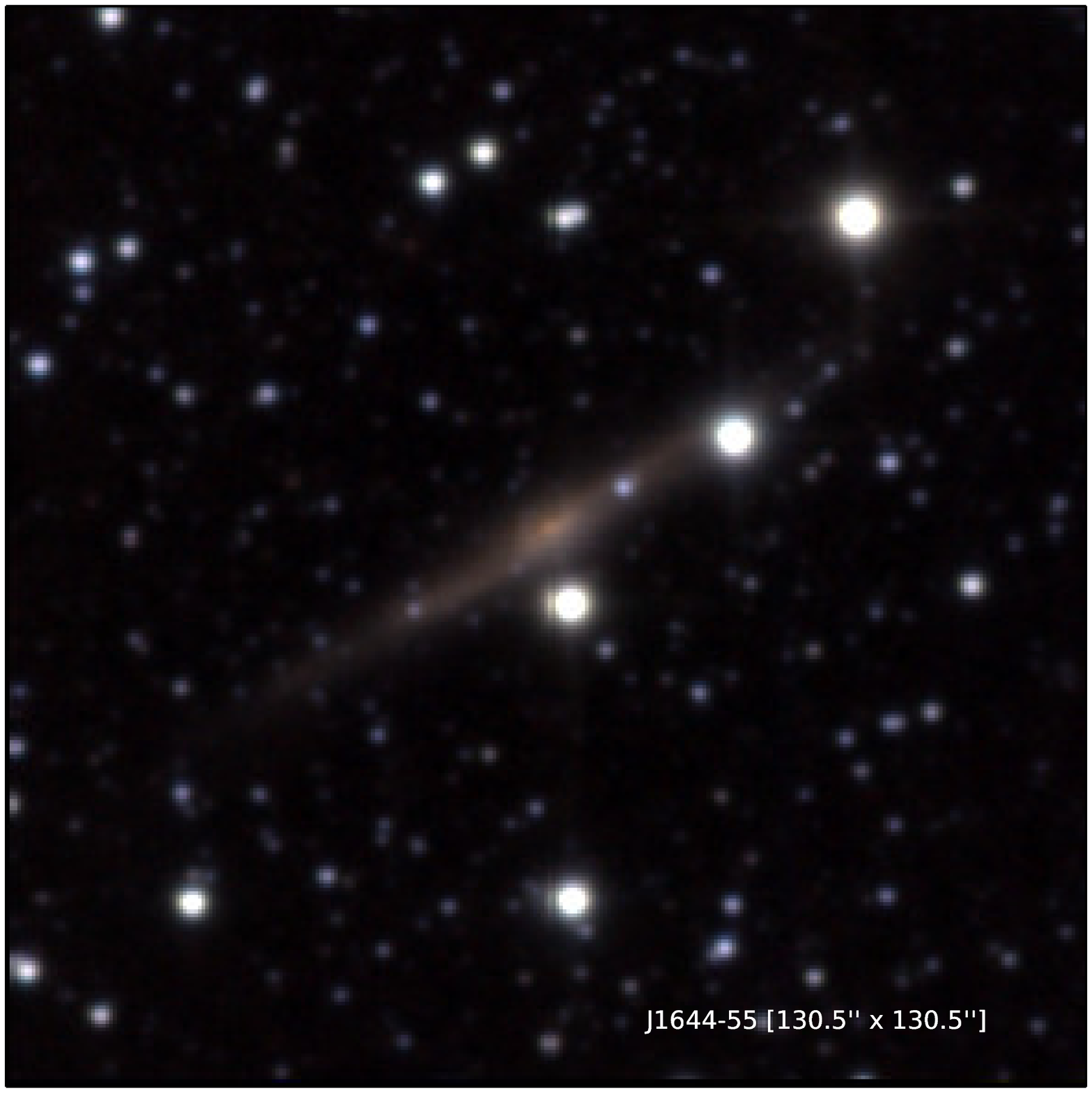}}
 & \subfloat{\includegraphics[scale=0.17]{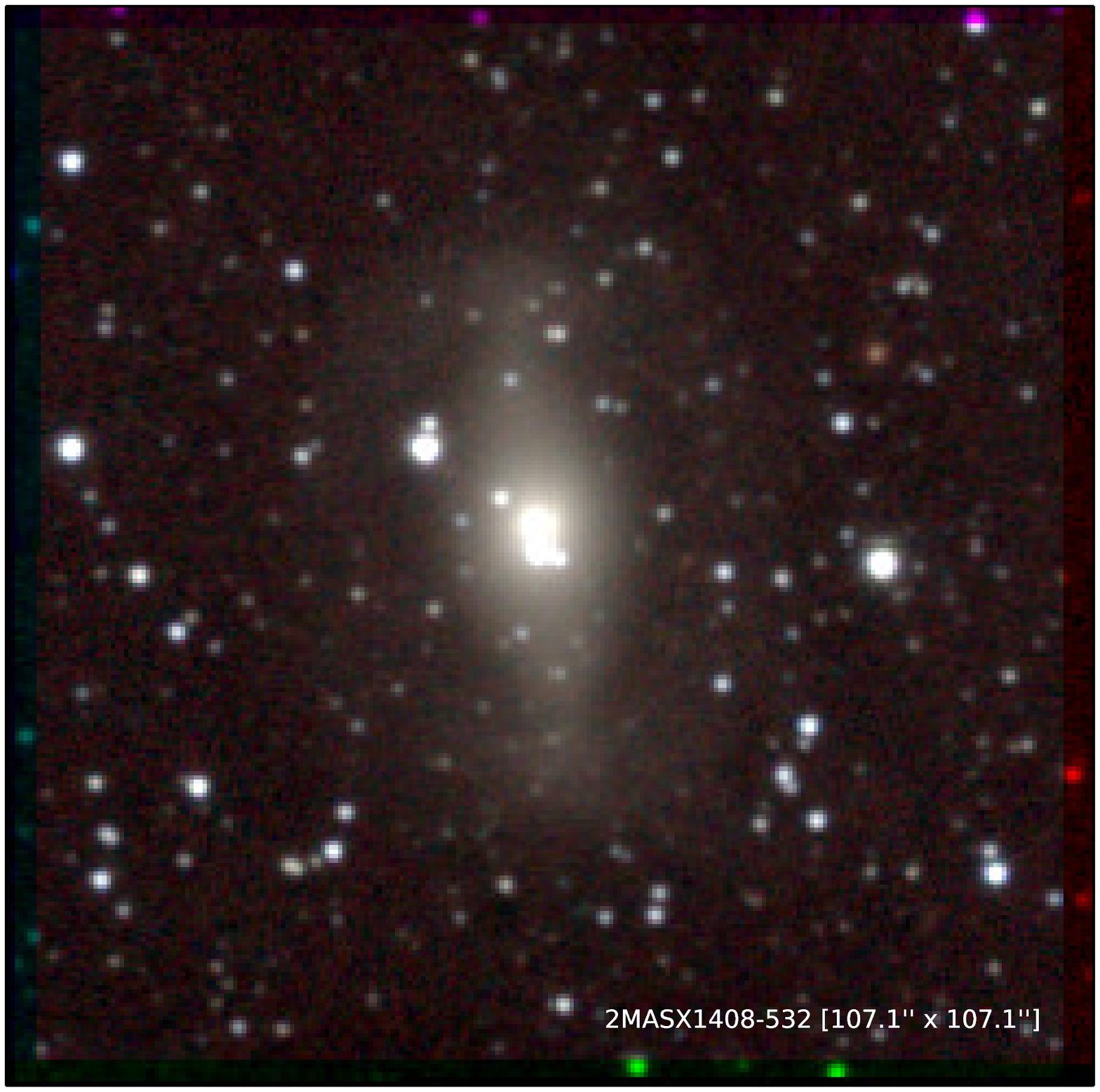}}
 & \subfloat{\includegraphics[scale=0.17]{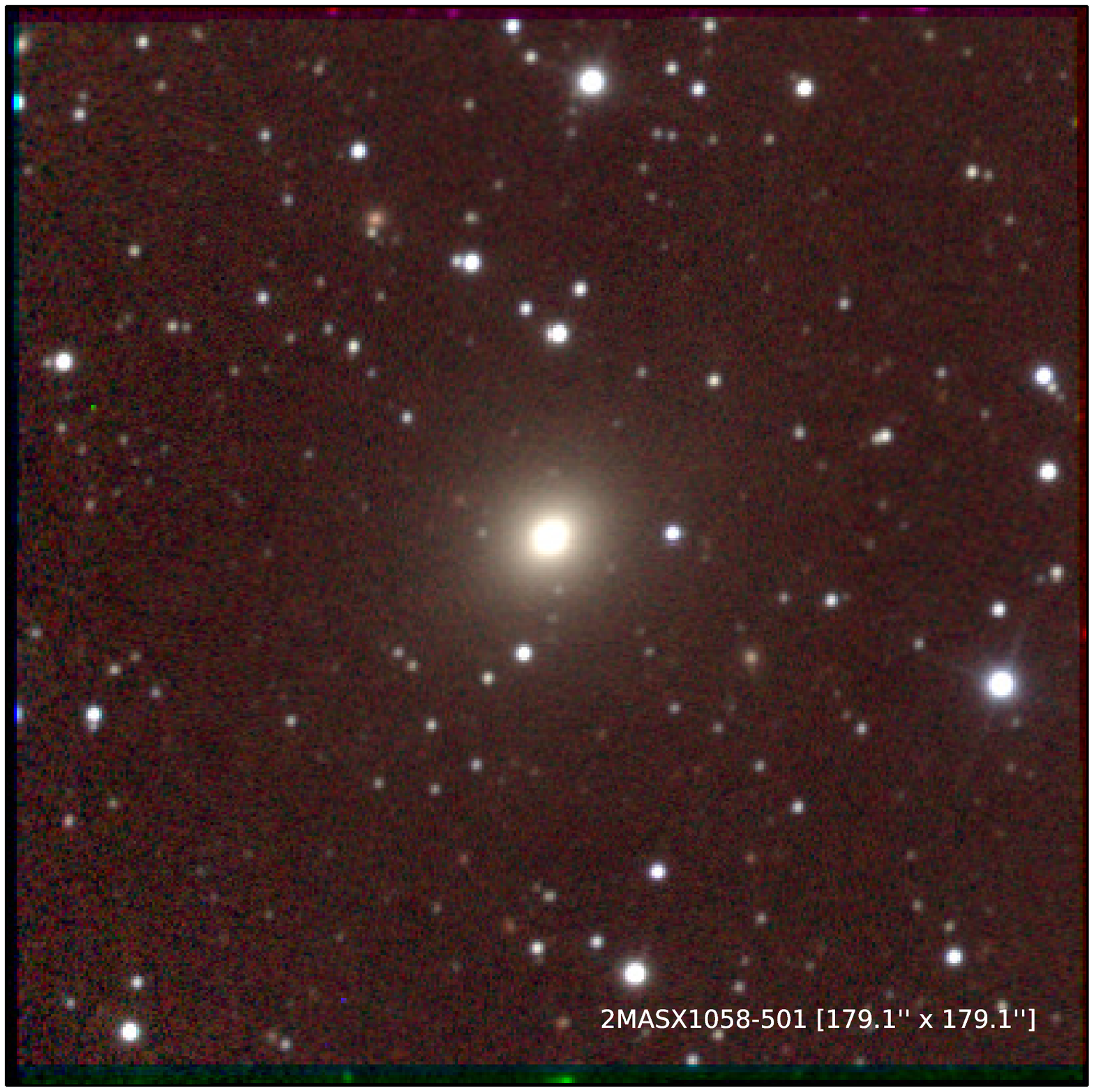}}\\
  \subfloat{\includegraphics[scale=0.17]{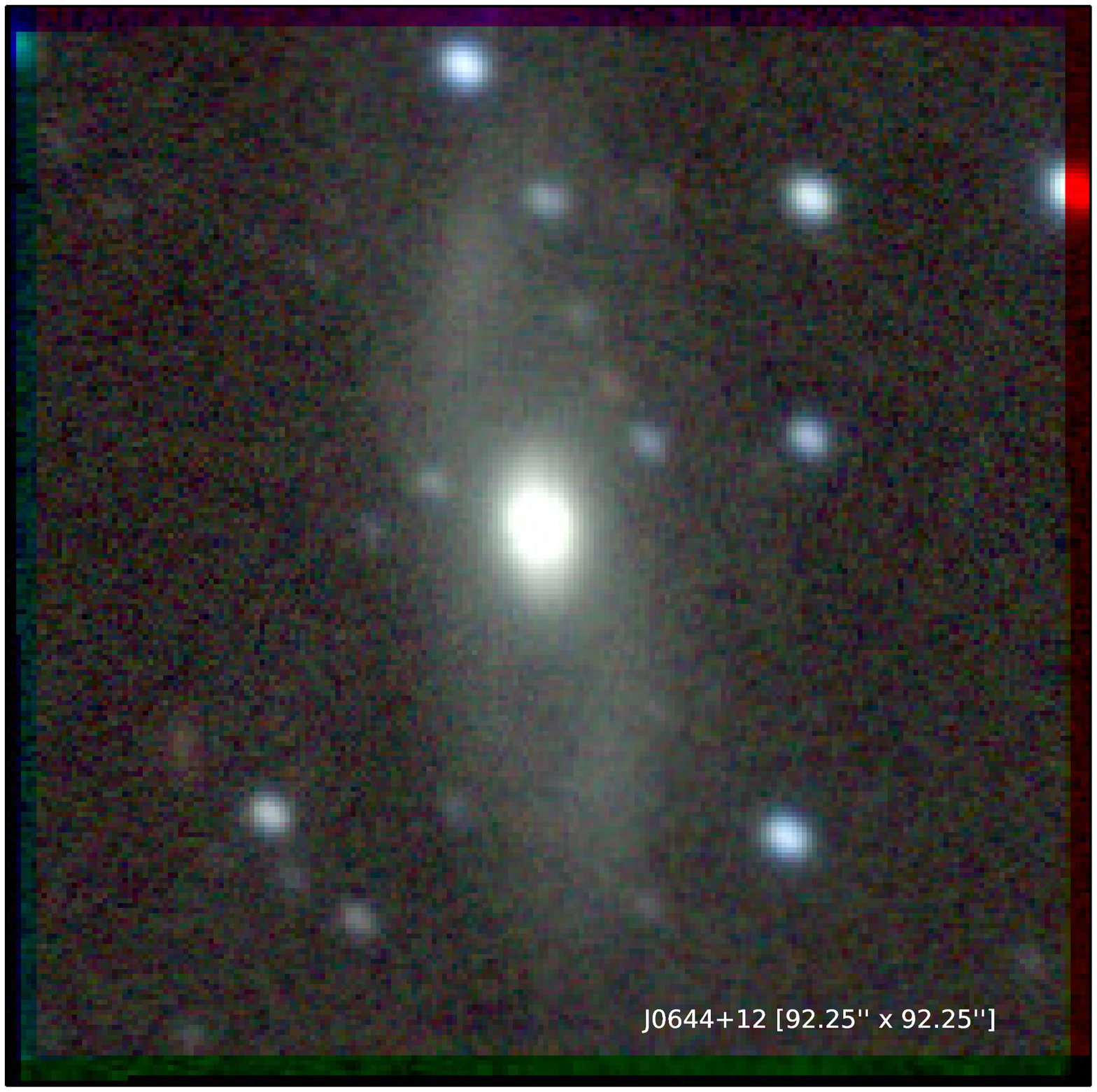}}
 & \subfloat{\includegraphics[scale=0.17]{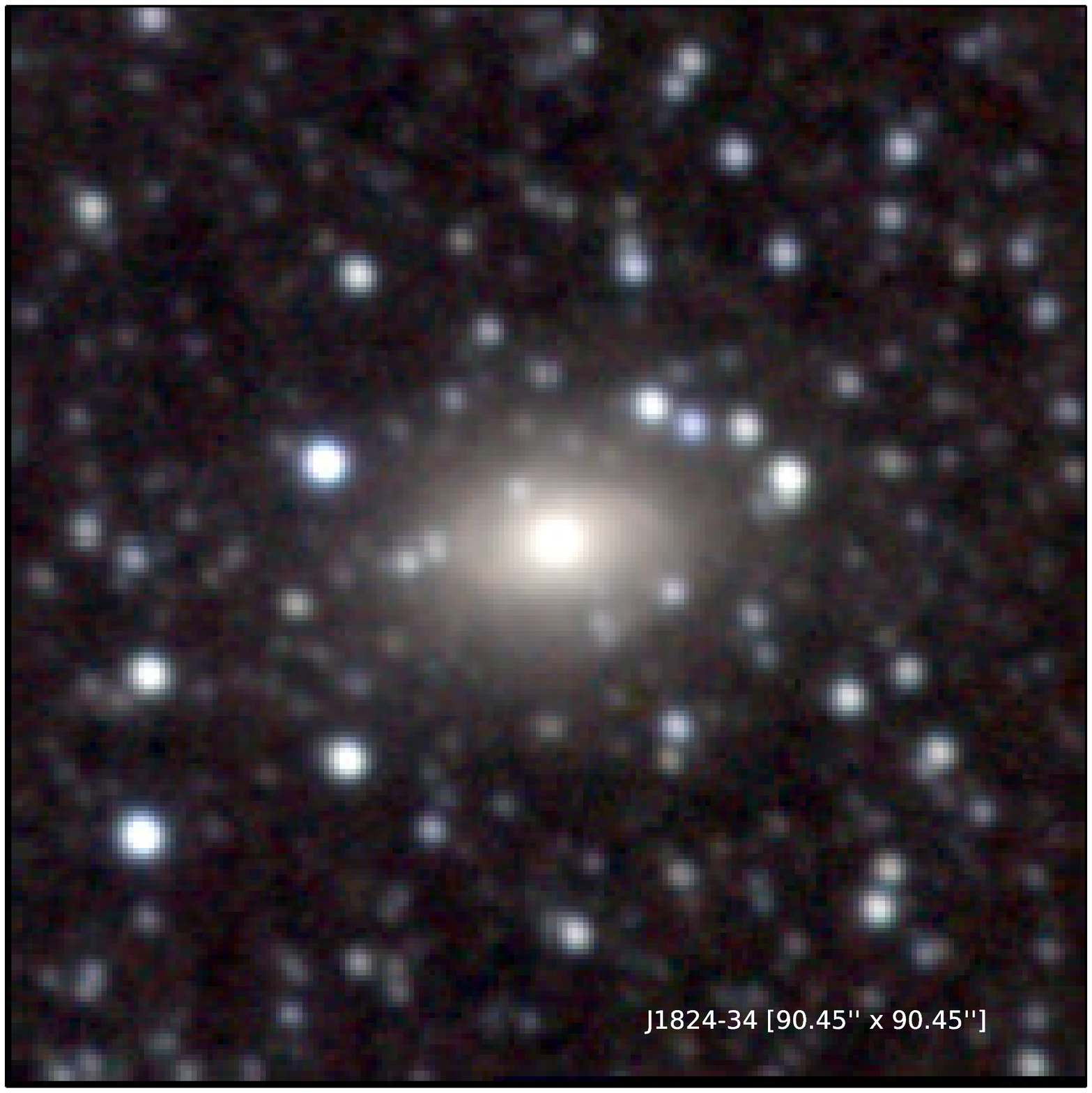}}
 & \subfloat{\includegraphics[scale=0.17]{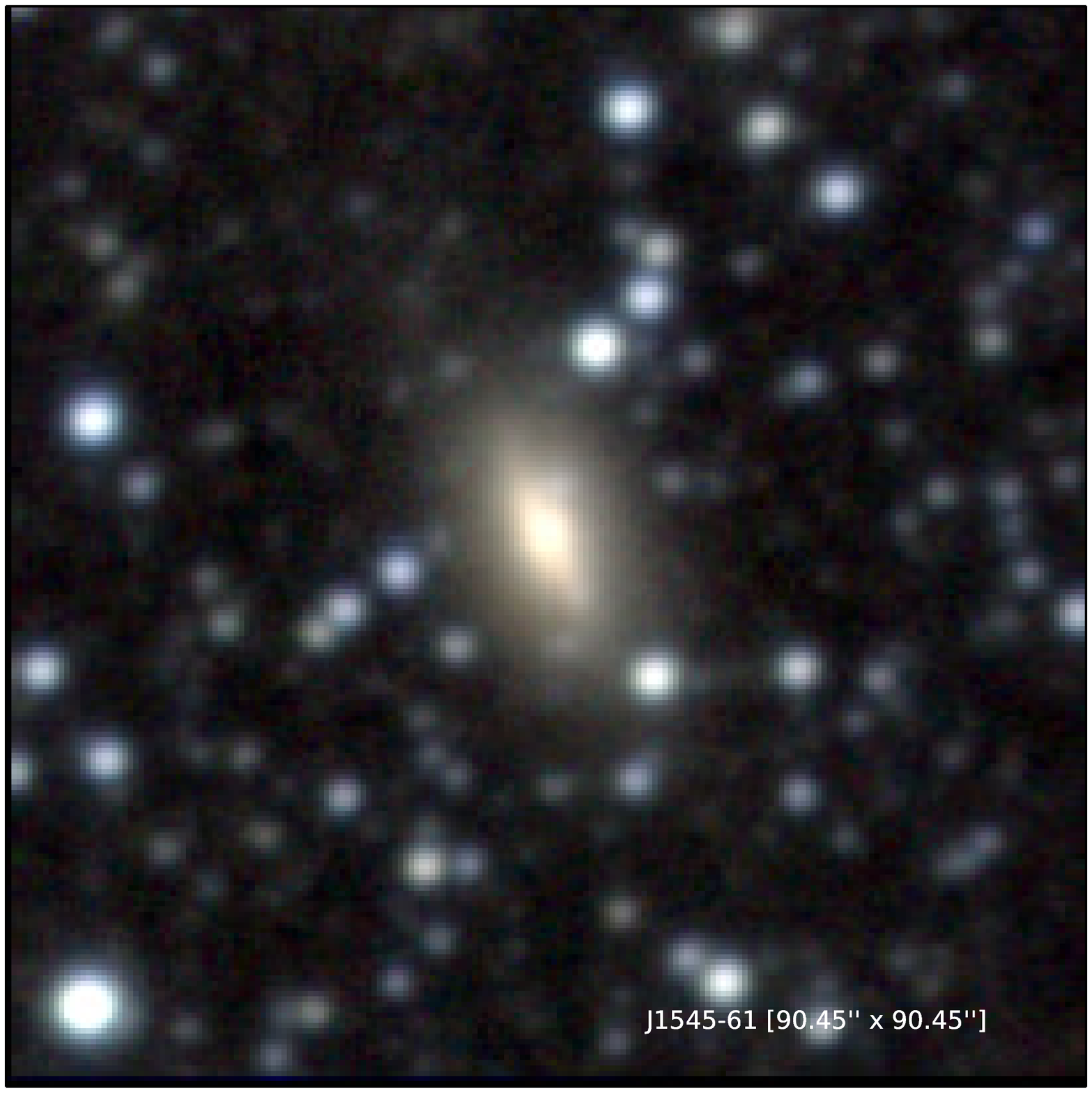}}
 & \subfloat{\includegraphics[scale=0.17]{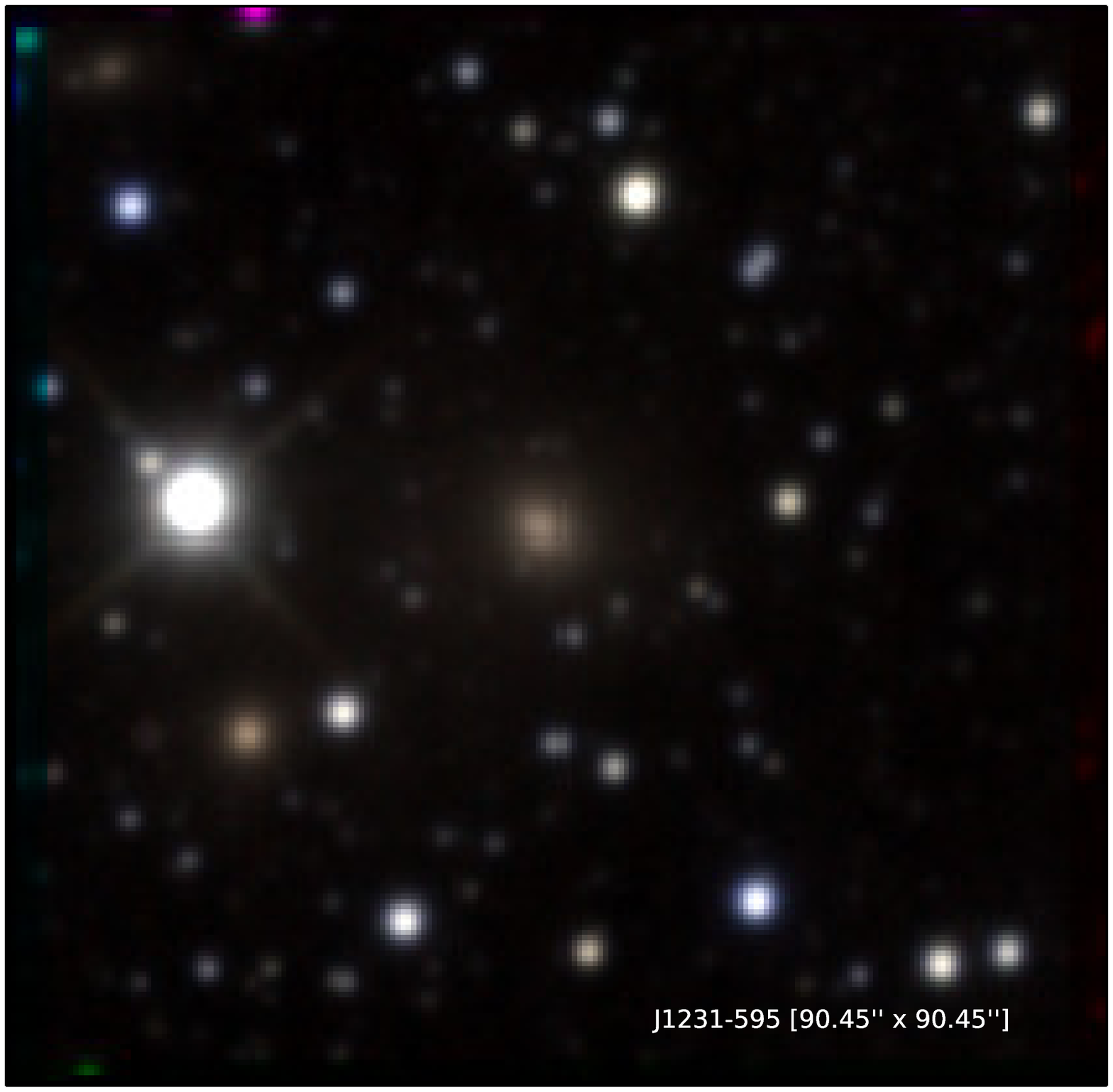}}
 & \subfloat{\includegraphics[scale=0.17]{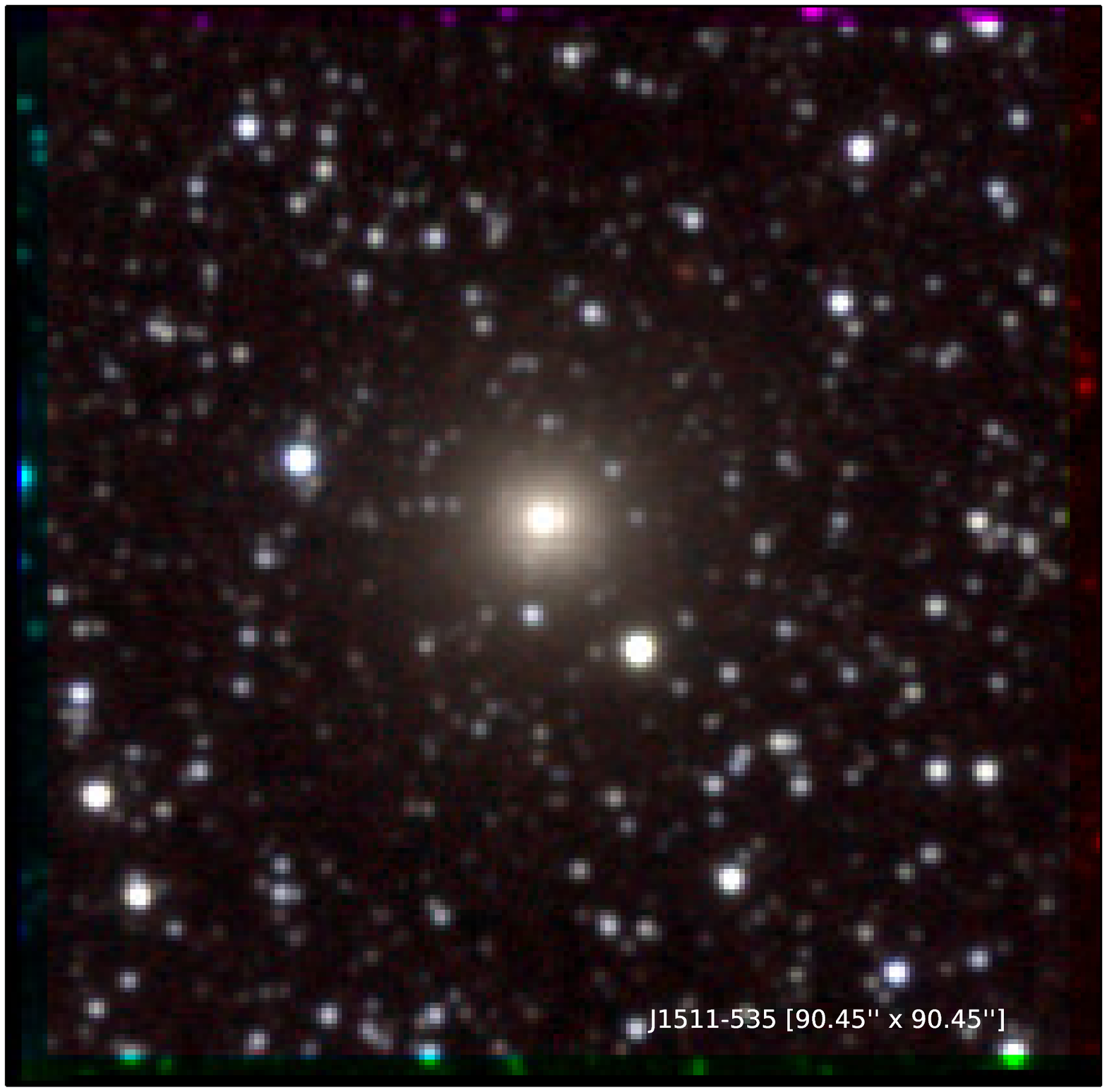}}\\
  \subfloat{\includegraphics[scale=0.17]{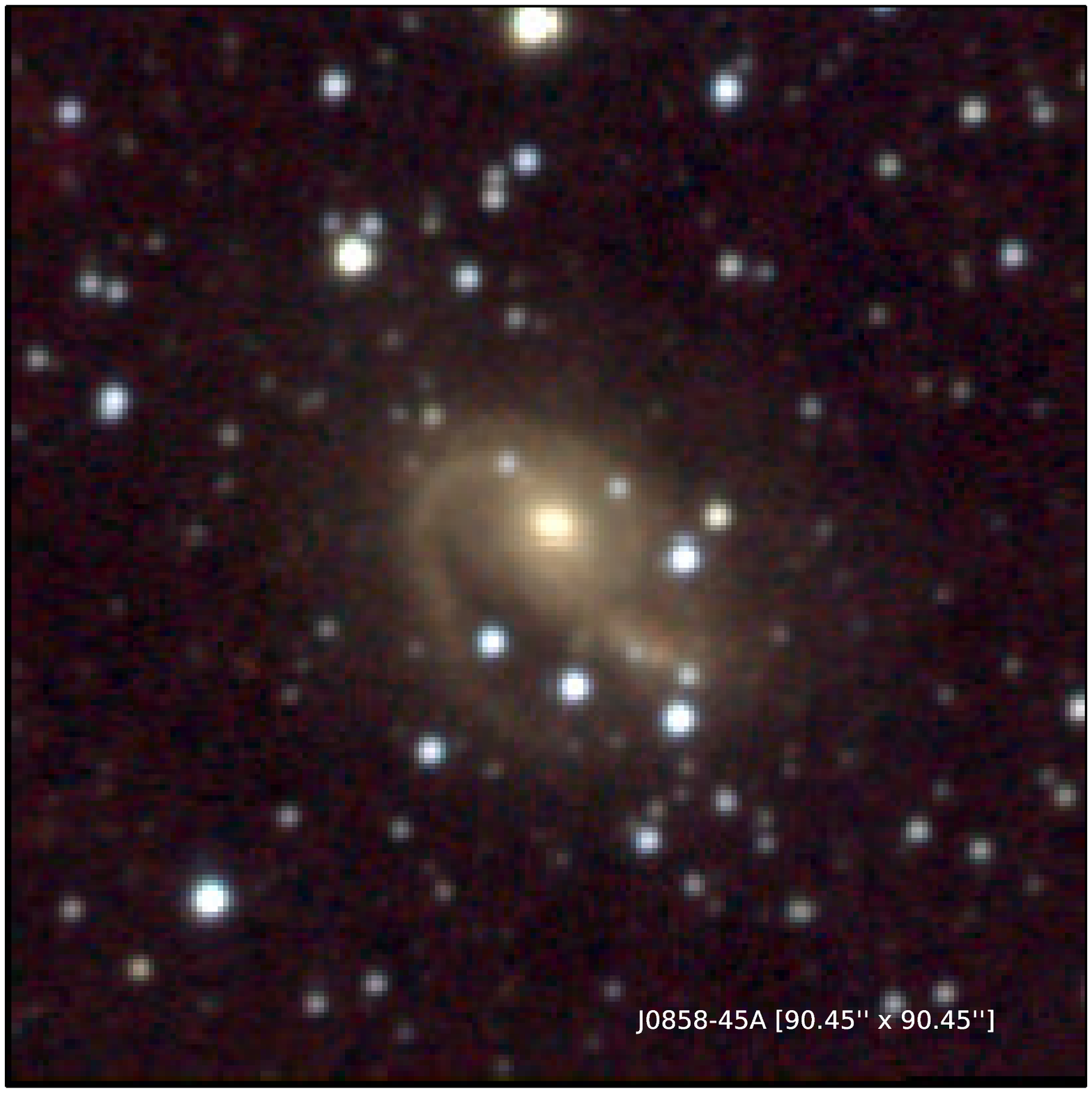}}
 & \subfloat{\includegraphics[scale=0.17]{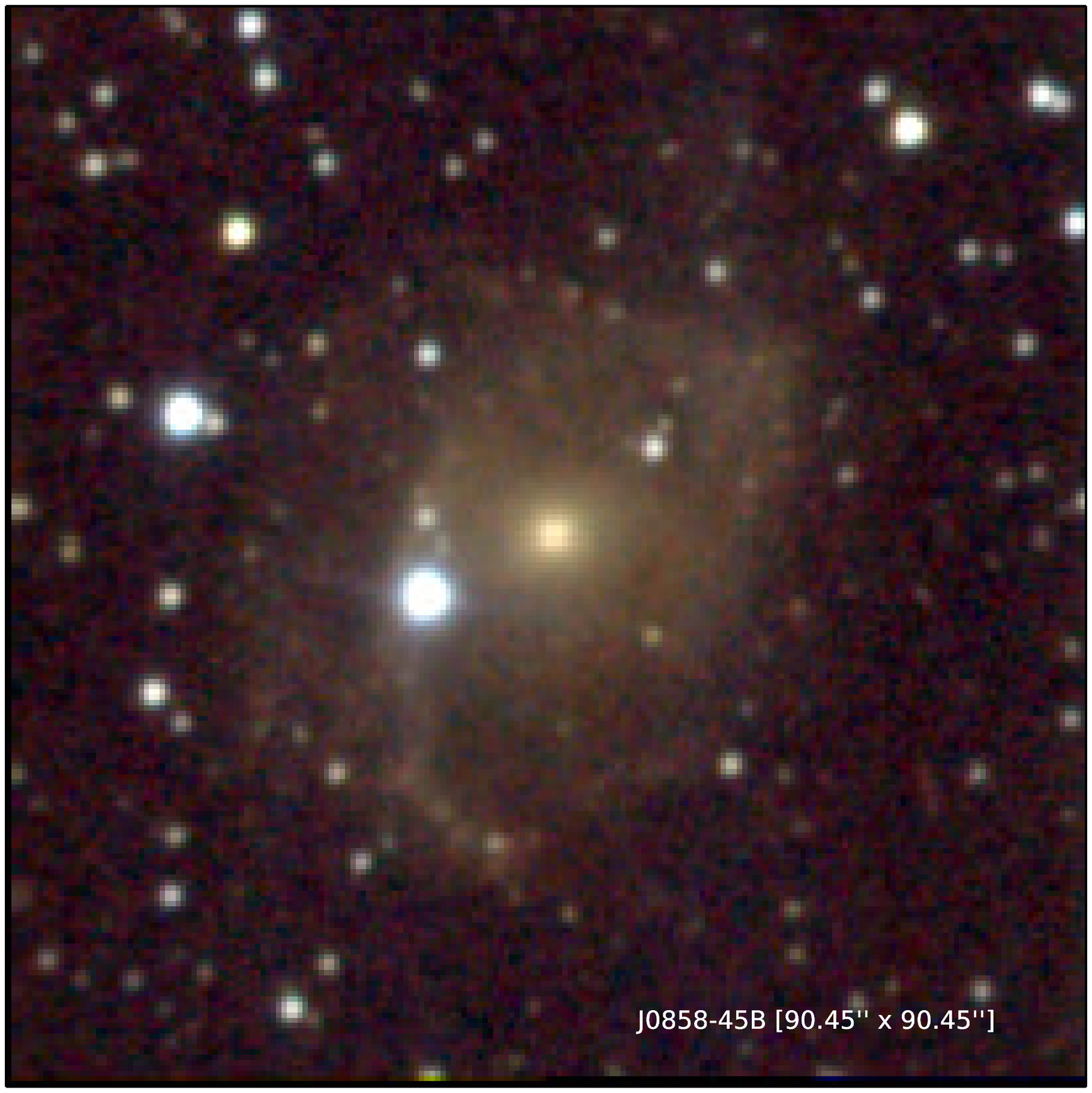}}
 & \subfloat{\includegraphics[scale=0.17]{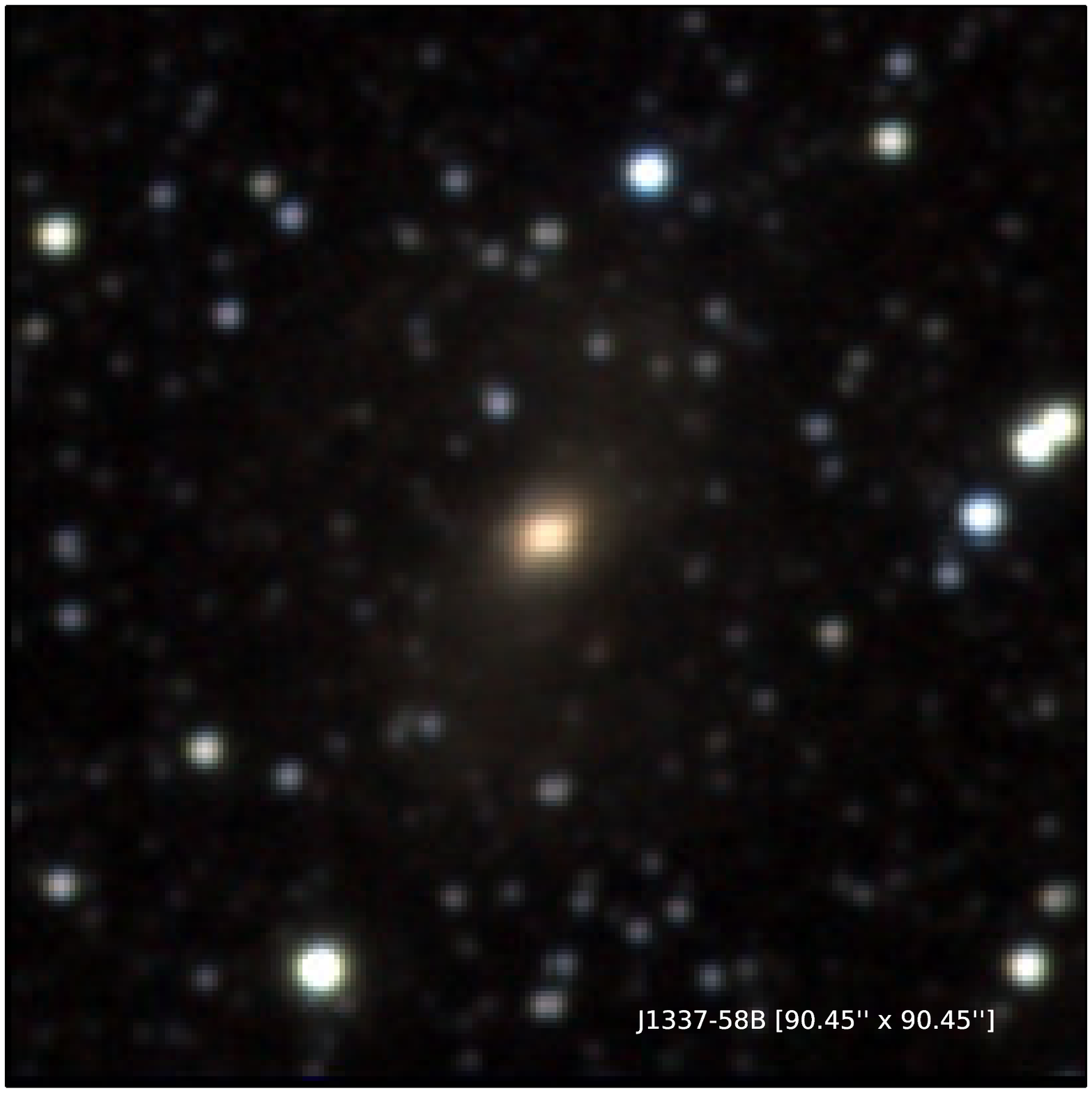}}
 & \subfloat{\includegraphics[scale=0.17]{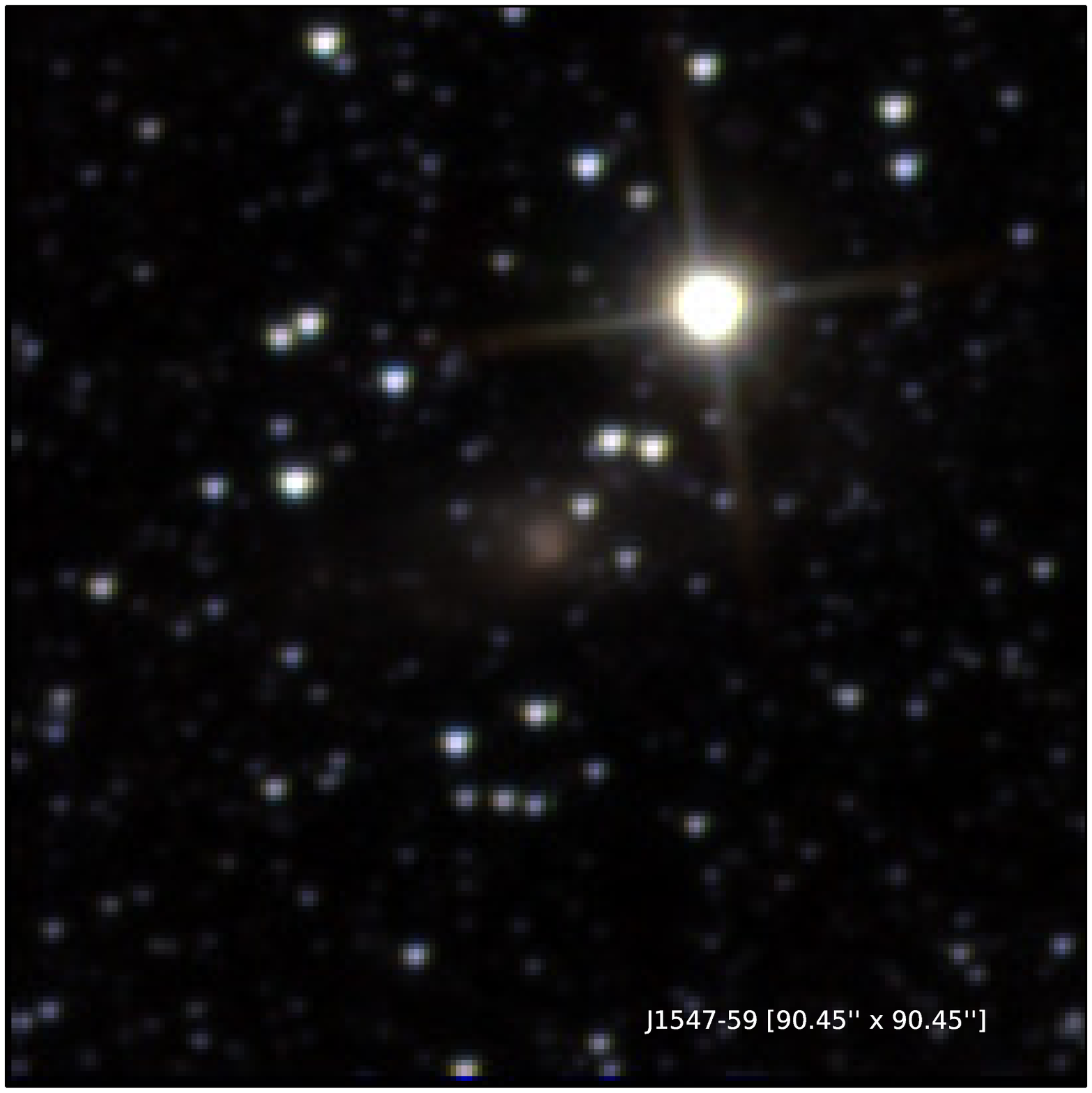}}
 & \subfloat{\includegraphics[scale=0.17]{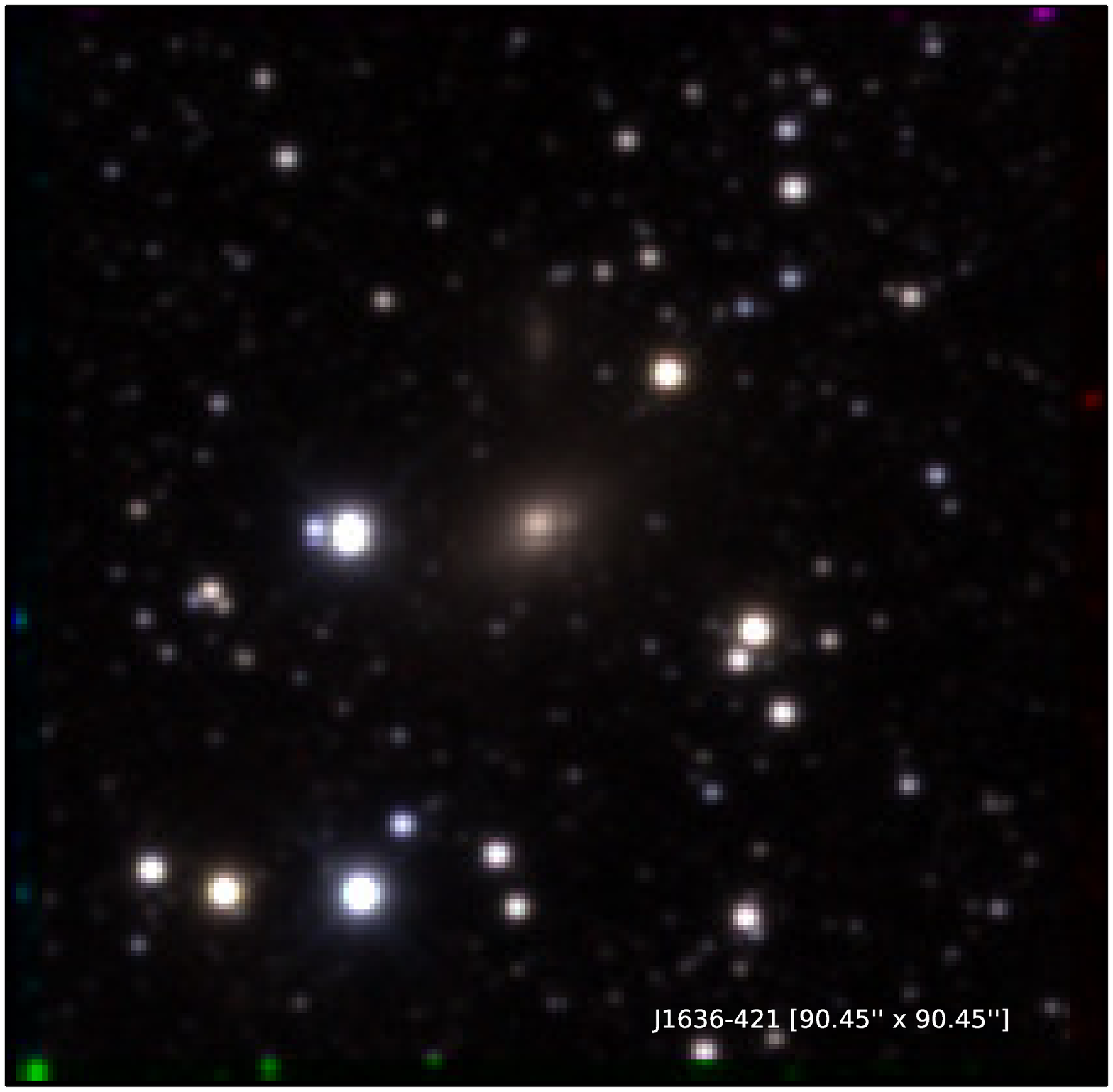}}\\
  \subfloat{\includegraphics[scale=0.17]{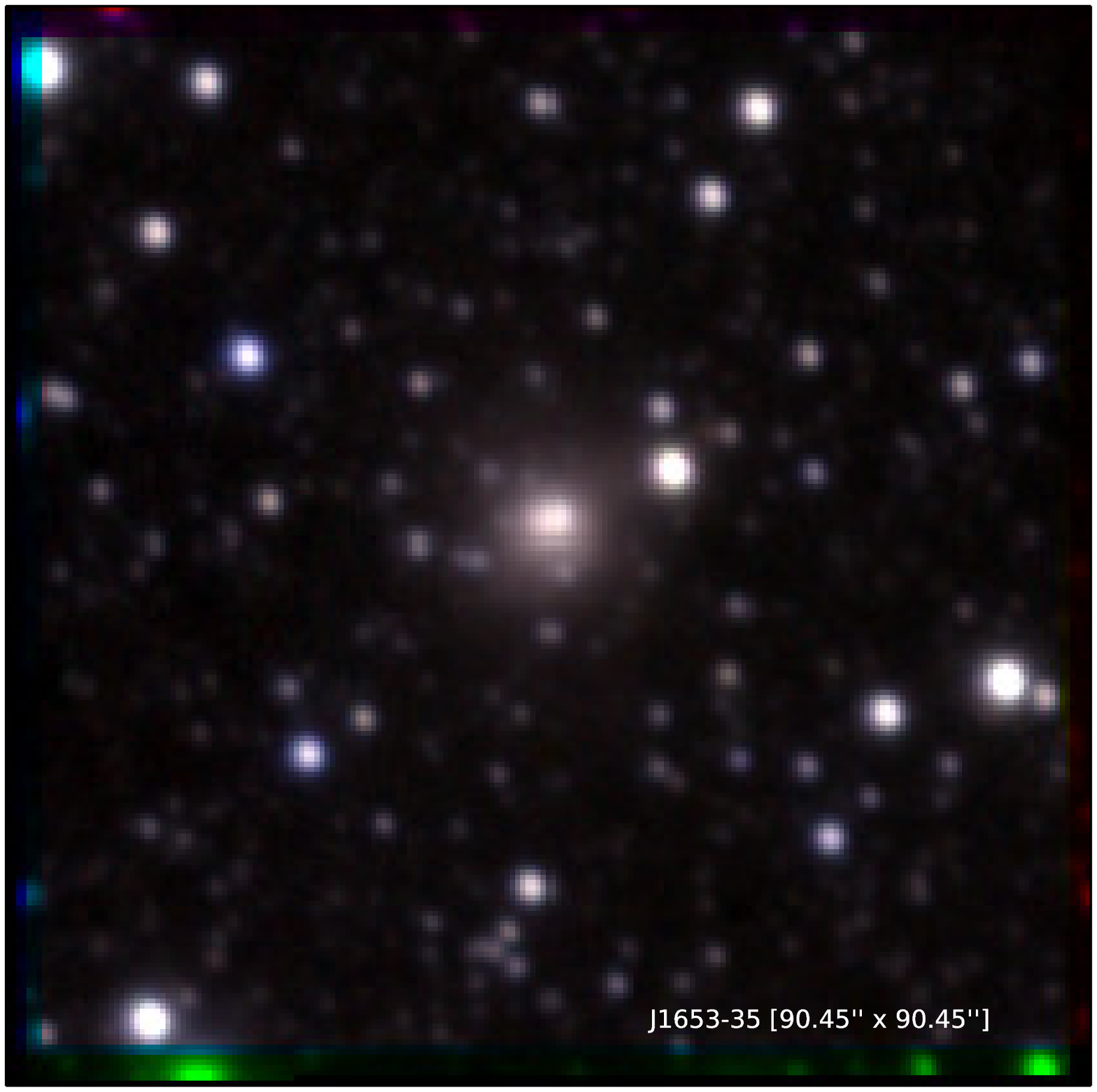}}
 & \subfloat{\includegraphics[scale=0.17]{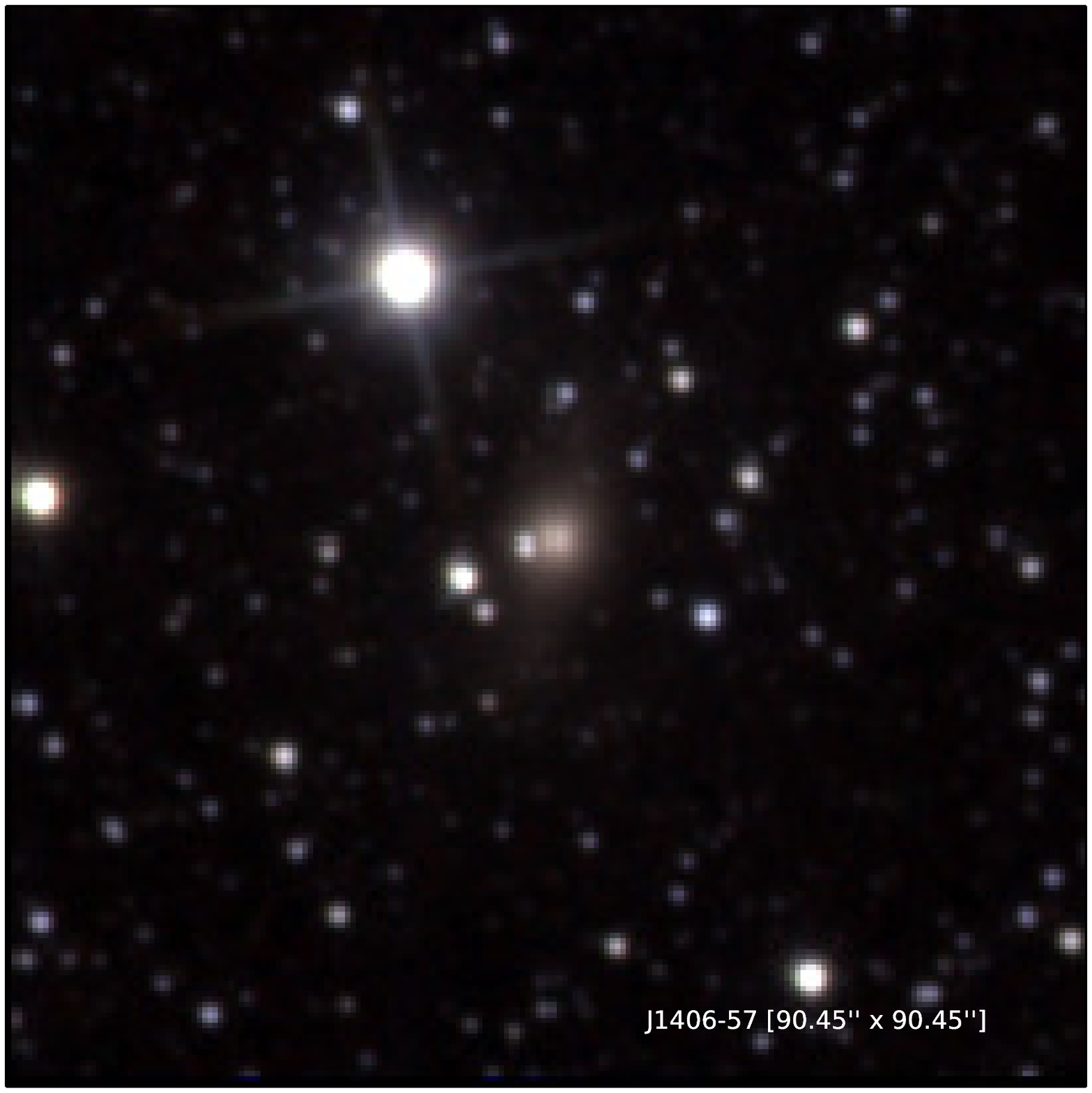}}
 & \subfloat{\includegraphics[scale=0.17]{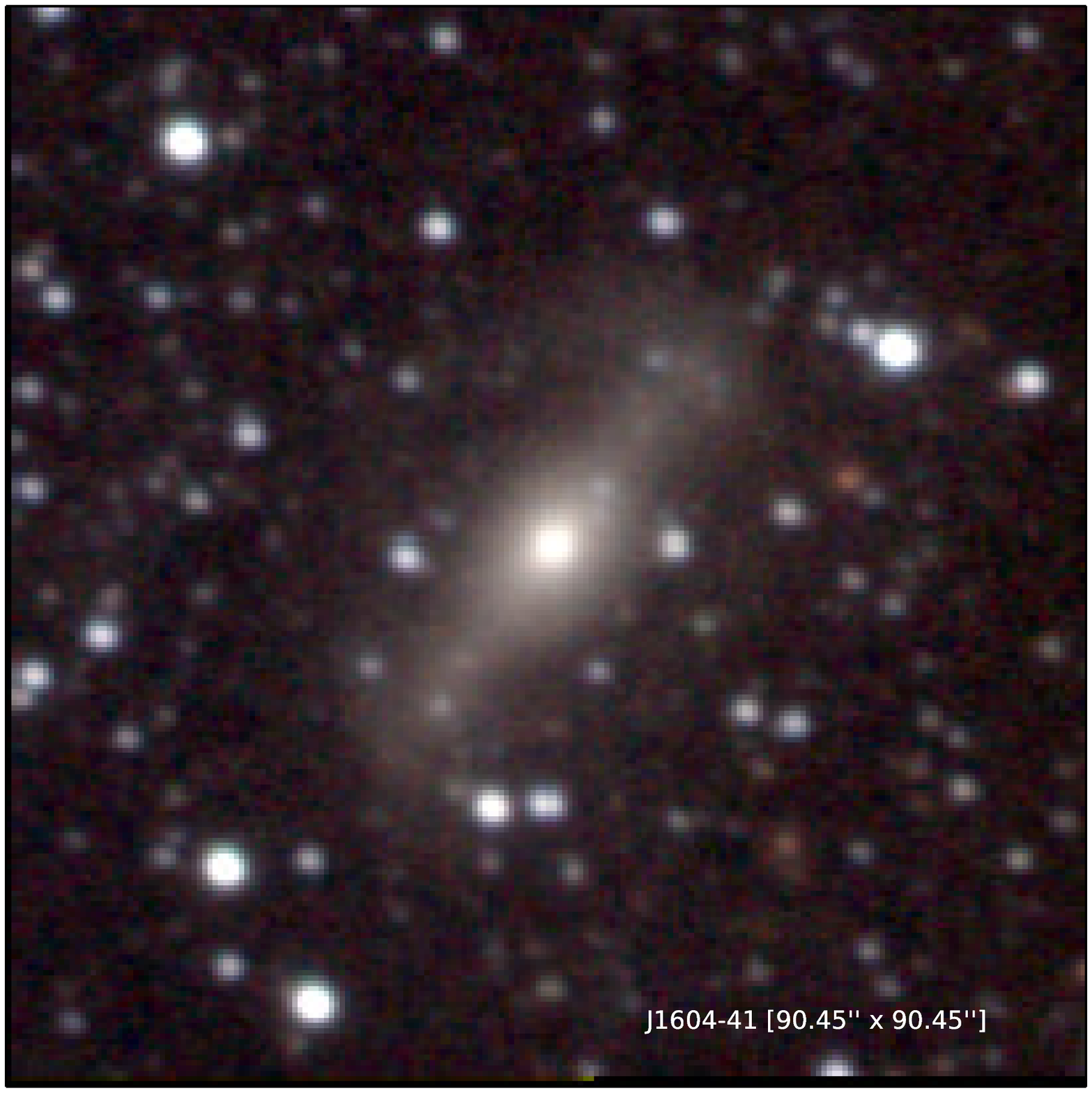}}
 & \subfloat{\includegraphics[scale=0.17]{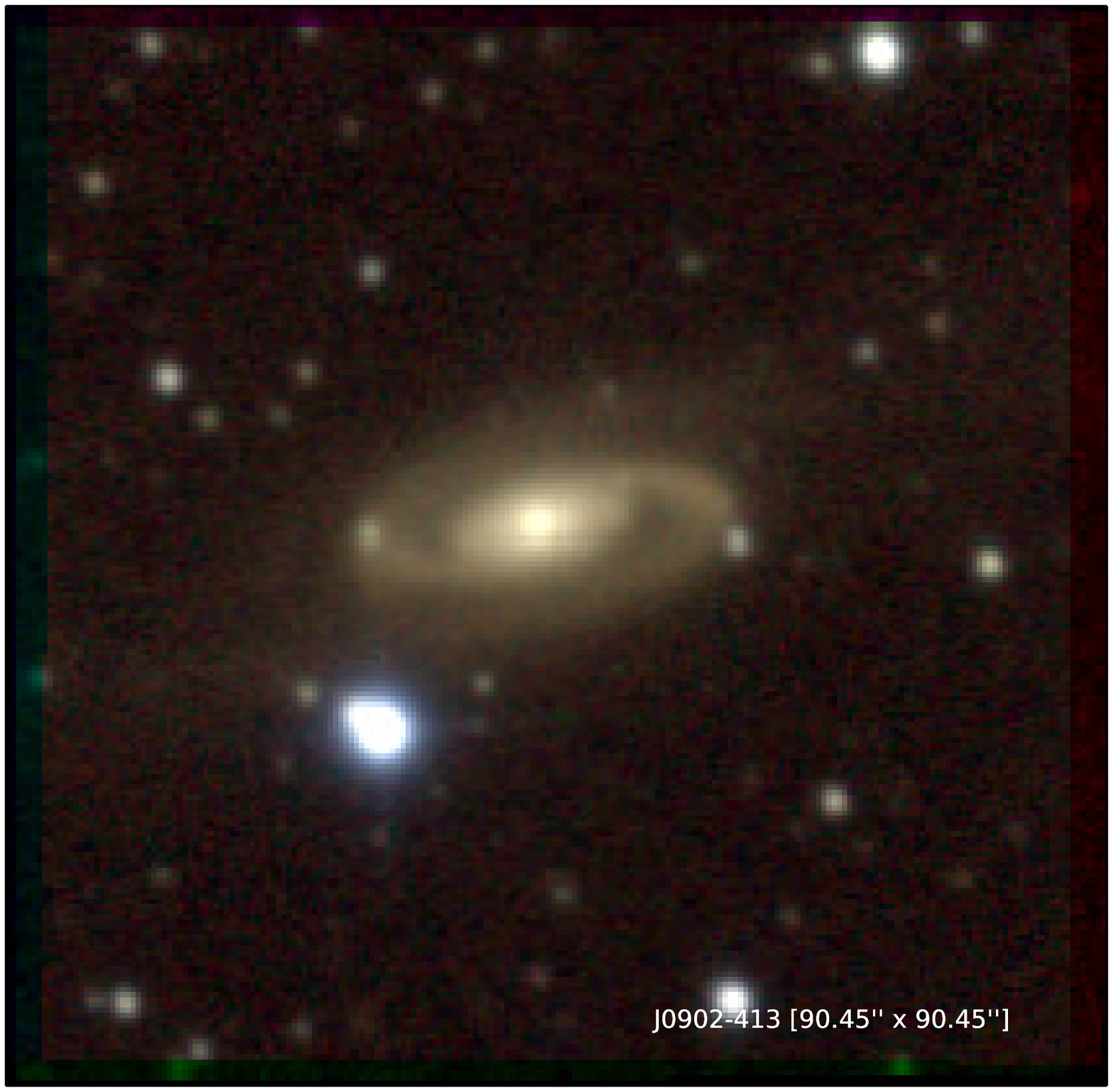}}
 & \subfloat{\includegraphics[scale=0.17]{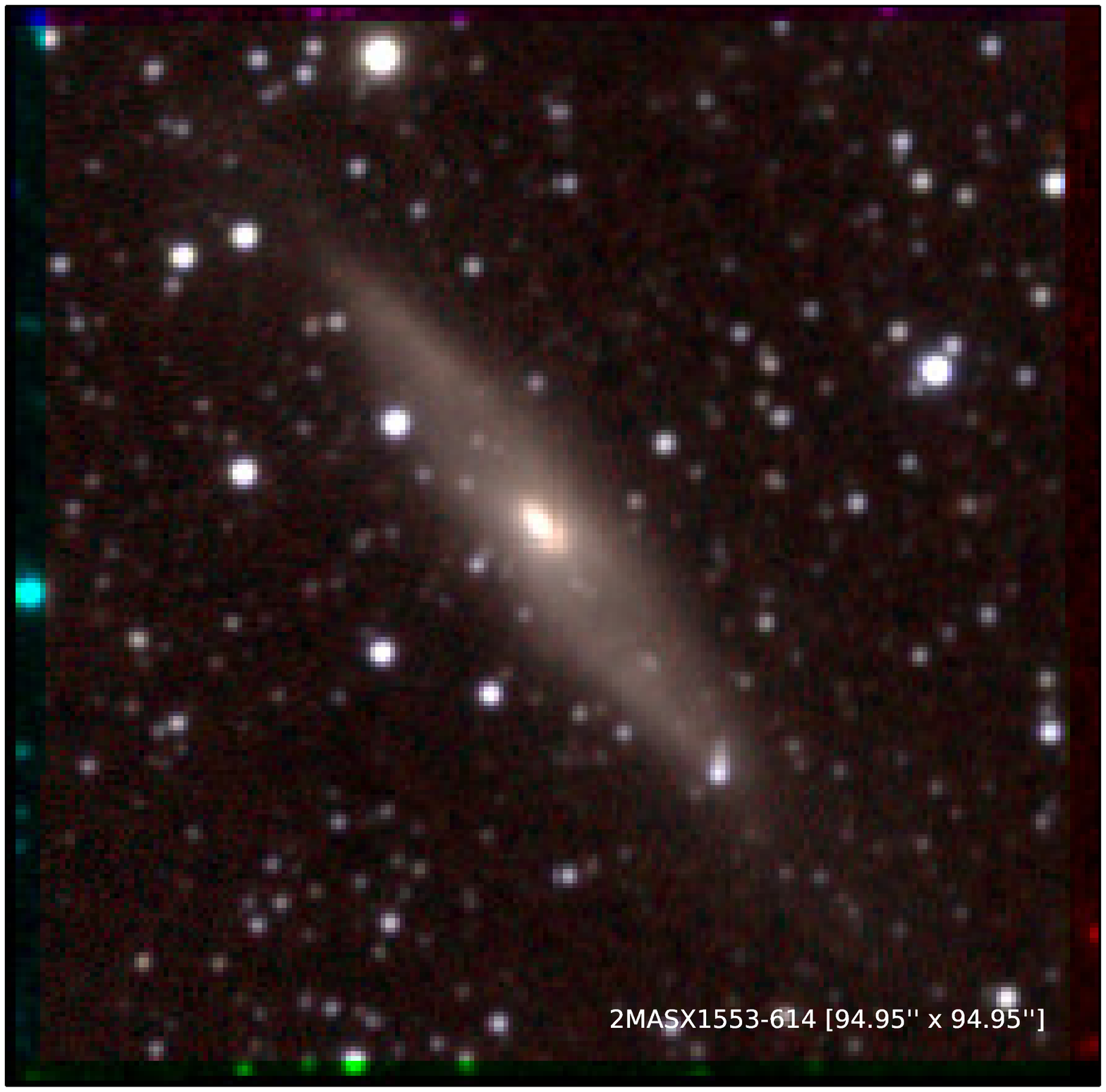}}\\
  \subfloat{\includegraphics[scale=0.17]{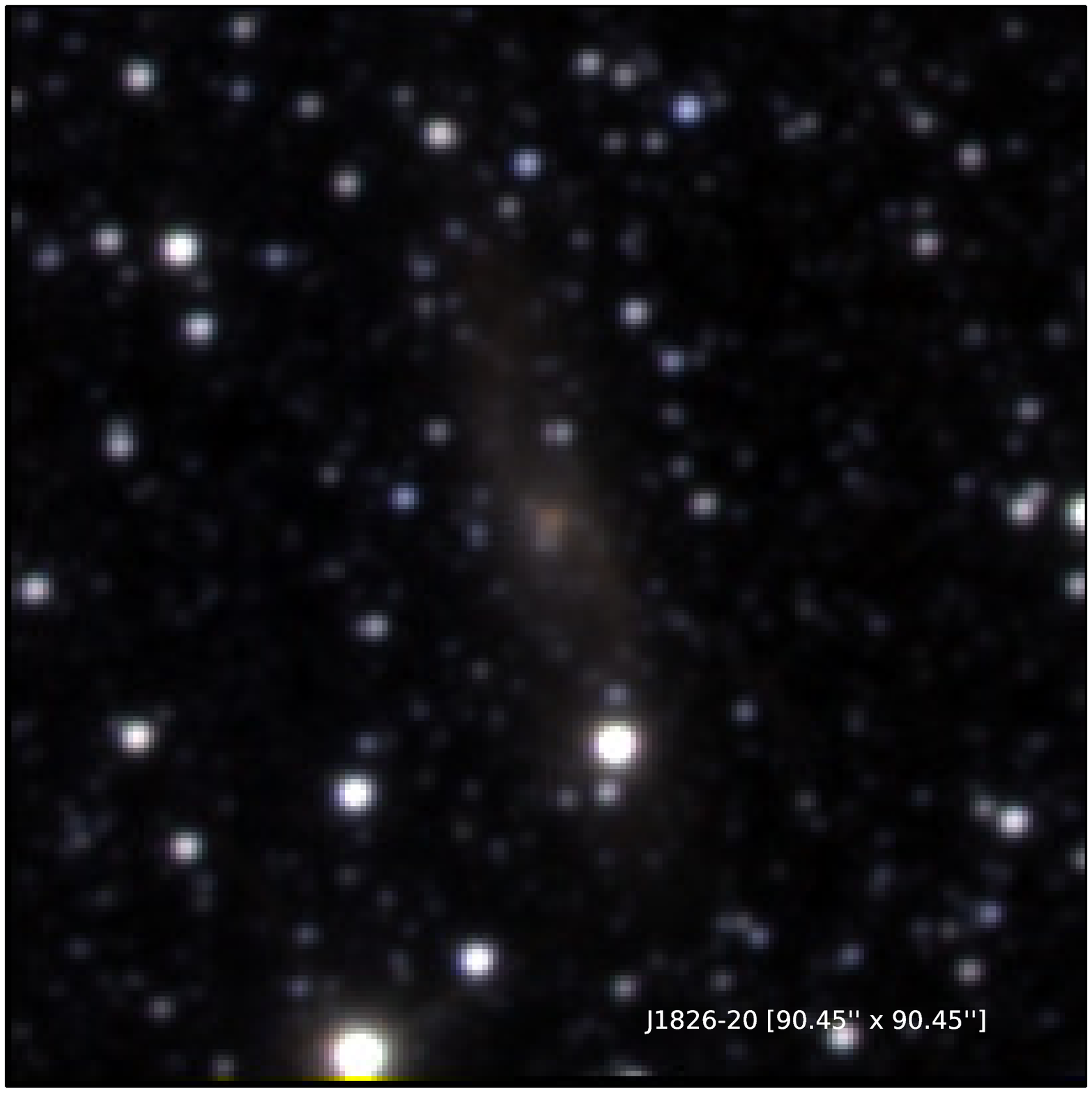}}
 & \subfloat{\includegraphics[scale=0.17]{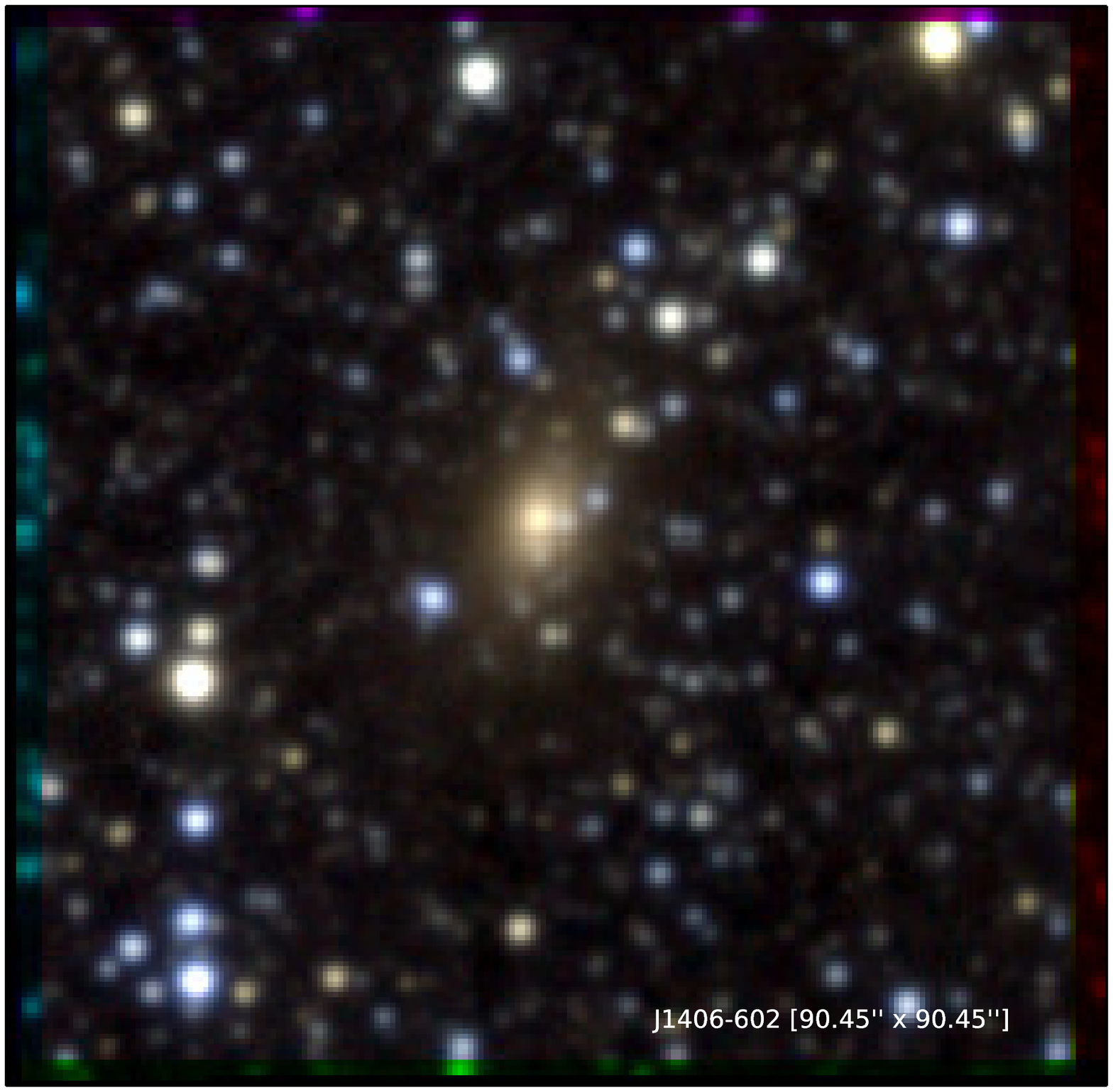}}
 & \subfloat{\includegraphics[scale=0.17]{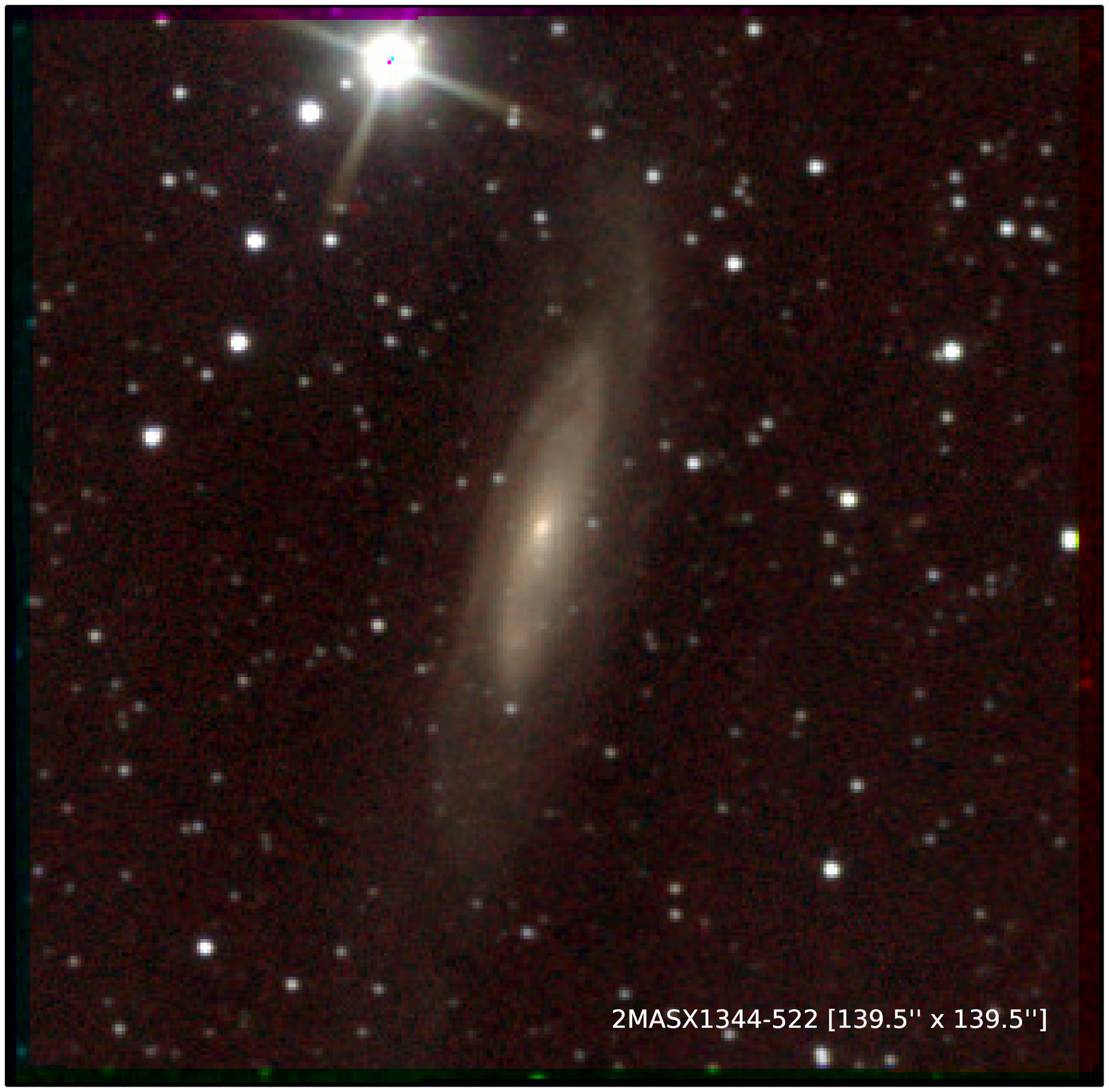}}
 & \subfloat{\includegraphics[scale=0.17]{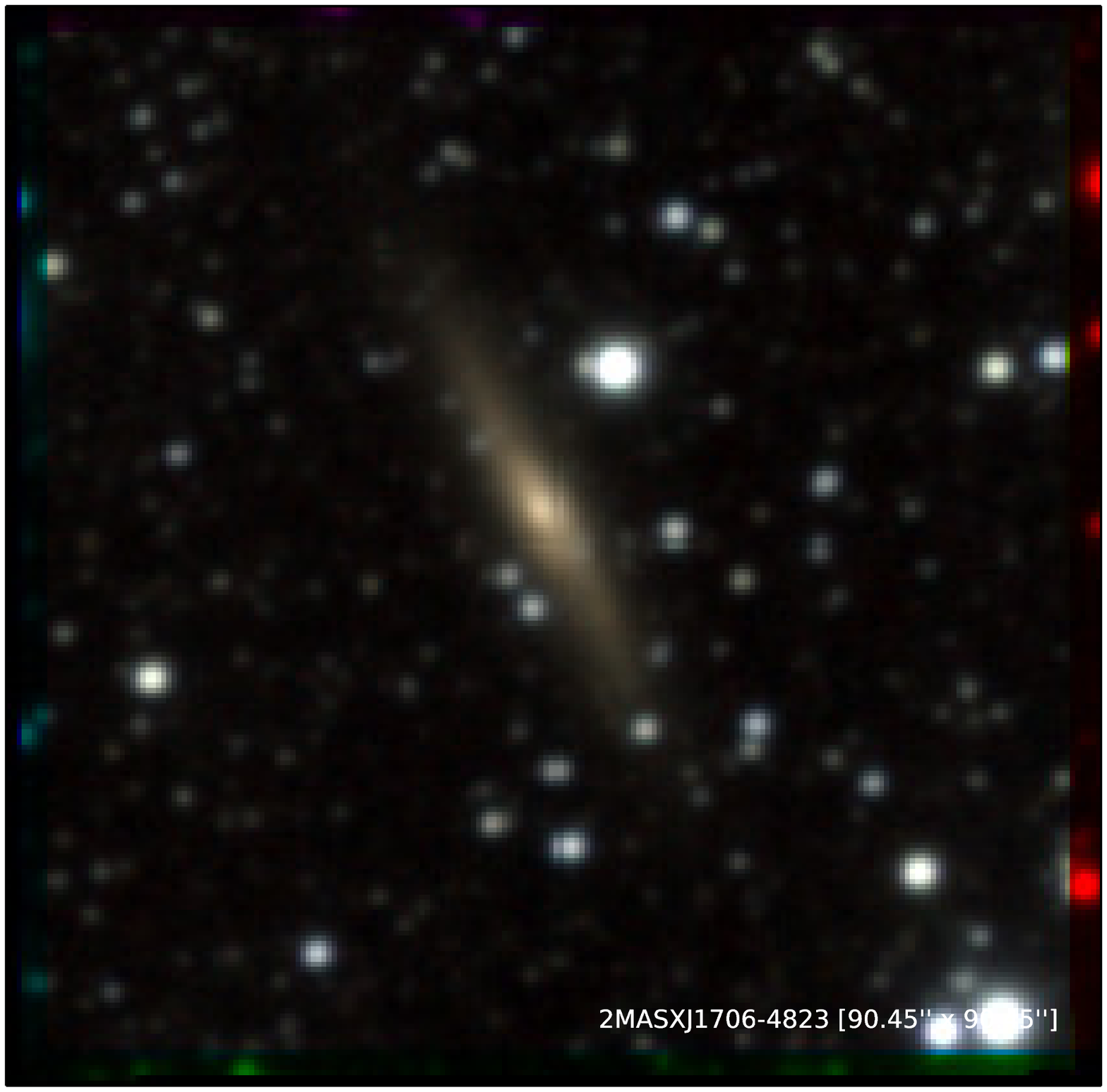}}
 & \subfloat{\includegraphics[scale=0.17]{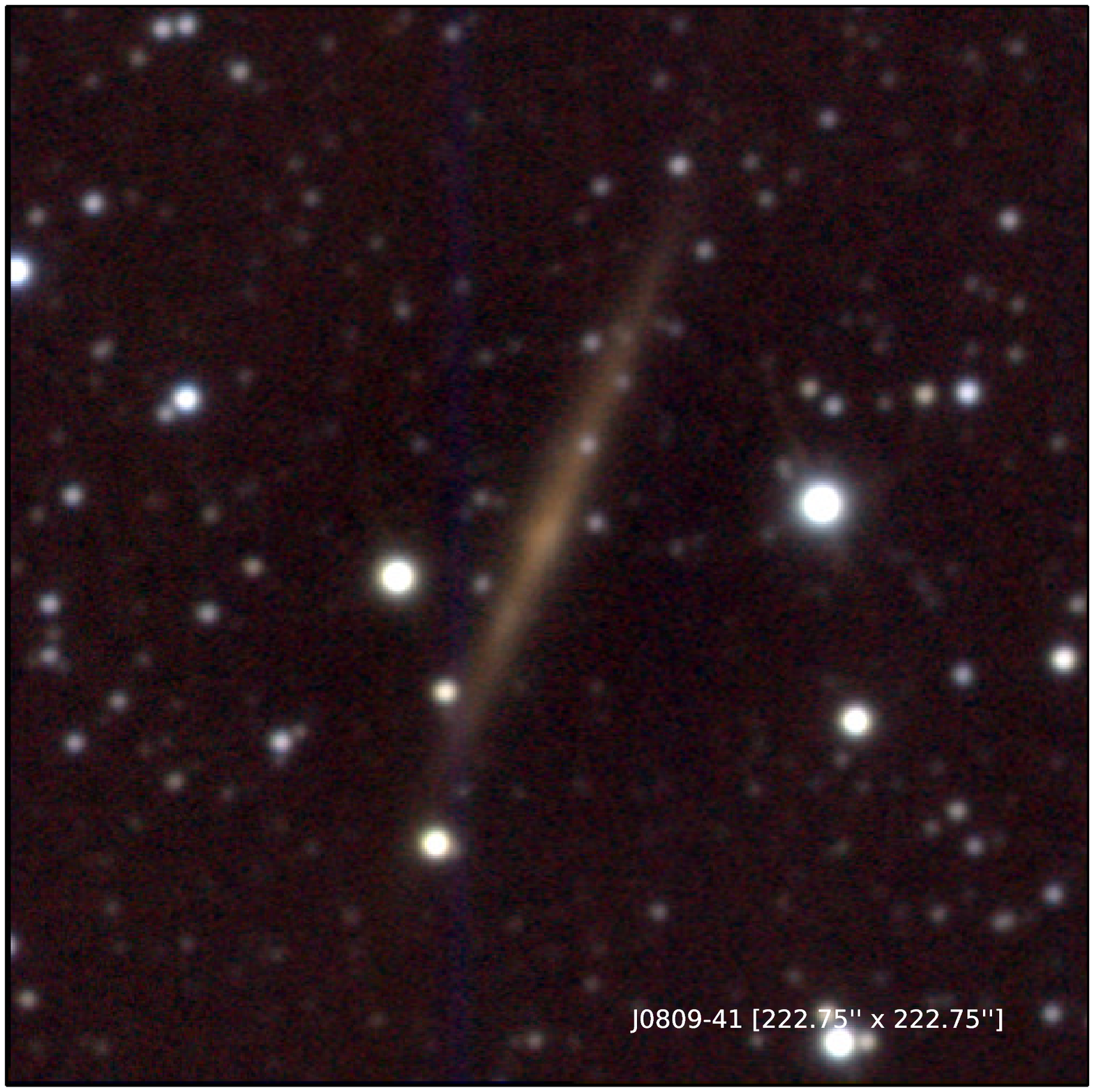}}\\
 \\
\\

 \end{tabular}
 \caption{Continued}
\end{figure*}

\begin{figure*}
\ContinuedFloat
\begin{tabular}{ccccc}
  \subfloat{\includegraphics[scale=0.17]{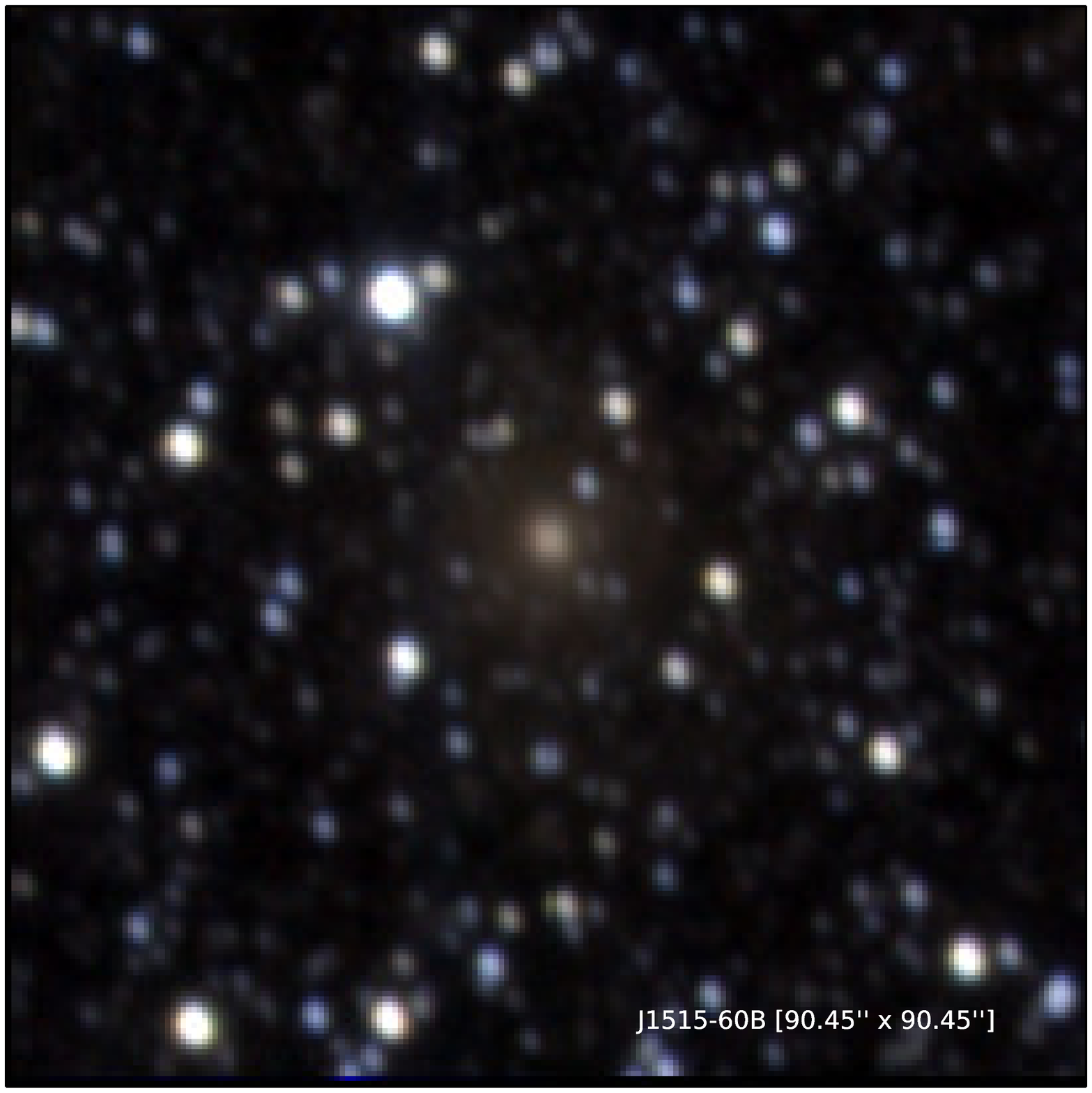}}
 & \subfloat{\includegraphics[scale=0.17]{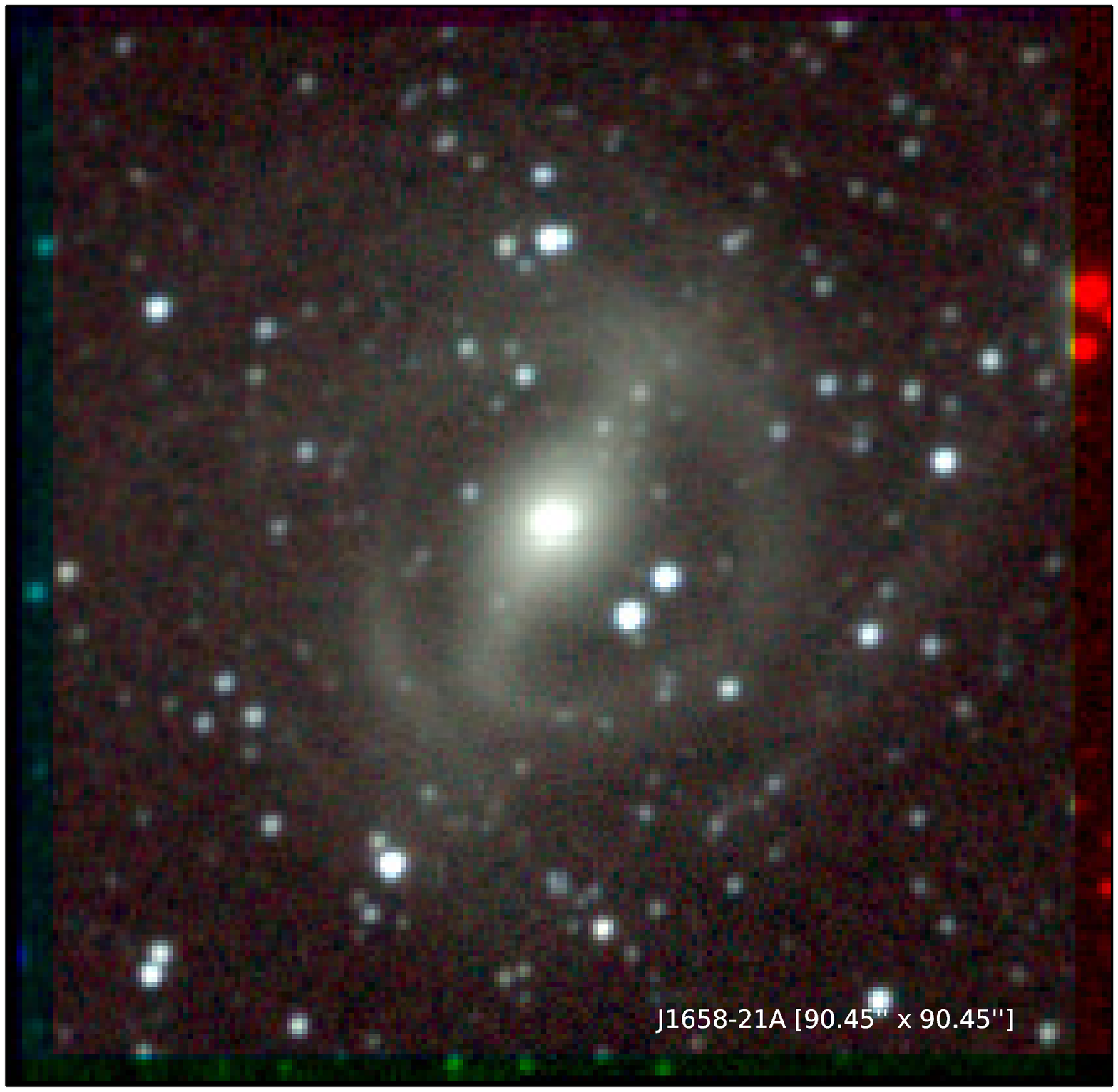}}
 & \subfloat{\includegraphics[scale=0.17]{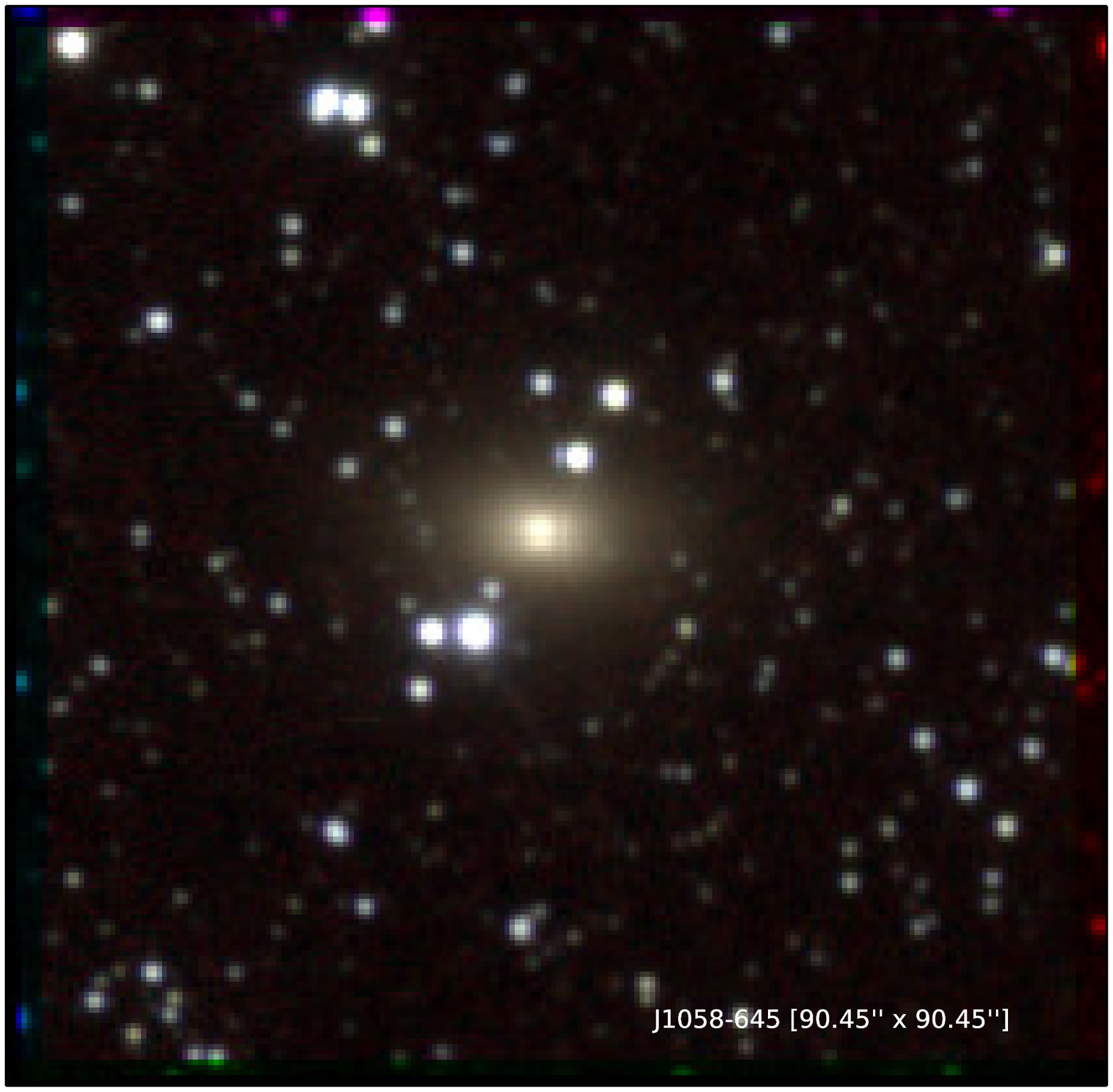}}
 & \subfloat{\includegraphics[scale=0.17]{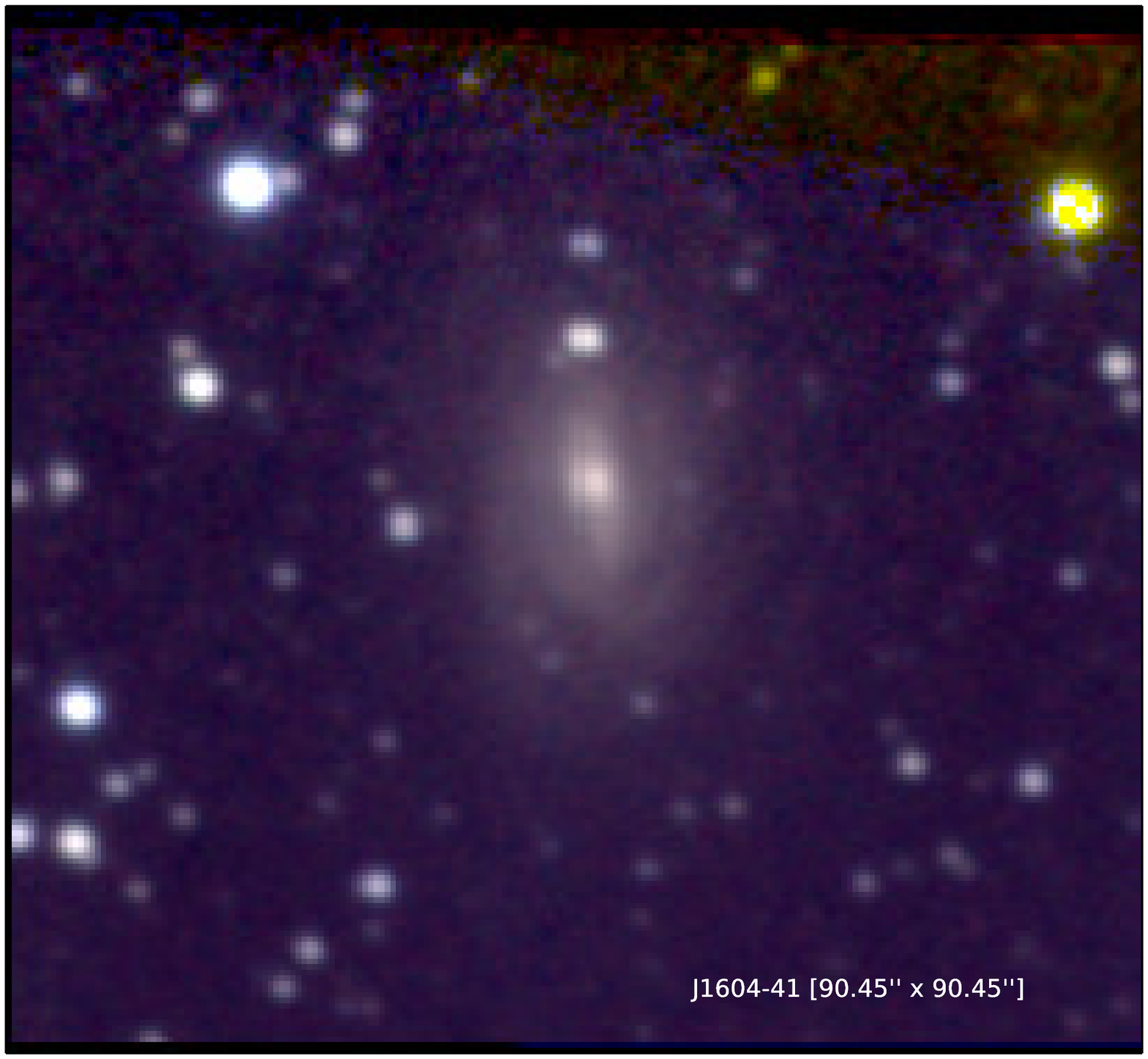}}
 & \subfloat{\includegraphics[scale=0.17]{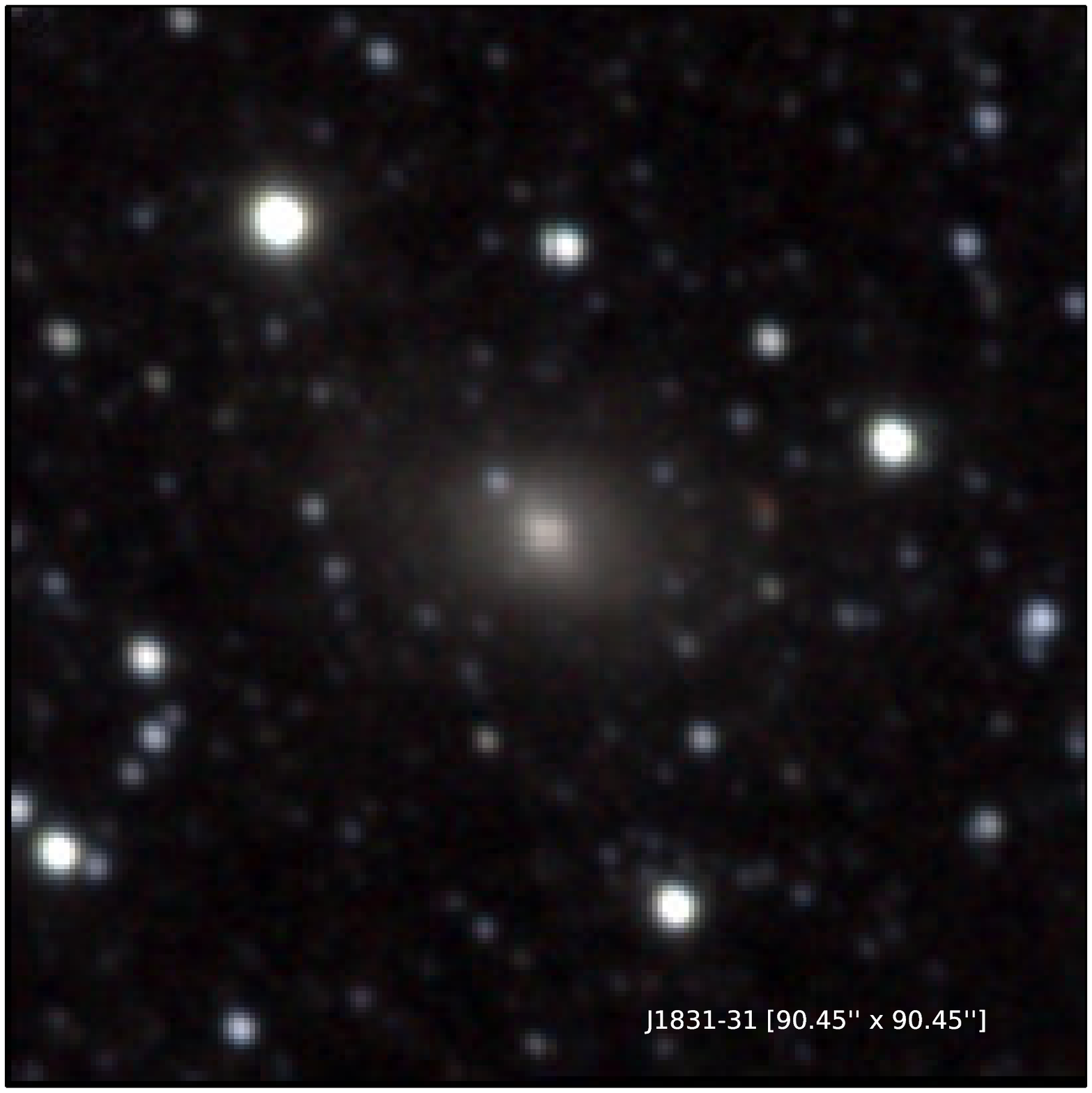}}\\
  \subfloat{\includegraphics[scale=0.17]{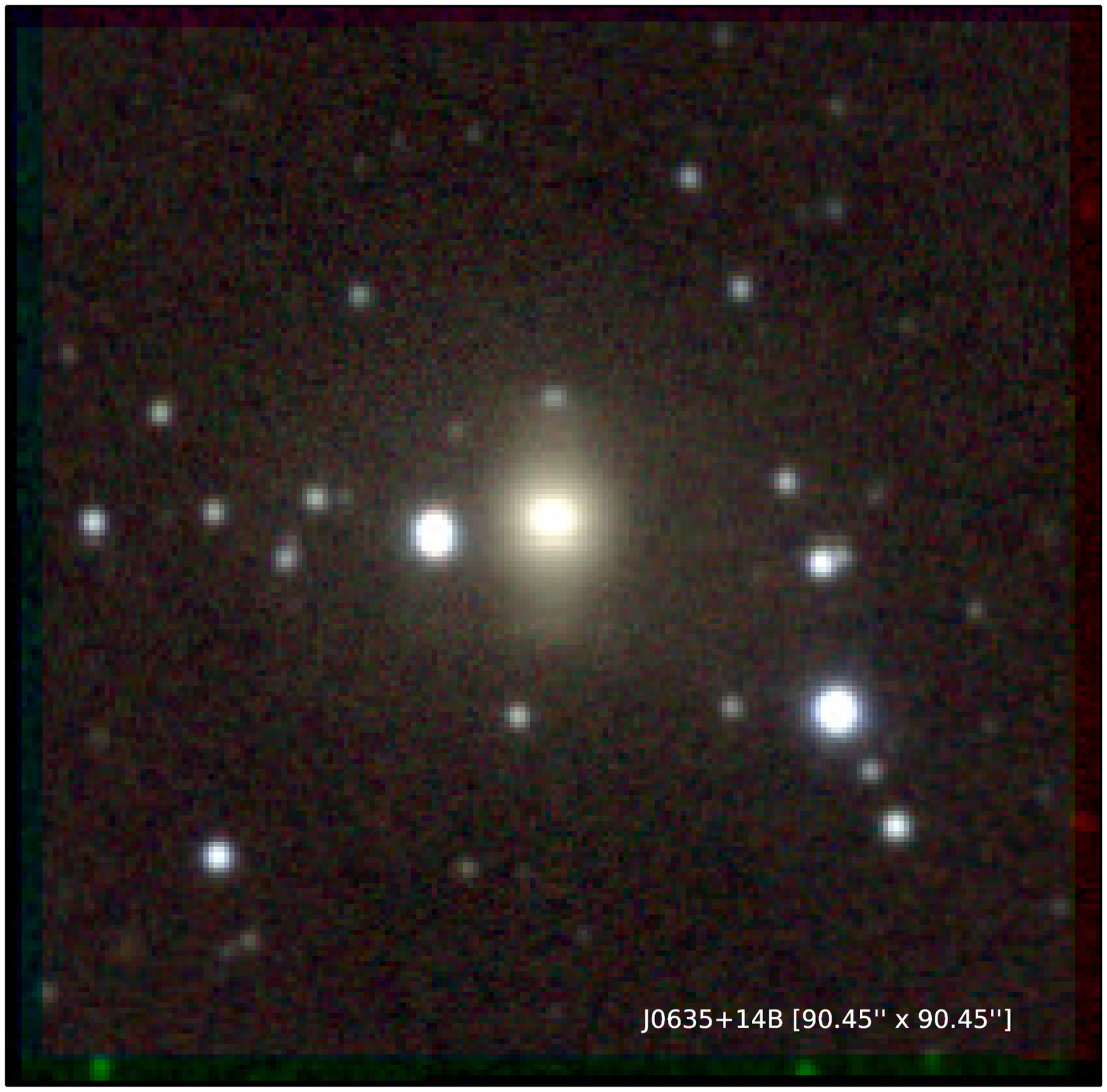}}
 & \subfloat{\includegraphics[scale=0.17]{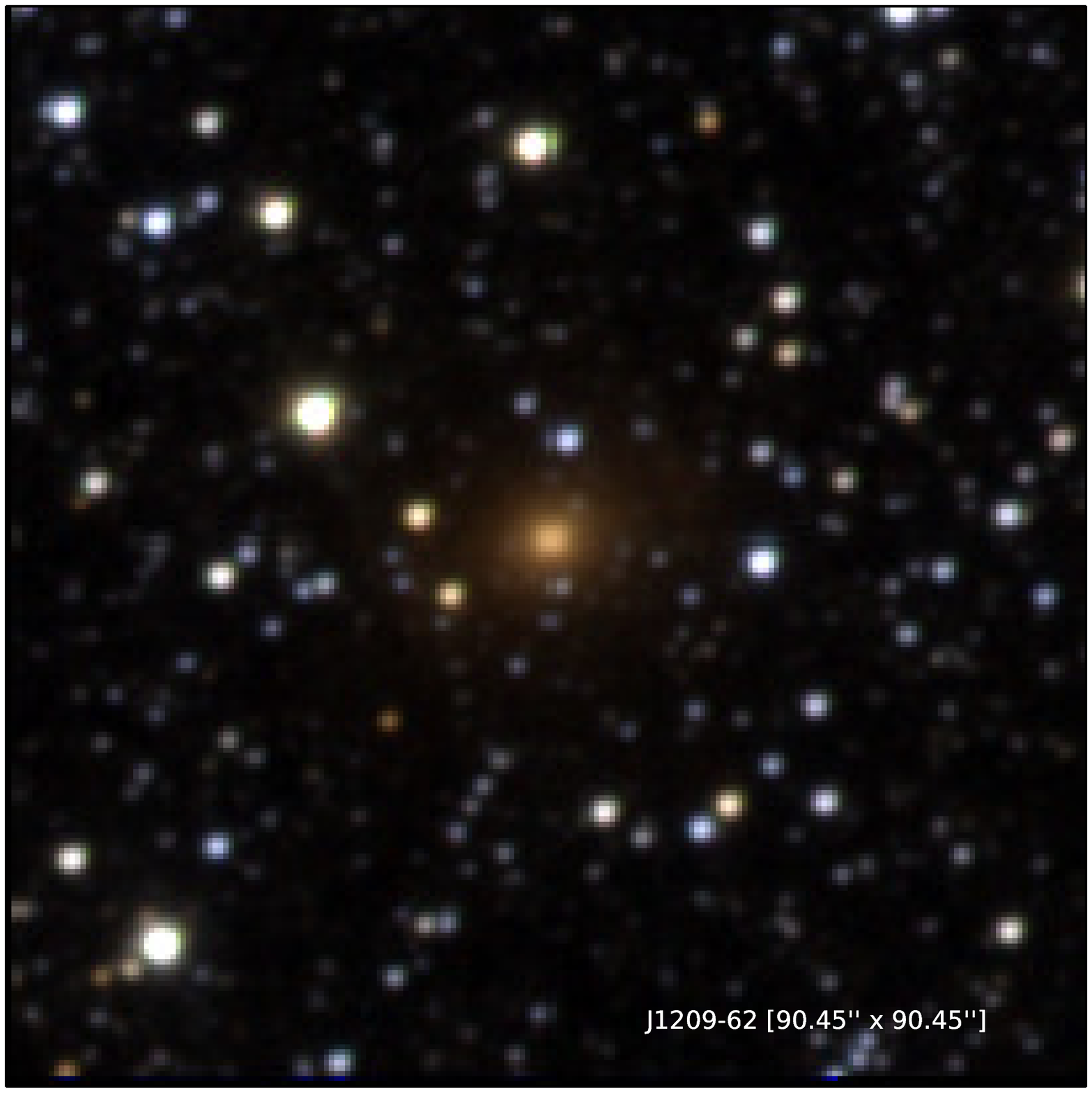}}
 & \subfloat{\includegraphics[scale=0.17]{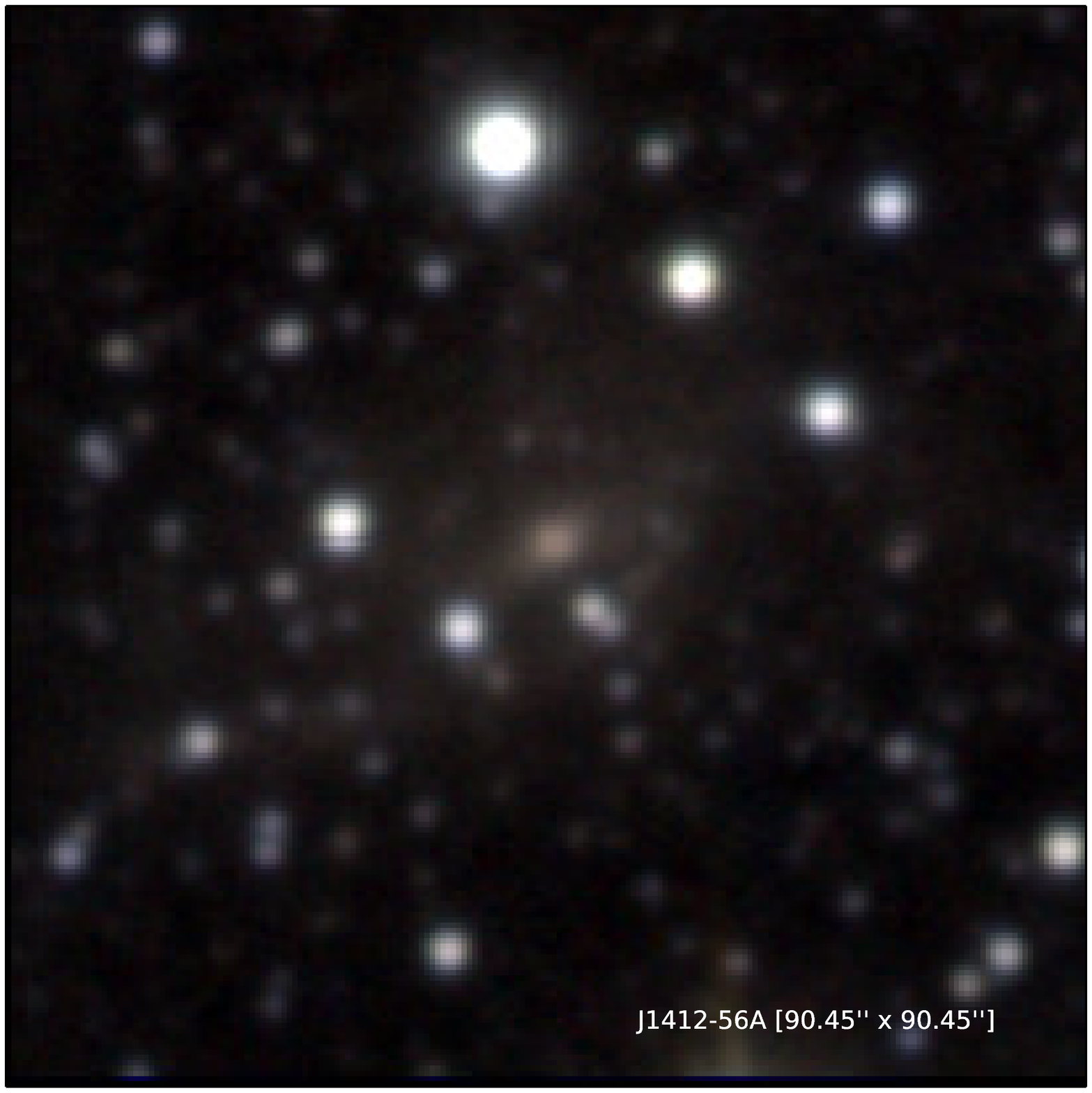}}
 & \subfloat{\includegraphics[scale=0.17]{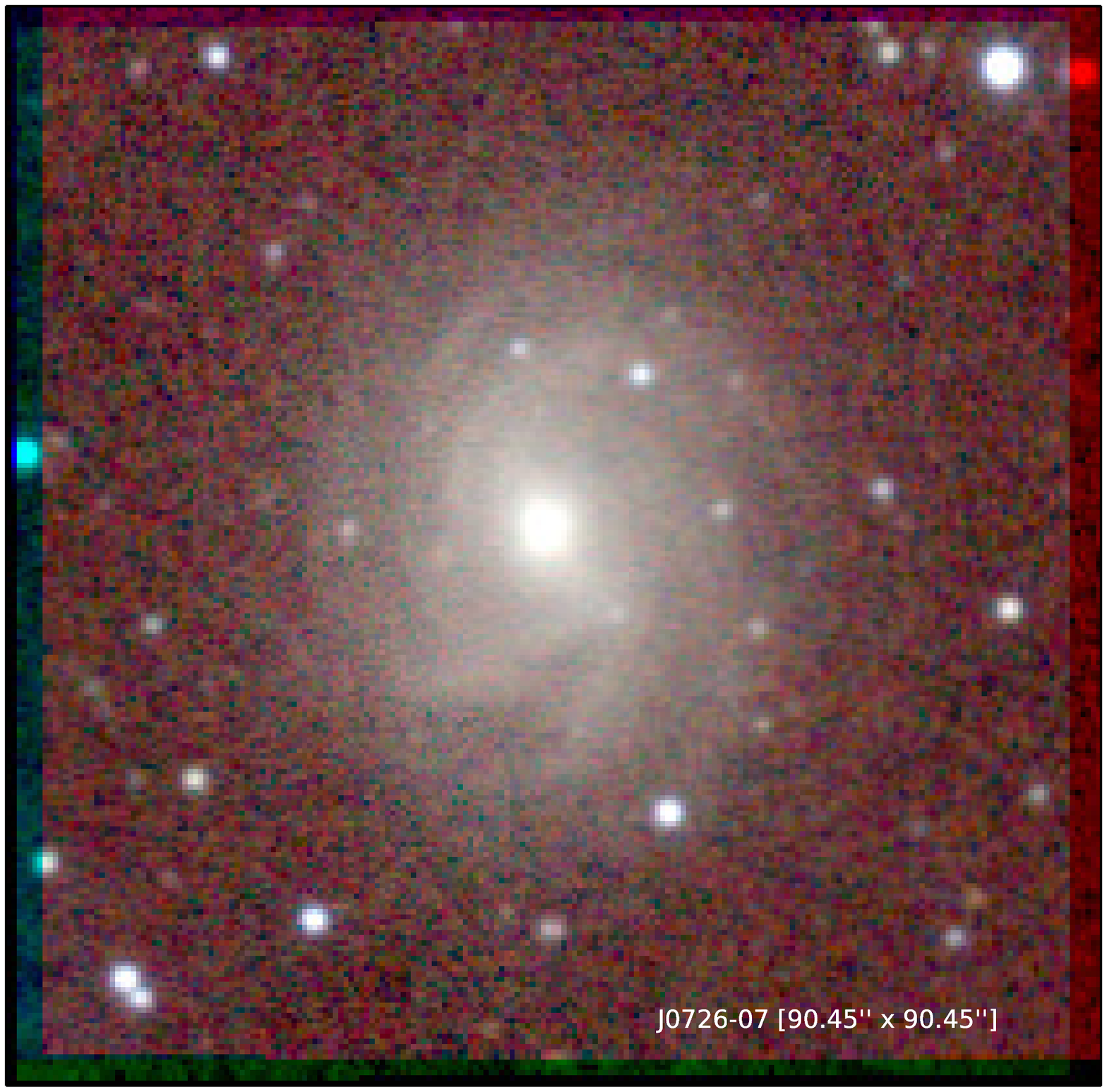}}
 & \subfloat{\includegraphics[scale=0.17]{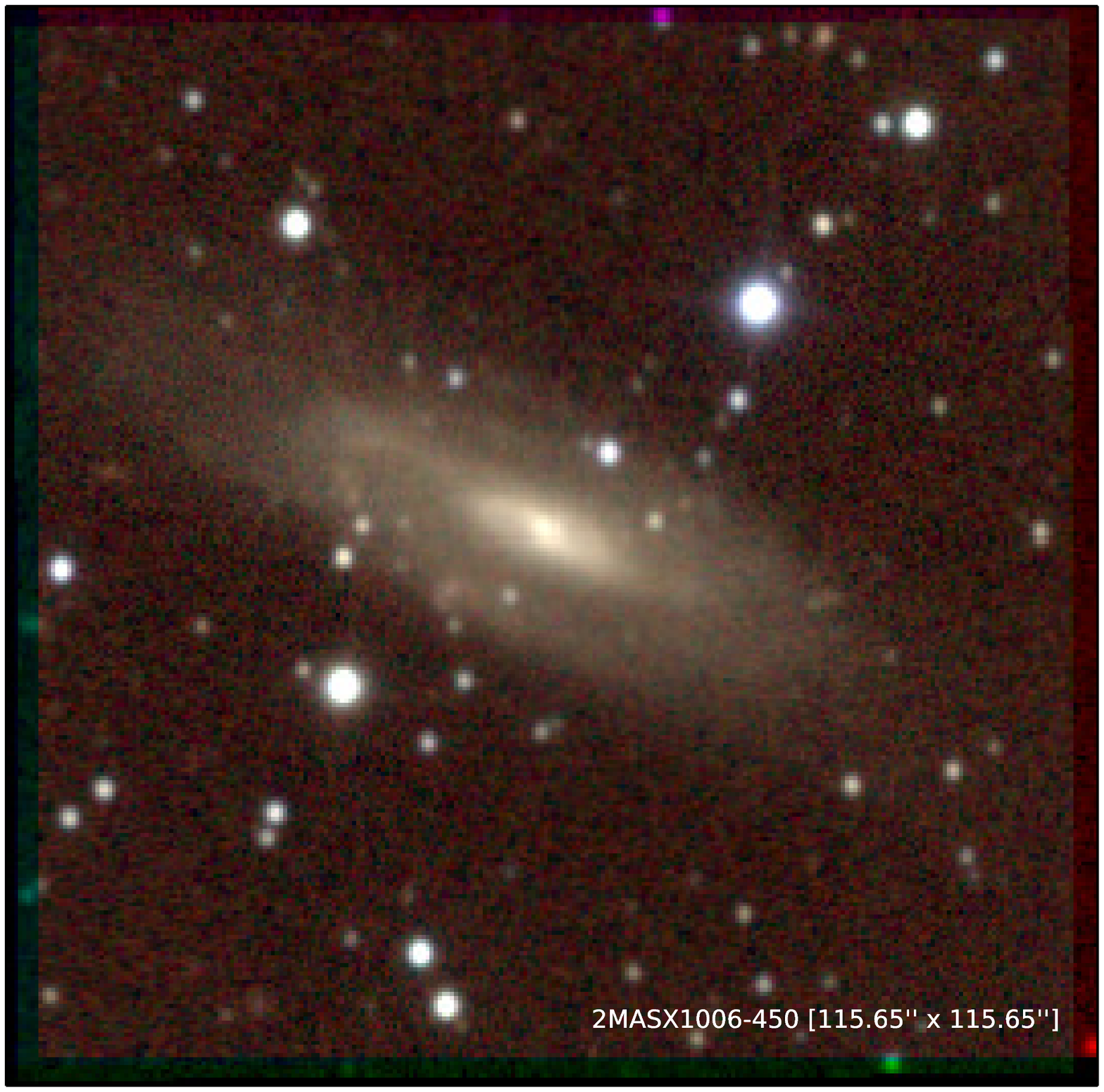}}\\
\end{tabular}
\caption[]{Continued}
\end{figure*}

\subsection{Data presentation}
\begin{figure*}
\begin{center}
\includegraphics[scale=0.5]{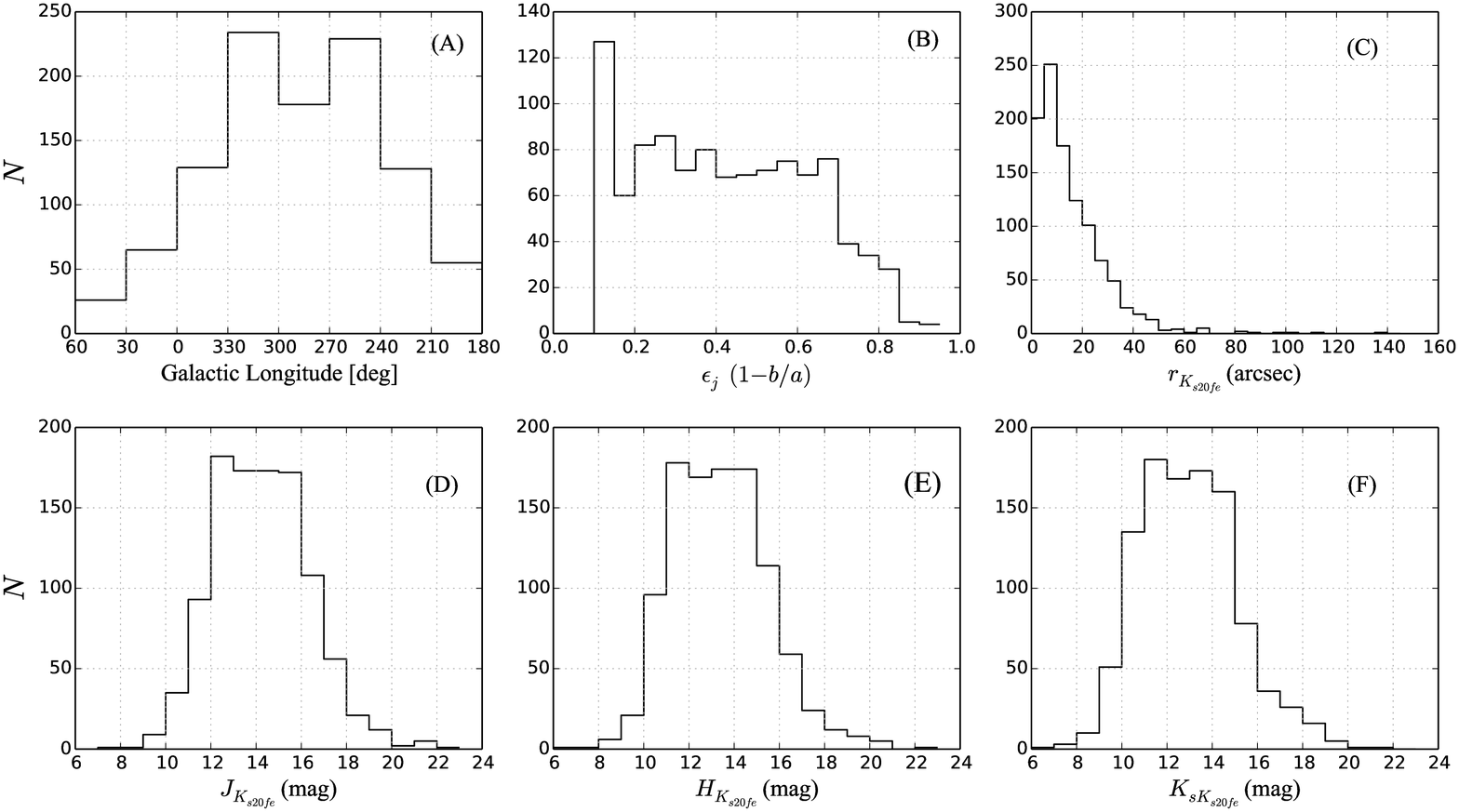}
\caption{Summary of the characteristic photometric properties of the catalogue. The top panels A, B, and C show the distribution as a function of Galactic longitude, the shape, represented by ($\epsilon_J$) the $J$-band ellipticity, and size, represented by ($r_{K_{s20fe}}$) $K_{s20}$ fiducial elliptical aperture semi-major axis, of 1044 galaxies in the catalogue. The bottom panels D, E, and F show the distributions of the $K_{s20}$ fiducial elliptical aperture magnitudes in the $J$, $H$, and $K_s$ bands, respectively.}
\label{summaryfig}
\end{center}
\end{figure*}

Table \ref{summarytab} and Fig. \ref{summaryfig} summarize the characteristic properties of the NIR catalogue. In Table \ref{summarytab} we list the mean, maximum and minimum for a number of parameters. These parameters are as follows: 
\begin{itemize}
\item $J$-band ellipticity ($\epsilon_J$)
\item $K_{s20}$ fiducial elliptical aperture semi-major axis ($r_{K_{s20fe}}$)
\item $J$-, $H$-, and $K_s$-band $K_{s20}$ fiducial elliptical aperture magnitudes ($J_{K_{s20fe}}$, $H_{K_{s20fe}}$, and $K_{sK_{s20fe}}$)
\item Galactic extinction as reported by \cite{2011ApJ...737..103S} ($E(B-V)$)
\item IRSF stellar density of stars brighter than 14 mag in the $K_s$ band ($SD$)
\end{itemize}
Fig. \ref{summaryfig} shows histograms of all parameters except $E(B-V)$ and stellar density which are discussed in detail in the next section as part of the completeness.

\begin{table}
\begin{center}
\caption[]{Summary of the characteristic properties of the catalogue.}
\begin{tabular}{l c c c}
\hline
\hline
Parameter & Mean & Max. & Min. \\ 
\hline
$\epsilon_J$    & 0.42 & 0.90 & 0.10 \\
$r_{K_{s20fe}}$ [arcsec]& 15.57 & 136.08 & 0.51 \\
$J_{K_{s20fe}}$ [mag] & 14.28 & 22.08 & 7.50  \\
$H_{K_{s20fe}}$ [mag] & 13.39 & 22.65 & 6.49  \\
$K_{sK_{s20fe}}$ [mag] & 13.02 & 23.02 & 6.09  \\
$E(B-V)$      [mag]  & 0.92 & 12.76 & 0.09  \\
$SD$            & 3.95 & 5.32 & 3.28  \\
\hline
\end{tabular}
\label{summarytab}
\end{center}
\end{table}

The top panels, A, B, and C, of Fig. \ref{summaryfig} show the distributions in Galactic longitude, shape and size. Panel A illustrates the success of this work in unveiling galaxies hidden behind the MW especially in over-dense regions like Puppis and the GA around $240^\circ < l < 270^\circ$ and $300^\circ < l < 330^\circ$ respectively. The drop in the number of sources toward the Galactic bulge is due to the LV \citep{2008glv..book...13K,2016AJ....151...52S}. Panel B shows a fairly flat distribution of galaxy ellipticities, which is consistent with the expectation of a random sample of disk galaxies. However, for our final TF sample we use only edge-on galaxies after applying the axial ratio correction from \cite{2015MNRAS.447.1618S}. In panel C we plot a histogram of the distribution of the $K_{s20}$ fiducial elliptical aperture semi-major axis. Panel C shows that only three galaxies in our sample have $r_{K_{s20fe}}$ larger than 100 arcsec and 21 galaxies are larger than 50 arcsec. The largest three galaxies are J1514-53, 2MASX1514-464, and J0730-22 which have $r_{K_{s20fe}}$ of 136, 114 and 104 arcsec, respectively.

In panels D, E, and F of Fig. \ref{summaryfig}, we show histograms of the $J$-, $H$-, and $K_s$-band $K_{s20}$ fiducial elliptical aperture magnitudes. Based on the deep NIR survey of the Norma Wall \citep{2010PhDT........33W,2011arXiv1107.1069K}, which used the same instrument with the same setup, we expect similar  completeness limits of $J^\circ=15.6$, $H^\circ=15.3$, and $K_s^\circ=14.8$ mag in the $J$, $H$, and $K_s$ bands, respectively. However, these limits are only valid for $A_{K_s} < 1.0$ mag and $\log(N_{(K_s<14)}/deg^2) < 4.71$. Compared to other NIR surveys, this IRSF survey is 1 mag deeper than 2MASS in the $J$ band and 2 mag deeper in the $K_s$ band. Moreover, it is only 1 mag shallower than the UKIDSS Galactic Plane Survey (GPS; \citealt{2008MNRAS.391..136L}) and VISTA Variables in the Via Lactea (VVV; \citealt{2012AJ....144..127A}) in the $K_s$ band. In Section \label{sect:ukidss} we present a full comparison with UKIDSS GPS galaxies. The three panels (D, E, and F) of Fig. \ref{summaryfig} show that the detection rate drops rapidly for galaxies fainter than 16 mag. However, this survey is not magnitude limited in any sense. Panel F shows that our survey has  63 galaxies brighter than 10 mag. 

\section{Completeness}
\label{sect:completeness}
\begin{figure*}
\begin{center}
\includegraphics[scale=0.55]{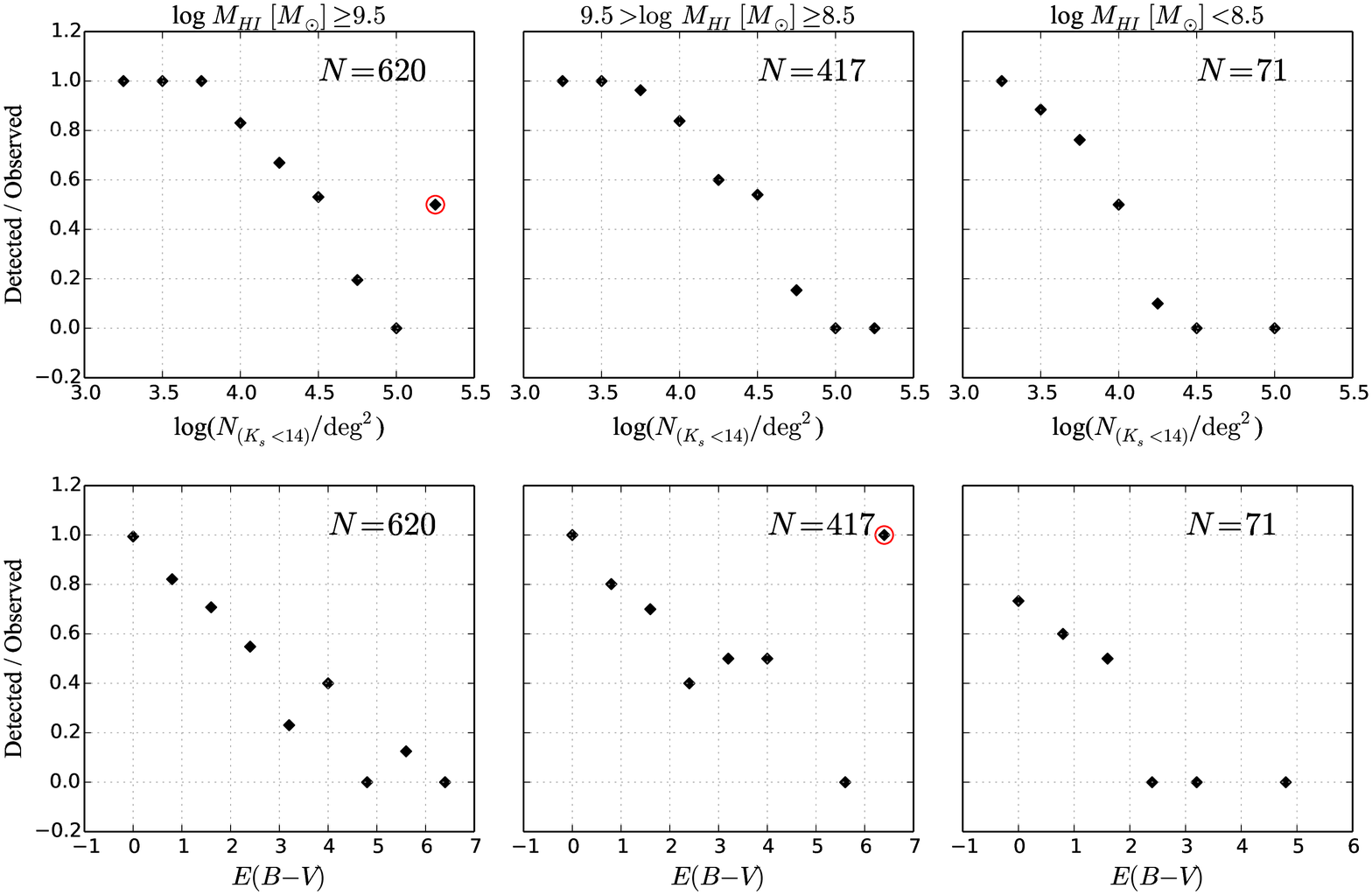}
\caption{The completeness as a function of stellar density and dust extinction for different HI mass ranges. The three columns present different HI mass ranges. The top panels show the completeness as a function of the logarithm of the stellar density of stars brighter than 14 mag in the $K_s$ band. The bottom panels show the completeness as a function of Galactic reddening along the line of sight. The outliers marked with the red circles are due to low numbers of galaxies in these two bins.}
\label{sd_dust_diff_mhi}
\end{center}
\end{figure*}

In this section we discuss the completeness as a function of dust extinction, stellar density and {\HI} mass. We divided the catalogue into three sub-samples according to their logarithm {\HI} mass reported by \cite{2016AJ....151...52S}. The first column in Fig. \ref{sd_dust_diff_mhi} shows galaxies with $\log M_{HI} [M_{\odot}] \geq 9.5$.
The second column presents galaxies within the range of $8.5 \leq \log M_{HI} [M_{\odot}] < 9.5$. The third column shows  galaxies with $\log M_{HI} [M_{\odot}] < 8.5$.

The top panels in Fig. \ref{sd_dust_diff_mhi} show the completeness as a function of the IRSF stellar density of stars brighter than 14 mag in the $K_s$ band for the three sub-samples. The first two columns in the top panel show that detection rate is 100 per cent for massive galaxies ($\log M_{HI} [M_{\odot}] > 8.5$) in regions with stellar density of $\log(N_{K_s<14}/\mathrm{deg}^{2}) \leq 4$. This detection rate  drops to 50 per cent for regions with $\log(N_{K_s<14}/\mathrm{deg}^{2}) > 4$. In contrast, the third column in the top panel shows that the detection rate of the least massive galaxies ($\log M_{HI} [M_{\odot}] < 8.5$) is 75 per cent complete in regions of $\log(N_{K_s<14}/\mathrm{deg}^{2}) \leq 4$ and only 30 per cent in regions with stellar density of $\log(N_{K_s<14}/\mathrm{deg}^{2}) > 4$. Most of these least massive galaxies are dwarfs and will be excluded from the TF analysis because they have the highest scatter in the TF relation.

Similar conclusions can be drawn from the bottom panels of Fig. \ref{sd_dust_diff_mhi} which show the completeness as a function of Galactic reddening along the line of sight \citep{2011ApJ...737..103S}. The first two columns in the bottom panel show that the detection rate of the massive galaxies  is nearly 90 per cent  in regions of $E(B-V)\leq1$ mag ($A_V \leq 3.1$ mag). Furthermore, we can still detect massive galaxies up to $E(B-V) = 7$ mag ($A_V=21.7$ mag). On the contrary, the third column in the bottom panel shows that the detection of  the least massive galaxies is not complete anywhere, not even in regions with $E(B-V)\leq1$ mag ($A_V \leq 3.1$ mag).

We note that the measured photometric parameters (e.g, magnitude, size and shape) will be also affected by these trends and should be corrected before use in the TF analysis. \cite{2010MNRAS.401..924R} discuss the effect of dust extinction on magnitude and size of galaxies and provide a correction model for extinction values up to $A_{K_s} = 3$ mag ($A_V = 25$ mag). While \cite{2015MNRAS.447.1618S} simulate the effect of dust extinction on the shape (ellipticity) of galaxies and also provide a correction model to reproduce the intrinsic axial ratio from the observed value up to extinction levels of $A_J = 3$ mag ($A_V = 11$ mag). These corrections will be used to correct the magnitudes, sizes and shapes of galaxies before use in the TF analysis.

\section{Counterparts and Comparisons}
\label{sect:counterparts}
In this section we discuss the counterparts of our survey. We present the confirmed HIZOA counterparts then check for counterparts in the 2MASX \citep{2000AJ....119.2498J} and UKIDSS GPS \citep{2008MNRAS.391..136L} surveys. We also present a comparison of our photometry with both the shallower 2MASX and the deeper UKIDSS GPS surveys.

\subsection{HIZOA counterparts}
The pixel size of the final HIZOA cubes of 4 arcmin  in RA and DEC makes the IRSF perfect for the follow-up observations given its $8.6 \times 8.6$ arcmin$^2$ field of view (after dithering). Centering the NIR camera on the HIZOA position has a high probability of locating the counterpart in the image. Thus, the detection of these {\HI} sources depends only on their {\HI} mass and the stellar density and dust extinction of the region in which they lie. Some NIR fields contain more than one possible counterpart to the {\HI} galaxy and therefore need more attention. All sources identified as {\HI} sources in the NIR fields were inspected by eye and information from their {\HI} profiles used to identify the {\HI} counterpart. A galaxy with double-horn {\HI} profile was normally identified as an edge-on galaxy in the NIR image, while a galaxy with a Gaussian profile was usually identified with a face-on galaxy in the NIR image. 

The final NIR catalogue contains counterpart galaxies to 674 sources from all three HIZOA catalogues (HIZOA-S; \citealt{2016AJ....151...52S}, HIZOA-N; \citealt{2005AJ....129..220D}, GB; \citealt{2008glv..book...13K}). A single counterpart was found for 527 galaxies, while more than one counterpart was found for 147 galaxies.

\subsection{2MASS counterparts}
\begin{figure}
\begin{center}
\includegraphics[scale=0.42]{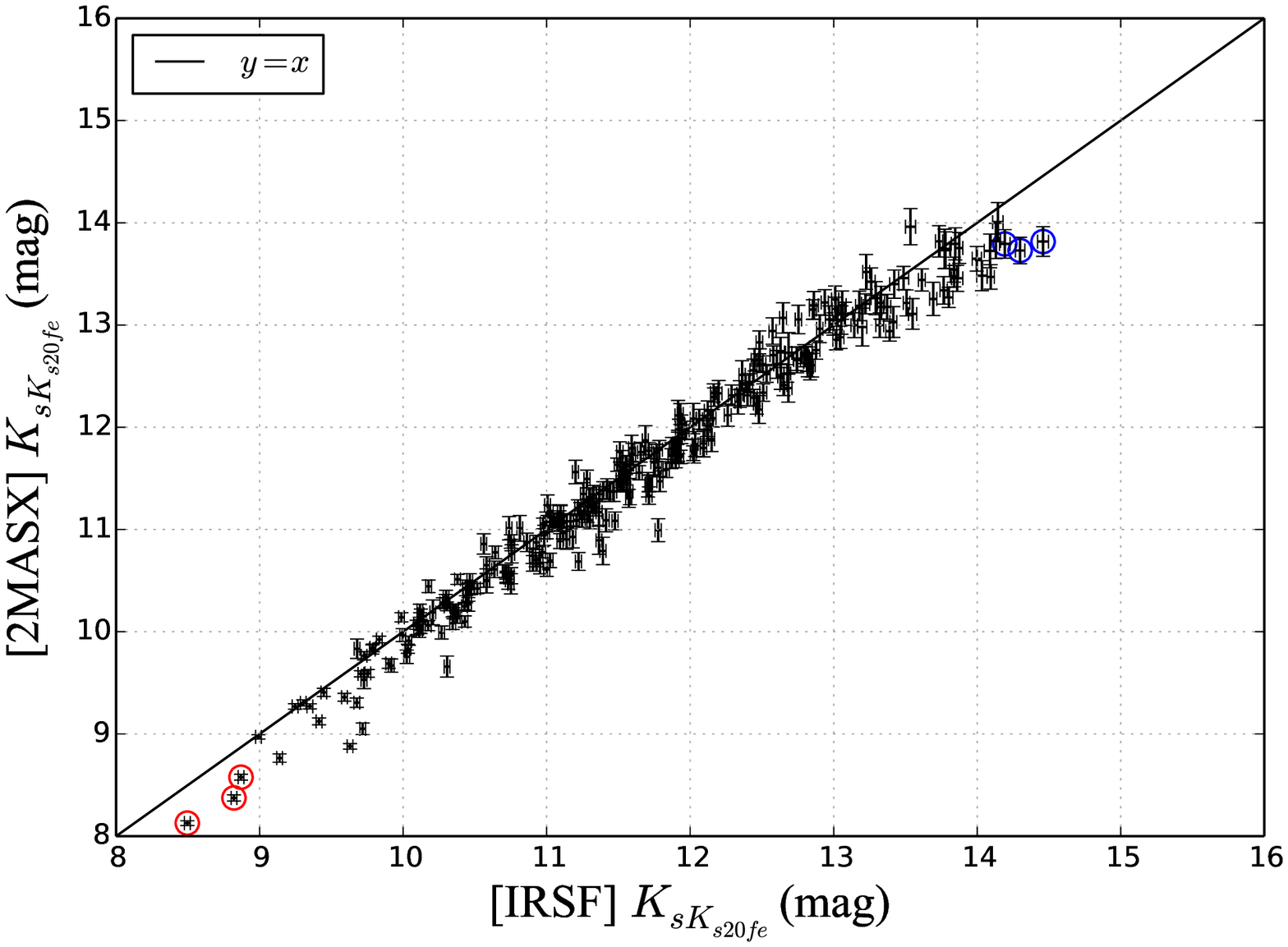}
\caption{A comparison between the $K_s$-band $K_{s20}$ fiducial elliptical aperture magnitudes in this catalogue and the same parameter reported in the 2MASX catalogue.}
\label{IRSF_vs_2MASX}
\end{center}
\end{figure}

Of the 1044 NIR galaxies in the final catalogue, 285 have counterparts in the 2MASX catalogue (180 HIZOA plus 105 2MASS). We used a search radius of only 1 arcsec because of the high positional accuracy of both 2MASS and the IRSF. In Fig. \ref{IRSF_vs_2MASX} we compare our measured $K_s$-band $K_{s20}$ fiducial elliptical aperture magnitudes with the same parameter from the 2MASX catalogue \citep{2000AJ....119.2498J,2003AJ....125..525J} for these 285 galaxies. 

Fig. \ref{IRSF_vs_2MASX} shows good agreement between the $K_{s20}$ measured for this catalogue and the same parameter reported in the 2MASX catalogue. Small systematic deviations are visible for both faint galaxies ($> 14$ mag) marked as blue circles and bright galaxies ($< 9$ mag) marked as red circles. The deviation for bright galaxies is due to the difference of the pixel scale between these two instruments. The IRSF has pixel scale of 0.45 arcsec pixel$^{-1}$ compared to 2.0 arcsec pixel$^{-1}$ for the 2MASX survey. Small, faint stars superimposed on bright galaxies can not be resolved by 2MASX. Thus, the magnitudes of these galaxies are over-estimated because these stars are not subtracted from the image before measuring the photometry. The high resolution of the IRSF instrument leads to more effective star-subtraction and thus more accurate photometry which is vital when working in the ZOA. The deviation for faint galaxies is a bias due to the completeness limit of 2MASX. The scatter should to be on both sides of the one-to-one line but there are no 2MASX galaxies fainter than 14 mag.

\subsection{UKIDSS counterparts}
\label{sect:ukidss}
The UKIDSS GPS \citep{2008MNRAS.391..136L} overlaps with the HIZOA survey in the northern extension published by \cite{2005AJ....129..220D}. We used the publically accessible \textit{UKIDSS DR8 plus} data release to search for counterparts. Given the high positional accuracy of both the IRSF and UKIDSS GPS, the minimum available search radius  of 3 arcsec was used. We found 30 confirmed counterparts in the UKIDSS GPS survey. A modified version of our IRSF photmetry pipeline was used to consistently measure the photometric parameters for these galaxies from the UKIDSS GPS images. 

\begin{figure}
\begin{center}
\includegraphics[scale=0.42]{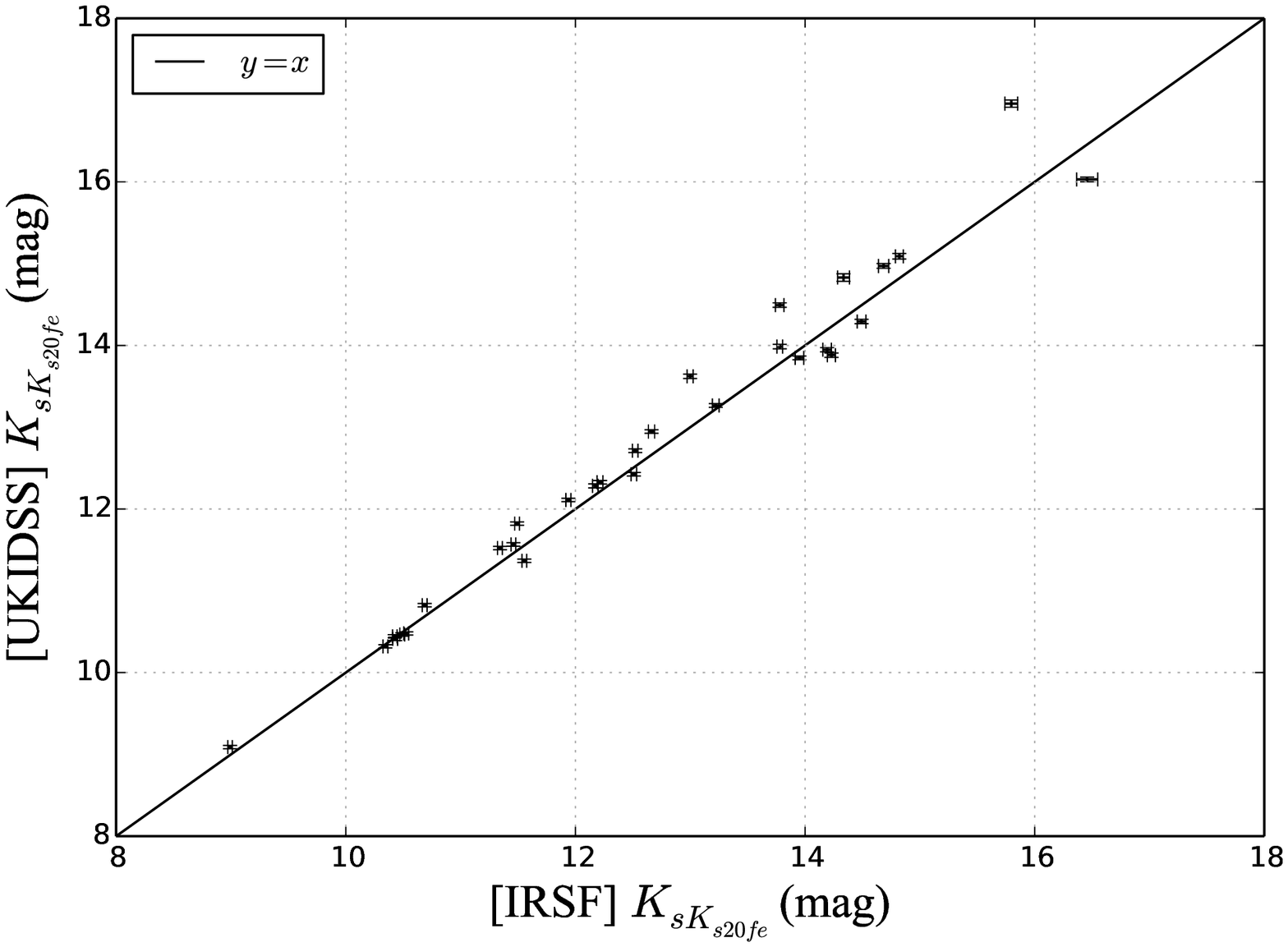}
\caption{A comparison between the $K_s$-band $K_{s20}$ fiducial elliptical aperture magnitudes measured from the IRSF images and the same parameter measured from the UKIDSS GPS images using a modified version of the IRSF photometry pipeline. The solid line is the one-to-one line.}
\label{IRSF_vs_UKIDSS}
\end{center}
\end{figure}

Fig. \ref{IRSF_vs_UKIDSS} shows the comparison between the $K_s$-band  $K_{s20}$ fiducial elliptical aperture magnitudes measured from the IRSF images and the same parameter measured from the UKIDSS GPS images.
The solid line in Fig. \ref{IRSF_vs_UKIDSS} is the one-to-one line. Excellent agreement between the IRSF and UKIDSS GPS photometry can be seen. The pixel scale for the UKIDSS GPS images is 0.4 arcsec pixel$^{-1}$, which is comparable to the IRSF pixel scale of 0.45 arcsec pixel$^{-1}$. The average seeing of 1.54 arcsec for the UKIDSS GPS images used in this comparison is similar to that of 1.38 arcsec for the IRSF survey. This agreement shows that the IRSF imaging does not suffer from foreground contamination, even after star removal, nor does it adversely under-estimate the isophotal flux of the ZOA galaxies.  It is satisfying to see this agreement between photometric parameters because we plan to extend the current TF project to  the northern ZOA using the UKIDSS GPS survey along with {\HI} data from the Nan\c{c}ay Radio Telescope.

\section{Summary}
\label{sect:summary}
In this paper, we present the observations, data reduction and final catalogue for 1044 NIR galaxies, in the $J$, $H$, and $K_S$ bands, in the southern ZOA. The observations were conducted between 2006 and 2013 using the IRSF, a 1.4-m telescope situated at the South African Astronomical Observatory site in Sutherland. This resulted in observations of all galaxies in the three blind systematic deep HIZOA surveys \citep{2005AJ....129..220D,2008glv..book...13K,2016AJ....151...52S}.

The quality of the images is discussed in detail. The survey has an average seeing and zero point magnitude of 1.38 arcsec and 20.1 mag in the $K_s$ band, respectively. These values agree well with thos from previous surveys done with the same instrument \citep{2010PhDT........33W}. The mean error of the measured isophotal magnitudes is 0.02 mag which is sufficient for the TF analysis. 

The completeness as a function of stellar density and dust extinction was found to be dependent on the {\HI} mass of each galaxy. The detection rate was found to be 100 per cent for massive galaxies ($\log M_{HI} [M_{\odot}] > 8.5$) in regions with stellar density of $\log(N_{K_s<14}/\mathrm{deg}^{2}) \leq 4$. However, for small galaxies ($\log M_{HI} [M_{\odot}] < 8.5$) the detection rate is 75 per cent in regions of $\log(N_{K_s<14}/\mathrm{deg}^{2}) \leq 4$. This detection rate  drops to 50 per cent and 30 per cent for regions within $\log(N_{K_s<14}/\mathrm{deg}^{2}) > 4$ for massive and small galaxies, respectively. The same conclusion was found for the detection rate as a function of dust extinction. Although the detection rate was high for massive galaxies in regions up to $E(B-V) = 1$ mag ($A_v=3.1$ mag), it was very low for small galaxies even in regions with very low dust extinction.

We identified 674 galaxies in the final NIR catalogue that have confirmed counterparts in the three HIZOA catalogues. Counterparts from similar NIR surveys are presented. We found 285 2MASX galaxies have counterparts in our final catalogue. However, only 30 galaxies in the UKIDSS Galactic Plane Survey have counterparts in our final NIR catalogue because UKIDSS GPS only overlaps with our survey in its northern extension. 

A comparison of our IRSF pointed observations with the 2MASX and UKIDSS GPS surveys was performed for these galaxies. We found good agreement between the $K_s$-band $K_{s20}$ fiducial elliptical aperture magnitude presented in this paper and the same parameter reported in the 2MASX catalogue. We detect only small deviations for both faint galaxies ($> 14$ mag) and bright galaxies ($< 9$ mag). The deviation for bright galaxies is due to the difference of the pixel scale between IRSF and 2MASX instruments. While the deviation for the faint galaxies is a bias due to the completeness limit of 2MASX.

Good agreement was found between the $K_s$-band $K_{s20}$ fiducial elliptical aperture magnitudes measured from the IRSF data and the UKIDSS GPS data. This agreement confirms that the IRSF images are of equal quality to the UKIDSS GPS images, which are one magnitude deeper. This indirectly implies that IRSF photometry does not suffer from foreground contamination, after star removal, nor does it under-estimate the isophotal flux of the ZOA galaxies. The measurement of UKIDSS photometry is regarded as a pilot project of our TF survey in the northern ZOA. 

This paper is the third in a series towards the full ZOA TF analysis. The data presented here will be used with the recently calibrated TF relation in \cite{2015MNRAS.447.1618S} as well as the {\HI} data presented in \cite{2016MNRAS.457.2366S} to derive distances and peculiar velocities for inclined spiral galaxies in the southern ZOA. An extension of this project into the northern ZOA already started last year with the {\HI} observations of bright inclined 2MASS galaxies using the Nan\c{c}ay Radio Telescope.

\section*{Acknowledgments}
This work is based upon research supported by the Science Faculty at University of Cape Town (UCT), the South African National Research Foundation and the Department of Science and Technology. Part of this research was conducted by the Australian Research Council Centre of Excellence for All-sky Astrophysics (CAASTRO), through project number CE110001020. WLW acknowledges support from the UK Science and Technology Facilities Council [ST/M001008/1]. We thank Dr. Tom Mutabazi and Hiroki Onozato for running some of the observations. We also thank Dr. Yasushi Nakajima, Dr. Takahiro Nagayama and Dr. Ihab Riad for many discussions regarding the reduction pipeline. The authors are very grateful to Dr. Christina Magoulas for her careful reading and comments on the paper.

\bibliographystyle{mn2e.bst}
\bibliography{774}

\label{lastpage}

\bsp

\end{document}